\documentclass[%
 aip,
 jmp,%
 amsmath,amssymb,
 reprint,%
]{revtex4-1}
\DeclareUnicodeCharacter{0301}{\'{e}}

\usepackage{graphicx}
\usepackage{dcolumn}
\usepackage{bm}

\usepackage{inputenc}
\usepackage{blindtext}
\usepackage{amsmath}
\usepackage{amsfonts}
\usepackage{amssymb}
\usepackage{setspace}
\usepackage{color}
\usepackage{float}
\restylefloat{table}
\usepackage{lettrine}
\usepackage{natbib}
\usepackage{blindtext}
\usepackage{titlesec}
\usepackage{caption}
\usepackage{subcaption}
\usepackage[export]{adjustbox}
\usepackage{wrapfig}
\usepackage{fullpage}
\usepackage{multirow}
\usepackage{filecontents}
\usepackage{epstopdf}
\usepackage{graphicx}
\usepackage{upgreek}
\usepackage{chemformula}
\usepackage[pdfencoding=auto, psdextra]{hyperref}

\makeatletter
\newcommand*{\rom}[1]{\expandafter\@slowromancap\romannumeral #1@}

\begin{document}

\preprint{AIP/123-QED}

\title[]{Report on laser-induced fluorescence transitions relevant for the microelectronics industry and sustainability applications}

\author{V. S. Santosh K. Kondeti}
\email{vkondeti@pppl.gov}
 \affiliation{Princeton Plasma Physics Laboratory, Princeton, New Jersey.}

\author{Shurik Yatom}
\affiliation{Princeton Plasma Physics Laboratory, Princeton, New Jersey.
}

\author{Ivan Romadanov}
\affiliation{Princeton Plasma Physics Laboratory, Princeton, New Jersey.
}

\author{Yevgeny Raitses}

\affiliation{Princeton Plasma Physics Laboratory, Princeton, New Jersey.}

\author{Leonid Dorf}
\affiliation{Applied Materials, Santa Clara, California.
}

\author{Andrei Khomenko}
\affiliation{Applied Materials, Santa Clara, California.
}


\date{\today}

\begin{abstract}

A wide variety of feed gases are used to generate low-temperature plasmas for the microelectronics and the sustainability applications. These plasmas often have a complex combination of reactive and non-reactive species which may have spatial and temporal variations in the density, the temperature and the energy. Accurate knowledge of these parameters and their variations is critically important for understanding and advancing these applications through validated and predictive modeling and design of relevant devices. Laser-induced fluorescence (LIF) provides both spatial and temporally resolved information about the plasma-produced radicals, ions, and metastables. However, the use of this powerful diagnostic tool requires the knowledge of optical transitions including excitation and fluorescence wavelengths which may not be available or scattered through a huge literature domain. In this manuscript, we collected, analyzed and compiled the available transitions for laser-induced fluorescence for more than 160 chemical species relevant to the microelectronics industry and the sustainability applications. A list of species with overlapping LIF excitation and fluorescence wavelengths have been identified. This summary is intended to serve as a data reference for LIF transitions and should be updated in the future.

%
\end{abstract}

\keywords{Laser-induced fluorescence, etching, microelectronics, sustainability}
\maketitle

%

\onecolumngrid
\section{Introduction}

According to the Moore's law, the number of transistors on a chip exponentially increase every year \cite{waldrop2016chips,lundstrom2022moore,shalf2020future,theis2017end}. The semiconductor industry has been able to keep up with this law by the use of the low-temperature plasma based processes to perform operations such as etching, deposition, or cleaning. The microelectronics industry widely uses plasma-based deposition techniques such as sputtering, plasma-enhanced chemical vapor deposition, oxidation, and planarization\cite{graves1994plasma}. Film and material removal techniques such as etching, photo resist stripping, and cleaning \cite{graves1994plasma} also involve plasma-material interactions. The underlying material science and the plasma-surface interactions are not completely understood, which poses a significant challenge to develop new tools for the industry. The fabrication of microelectronics in the present era requires sub-nm precision accuracy to enable technologies such as the 3 nm and the 2 nm nodes \cite{adamovich20222022}. Modern microelectronics industry uses processes such as atomic layer etching (ALE), atomic layer deposition (ALD), area selective deposition and high aspect ratio processing to achieve such precision \cite{parsons2020area, knoops2019status,oehrlein2015atomic,dorf2016atomic,jagtiani2016initial,rauf2017three}. Achieving such precision without damage to the underlying material layer will require meticulous control over the ion energy (or velocity) distribution function (IEDF/IVDF), the fluxes of ions and radicals and the chemical composition near the material surface \cite{dorf2022plasma1, dorf2022plasma2}.

Since their inception, the standard diameter of the silicon wafers used in the microelectronics industry has been increasing to enhance the throughput and reduce the operational cost. Starting from a wafer diameter of 12.5 mm in the year 1957, the industry currently uses a size of 300 mm and proposes to use a size of 450 mm in the future\cite{zhang2014silicon}. This increase in wafer size necessitates larger plasma chambers. Radio-frequency capacitively coupled plasmas are known for generating large-scale self-organized patterns, leading to spatial non-uniformities that become particularly critical in larger chambers\cite{bera2021self}. Etching and deposition occurs on the silicon wafer surface and it is essential to know the chemical and plasma properties close to the wafer surface. The control over the processes occurring on the wafer surface can be achieved by understanding the chemical stoichiometry close to the wafer surface\cite{cunge2005plasma}. The measurement of the spatial variation in the properties of the radicals, ions and metastable species will help in determining the operational parameters that generate a spatially uniform plasma.

Material processing is often controlled by using a pulsed plasma with a duty cycle that allows for periodic intervals where ions or radicals are acting on the material. The volatile products from the material surface are removed during the plasma-off phase. Ions are accelerated towards the material surface by applying a bias voltage on the material surface. By varying the duty cycle of the pulsed plasma, a reproducible high aspect ratio anisotropic etching can be achieved \cite{dorf2022plasma1, dorf2022plasma2}. The measurement of the temporal variation of the radicals, ions and metastable species in a pulsed plasma will validate the simulation models, that are widely used in the microelectronics industry.  

The emergence of alternative materials with unique properties such as a single atomic layer or thin films, two-dimensional materials, materials with controlled defects, and nanocrystal assemblies, the semiconductor industry is entering a new post-silicon era\cite{badaroglu2017international}. These materials include group \rom{3}, \rom{4} materials, MoS$_{2}$, WSe$_2$, graphene, TiO$_{2}$, VO$_{2}$, SmNiO$_{2}$ and others\cite{atature2018material,zwanenburg2013silicon}. The post-silicon era materials are of interest for quantum computing, quantum electronics, and quantum photonics \cite{ladd2010quantum}. 


Similar to low-pressure plasmas used in the microelectronics industry, atmospheric pressure plasmas which are widely used for sustainability applications, can also involve a complex chemistry \cite{samukawa20122012,adamovich20172017,adamovich20222022}. These plasmas are used for a wide variety of existing and emerging applications such as plasma medicine \cite{kong2009plasma}, plasma agriculture\cite{ranieri2021plasma}, plasma synthesis of nanomaterials\cite{chen2015review}, plasma catalysis\cite{bogaerts20202020,whitehead2016plasma}, thermal plasmas for welding, cutting and spraying\cite{vardelle2015perspective}, plasma-assisted combustion\cite{starikovskaia2006plasma}, environmental applications\cite{huang2015recent,nijdam2012introduction}, pollution control\cite{byeon2017ballast,trenchev2020dual}, additive manufacturing\cite{sui2020plasmas} and others. These plasmas operate at atmospheric pressure (1 atm) and do not require vacuum hardware to generate the plasma. All these applications require the interaction of plasmas with solids and liquids \cite{bruggeman2016plasma,oehrlein2011plasma}. 



While the plasmas used in the microelectronics industry can be tens of cms in dimensions, the plasmas used in sustainability applications vary in size from microplasmas ($\mu$m - mm size)\cite{becker2006microplasmas} to large volume reactors (several cm's)\cite{kusano2014atmospheric}. Microplasmas have large gradients in the density of ions, electrons and radical species over a dimension of a mm. Large-volume reactors can result in a spatially non-uniform discharge. To avoid forming filamentary discharges, the atmospheric pressure plasmas are often generated using pulsed voltage waveforms with a duty cycle. So, the plasma is on only for a fraction of the time. Similar to the use of pulsed plasmas for processing of microelectronics, it results in an intermittent plasma with a temporal variation of the density of electrons, ions, and radicals generated by the plasma.

The predictive design capabilities of the plasma sources used for the microelectronics and the sustainability applications can be developed by the comprehensive modeling of the plasma reactors involving hybrid fluid simulations. Modeling is often plagued by the lack of the chemical kinetics information and the experimental data to validate the simulation results. Measurement of the spatial and temporal variation in the IEDF/IVDF, ion/radical/metastable density distributions and gas temperatures can inform and improve the predictive capabilities of the simulations. Non-intrusive optical diagnostic techniques such as optical emission spectroscopy and absorption spectroscopy are line-of-sight integrated techniques and obtaining spatially resolved information requires the assumption of a spatial profile. 

A laser-based technique such as laser-induced fluorescence can resolve both the spatial and the temporal information, while being non-intrusive. Being a resonant laser-based technique, it can detect species even if there is no detectable optical emission from the plasma, in addition to when there is optical emission from the plasma. The LIF transitions for various species such as ions, atoms and molecules is spread over different scientific papers from various fields. It is often not straightforward to find the different species that can be detected by LIF and the various reported transitions previously reported for a particular species. In this work, we reviewed the basic principles of LIF and the different types of LIF applicable for the microelectronics industry and the sustainability applications. We compiled a LIF reference database with the available LIF transitions that will enable the measurement and the detection of the various plasma-produced ions, radicals, and metastables that are relevant for microelectronics and sustainability applications. We describe the principle, the required instrumentation, calibration and the applications of LIF in section \ref{Methods:LIF}. Section \ref{Section: LIF/TALIF reference data} provides the compilation of the LIF transitions and we summarize in section \ref{Summary}.

\section{Laser induced fluorescence}
\label{Methods:LIF}

\subsection{Introduction and principle}
\label{Methods: LIF introduction and principle}
Laser-induced fluorescence can provide both spatial and temporal measurements of the plasma-produced neutral, ionic, and metastable species. In LIF, laser irradiation resonantly excites the plasma-generated species to a higher electronic state. The excited species emit fluorescence upon relaxation to a lower level. This fluorescence intensity can be used to quantify the species density and the velocity distribution functions \cite{amorim2000laser, dobele2005laser, donnelly1982laser, konthasinghe2015laser, lee2007measurements, tachibana1991measurement,herring1988two,selwyn1987detection,van1989laser}. When the wavelength required for a resonant transition in LIF falls in the vacuum ultraviolet (VUV) region that is readily absorbed by the ambient air and water vapor, a two-photon excitation scheme is used and the technique is called two-photon absorption laser-induced fluorescence (TALIF). In this method, the energy required for exciting to the upper level is provided by two photons. A schematic of the excitation schemes for LIF and TALIF is shown in Figure \ref{fig: Pulsed LIF (a) Transitions schematic (b) Experimental schematic}(a). 

LIF has been used for measuring relative/absolute densities of neutral radical species \cite{dobele2005laser}, velocity distribution functions (VDF) \cite{diallo2015time} and rotational temperatures of molecules \cite{bruggeman2014gas}. Densities as low as 10$^{8}$ cm$^{-3}$ have been successfully measured by LIF when using sensitive modern optical sensors such as a photomultiplier tube (PMT) or an intensified charge-coupled device (iCCD) camera \cite{yatom2023measurement}. The spatial resolution of LIF is about 100-200 $\mu$m. Absolute densities of neutral radicals can be obtained through calibration of the fluorescence signal. Rayleigh scattering from a known concentration of air coupled with a multi-level model is typically used for obtaining absolute densities for one-photon LIF schemes. Fluorescence from the two-photon excitation of a known concentration of a neutral gas is used for the calibration of TALIF schemes. Details of the calibration methods have been summarized in section \ref{Methods: LIF calibration general}. For obtaining absolute densities, the laser energy in the LIF measurements is maintained in the linear regime, i.e., when the LIF intensity is proportional to the laser energy. This ensures that the ground state density does not get depleted by the laser beam and the laser energy is spent in the excitation from the lower level to the higher level rather than from the stimulated emission from the higher level to the lower level. Fully saturated laser induced fluorescence has also been used for detecting species using LIF. In the fully saturated LIF regime, a high laser energy is used that ensures that the stimulated emission dominates over the spontaneous emission and the collisional quenching of the excited state \cite{forster1995high,mrkvivckova2022dealing}. Further, saturation will broaden the LIF absorption line profiles and modify the line profile to a Lorentzian shape\cite{goeckner1993saturation, mazouffre2016laser}. Saturation broadening needs to be accounted for deducing the velocity distribution function. While fully saturated LIF has been used to get a better signal to noise ratio, it is difficult to determine the fully saturated regime and maintain such saturation over the entire interrogation volume\cite{forster1995high,mrkvivckova2022dealing}. For TALIF measurements, the used laser energy is in the quadratic regime where the TALIF intensity is proportional to the square of the laser energy to avoid the consequences of saturation effects described above. LIF signal intensities can be reduced due to quenching of the excited state by the feed gases, humidity, air, and molecular gases such as the ones used in the microelectronics industry. A correction of the fluorescence intensity due to the presence of such quenchers needs to be accounted to obtain absolute densities \cite{kondeti2020h, yatom2017nanosecond, kondeti2017ag+}. Resonant transitions from the ground state are often not available for all atomic and molecular species. Excitation from the metastable states of neutrals and ions has been used as a proxy for ground state species by assuming thermal equilibrium between the metastable and the ground states. This allows to obtain useful information about the species velocities that are important for applications with plasma flows such as ion sources and plasma thrusters \cite{svarnas2018laser,mazouffre2016laser, pietzonka2023laser, vinci2022laser}. The metastables in a plasma are predominantly produced by the presence of electrons with the appropriate energy. The absence of electrons with such energy, quenching due to collisions or ionization can deplete or reduce the density of density of the metastable states below the detection limit of LIF. Hence, the application of LIF for metastable excitation may not be applicable to a wide variety of discharge environments.

\begin{figure}[H]
    \centering
    \includegraphics[width=1\linewidth]{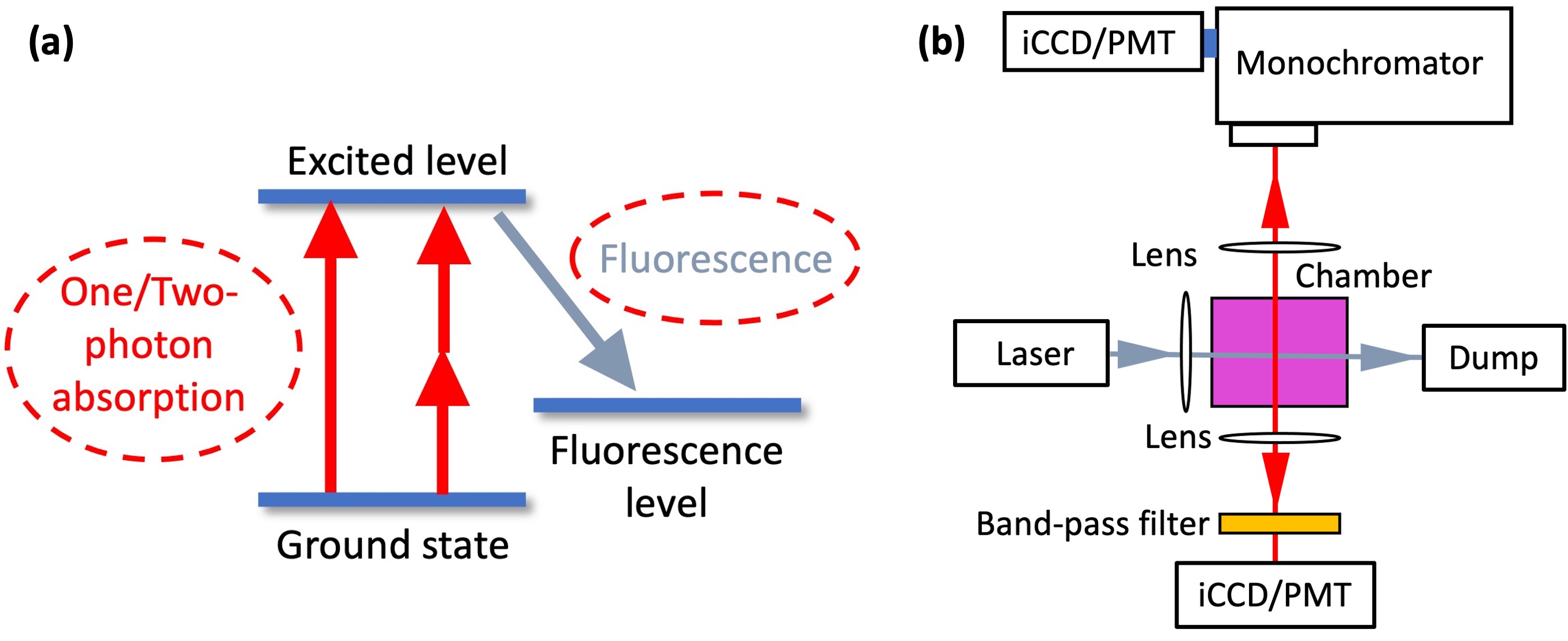}
    \caption{(a) Transitions schematic and (b) Experimental schematic of laser-induced fluorescence}
    \label{Pulsed_LIF_schematica-Transition_scheme,b-Experimental_setup.jpg}
\end{figure}

\subsection{Implementation and instrumentation}
\label{Methods: LIF implementation and instrumentation}

The laser beam is focused on to the plasma and the fluorescence is typically detected perpendicular to the laser beam propagation by a detector (Figure \ref{fig: Pulsed LIF (a) Transitions schematic (b) Experimental schematic}(b)). The spatial resolution in LIF is limited by the focus size of the laser launch and the collection optics. The laser beam is imaged onto the detector by using a lens. Using a PMT results in a point measurement and the spatial distribution can be obtained by using diaphragms and positioning the PMT to image different regions of the laser beam. The wavelength of fluorescence is typically different from the laser excitation wavelength. However, the fluorescence detection at the same wavelength as the excitation wavelength has also been reported\cite{tachibana1991measurement}. In such situations, fluorescence is detected with a certain delay after the laser beam. This technique relies on the longer lifetime of the laser excited level than the pulse width of the laser beam. When the laser excitation wavelength is different from the fluorescence wavelength, an appropriate narrow band-pass filter or a cut-off filter is installed in front of the detector to transmit only the fluorescence onto the detector or block the Rayleigh scattering of the laser beam. When a fluorescence spectrum is desired, the laser beam is imaged onto the entrance slit of a monochromator that is coupled with an iCCD camera or a PMT. The shaping of the laser beam into a sheet provides a two-dimensional distribution of the imaging species and this technique is called planar laser-induced fluorescence (PLIF) \cite{brackmann2016strategies,hirano1994visualization,kirby2000imaging,sappey1992planar}. PLIF can provide the spatial distribution of non-uniformities in the discharge with a single measurement (Figure \ref{fig: Pulsed LIF (a) Transitions schematic (b) Experimental schematic}(b)). Examples of the measured OH density using LIF and PLIF are shown in Figure \ref{fig: LIF, PLIF examples (a) OH LIF example (b) OH PLIF example}(a) and \ref{fig: LIF, PLIF examples (a) OH LIF example (b) OH PLIF example}(b). For obtaining time-resolved information on the plasma-produced species in a pulsed plasma, a pulsed laser is typically used. The temporal resolution is limited by the pulse width of the laser beam. An example of the time resolved OH density measured in a pulsed plasma is shown in Figure \ref{fig: LIF, PLIF examples (a) OH LIF example (b) OH PLIF example}(a). Nanosecond pulsed laser beams are a popular and a cost-effective way for executing LIF, while the relatively more expensive picosecond and femtosecond lasers have also been used for performing LIF \cite{niemi2001absolute,kulatilaka2009interference,dogariu2021non}. 

\begin{figure}[H]
    \centering
    \includegraphics[width=1\linewidth]{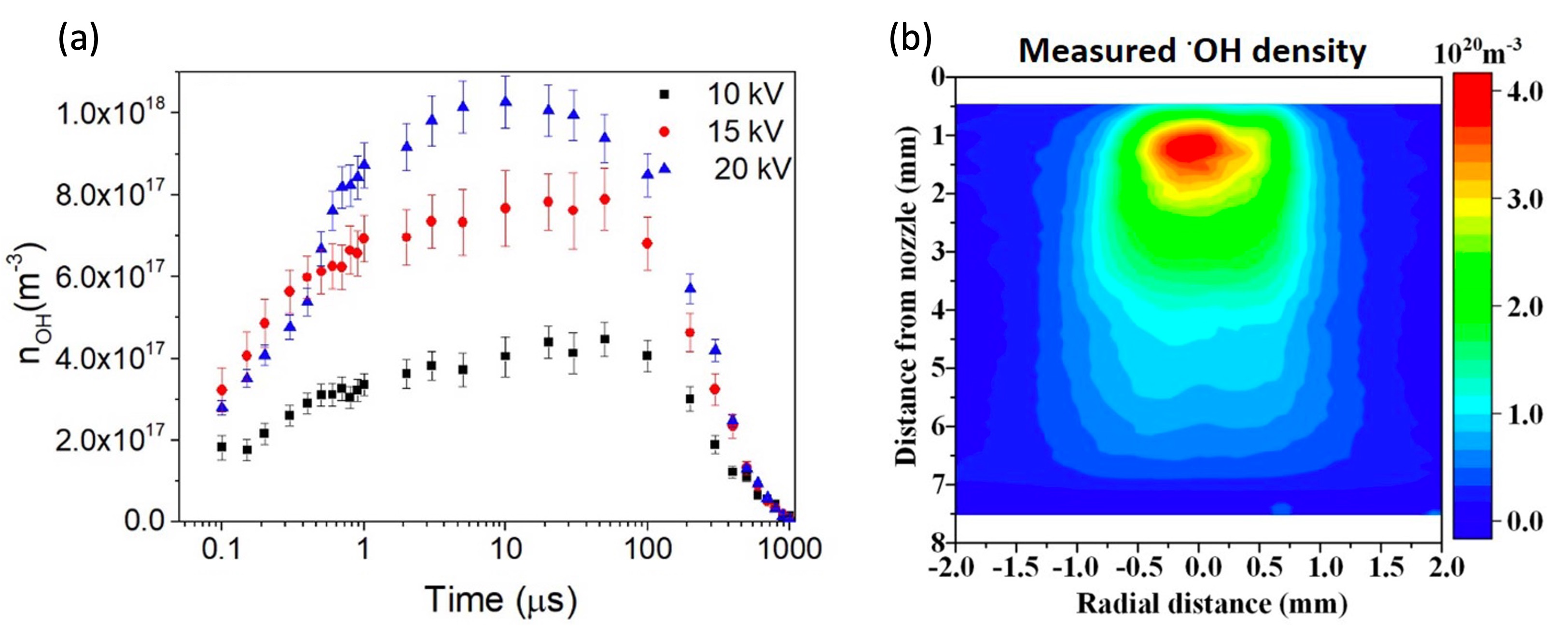}
    \caption{Examples of absolute OH density measured using (a) LIF of a pulsed dielectric barrier discharge\cite{yatom2022examination} and (b) PLIF of an RF plasma jet (reproduced from Ref.\cite{kondeti2020h} with permission from the American Vacuum Society (AVS)).}
    \label{fig: LIF, PLIF examples (a) OH LIF example (b) OH PLIF example}
\end{figure}

Depending on the application, different types of lasers have been used to perform LIF measurements. These include solid state lasers, dye lasers, diode lasers, quantum cascade lasers and optical paramagnetic oscillators (OPO) \cite{trager2012springer,renk2012basics}. Diode lasers can have a line-width down to the fm range, while dye lasers typically have line-widths in the pm range and commercial OPO lasers have line-widths in the range of tens to hundreds of pm. The narrow line-width of diode lasers allows the accurate resolution of the narrow spectral features of atomic, ionic and molecular transitions. These lasers can be applied to detect precise Doppler shifts in the transition. These shifts can be caused by velocity changes as small as tens of m/s along the direction of the laser beam propagation (Section \ref{Methods: VDF}). The inherent large line-width of the dye lasers and the OPO lasers compared to the diode lasers limits their application to measuring only large Doppler shifts that are produced by very high ion or atom velocities \cite{vekselman2010high}. The dye lasers and the OPO lasers offer the ability to generate a broad range of wavelengths ranging from the deep ultraviolet (DUV) to the infrared region. Solid state lasers, diode lasers and quantum cascade lasers often generate radiation only in a narrow spectral region and do not have the versatility to generate a broad range of wavelengths.

\subsection{Calibration method for obtaining absolute density for LIF}
\label{Methods: LIF calibration general}

The LIF signal can be calibrated to obtain absolute densities of the ground state species. The LIF signal can be described by the following equation\cite{verreycken2013absolute,yatom2023measurement}: 
 \begin{equation}
     I_{LIF} = \frac{1}{4\pi}\int(\eta_{LIF} \frac{hc}{\lambda_{fl}})n_{exc}(x,y,z,t) A_{fl} dx dy dz dt
     \label{Eq: I_LIF}
 \end{equation}

Where $\eta_{LIF}$ is the instrumental factor that is the product of quantum efficiency of detector, transmission of the optical setup and the solid angle of signal collection, h is the Planck's constant, c is the speed of light, n$_{exc}$(x,y,z,t)(m$^{-3}$ ) is the density of the excited level, from which the fluorescent photons are emitted, $\lambda_{fl}$ is the fluorescence wavelength (m), A$_{fl}$ - Einstein coefficient of the fluorescent transition (s$^{-1}$). The unknown parameters in this equation are $\eta_{LIF}$ and $n_{exc}$. This section describes a method to calculate the ground state density $n$ by using Rayleigh scattering and a population kinetics model.

 \subsubsection{Rayleigh scattering}
 \label{Methods: LIF calibration Rayleigh}

The experimental setup for implementing this calibration method is similar to the experimental setup used for executing LIF (Figure \ref{fig: Pulsed LIF (a) Transitions schematic (b) Experimental schematic} (b)). The plasma is turned off and the region of interest is filled with a gas such as air. The spectral filter that was used for LIF is removed and the laser beam is imaged on to the detector to collect the Rayleigh scattering signal from air. If a monochromator was used for LIF, the wavelength window of the mondochromator is adjusted to detect the Rayleigh scattering laser beam. The position of the detector is not changed compared to the LIF setup to ensure that the solid angle of detection is the same as that for the LIF measurements. 


The Rayleigh scattering signal (I$_{R}$) can be described as\cite{miles2001laser,verreycken2013absolute,yatom2023measurement}:
\begin{equation}
    I_{R} = \eta_{R}N_{n} (\frac{\partial\sigma_{0}}{\partial\Omega})E_{L}\Delta x
    \label{Eq: I_Rayleigh}
\end{equation}

Where N$_n$ is the scattering particles density, E$_L$ is the laser pulse energy, $\Delta$x is the the detection volume length, $\eta_R$ is the instrumental factor for the wavelength of the Rayleigh scattering, and $\frac{\partial\sigma_{0}}{\partial\Omega}$ is the Rayleigh scattering differential cross-section. The differential Rayleigh scattering cross-section depends on the polarization direction of the laser beam and the angle at which the scattering light is collected. A convenient configuration is when the polarization of the laser beam is perpendicular to the laser propagation direction and the Rayleigh scattering collection is at angle of 90$^{o}$ with respect to the laser propagation direction and the polarization direction. Assuming that the detector collects all polarizations of the scattered beam, the differential Rayleigh scattering cross-section is  defined as\cite{miles2001laser}:

\begin{equation}
    \frac{\partial\sigma_{0}}{\partial\Omega} = \frac{3\sigma}{8\pi} \frac{2}{(2+\rho_{0})}
    \label{Eq: Rayleigh_cross section}
\end{equation}

Where $\sigma$ is the total Rayleigh scattering cross-section (m$^2$), $\rho_{0}$ is the ratio of the horizontally-to-vertically polarized light in an unpolarized laser beam. The subscript $0$ corresponds to the detection of both horizontal and vertical polarization of the scattering. The values of $\sigma$, $\rho_{0}$ for the commonly used scattering from air are tabulated as functions of wavelength in the review by R. Miles \textit{et al.}\cite{miles2001laser}. Calibration using Rayleigh scattering is usually performed at a known pressure and temperature. The scattering particles density can be obtained by using the ideal gas law\cite{kauzmann2012kinetic}: 

\begin{equation}
    N_{n} = \frac{P}{k_{B}T}
    \label{Eq: Ideal gas law}
\end{equation}

\begin{figure}[H]
    \centering
    \includegraphics[width=0.5\linewidth]{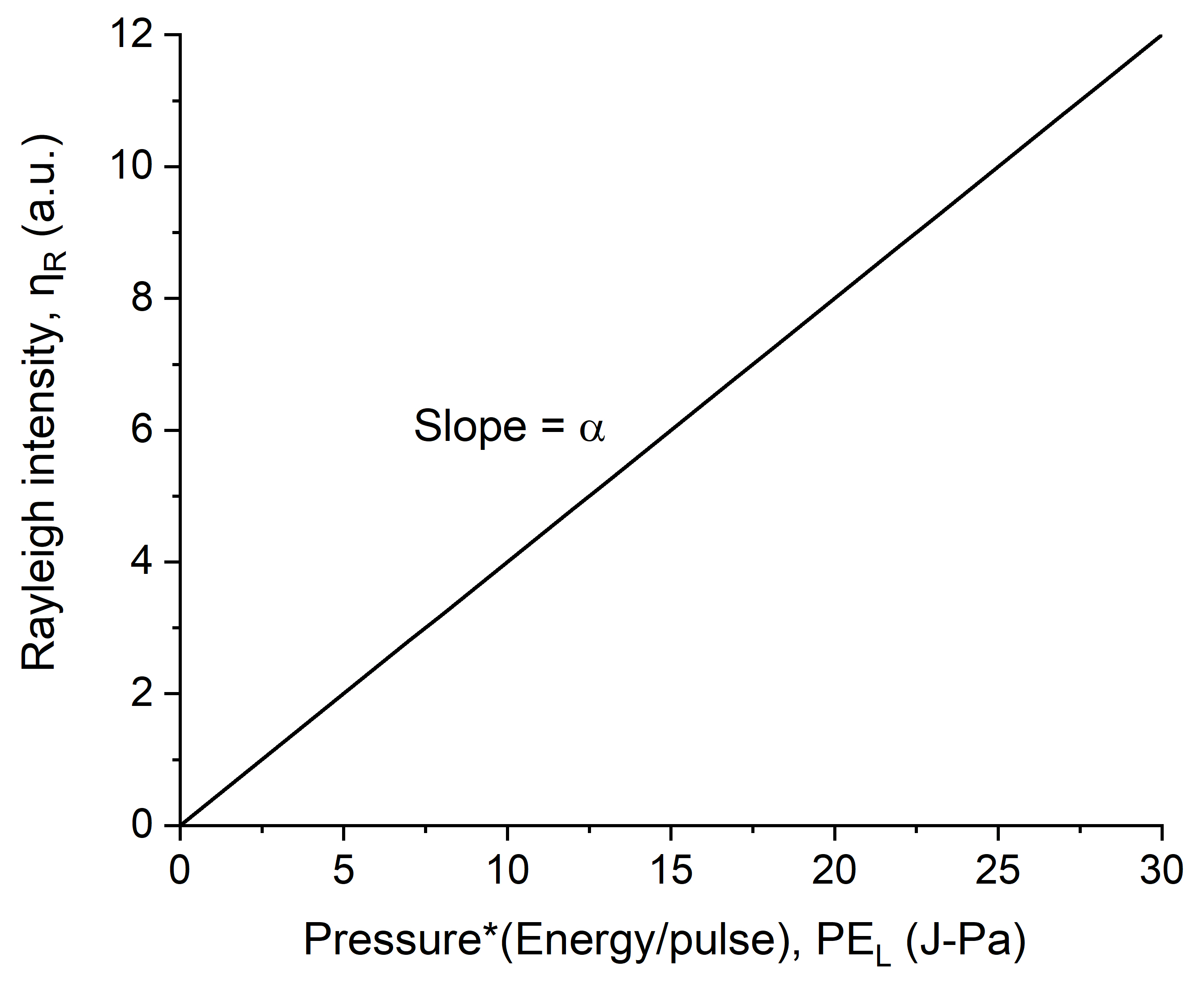}
    \caption{An example to determine the instrumental factor by Rayleigh scattering.}
    \label{fig: eta calculation Rayleigh scattering}
\end{figure}

Where k$_{B}$ is the Boltzmann constant. To calculate the instrumental factor, $\eta_{R}$ in equation \ref{Eq: I_Rayleigh}, the Rayleigh scattering intensity is measured for several laser energies at a fixed pressure. The slope of the straight line curve ($\alpha$) is calculated by plotting the Rayleigh scattering intensity (I$_R$) as a function of the product of the laser pulse energy, E$_L$ and the pressure, P (Figure \ref{fig: eta calculation Rayleigh scattering}).

\begin{equation}
    \alpha = \frac{I_{R}}{P E_{L}} = \frac{\eta_{R}\Delta x}{k_{B}T}(\frac{\partial\sigma_{0}}{\partial\Omega})
\end{equation}

The instrumental factor reduces to:

\begin{equation}
    \eta_{R} = \frac{\alpha k_{B}T}{(\frac{\partial\sigma_{0}}{\partial\Omega})\Delta x}
\end{equation}

The wavelength of the LIF signal is different from the laser excitation wavelength. The instrumental factor depends on the efficiency of the detector at the wavelength of detection. Hence, the instrumental factor for LIF needs to be corrected for the detector efficiency at the two different wavelengths. So, $\eta_{LIF}$ = $\eta_R\frac{\epsilon_{LIF}}{\epsilon_{R}}$, where $\epsilon_{LIF}$ and $\epsilon_{R}$ are the products of of the transmission factor of the optics and the quantum efficiency of the detector at the LIF and Rayleigh scattering wavelengths respectively. This correction can be avoided by setting the laser wavelength to the LIF wavelength for Rayleigh scattering measurements. However, a large difference between the LIF wavelength and the laser excitation wavelength for such a measurement might shift the laser beam path, the energy density of the focused laser beam can be different and such large wavelength shift may not be in the tunable range of the laser being used. 


\subsubsection{Laser parameters and population kinetics model}

Several parameters are required to develop for the population kinetics model and we define them here. The time evolution of the laser irradiance ($I_L(t)$) is defined as\cite{verreycken2013absolute,yatom2023measurement}:
\begin{equation}
I_L(t) = \frac{E_L\Gamma^{<1>}}{\Delta\nu_L\tau_LA_L}L(t)
    \label{Eq: Laser irradiance}
\end{equation}
Where $\Gamma^{<1>}$ is the dimensionless one-photon overlap integral, $\Delta\nu_L$ is the laser line-width (m$^{-1}$), $A_L$ is the area of the laser beam at the detection location (m$^2$) and $L(t)$ is a normalized function that describes the temporal evolution of the laser pulse. $\Delta\nu_L$ can be obtained from the specifications sheet of the laser. $A_L$ can be measured by a laser camera. Assuming the beam profile of the laser beam, it can also be calculated by measuring the residual laser energy by traversing a knife edge at the measurement location. $I_L$ can be measured by measuring the Rayleigh scattering intensity with a short gate time at different delay times with respect to the laser beam. The dimensionless one-photon overlap integral is defined as\cite{fiechtner2001absorption}:

\begin{equation}
    \Gamma^{<1>} = \Delta\nu_{L}\int_{-\infty}^{+\infty}P_{abs}(\nu)P_{L}(\nu)d\nu
    \label{Eq: One photon overlap integral}
\end{equation}

Where $P_{abs}$ is the absorption transition spectral profile (m) and $P_L(\nu)$ is the laser spectral profile (m). $P_{abs}$ can be calculated by measuring the absorption profile and accounting for the different broadenings such as Van der Waals broadening and Doppler broadening\cite{yatom2023measurement,verreycken2013absolute,yatom2022examination}. $P_L(\nu)$ can be taken from documentation of the laser manufacturer.

\begin{figure}[H]
    \centering
    \includegraphics[width=0.5\linewidth]{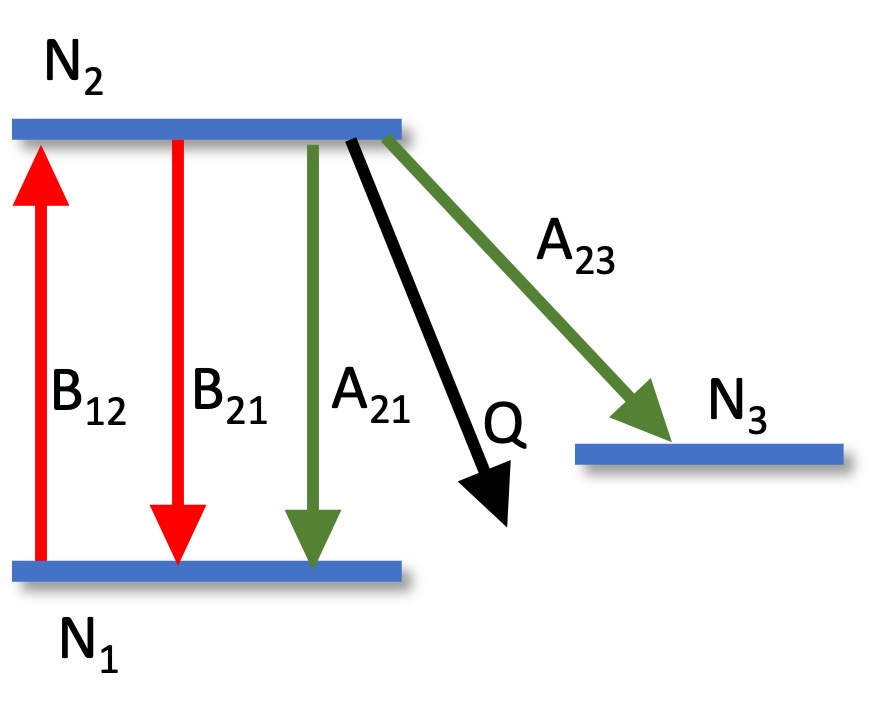}
    \caption{A three-level model of LIF. A$_{ij}$: spontaneous optical transitions, B$_{ij}$: Laser-induced transitions, Q: Collision induced transitions.}
    \label{fig: LIF 3-level model}
\end{figure}

The LIF signal is related to the population density of the excited state state species $n_{exc}$ (equation \ref{Eq: I_LIF}). A population kinetics model is used to deduce the ground state density. We illustrate a 3-level model here (Figure \ref{fig: LIF 3-level model}). A 4, 5 and 6-level model has also been implemented\cite{verreycken2013absolute,luque1996absolute,dobele2005laser,nemschokmichal2012metastable,yatom2023measurement}. Upon laser irradiation, several processes occur that result in the measured LIF signal. These include laser excitation from the ground state to the excited state, laser de-excitation from the excited state to the ground state, spontaneous relaxation from the excited state to any lower energy level and collisional quenching of the excited state. While there can be an additional source of production that can increase the density of these different levels, it is often neglected in the analysis. This is done by assuming the level to be in a quasi-steady state, where the laser excitation and its subsequent de-excitation is the most dominant process that results in the changes of the population of these states. A set of time dependent differential equations are solved to obtain the ground state density\cite{verreycken2013absolute,luque1996absolute,dobele2005laser,nemschokmichal2012metastable,yatom2023measurement}. 

The time dependent differential equations for a 3-level model are described below.

\begin{equation}
    \frac{dN_1}{dt} = I_L(t)(B_{21}N_2f_B^2 - B_{12}N_1f_B^1) + A_{21}N_2
    \label{Eq LIF 3-l3vel model: dN1/dt}
\end{equation}

\begin{equation}
    \frac{dN_2}{dt} = I_L(t)(B_{12}N_1f_B^1 - B_{21}N_2f_B^2) - QN_2
     \label{Eq LIF 3-l3vel model: dN2/dt}
\end{equation}

\begin{equation}
    \frac{dN_3}{dt} = A_{23}N_3
     \label{Eq LIF 3-l3vel model: dN3/dt}
\end{equation}

Where $N_{i}$ is the density of the level $i$ normalized to the ground state species density $n$, $I_L(t)$ is as defined in equation \ref{Eq: Laser irradiance}, $B_{ij}$ is the  Einstein B coefficient from level $i$ to level $j$ (mJ$^{-1}$), $A_{ij}$ is the  Einstein A coefficient from level $i$ to level $j$ (m$^3$s$^{-1}$), $f_{B}^{i}$ is the temperature-dependent Boltzmann factor of the level $i$ and Q is the quenching rate of the excited level. The Einstein A coefficient, the Einstein B coefficient and the Boltzmann factor are available from literature for several species \cite{NIST_ASD}. This information is not readily available for all the species of interest. The Einstein A coefficient for atomic species is available from the NIST database\cite{NIST_ASD}, while it is available for several molecular species in the cited references in the tables\cite{huber2013molecular}. The Einstein B coefficients are available for well studied species in the references provided in the tables. The Boltzmann factor needs to be calculated for the specific energy levels involved in the transitions\cite{kaminski1995absolute,luque1997spatial}. For the ground state and the metastable level atomic species, the Boltzmann factor can be assumed to be unity. However, for resonant atomic levels, the population of the resonant levels need not be larger than the other resonant levels involved in the transitions. For example, LIF can be performed by exciting the Ar(1s$_2$) and the Ar(1s$_4$) resonant levels, which are the ground states for excitation of these species. The validity of $f_B$ being taken as unity needs to carefully examined for each case. $Q$ can be estimated by measuring the fluorescence decay time constant after accounting for the collisions with different species\cite{yatom2023measurement}. These equations are solved to obtain the ground state density of the species.

\subsection{Calibration using a known concentration of a gas for TALIF}
\label{Methods: TALIF calibration}

TALIF is a non-linear two-photon excitation process\cite{niemi2001absolute}. This does not allow the use of Rayleigh scattering to calibrate the fluorescence intensity to deduce the absolute density of the species discussed in section \ref{Methods: LIF calibration Rayleigh}. The experimental setup for TALIF calibration is the same as the setup used for collecting the TALIF signal (Figure \ref{fig: Pulsed LIF (a) Transitions schematic (b) Experimental schematic}(b)). In this method, a known concentration of a reference gas is excited to a higher level using two-photon excitation by a laser. The reference gas is chosen such that the two-photon excitation wavelength of the reference gas is spectrally similar to the excitation wavelength of the species whose absolute density needs to be determined. The position of the detector for the TALIF of the measured species and the reference gas setup is the same. The TALIF signal from the reference gas can be collected by using an appropriate spectral filter or by setting a monochromator to transmit the desired wavelength. The used laser energy is in the quadratic regime for both the reference gas and the detection species to avoid correction for the complicated saturation effects. The TALIF signal ($I_{TALIF}$) can be defined as\cite{niemi2001absolute,yatom2017nanosecond, kondeti2020h}:

\begin{equation}
I_{TALIF} = T\frac{\sigma^{(2)}a \Gamma^{<2>}n}{(h\nu)^2}
    \label{Eq: TALIF definition equation}
\end{equation}

Where $T$ is the instrumental factor, $\sigma^{(2)}$ is the two-photon absorption cross-section, $h\nu$ is the photon energy, $\Gamma^{<2>}$ is the dimensionless two-photon overlap integral between the laser spectral profile and the two-photon absorption profile\cite{fiechtner2001absorption} and $n$ is the species density. $\sigma^{(2)}$ is available for different species in the cited references of the tables. The branching ratio (\textit{a}) is defined as\cite{niemi2001absolute,yatom2017nanosecond}:
 
 \begin{equation}
     a_{i} = \frac {A_{ik}}{A_{i} + \sum_{q}k_{q}n_{q}}
     \label{Eq. branching ratio}
 \end{equation}
 
 where the subscript \textit{q} refers to the quenching species, \textit{A$_{ik}$} is the spontaneous emission coefficient of the observed fluorescence line from level $i$ to level $k$, \textit{A$_{i}$} is the total spontaneous emission rate from the excited level $i$, \textit{k$_{q}$} is the quenching coefficient and \textit{n$_{q}$} is the density of the quenching species assuming that the 3-body collisional quenching is negligible\cite{schmidt2015ultrashort}. 

 The dimensionless two-photon overlap integral is the two-photon analogue to the one-photon overlap integral defined in equation \ref{Eq: One photon overlap integral}. It is defined as\cite{fiechtner2001absorption}: 

\begin{equation}
    \Gamma^{<2>} = \Delta\nu_L\int_{-\infty}^{+\infty}\int_{-\infty}^{+\infty}P_{abs}(2\nu)P_{L}(\nu)P_{L}(\nu)d\nu d\nu
    \label{Eq: Two photon overlap integral}
\end{equation}
 
The TALIF signal from both the reference gas ($I_r$) and main species ($I_m$) are collected. The density of the main species reduces to\cite{niemi2001absolute}:
 
 \begin{equation}
     n_{m} = \frac{T_{r}}{T_{r}}\frac{\sigma_{r}^{(2)}}{\sigma_{m}^{(2)}}\frac{a_{r}}{a_{m}}(\frac{h\nu_{m}}{h\nu_{r}})^{2}\frac{I_{m}}{I_{r}}\frac{\Gamma^{<2>}_{r}}{\Gamma^{<2>}_{m}}n_{r} 
     \label{Eq. TALIF absolute density equation}
 \end{equation}
 
 where the subscripts $m$ and $r$ refer to the main species and the reference gas respectively\cite{schmidt2015ultrashort, niemi2001absolute}.

\subsection{Laser-induced fluorescence for velocity distribution functions}
\label{Methods: VDF}
LIF is a crucial optical measurement technique for characterizing the flows and studying the kinetic processes in plasmas \cite{stern1975plasma, mazouffre2016laser, vekselman2009laser}. LIF allows for measurements of the VDF, which is essential for understanding the thermodynamic state of a medium. This function allows us to derive important parameters such as the temperature, the mean velocity, and the most probable velocity from the fluorescence profile. Moreover, LIF can directly measure the electric and the magnetic fields in plasmas through the Stark\cite{gavrilenko2000measurement} and the Zeeman effects\cite{ngom2008numerical} respectively. It is also possible to indirectly determine the electric fields by analyzing the moments of the VDFs\cite{Pérez-Luna_2009}. LIF utilizes the Doppler effect, where the absorption of the light occurs at a frequency shifted from the resonant transition frequency, a shift that depends on the velocity of the moving atom or the ion. Thus, it is sometimes referred to as Doppler-shift LIF.

\begin{figure}[H]
    \centering
    \includegraphics[width=0.2\linewidth]{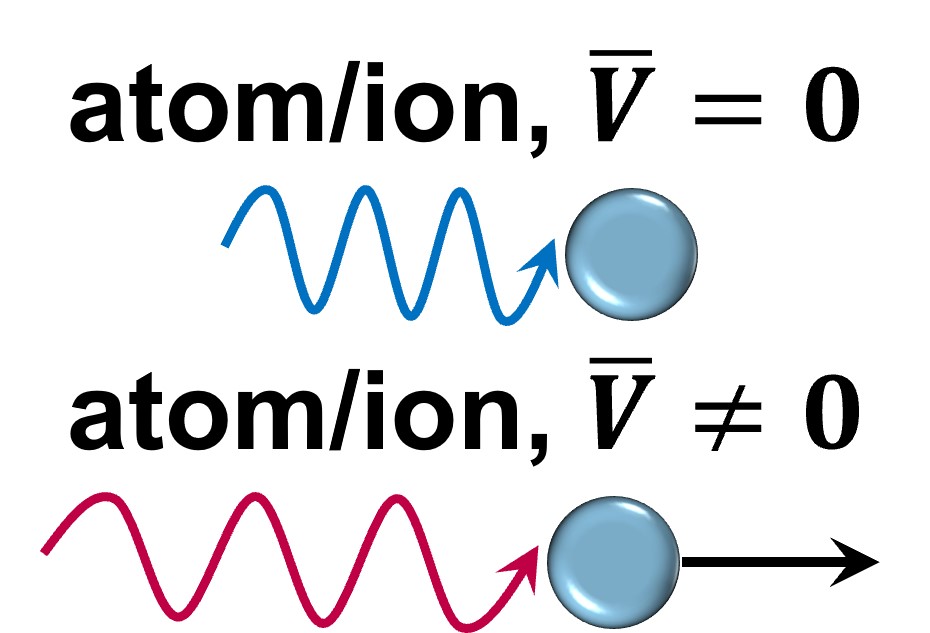}
    \caption{A schematic of the Doppler effect.}
    \label{fig: Doppler effect}
\end{figure}

The LIF process consists of two steps, as illustrated in Figure \ref{fig: Pulsed LIF (a) Transitions schematic (b) Experimental schematic}(a). In the first step, the species in the plasma are often in the excited state due to collisions with energetic electrons or other species. These states, known as metastable states, have long lifetimes — up to seconds, compared to the fluorescence states. It is important to note that when a metastable state is excited to deduce the properties of the ground state species, an assumption about the thermal equilibrium between this state and the ground or other states of the probed species in the plasma is necessary. When a laser beam is introduced, these excited atoms or ions absorb photons and are further excited to a higher energy state. The selection of this transition is guided by rules based on various quantum numbers\cite{cohen1977quantum} and must be sufficiently populated to ensure a detectable signal\cite{romadanov2018limitations}. Plasma parameters, such as the electron temperature and the electron density, influence the excitation rate and the de-excitation mechanisms such as quenching (e.g. collisions). To facilitate further excitation to a higher state, the photon energy (wavelength) of the laser must correspond precisely with the energy required for a resonant transition. However, due to thermal or directional motion of the probed species in the plasma, they perceive the incoming laser light wavelength Doppler-shifted (see Figure \ref{fig: Doppler effect}). This requires tuning of the laser wavelength to the target species moving at specific velocities.

\begin{figure}[H]
    \centering
    \includegraphics[width=1\linewidth]{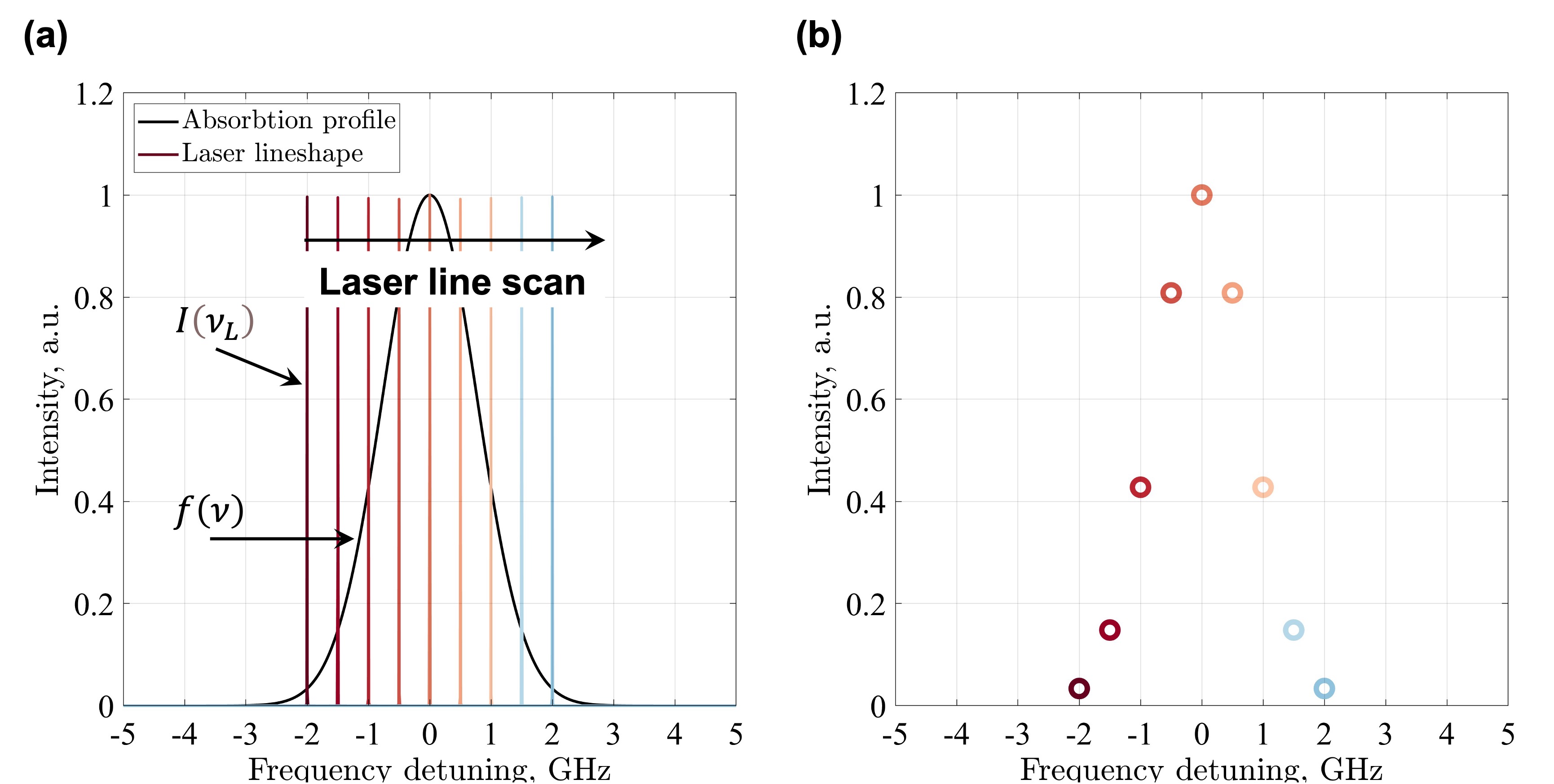}
    \caption{Schematic of LIF measurements. a) Laser wavelength ($I(\nu_L)$) is scanned across the Doppler broadened absorption profile ($f(\nu)$). b) The recovered signal, which is proportional to the integral: $\int f(\nu) I(\nu_L) d\nu $. $\nu_L$: laser wavelength.}
    \label{fig: Doppler LIF schematic}
\end{figure}

In the second step of the process, the excited species emit light through spontaneous emission, collisions, or stimulated emission, transitioning to a lower energy state. This transition can be resonant, in case if the species return to their initial energy state, and non-resonant if decay leads to a different level. Non-resonant emission is typically utilized in Doppler-shift LIF diagnostics because it allows the emitted wavelength to be easily distinguished from the intense laser light. The intensity of the emitted light is directly proportional to the number of species moving at the targeted velocity. By measuring emissions at various excitation frequencies, one can reconstruct the VDF, yielding detailed parameters of the ions or the neutrals. A schematic of this process is shown in Figure \ref{fig: Doppler LIF schematic}(a). When the laser frequency is scanned across the Doppler broadened absorption profile, the excited species then emit the fluorescence light. This signal is detected by various methods described in section \ref{Section: Doppler LIF optical setup}, and the profile of the VDF can be reconstructed (Fig. \ref{fig: Doppler LIF schematic}(b)). Doppler-shift LIF can be characterized with a high spatial resolution of 100’s of $\upmu$m\cite{Mazouffre_2013}, a temporal resolution of 10's of $\upmu$s\cite{chaplin2020time, vaudolon2014time, Young_2018}, and a spectral resolution which in terms of velocity translates to as low a few m/s\cite{aramaki2009high}.

\subsubsection{Line shape}

While Doppler broadening is the primary effect that allows us to extract information about the shape of the VDF, it is crucial to consider other mechanisms that might affect the measured line shape profile. In this subsection, we will discuss several physical mechanisms that contribute to the broadening of the line shape. Typically, the profile is influenced by broadening mechanisms such as hyperfine structure, isotopic splitting, Zeeman splitting, lifetime broadening, or Stark broadening. It is crucial to note that VDF measurements are feasible only under conditions where pressure broadening is not significant compared to the measured line-width of the LIF transition. This depends on the transition being probed and the operating conditions, for example at a pressure $<$ 10 Torr\cite{boivin2003laser}. The measured LIF profile, denoted as $f_{LIF}(\nu_L)$, results from the convolution of the Doppler line shape with the effects of broadening mechanisms and the laser intensity. This relationship can be represented mathematically as follows:

\begin{equation}
    f_{LIF}(\nu_L) \propto f(\nu_L) \otimes \phi_L(\nu_L) \otimes \phi_b (\nu_L)
    \label{VDF_broad}
\end{equation}


where $f(\nu_L)$ represents the amplitude of the Doppler line shape, $\phi_L(\nu_L)$ is the laser intensity profile (typically negligible), and $\phi_b(\nu_L)$ includes all broadening mechanisms and is referred to as the Doppler-free line shape.

Doppler broadening arises from the thermal and the directional velocities of the species being probed. Consider a laser with a frequency of $\nu_L$ and a wave-vector of $\mathbf{k}$. For a moving particle with velocity $\mathbf{v}$, the shift in the laser frequency, $\Delta \nu$ in the specie's frame is given by:

\begin{equation}
	\dfrac{\Delta \nu}{\nu_L} = -\dfrac{\mathbf{v}\cdot\mathbf{k}}{c}
 \label{Eqn: VDF_Laser frequency}
\end{equation}

where $c$ is the speed of light. The transition frequency $\nu_{T}$, at which the particle absorbs the laser, then becomes:
\begin{equation}
	\nu_{T} = \nu_L + \Delta\omega = \omega_L\Big(1 - \dfrac{v_k}{c}\Big)
 \label{Eqn: VDF_Transition frequency}
\end{equation}
where $(v_k)$ is the velocity component along $\mathbf{k}$. This establishes a direct correlation between the laser frequency and the particle velocities, facilitating the measurement of the VDF.
LIF diagnostics involve scanning laser frequencies near the target transition and recording the fluorescence. Each frequency corresponds to a velocity group $(v_k)$ of the species, allowing the velocity distribution to be mapped as a function of the laser frequency:
\begin{equation}
	f_{LIF}(v_k) \sim  F\Bigg(\dfrac{\nu_{T}}{1 - \dfrac{v_k}{c}}\Bigg)
    \label{ch3:VDF_LIF}
\end{equation}
This approach provides a one-dimensional representation of the VDF along the laser beam direction.

The broadening of the line shape due to the hyperfine structure of the energy levels and the isotope splitting significantly impacts the spectral profile. The hyperfine energy levels are determined by the total angular momentum $\mathbf{F}$, which results from the coupling between the total electron angular momentum $\mathbf{J}$ and the nuclear spin $\mathbf{I}$, represented as $\mathbf{F} = \mathbf{J} + \mathbf{I}$. For example, among the nine isotopes of xenon, seven have zero nuclear spin due to an even number of neutrons and thus do not contribute to hyperfine structures, leaving only two isotopes that affect the line shape. This complexity in the hyperfine structure\cite{huang2009obtaining}, can complicate the spectral analysis, especially since not all transitions have well-documented hyperfine structures. However, the most probable velocities can be discerned from the location of the LIF signal peak, which remains unaffected by this broadening mechanism. This allows for the characterization of flows and their variations. However, for more comprehensive analysis it is important to know the hyperfine structure and the isotope splitting of the species.

The lifetime broadening mechanism is related to the Heisenberg uncertainty principle, which posits that certain pairs of physical properties, like position and momentum or energy ($E$) and time ($t$), cannot be precisely measured simultaneously. This principle is mathematically expressed as:
\begin{equation}
\Delta E\Delta t \geq \hbar.
\end{equation}
The uncertainty in the photon energy, $\Delta \omega$, is influenced by the lifetime of the upper energy level, $\tau_p$. This relationship modifies the line shape into a Lorentzian function, as outlined in the literature \cite{kunze2009introduction}:
\begin{equation}
    \phi(\nu) = \dfrac{\Delta \nu}{(\nu - \nu_0)^2 + (\Delta \nu/2)^2}, \\
    \Delta \nu = \dfrac{1}{\tau_p}.
\end{equation}
For instance, for the metastable level $5d^2[4]_{7/2}$ of Xe II, the lifetime ranges from 7 to 9 ns. The resultant broadening is significantly smaller by one or two orders of magnitude than that caused by the Doppler broadening, and thus can generally be neglected in analysis.

Two other mechanisms responsible for spectral line splitting are the Stark and the Zeeman effects. The Stark effect involves the line splitting due to the strong electric fields, which can induce additional angular momentum between the nucleus and the electron cloud. However, the electric fields must be considerably strong to have a measurable impact. For instance, it was demonstrated that the Stark effect is negligible under typical plasma experimental conditions due to the field strengths involved\cite{hubner2013density}. On the other hand, the Zeeman effect, which results from the influence of the magnetic fields, is more significant and can substantially alter the LIF line shape\cite{boivin2003laser}. This effect allows for the simulation of the spectral line splitting and the deduction of local magnetic field values, providing valuable insights into the magnetic environment within the plasma.

\subsubsection{Laser systems}

A tunable laser is essential for scanning the absorption profile in LIF experiments. It is important to distinguish between the continuous and the pulsed modes of the laser operation \cite{renk2012basics, trager2012springer}. Continuous wave (CW) lasers, such as laser diodes, have extremely narrow bandwidths, typically in the femtometer range or below. They allow for the direct observation of the species VDF in LIF measurements. Pulsed lasers, in contrast are broader in their line-width and deliver a large amount of energy per pulse, with pulse widths as short as a few femtoseconds, making them more suitable for time-resolved measurements.

\begin{figure}[H]
    \centering
    \includegraphics[width=0.33\linewidth]{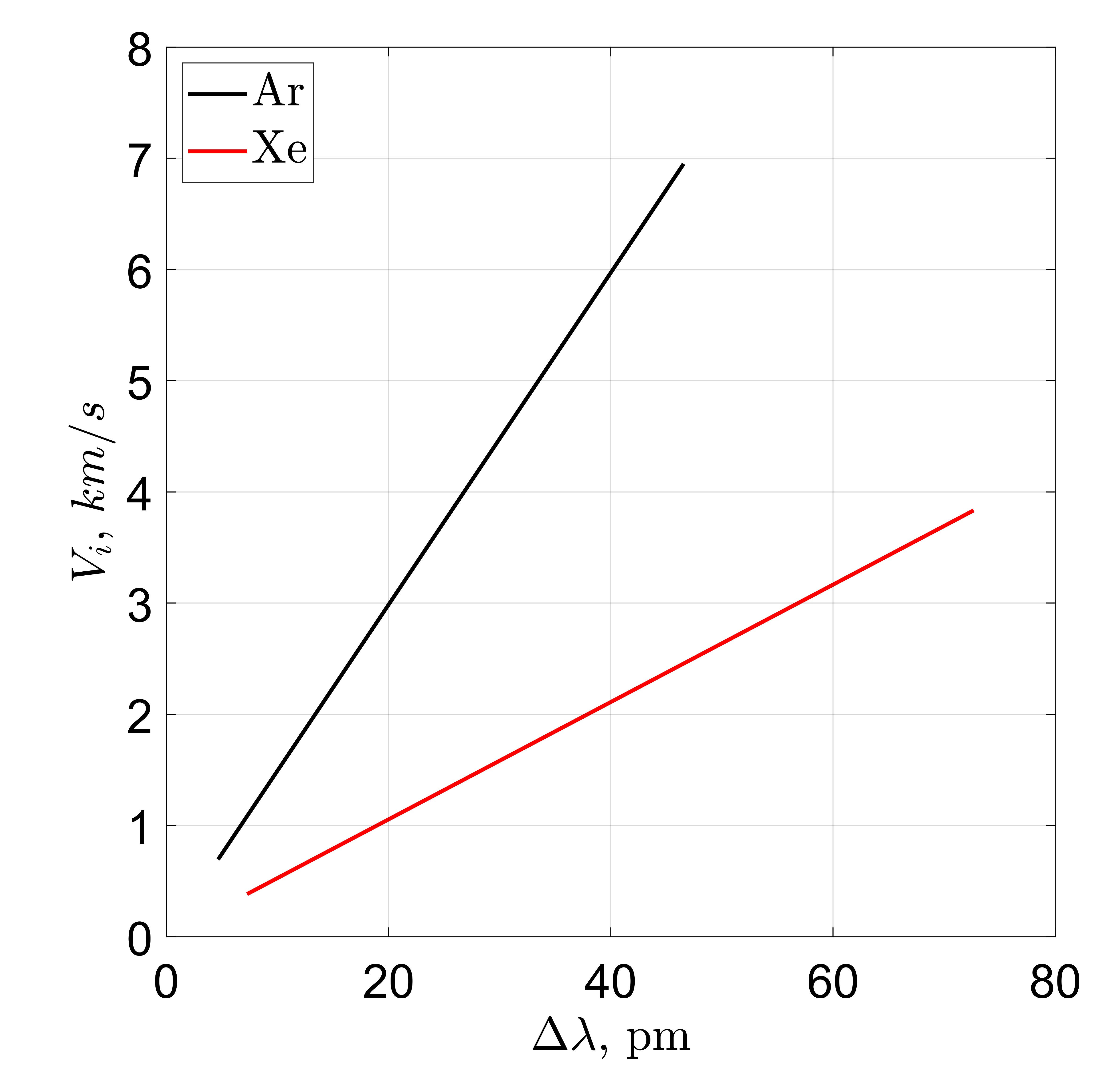}
    \caption{The detectable argon and xenon ion velocities $V_i$ as a function of the laser line-width, $\Delta\lambda$. The wavelength considered for the excitation of the Ar and Xe ions are 667 nm and 834 nm respectively.}
    \label{fig: Velocity to linewidth}
\end{figure}

The measurable VDFs vary with the type of the laser: diode lasers measure in the range of tens of m/s, while dye lasers and optical parametric oscillators (OPO) can detect velocities up to a few km/s and tens of km/s, respectively. As an example, the dependence of the laser line-width ($\Delta \lambda$) in $pm$ on the ion velocity for argon and xenon ions has been shown in Fig. \ref{fig: Velocity to linewidth}. Diode lasers are particularly effective for measuring the profiles of the velocity distributions, whereas dye and OPO lasers are better suited for detecting velocity drifts in systems with high-energy flows. However, it is important to note that in the case of high-resolution measurements, the limiting factor is usually a resolution of measurement system that detects the current laser wavelength, e.g. a wavemeter.

CW diode lasers are commonly used in Doppler-shift based LIF, especially for noble gases. They offer wide tunability (up to 10 nm, depending on the design) and very narrow line-width. Additionally, they facilitate various high-frequency laser light modulations, such as amplitude and wavelength (or frequency) modulation, which are crucial for implementing detection methods based on homodyne\cite{scime2005laser}, heterodyne\cite{diallo2015time}, or photon counting principles\cite{pelissier1996time, vaudolon2013photon}.

\subsubsection{Optical setup}
\label{Section: Doppler LIF optical setup}

\begin{figure}[H]
    \centering
    \includegraphics[width=1\linewidth]{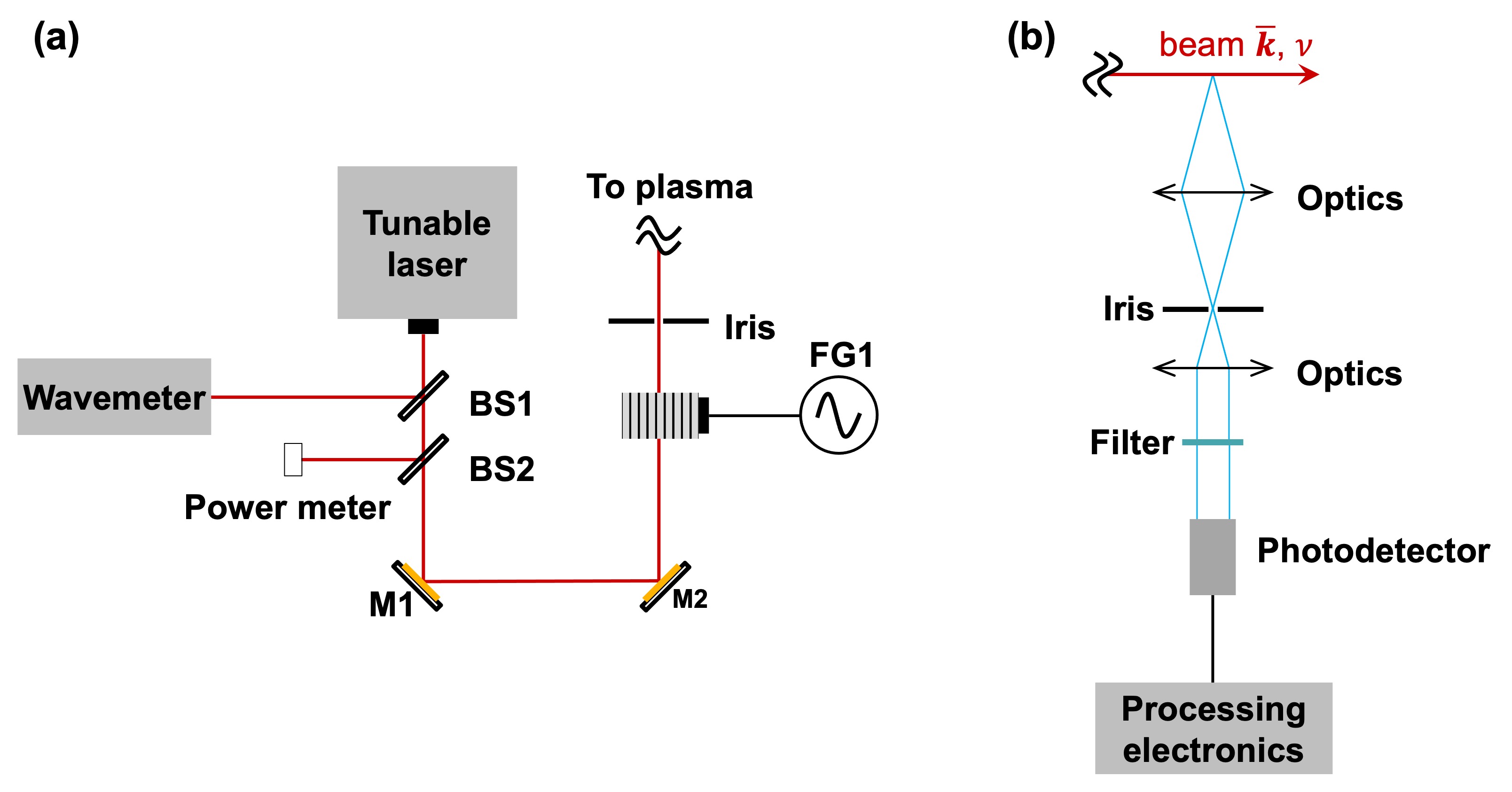}
    \caption{Generalized LIF setup. a) Laser launch brunch, BS1,2 - are beam splitters, and M1,2 - mirrors. b) Detection branch. "Optics" can represent various combination of lenses arranged to collect fluorescent light.}
    \label{fig: LIF setup}
\end{figure}

The typical optical setup for the Doppler LIF measurements can be divided into two main components: the laser launch branch and the detection branch, as illustrated in Figure \ref{fig: LIF setup}(a) and \ref{fig: LIF setup}(b). The laser launch branch consists of optical elements tasked with the beam conditioning, directing, light modulation, and the measurement of the laser power and the laser wavelength. The beam conditioning involves a series of mirrors and lenses that direct and focuses the laser beam into the interrogation volume. Optical fibers may also be used to transport the laser beam. In some setups, the laser beam is shaped into a laser sheet for planar LIF measurements\cite{jacobs2007laser}. The laser wavelength is controlled using a calibrated wavemeter. Additionally, a Fabry-Perot interferometer can be employed to monitor the quality of the laser mode in real-time and to detect any mode hops. The power of the beam is typically monitored continuously with a photodiode. The modulation of the laser wavelength or amplitude is an essential aspect of the setup. Amplitude modulation can be achieved using mechanical choppers, acousto-optic modulators (AOM), or electro-optic modulators (EOM). Wavelength modulation can be performed by modulating the laser diode current, which allows for the high-frequency modulation but also results in the amplitude modulation, or by using the piezo-driven modulation, which supports the lower-frequency modulations without affecting the laser amplitude. Wavelength modulation is particularly useful for implementing the derivative spectroscopy techniques\cite{romadanov2024wavelength, dixit1985quantitative}.

The detection branch of the optical setup consists of a collecting lens that directs light through a series of lenses and a spatial filter (pinhole) into the detector. This lens is typically positioned so that its optical axis is perpendicular to the laser beam's wavevector, which enhances spatial resolution. The collected light is filtered around the fluorescence wavelength, either by a narrow bandpass filter or by a monochromator. Various detectors such as a photomultiplier tube (PMT), a photodiode, or a CCD camera can be used depending on the measurement requirements. For point measurements, a PMT is preferred due to its high sensitivity. A CCD camera is more suitable for planar LIF measurements. Additionally, the detection branch can be mounted on a movable stage, allowing measurements along the laser beam and enabling the acquisition of 1D distributions of parameters such as velocity and temperature.

Signal processing for the LIF measurements can be performed in various ways to achieve either time-averaged or time-resolved information. A lock-in amplifier, synchronized with the laser modulation frequency, is commonly used. For more advanced techniques, utilizing beating frequencies help in the extraction of time-resolved information effectively. The photon-counting technique is also widely employed for this purpose. When the laser wavelength is modulated, the use of a lock-in amplifier to extract higher harmonics facilitates the derivation of the VDF profile. This technique provides a more sensitive analysis of the true VDF shape, particularly in cases of complex signals that deviate from a Maxwellian distribution.

The placement of the detection branch perpendicular to the laser beam typically necessitates at least two optical ports in the plasma chamber: one for the entry of the laser and another for collecting the LIF signal. However, modern industrial-scale plasma reactors are fully enclosed to ensure uniform processing of the materials and having more than one viewport can disrupt the plasma processes and alter the processing of the material. Additionally, space constraints often limit access to only one location around a plasma reactor. In such cases, confocal LIF is a suitable alternative. This technique allows both the input laser beam and the collected LIF signal to pass through the same axis into the plasma discharge, minimizing the need for multiple ports. Originally widespread in the fields of biology and medicine \cite{amos2003confocal}, confocal LIF has been adapted for TALIF and single LIF measurements across various scenarios\cite{vandervort2014optimization, thompson2017confocal, Kajiwara_1990, caron2023ion}. Recently, this method has been enhanced with structured laser light, specifically Laguerre-Gaussian beams, to improve the signal-to-noise ratio. This modification has been successfully implemented on a plasma source to measure the velocity of argon ions \cite{romadanov2023confocal}.

\section{LIF/TALIF reference data }
\label{Section: LIF/TALIF reference data}

The microelectronics industry and sustainability applications use a combination of a wide variety of precursor gases such as CH$_4$, CF$_4$, CH$_2$F$_2$, CHF$_3$, C$_2$F$_4$, C$_3$F$_6$, C$_4$F$_6$, C$_4$F$_8$, NF$_3$, HBr, Cl$_2$, COS, SF$_6$, H$_2$, O$_{2}$, H$_{2}$O, N$_{2}$, He and Ar, among others to generate the plasma \cite{lee2018relationships,foad1992ch4,bell1994investigation,li2002fluorocarbon,cho2012angular,rhee2008comparison,d1981plasma,efremov2020kinetics,kim2013study,jiang2004dry,donnelly1984anisotropic,nishino1993damage,kondeti2020h,hefny2016atmospheric,yatom2023diagnostics}. It results in the generation of a combination of a wide variety of ions, radicals and metastable species that play different roles in the overall chemical process. The LIF/TALIF transitions to detect such species should be carefully selected such that there is no overlap in the excitation and fluorescence wavelengths of the species of interest with other species present in the detection volume. There are several species that have coincident excitation and fluorescence wavelengths. Depending on the species present in the interrogation volume, the excitation and fluorescence wavelengths should be carefully examined to determine if a spectral filter or a monochromator is required for collecting fluorescence. While using a spectral filter for collecting fluorescence often provides a higher optical efficiency, the presence of overlapping fluorescence might necessitate the use of a monochromator to obtain the wavelength of the fluorescence and differentiate between the interfering species. For example, in an industrial scale silicon etching CCP discharge with argon and oxygen admixture, we found that both the excitation and the fluorescence wavelengths of O$_{2}^{+}$ ions and SiO radicals coincide with one another inhibiting the differentiation between the two species even with the use of a monochromator. This reference data is intended to assist in the selection of the LIF/TALIF transitions that do not overlap with other species present in the interrogation volume. Some groups of species with overlapping excitation and fluorescence wavelengths that may not be differentiable by the use of a spectral filter are listed below:
\begin{enumerate}
    \item NO, NS, O$_2$, O$_2^+$ and SiO.
    \item CH$_3$O, OH, SiCl, SiF, Ti and W.
    \item CF$_3$O, CN$_2$, C$_2$H$_3$O, C$_2$H$_5$O,  HC$_{2}$O, N$_{2}$O$^{+}$ and SH.
    \item CH$_{2}$O and CN.
    \item C$_2$H$_5$S, C$_{3}$H$_{7}$S and N$_{2}^{+}$.
    \item C$_2$H$_5$S, C$_{3}$H$_{7}$S and SiN.
    \item C$_2$H$_5$S, C$_{3}$H$_{7}$S, CS$_2^+$, HC$_2$S$_2$ and HSiF.
    \item Br$_2^+$, CuSH and Na$_2$.
    \item NO$_{3}$ and RuO. 
    \item BO$_2$ and Br$_2$.
    \item CH$_2$, C$_{2}$O and NH$_{2}$.
    \item SiCCl and SiCF. 
    \item SiCCl and TiC. 
    \item CH$_2$ and C$_2$O.
    \item C$_2$O and HNO.
    \item C$_2$P, C$_2$S, HBF, SiCl$_3$O and SiC$_3$H.
    
\end{enumerate}

A complied list of available LIF/TALIF transitions for atomic, molecular, ionic, and metastable species relevant to the microelectronics industry and sustainability applications are shown in tables \ref{Table: Neutral molecules LIF1: A to G}, \ref{Table: Neutral molecules LIF 2: C2 to H}, \ref{Table: Neutral molecules LIF 3: I to P}, \ref{Table: Neutral molecules LIF 4: Q to S}, \ref{Table: Neutral molecules LIF 5: S to Z}, \ref{Table: Neutral atoms LIF}, \ref{Table: Ions LIF} and \ref{Table: Neutral metastables LIF}. Several of these transitions were studied a few decades ago when the availability and the use of equipment such as iCCD cameras or a monochromator were unavailable. When the transition/fluorescence information was unavailable, a "not-reported" ($NR$) indication was mentioned in the tables. Certain neutral molecules cannot be detected by excitation from the ground state and require the presence of higher vibrational states for their detection. They have been marked in the tables. While we have put efforts into reporting all the relevant species that have been reported in the literature, it is possible that the reported tables inadvertently omitted relevant species/transitions that have been previously reported.


\newpage

\renewcommand{\arraystretch}{1.4}
\newcounter{magicrownumbers}
\newcommand\rownumber{\stepcounter{magicrownumbers}\arabic{magicrownumbers}}

\begin{table*}[b!]
    \centering
    \resizebox{\textwidth}{!}
    {
    \begin{tabular}{|c|c|c|p{4.5 cm}|p{4.5 cm}|c|}
    \hline
    Sl. No.  & Species & Transition & Excitation (nm)  & Fluorescence (nm) & Reference\\
    \hline
   \rownumber & AlO & B$^2\Sigma^{+}$ -- X$^2\Sigma^{+}$ & 464--470 & 507--519 & \cite{colibaba2000detection,honma2003reaction,li2019laser}\\
    \hline
    \rownumber & AlS & A$^2\Sigma^{+}$ -- X$^2\Sigma^{+}$ & 418 -- 430 & 440 & \cite{he1993production}\\
    \hline
    \multirow{2}{*}{\rownumber} & \multirow{2}{*}{As$_2$} & A$^{1}\Sigma^{+}_{u}$ -- X$^{1}\Sigma^{+}_{g}$ & 240 -- 241 & 275 -- 305 & \cite{alstrin1992vibrational,donnelly1982development,smilgys1990state}\\
    \cline{3-6}
    & & A$^{1}\Sigma^{+}_{u}$ -- X$^{1}\Sigma^{+}_{g}$ & 248 & 289.3 & \cite{donnelly1982development} \\
    \hline
    \multirow{2}{*}{\rownumber} & \multirow{2}{*}{BC} & B$^{4}\Sigma^{-}$ -- X$^{4}\Sigma^{-}$ & 557.4 -- 559.2 & 596.9 & \cite{ng2011laser,sunahori2015optical}\\
    \cline{3-6}
    & & E$^{4}\Pi$ -- X$^{4}\Sigma^{-}$ & 290.9 -- 291.5 & $NR$ & \cite{sunahori2015optical}\\
    \hline
    \rownumber & BCl & A$^1\Pi$ -- X$^1\Pi^{+}$ & 272 & 278 & \cite{fleddermann1998measurements}\\
    \hline
    \rownumber & BH & A$^{1}\Pi$ -- X$^{1}\Sigma^{+}$ & 396 & 429 -- 434 & \cite{rice1989gas, harrison1988kinetics}\\
    \hline
    \rownumber &   BH$_{2}$ & $\tilde{A}^{2}$B$_{1}$($\Pi_{u}$) -- $\tilde{X}^{2}$A$_{1}$ & 727.2 -- 742.4 & $NR$ & \cite{sunahori2015bh2} \\
    \hline
    \rownumber & BO & A$^{2}\Pi$ -- X$^{2}\Sigma^{+}$ & 313.5 -- 316.5, 422 -- 428, 434.5 & $NR$ & \cite{clyne1980laser,maksyutenko2011lif}\\
    \hline
    \multirow{3}{*}{\rownumber} & \multirow{3}{*}{BO$_2$} & A$^{2}\Pi_{u}$ -- X$^{2}\Pi_{g}$ & 434 -- 434.3 & $NR$ & \cite{clyne1980laser}\\
    \cline{3-6}
    & & $NR$ & 545.6, 547.1 & 574 -- 586 & \cite{schneider1988application} \\
    \cline{3-6}
    & & A$^{2}\Pi_{u}$ -- X$^{2}\Pi_{g}$ & 579 -- 583 & 538--554, 571 -- 587, 611 -- 627, 630 -- 646 & \cite{weyer1980laser} \\
    \hline
    \rownumber & BS$_2$ & $\tilde{A}^{2}\Pi_{u}$ -- $\tilde{X}^{2}\Pi_{g}$  & 530 -- 760 & 424 -- 486 & \cite{he2003study} \\
    \hline
    \rownumber & Br$_2$ & B$^3\Pi_{0^{+}u}$ -- X$^1\Sigma^+_g$  & 548 -- 595 & 618 -- 621, 714 -- 1110 & \cite{focsa2000characterization,bullman1981laser} \\
    \hline
    \rownumber & c--C$_6$H$_{7}$ & $\tilde{A}^{2}$A$_{2}$ -- $\tilde{X}^{2}$B$_{1}$ & 549.5 -- 549.9 & $NR$ & \cite{nakajima2007rotationally}\\
    \hline
    \rownumber & CBr$_2$ & $\tilde{A}^{1}$B$_{1}$ -- $\Tilde{X}^{1}$A$_{1}$ &  561 -- 575 & 588 -- 672 & \cite{zhou1990gas,lee2003electronic}\\
    
    \hline
    \rownumber & CCl & A$^{2}\Delta$ -- X$^{2}\Pi$ & 278 & $NR$ & \cite{hack1985detection,tiee1980reactions}\\
    \hline
    \rownumber & CCl$_{2}$ & A$^{1}$B$_{1}$ -- X$^{1}$A$_{1}$ & 416 -- 556 & 391 -- 500, $>$580 & \cite{huie1977laser,gomez2013kinetic,liu2004reaction,lu1991laser,tiee1979laser,guss2005laser}\\
    \hline
    \rownumber & CCl$_{2}$S & $\tilde{B}$ -- $\tilde{X}$ & 274 -- 297 & $NR$ & \cite{ludwiczak1991bx}\\
    \hline
    \rownumber & CF & A$^2$ $\Sigma$ $^{+}$ -- X$^2\Pi$ & 223.8 & 230.6 & \cite{booth1989spatially}\\
    \hline
    \rownumber & CF$_2$ & $\tilde{A}$ -- $\tilde{X}$ & 260 -- 276 & 260 -- 390 & \cite{king1979spectroscopy,mcmillin1997two,rubio2001laser}\\
    \hline
    \rownumber & CF$_3$O & $\tilde{A}^{2}$A$_1$ -- $\tilde{X}^{2}$E  & 332 -- 352 & 355 -- 400 & \cite{yang1997rotational,li1991laser}\\
    \hline
    \rownumber & CF$_3$S & $\tilde{A}^{2}$A$_{1}$ -- $\tilde{X}^{2}$E$_{3/2}$   & 373.8 -- 374, 377.7 -- 377.8, 365.8 --366 & $NR$ & \cite{yang1997rotational} \\
    \hline
    \rownumber & CFBr & $\tilde{A}^{1}$A$^{''}$ -- $\tilde{X}^{''}$A$^{''}$ & 422.6 -- 424.8 & 416 -- 556 & \cite{purdy1980laser, truscott2009reanalysis}\\
    \hline
    \rownumber & CFCl & $\tilde{A}$ -- $\tilde{X}$ & 363.3, 367.3, 372.7, 378.3, 384 & 330 -- 395 & \cite{huie1977laser, tiee1979laser,guss2001electronic}\\
    \hline
    \multirow{3}{*}{\rownumber} &  \multirow{3}{*}{CH} & A$^2$ $\Delta$ -- X$^2$ $\Pi$ & 431 & 489 & \cite{raiche1993laser}\\
    \cline{3-6}
    
     & & A$^2$ $\Delta$ -- X$^2$ $\Pi$ & 427.9 -- 428.1, 435.4 -- 435.8 & 485 & \cite{luque1997spatial,luque1997absolute}\\
     \cline{3-6}
       &  & B$^{2}\Sigma^{-}$ -- X$^{2}\Pi$ & 363.3 -- 363.6, 387 -- 388 & 385 -- 415, 423 -- 435, 443 -- 460 & \cite{engelhard1995new,luque1997absolute,luque2002quantitative,luque2002quantitative}\\
    \hline
    
    \end{tabular}
     }
     \caption{Laser-induced fluorescence transitions of neutral molecular species - 1. $NR$: not reported.}
    \label{Table: Neutral molecules LIF1: A to G}
\end{table*}

\begin{table*}[p]
    \centering
    \resizebox{\textwidth}{!}
    {
   \begin{tabular}{|c|c|c|p{4.5 cm}|p{4.5 cm}|c|}
    \hline
    Sl. No. & Species & Transition & Excitation (nm)  & Fluorescence (nm) & Reference\\
    \hline
    \rownumber  & CHF & $\tilde{A}$ $^1$ A$^{\prime\prime}$ -- $\tilde{X}$ $^{1}$ A$^{\prime}$ & 492 -- 493, 514.5, 574 --582 & 515 -- 521, 550 -- 570, 600 -- 620, 650 -- 670 & \cite{hakuta1984vibration,l1997detection,kakimoto1981doppler,qiu1987laser,ashfold1980laser,hancock1982chf,schmidt1999characterization}\\

    \hline
      \rownumber & CH$_2$ & $\tilde{b}$ $^1$B$_{1}$ -- $\tilde{a}$ $^1$A$_1$ & 589 -- 595 & 635 -- 651 & \cite{danon1978laser}\\
     \hline
\rownumber & CH$_{2}$O & $\tilde{A}^{1}$A$_{2}$ -- $\tilde{X}^{1}$A$_{1}$ & 352 -- 357 & 380 -- 550 & \cite{brackmann2005strategies,harrington1993laser} \\
     \hline   

     \multirow{2}{*}{\rownumber} & \multirow{2}{*}{CH$_{3}$O} & A$^{2}$A$_{1}$ -- X$^{2}$E & 270 -- 330 & 300 -- 400 & \cite{gutman1982kinetics,ebata1982a,inoue1979laser,inoue1980spectroscopy,kappert1989rotationally}\\
     \cline{3-6}
      & & $\tilde{B}$ -- $\tilde{X}$ & 357 -- 376 & 316 -- 375 & \cite{zu2004jet}\\
      \hline
    \rownumber &  CH$_3$S & $\tilde{A}^{2}$A$_{1}$ -- $\tilde{X}^{2}$E & 364 -- 378 & 370 -- 500 & \cite{suzuki1984laser, black1986laser}\\
    \hline
   
       \rownumber  &  CN & B$^2\Sigma$ -- X$^2\Sigma$ & 356.5, 381 -- 388.5, 565 -- 610 & 388.8, 389.5 & \cite{hirano1994visualization,conley1980laser,cody1977laser,bonczyk1979measurement}\\
       \hline
      \rownumber & CN$_{2}$ & $^{3}\Pi_{1}$ -- $^{3}\Sigma^{-}$, $\tilde{A}^{3}\Pi$ -- $\tilde{X}^{3}\Sigma^{-}$ & 327 -- 336 & 330 -- 440 & \cite{sutton2008laser,smith1989electronic,curtis1988laser}\\
    \hline 
    \rownumber &  CO & B$^{1}$ $\Sigma$ $^{+}$ -- X$^{1}$ $\Sigma$ $^{+}$ & 2 $\times$ 230.1 & 484 & \cite{dally2003two} \\
    \hline
       \rownumber &  CO$_{2}$ & (00$^{0}$0) -- (10$^{0}$01) & 2005, 2700 & 4260 -- 4270 & \cite{zetterberg2012situ,alwahabi2007high} \\
    \hline
       \rownumber &  CS & A$^{1}\Pi$ -- X$^{1}\Sigma^{+}$ & 257.5 -- 258.3 & 250 -- 290 & \cite{clough1980laser,hynes1979laser,martin1982two} \\
    \hline
      \rownumber & CS$_2$ & V$^{1}$B$_{2}$ -- $\tilde{X}^{1}\Sigma^{+}_{g}$  & 280 -- 338 & 370, 380, 403.7 & \cite{ochi1987rotationally,weyh1996lifetime,martinez1997laser,sisk1999influence}\\
   \hline
       \rownumber &  CuSH & $\tilde{A}^{1}$A$^{''}$ -- $\tilde{X}^{1}$A$^{'}$ & 472 -- 515 & 480 -- 506 & \cite{sunahori2006electronic} \\
       \hline
     \multirow{2}{*}{\rownumber} & \multirow{2}{*}{Cu$_{2}$} & A$^{1}\Sigma^{+}_{u}$ -- X$^{1}\Sigma^{+}_{g}$ & 490.27 & 495 -- 499 & \cite{sappey1992planar} \\
    \cline{3-6} 
     & & B$^{1}\Pi_{u}$ -- X$^{1}\Sigma^{+}_{g}$ & 449.8 & 460.7 & \cite{dreyfus1991cu0}\\
     \hline
  \multirow{4}{*}{\rownumber} & \multirow{4}{*}{C$_{2}$} & d$^3$ $\Pi$ -- a$^3$ $\Pi$ & 516 & 563 & \cite{kaminski1995absolute}\\
    \cline{3-6}
     &  & d -- a & 470 & 425 -- 438 & \cite{vekselman2018quantitative,luque1997spatial} \\
     \cline{3-6}
     &  & A$^{1}\Pi_{u}$ -- X$^{1}\Sigma_{g}^{+}$ & 690.9 & 790.8 & \cite{reisler1980kinetics,pitts1982temperature,jones1976evaluation}\\
     \cline{3-6}
     &  & d$^{3}\Pi_{g}$ -- a$^{3}\Pi_{u}$ & 516.5 & 559 & \cite{reisler1980kinetics,tatarczyk1976lifetime}\\ 
     \hline
     \rownumber & C$_{2}$H$_{2}$ & $\tilde{A}$ -- $\tilde{X}$ & 225 -- 235 & 250 -- 360 & \cite{williams2002laser}\\
    \hline
    \rownumber & C$_2$H$_3$O &  $\tilde{B}^{2}$A$^{''}$ -- $\tilde{X}^{2}$A$^{''}$ & 330 -- 350 & 340 -- 420 & \cite{inoue1981laserC2H3O,kleinermanns1981laser,dimauro1984laser}\\
    \hline
    \rownumber & C$_2$H$_3$S & $\tilde{B}$ -- $\tilde{X}^{2}$A$^{''}$  & 458.2 -- 458.5 & 464 -- 531 & \cite{nakajima2007laser}\\
    \hline
    \rownumber & C$_2$H$_5$O &  $\tilde{A}$ -- $\tilde{X}$ & 310 -- 350 & 340 -- 430 & \cite{gutman1982kinetics,inoue1981laser,ebata1982a}\\
    \hline
    \multirow{2}{*}{\rownumber} & \multirow{2}{*}{C$_2$H$_5$S} & $\tilde{A}$ -- $\tilde{X}$  & 390 -- 450 & 420 -- 580 & \cite{black1987laser}\\
    \cline{3-6}
    & & $\tilde{B}^{2}$A$^{'}$ -- $\tilde{X}^{2}$A$^{''}$ & 397 -- 427 & 352 -- 425 & \cite{hung1996vibronic}\\
    \hline
    \rownumber & C$_2$N & $\tilde{A}^{2}\Delta$ -- $\tilde{X}^{2}\Pi$  & 465.8, 470 -- 471.5 & 466 -- 580 & \cite{kakimoto1982doppler, hakuta1983laser}\\
    \hline
     \rownumber & C$_{2}$O &  $\tilde{A}^{3}\Pi_{i}$ -- $\tilde{X}^{3}\Sigma^{-}$ & 588 -- 685 & 650 -- 780 & \cite{pitts1981c2o}\\
    \hline
      \rownumber & C$_{2}$P &  $^{2}\Delta_{i}$ -- $\tilde{X}^{2}\Pi_{r}$ & 596 -- 633 & 556 -- 627 & \cite{sunahori2008electronic}\\
    \hline
    \rownumber  & C$_{2}$S & $\tilde{A}^{3}\Pi_{i}$ ← $\tilde{X}^{3}\Pi^{-}$ & 675 -- 690, 604 -- 614 & 553 -- 614 & \cite{schoeffler2001laser}\\
    \hline
    \multirow{2}{*}{\rownumber} &  \multirow{2}{*}{C$_{3}$} & A$^{1}\Pi_{u}$ -- X$^{1}\Sigma^{+}_{g}$ & 405.3 & 398 -- 411 & \cite{ikegami2001spatial}\\
     \cline{3-6}
     & & A -- X & 425 -- 430 & 405 & \cite{luque1997spatial}\\
     \hline
     \rownumber &  C$_{3}$H$_{7}$S & $\tilde{A}$ -- $\tilde{X}$ & 395 -- 430 & 420 -- 520 & \cite{black1987laserC3H7S}\\
     \hline
     \rownumber &  C$_{3}$N & $\tilde{B}^{2}\Pi_{i}$ -- $\tilde{X}^{2}\Sigma^{+}$ & 343 -- 350 & 284 -- 336 & \cite{hoshina2007laser}\\
     \hline
    \rownumber &  C$_{4}$H & $\tilde{B}^{2}\Sigma^{+}$ -- $\tilde{X}^{2}\Sigma^{+}$ & 400 -- 416.7 & $>$ 420 & \cite{hoshina1998laser}\\
     \hline
     \rownumber &  FCO & $\tilde{A}^{2}\Pi$ -- $\tilde{X}^{2}$A$^{'}$ & 307 -- 345 & 333 -- 477 & \cite{williams1997laser}\\
     \hline
     \rownumber & GaCl & A$^{3}\Pi_{0}^{+}$ -- X$^{1}\Sigma^{+}$ & 332 -- 346 & 364 & \cite{donnelly1982development}\\ 
    \hline
    \rownumber & HBF & $\tilde{A}^{2}$A$^{\prime\prime}$ $\Pi$ -- $\tilde{X}^{2}$A$^{\prime}$  & 602 -- 666 & 535 -- 628 & \cite{sunahori2009electronic}\\
    \hline
    \rownumber & HBr & M$^{1}\Pi$  -- X$^{1}\Sigma^{+}$  & 115.2 -- 116.2  & $NR$ & \cite{callaghan1989single}\\
    \hline
    \rownumber & HCBr & $\tilde{A}$ -- $\tilde{X}$2$^{7}_{0}$  & 560 -- 561.3 & 468 -- 528  & \cite{lee2003electronic}\\
    \hline

\end{tabular}
     }
     \caption{Laser-induced fluorescence transitions of neutral molecular species - 2. $NR$: not reported.}
    \label{Table: Neutral molecules LIF 2: C2 to H}
\end{table*}

\begin{table*}[t!]
    \centering

    \resizebox{\textwidth}{!}
    {
    \begin{tabular}{|c|c|c|p{4.5 cm}|p{4.5 cm}|c|}
    \hline
    Sl. No. & Species & Transition & Excitation (nm)  & Fluorescence (nm) & Reference\\
    \hline
    \rownumber & HCl & V$^{1}\Sigma^{+}$ -- X$^{1}\Sigma^{+}$  & 115.7 -- 115.9 & $NR$ & \cite{callaghan1989single}\\
    \hline
    \rownumber & HCO & $^{2}\tilde{A}^{'}$ -- $^{2}\tilde{A}^{''}$  & 613 -- 618 & 638 -- 664 & \cite{konig1983laser}\\
    \hline
    \rownumber & HC$_{2}$O & X$^{2}$A$^{'}$ -- $^{2}$A$^{''}$  & 310 -- 360 & 350 -- 460 & \cite{inoue1986laser}\\
    \hline
    \rownumber & HC$_2$S$_2$ &  $\tilde{B}^{2}$A$^{''}$ -- $\tilde{X}^{2}$A$^{''}$ & 435 -- 458 & 458 -- 530 & \cite{nakajima2005gas} \\
    \hline
   \rownumber & HC$_4$S & $\tilde{A}^{2}\Pi_{3/2}$ -- $\tilde{X}^{2}\Pi_{3/2}$  & 487 -- 501 & 426 -- 488 & \cite{nakajima2002laser,reilly2006experimental}\\
    \hline
    \rownumber & HC$_6$S & $^{2}\Pi_{3/2}$ -- $^{2}\Pi_{3/2}$  & 572 -- 590 & $NR$ & \cite{nakajima2002laserHC6S} \\
    \hline  
    \rownumber & HNO & $\tilde{A}^{1}$A$^{''}$ -- $\tilde{X}^{1}$A$^{''}$  & 623 -- 626, 641.6 -- 643 & 685 -- 780 & \cite{crosley1986laser,mayama1989laser,taylor2024laser} \\
     \hline 
    \rownumber & HSiBr & $\tilde{A}^{'}$A$^{''}$ -- $\tilde{X}^{1}$A$^{1}$  & 458 -- 503 &  $NR$ & \cite{harjanto1996resolution}\\
    \hline
    \rownumber & HSiCl & $\tilde{A}^{'}$A$^{''}$ -- $\tilde{X}^{'}$A$^{'}$  & 444 -- 485, 538 -- 550 & 471, 483, 489, 503, 524 & \cite{ho1986kinetics,de2000induced,harper1997reinvestigation}\\
    \hline
    \rownumber & HSiF & $\tilde{A}$ $^{1}$A$^{\prime\prime}$ -- $\tilde{X}$ $^{1}$A$^{\prime}$  & 430, 446.5 & 425 -- 530 & \cite{lee1983laser,dixon1985rotational}\\
    \hline
    \rownumber & HSiNC & $\tilde{A}^{1}$A$^{''}$ -- $\tilde{X}^{1}$A$^{'}$ & 499.75 -- 502 & 428.2, 444.3, 461.9, 480.6  & \cite{evans2009spectroscopic}\\
    \hline
    
    \rownumber & HSiNCO & $\tilde{A}^{1}$A$^{''}$ -- $\tilde{X}^{1}$A$^{'}$ & 490 -- 492 & 379.6 -- 477.51  & \cite{dover2009spectroscopic}\\
    \hline
    \rownumber & HSO &  $\tilde{A}$ -- $\tilde{X}$ & 539 --  540, 585, 600 -- 615, 625 -- 645 & $>$ 665 & \cite{lee1994kinetics,kawasaki1987laser,yoshikawa2009laser,crosley1986laser} \\
    \hline
    \rownumber & HSnCl & $\tilde{A}^{1}$A$^{''}$ -- $\tilde{X}^{1}$A$^{'}$ & 455 -- 500 & 389 -- 482  & \cite{rothschopf2022barely}\\
    \hline
    \multirow{2}{*}{\rownumber} & H$_2$ & E, F $^{1}\Sigma^{+}_{g}$ -- X$^{1}\Sigma^{+}_{g}$ & 2 $\times$ 193.3 & 750, 830 & \cite{lempert1991two}\\
    \cline{2-6}
    & H$_{2}$ $^{\dag\dag}$ & B$^{1}\Sigma^{+}_{g}$ -- X$^{1}\Sigma^{+}_{g}$ & 106.2 -- 106.7 & $NR$ & \cite{northrup1984vuv, mosbach2000situ, gabriel2008formation}\\
    \hline    
    \rownumber & InCl & A$^{3}\Pi_{0}^{+}$ -- X$^{1}\Sigma^{+}$, B$^{3}\Pi_{1}$ -- X$^{1}\Sigma^{+}$  & 340 -- 360 & 363.7 & \cite{donnelly1982development}\\
    \hline
    \rownumber & IO & A$^{2}\Pi_{3/2}$ -- X$^{2}\Pi_{3/2}$ & 435 -- 468 & 500 -- 600 & \cite{turnipseed1995lif,bloss2003application}\\
    \hline
    \rownumber & I$_{2}$ & B$^{3}\Pi^{+}_{0u}$ -- X$^{1}\Sigma^{+}_{g}$ & 501.7, 514.5 & 515 -- 830 & \cite{hiller1990properties,ezekiel1968laser, zucco2013laser}\\
    \hline
    \rownumber & KH & A$^1\Sigma^+$ -- X$^1\Sigma^+$ & 502.4 -- 504 & 548.5 -- 557.5 & \cite{lin1989state}\\
    \hline
    \multirow{2}{*}{\rownumber} & \multirow{2}{*}{K$_2$} & B$^1\Pi_u$ -- X$^1\Sigma^+_g$  & 632.8 & 633 -- 650 & \cite{tango1968spectroscopy}\\
    \cline{3-6}
    & & C$^{1}\Pi_{u}$ -- X$^{1}\Sigma_{g}^{+}$ &420 -- 435, 457.9 & 420 -- 475, 550 -- 580 & \cite{milosevic1987study,meiwes1982predissociation} \\
    \hline
    \rownumber & NaK & D$^{1}\Pi$ -- X$^{1}\Sigma^{+}$ & 476 -- 529 & 1052 -- 2500 & \cite{ross1986laser}\\
    \hline
    \rownumber & Na$_2$ & B$^{1}\Pi_{u}$ -- X$^{1}\Sigma_{g}^{+}$ & 465.7 -- 514.5 & 470 -- 540 & \cite{demtroder1969spectroscopy}\\
    \hline
    \rownumber & NCO & A -- X & 420 -- 440 & 465 & \cite{copeland1985laser}\\
    \hline
    \multirow{2}{*}{\rownumber} & \multirow{2}{*}{NC$_{3}$O} & $^{2}\Sigma$ -- $^{2}\Sigma$ & 365.7 -- 365.9 & 329 -- 365 & \cite{yoshikawa2008laser}\\
    \cline{3-6}
     & & $^{2}\Pi$ -- $^{2}\Pi$ & 365.6 -- 365.8 & 329 -- 365 & \cite{yoshikawa2008laser}\\
     \hline
    \rownumber & NC$_{3}$S & $\tilde{A}^{2}\Pi_{3/2}$ -- $\tilde{X}^{2}\Pi_{3/2}$ & 463.9 -- 464.1   & 408 -- 455 & \cite{nakajima2004laser}\\
    \hline

    \rownumber & NF & b$^{\prime}$ $\Sigma$ -- X$^3\Sigma$ $^{-}$& 530 & 528.9 & \cite{heidner1989direct}\\
    \hline
     \rownumber & NH & A$^{3}\Pi$ -- X$^3\Sigma^{-}$ & 303 -- 305, 329 -- 336 & 330 -- 345 & \cite{wang2023simultaneous,copeland1989time,brackmann2016strategies,umemoto1996production}\\
    \hline   
    \rownumber & NH$_{2}$ & $\tilde{A}^{2}$A$_{1}$ -- $\tilde{X}^{2}$B$_1$ & 571 -- 662.2  & 516 -- 824 & \cite{copeland1985laser,halpern1975laser,hancock1975laser} \\
    \hline
    \rownumber & NH$_{3}$ & X -- C$^{\prime}$ & 2 $\times$ 305 & 550 -- 575, 720 & \cite{westblom1990laser} \\
    \hline
    \rownumber & N$_{3}$ & $\tilde{A}$ $^2$ $\Sigma$ $^{+}$ -- $\tilde{X}$ $^2\Pi_{3/2}$ & 271.9  & 271.1 - 272.8 & \cite{beaman1987observation} \\
    \hline

    \end{tabular}
     }
     \caption{Laser-induced fluorescence transitions of neutral molecular species - 3. $NR$: not reported. $^{\dag\dag}$: Laser excitation is from a vibrationally/rotationally excited state of a neutral gas molecule.}
    \label{Table: Neutral molecules LIF 3: I to P}
\end{table*}

\begin{table*}[h]
    \centering
    
    \resizebox{\textwidth}{!}
    {
    \begin{tabular}{|c|c|c|p{4.5 cm}|p{4.5 cm}|c|}
    \hline
    Sl. No. & Species & Transition & Excitation (nm)  & Fluorescence (nm) & Reference\\
    \hline 
     \multirow{3}{*}{\rownumber} & \multirow{2}{*}{NO} & A$^2$ $\Sigma$ $^{-}$ -- X$^2$ $\Pi$ & 226 -- 227 & 248 & \cite{brackmann2017quantitative,van2013temperature,lee1993temperature}\\
     \cline{3-6}
     & & D$^2$ $\Sigma$ $^{-}$ -- X$^2$ $\Pi$ & 193 & 196 -- 226 & \cite{wodtke1988high}\\
     \cline{2-6}
     & NO $^{\dag\dag}$& C$^2$ $\Sigma$ $^{-}$ -- X$^2$ $\Pi$ & 193 & 208 & \cite{wodtke1988high}\\
    \hline
    \rownumber & NO$_{2}$ & $\tilde{X}^{2}$A$_{1}$ -- $\tilde{A}^{2}$B$_{2}$ & 434.9, 430.6, 435.1  & 638 -- 663 & \cite{delon1991laser}\\
    \hline 
     \rownumber & NO$_{3}$ &  $\tilde{B}^{2}$E$^{'}$ -- $\tilde{X}^{2}$A$^{'}_{2}$ & 500 -- 680 & 464 -- 662 & \cite{kim1992no3,fukushima2022laser} \\
    \hline
     \multirow{2}{*}{\rownumber} & \multirow{2}{*}{NS} & C$^{2}\Sigma^{+}$ -- X$^{2}\Pi$ & 230 -- 232 & 236 -- 238  & \cite{jeffries1986laser}\\
    \cline{3-6}
    &  & $NR$ & 300 -- 310, 360 -- 370 & $NR$ & \cite{ongstad1992photodissociation}\\
    \hline
    \rownumber & O$_{2}$ $^{\dag\dag}$ & B$^{3}\Sigma^{-}_{u}$ -- X$^{3}\Sigma^{-}_{g}$ & 223 -- 226 & 240 -- 440 & \cite{goldsmith1986laser}\\
    \hline
    \rownumber & OH & X -- A & 280 -- 286 & 304 -- 314 & \cite{verreycken2013absolute,irvine1990laser}\\
    \hline
    \multirow{2}{*}{\rownumber} & \multirow{2}{*}{PbF} & A$^{2}\Sigma^{+}_{1/2}$ -- X$_{1}^{2}\Pi_{1/2}$ & 260 -- 285 & $NR$ & \cite{zhu2022fine,chen2022spectroscopic}\\
    \cline{3-6}
    & & B$^{2}\Sigma^{+}$ -- X$^{2}\Pi_{1/2}$ & 426 -- 450 & 435 -- 500, 700 -- 800, 1200 -- 1300 & \cite{shestakov1992radiative}\\
    \hline
    \rownumber & P$_2$ & $NR$ & 193 & 322.2 & \cite{donnelly1982development}\\
    \hline
    \rownumber & PH$_2$ & $\tilde{A}^{2}$A$_{1}$ -- $\tilde{X}^{2}$B$_{1}$  & 419 -- 547 & 430 -- 555 & \cite{chen1994laser,huie1978laser, xuan1990dynamics}\\
    \hline
    \multirow{3}{*}{\rownumber} & \multirow{3}{*}{PO} & B$^{2}\Sigma^{+}$ -- C$^{2}\Pi$  & 324 & $NR$ & \cite{clyne1981laser}\\
    \cline{3-6}
    & & A$^{2}\Sigma^{+}$ -- X$^{2}\Pi_{1/2}$ & 245.6 -- 246.6 & 239 --279, 319 -- 359 & \cite{wong1986radiative} \\
    \cline{3-6}
     & & A$^{2}\Sigma^{+}$ -- X$^{2}\Pi_{3/2}$ & 247 -- 247.8 & 239 --279, 319 -- 359 & \cite{wong1986radiative} \\
\hline
         \rownumber & RuB & [18.4]2.5–X$^2\Delta_{5/2}$ & 522 -- 524, 542 -- 543 & $NR$ & \cite{wang2012laser} \\
    \hline
    
         \rownumber & RuC & [18.1]$^1\Pi_{1}$ -- X$^{1}\Pi^{+}$, [5.7]$^{1}\Delta_{2}$ & 431 -- 439, 552.9 & 553, 806 & \cite{dabell2001electronic} \\
    
    \hline
    \multirow{2}{*}{\rownumber}& \multirow{2}{*}{RuCl} & $^{4}\Gamma_{5.5}$ -- X$^{4}\phi_{4.5}$ & 508 -- 557 & 499 -- 546 & \multirow{2}{*}{\cite{zarringhalam2022high}} \\
    \cline{3-5}
     & & $^{4}\phi_{4.5}$ -- X$^{4}\phi_{4.5}$ & 426 -- 569 & 430 -- 442 &  \\
    \hline
    
    \multirow{3}{*}{\rownumber}& \multirow{3}{*}{RuF} & \multirow{3}{*}{[18.2]$^{4}\Gamma_{4.5}$ -- X$^{4}\phi_{3.5}$} & 450 -- 452 & 410 -- 441 & \multirow{3}{*}{\cite{zarringhalam2022high}} \\
    \cline{4-5}
     & & & 545 -- 548 & 470 -- 533 &  \\
    \cline{4-5}
    & & & 758 -- 764 & 697 -- 732 &  \\
    \hline
     \rownumber & RuN & (0,0)F$^{2}\Sigma^{+}$ -- X$^{2}\Sigma^{+}$ & 529.6 -- 532 & $NR$ & \cite{steimle2003permanent} \\
    \hline

    \rownumber & RuO & [18.1]4 -- X$^{5}\Delta_{4}$ & 529.5 -- 531 & 463 -- 503 & \cite{zarringhalam2022high} \\
    \hline

    \rownumber & SH & $^{2}\Sigma^{+}$ -- $^{2}\Pi_{3/2,1/2}$ & 323 -- 330 & 320 -- 362 & \cite{tiee1981spectroscopy,hawkins1980193,becker1974lifetime}\\
    \hline
    \rownumber & SiBr & B$^{2}\Sigma$ -- X$^{2}\Pi_{r}$ & 308 & 296 -- 304 & \cite{herman1996surface,cheng1995competitive}\\
    \hline
    \rownumber & SiCCl & $\tilde{A}^{2}\Sigma^{+}$ -- $\tilde{X}^{2}\Pi$ & 550 -- 615 & 545 -- 650 & \cite{rothschopf2019high}\\
    \hline
    \rownumber & SiCF & $\tilde{A}^{2}\Sigma^{+}$ -- $\tilde{X}^{2}\Pi$ & 574 -- 604 & 547 -- 596 & \cite{rothschopf2018laser}\\
    \hline
    \rownumber & SiCH & $\tilde{A}^2\Sigma^{+}$ -- $\tilde{X}^2\Pi_{i}$ & 650 -- 680, 700 -- 740 & $NR$ & \cite{smith2000electronic}\\
    \hline

    \end{tabular}
     }
     \caption{Laser-induced fluorescence transitions of neutral molecular species - 4. $NR$: not reported. $^{\dag\dag}$: Laser excitation is from a vibrationally/rotationally excited state of a neutral gas molecule..}
    \label{Table: Neutral molecules LIF 4: Q to S}
\end{table*}

\begin{table*}[h]
    \centering
    
    \resizebox{\textwidth}{!}
    {
    \begin{tabular}{|c|c|c|p{4.5 cm}|p{4.5 cm}|c|}
    \hline
    Sl. No. & Species & Transition & Excitation (nm)  & Fluorescence (nm) & Reference\\
    \hline
    \rownumber & SiCl & B$^2$ $\Sigma$ -- X$^2$ $\Pi$ & 275, 295, 308 & 280, 320, 290 -- 307 & \cite{singleton1992vibrational,herman1996surface}\\
    \hline
     \rownumber & SiCl$_{2}$ & $\tilde{A}^{1}$B$_{1}$ -- $\tilde{X}^{1}$A$_{1}$ &  309 -- 333 & 331 -- 390 & \cite{suzuki1986laser}\\
    \hline
    \rownumber & SiCl$_{3}$O & $\tilde{A}^{2}$A$_{1}$ -- $\tilde{X}^{2}$E &  590 -- 650 & 617 -- 653 & \cite{smith2020identification}\\
    \hline
    \rownumber & SiC$_{3}$H & $\tilde{A}^{2}\Sigma^{+}$ -- $\tilde{X}^{2}\Pi_{i}$ &  613 -- 681 & 578 -- 663 & \cite{umeki2015laser}\\
    \hline
    \rownumber & SiF & X -- B & 288 & 308 -- 315 & \cite{hebner2001spatially}\\
    \hline
    \rownumber & SiF$_{2}$ & X -- A & 221.6 & 234 -- 244 & \cite{hebner2001spatially}\\
    \hline
    \rownumber & SiH & A$^2$ $\Delta$ -- X$^2\Pi$ & 409 -- 416 & 413 & \cite{tachibana1991measurement,nozaki2000identification,hertl1998laser,matsumi1986laser}\\
    \hline
    \rownumber & SiH$_{2}$ & $\tilde{A}$ $^1$B -- $\tilde{X}$ $^1$A & 579.2 -- 580.4 & 618 & \cite{hertl2000laser,kono1993laser,kono1995laser}\\
    \hline
    \rownumber & SiN & $\tilde{B}$ $^2$ $\Sigma$ -- $\tilde{X}$ $^2\Sigma$ & 396 & 414 & \cite{walkup1984laser}\\
    \hline 
    \rownumber & SiO & A $^1\Sigma$ -- X$^1\Sigma$ & 220 -- 237  & 214 -- 350 & \cite{van1989laser,chrystie2017sio,chrystie2019absolute}\\
    \hline
    \rownumber &  SO & $NR$  & 248.1, 248.7 & 250 -- 450 & \cite{greenberg1990laser}\\
    \hline
    \multirow{2}{*}{\rownumber} & \multirow{2}{*}{SO$_2$}  & $NR$ & 248.1, 248.7 & 250 -- 450 & \cite{greenberg1990laser}\\
    \cline{3-6}
    & & $\Tilde{X}$($^1$A$_1$) -- $\Tilde{C}$($^1$B$_2$), $\Tilde{B}$($^1$B$_1$), $\Tilde{A}$($^1$A$_2$)  & 266 & 250 -- 350 & \cite{weng2019quantitative}\\
    \hline
    \rownumber & S$_{2}$ & B$^{3}\Sigma_{u}^{-}$ -- X$^{3}\Sigma_{g}^{-}$ & 325, 337 & 310 -- 410 & \cite{smith1981fluorescence}\\ 
    \hline
    \rownumber & S$_{2}$O & $\tilde{C}^{1}$A$^{'}$ -- $\tilde{X}^{1}$A$^{'}$  & 340 & $NR$ & \cite{crosley1986laser,zhang1995laser} \\
    \hline
    \rownumber & TaC & [18.61]$^2\Pi_{3/2}$ -- X$^2\Sigma^{+}$, $NR$ & 537.1 -- 537.8, 505.2 & 401 -- 509 & \cite{nakhate2021jet}\\
    \hline
    \multirow{4}{*}{\rownumber} & \multirow{4}{*}{TiC} & $^1\Pi$ -- $X^{3}\Sigma^{+}$  & 606.9 -- 608.3 & 538 -- 603 & \multirow{4}{*}{\cite{nakhate2017laser}} \\
    \cline{3-5}
    & & $^1\Pi$ -- $a^{1}\Sigma^{+}$  & 619.8 -- 621.6 & $NR$ & \\
    \cline{3-5}
        & & $^1\Pi$ -- $b^{1}\Sigma^{+}$  & 690.4 -- 691.5 & $NR$ & \\
    \cline{3-5}
        & & $^1\Pi$ -- $c^{1}\Sigma^{+}$  & 664.5 -- 665.5 & $NR$ & \\ 
    \hline
       \multirow{2}{*}{\rownumber} & \multirow{2}{*}{TiF} & [37.8]$^4\Phi$ -- $X^{4}\Phi$  & 250.9 -- 255.4 & $NR$ & \cite{zhang2009laser} \\
    \cline{3-6}
     & & $^4\Delta$ -- $X^{4}\Phi$  & 249.5 -- 250.1 & $NR$ & \cite{zhang2009laser} \\
    \hline
    \end{tabular}
     }
     \caption{Laser-induced fluorescence transitions of neutral molecular species - 5. $NR$: not reported.}
    \label{Table: Neutral molecules LIF 5: S to Z}
\end{table*}

\setcounter{magicrownumbers}{0}
\begin{table}[h]
    \centering
    
    \begin{tabular}{|c|c|c|p{3 cm}|p{3 cm}|c|}
    \hline
    Sl. No. & Species & Transition & Excitation (nm)  & Fluorescence (nm) & Reference\\
    \hline
    \multirow{2}{*}{\rownumber} & \multirow{2}{*}{Al} & $^2$P$^{0}_{3/2}$ -- $^2$D$_{5/2}$ & 309.3 & 394.4, 396.2  & \cite{zhao2019experimental} \\
    \cline{3-6}
    & & 3$^2$P$_{1/2}$ -- 4$^2$S$_{1/2}$ & 394.4 & 396.2 & \cite{dreyfus1986laser} \\
    \hline
    \rownumber & As & 4s$^3$ $^4$S$_{3/2}$ -- 5s $^4$P$_{3/2}$ & 193.8 & 245  & \cite{selwyn1987atomic} \\
    \hline
    \multirow{2}{*}{\rownumber} & \multirow{2}{*}{B} & 2s$^2$2p $^2$P$^0_{3/2}$ -- 2s$^2$3s $^2$S$_{1/2}$ & 249.7, 249.8 & 209, 206.7 & \cite{li2016determinations} \\
    \cline{3-6}
     & & 2s$^2$2p $^2$P$^0_{3/2}$ -- 2s2p$^2$ $^2$D$_{3/2, 5/2}$ & 208.96 & 208.96 & \cite{aubert2017comparison} \\
    \hline
    \rownumber & Br & 4p$^5$ $^2$P$^0_{3/2}$ -- 5p $^4$S$^0_{3/2}$ & 2 $\times$ 252.59 & 635 & \cite{sirse2014ground} \\
    \hline
    \rownumber & C & 2p3p $^3$P --  2p$^2$ $^3$P & 2 $\times$ 280 & 910 & \cite{alden1989detection}\\
    \hline
    \rownumber & Cl & 3p$^5$ $^2$P$^0$ -- 4p $^4$ S$^0$ & 2 $\times$ 233.2 & 725 -- 775 & \cite{selwyn1987detection,booth2012absolute,heaven1982two}\\
    \hline
    \rownumber & Cr & 3d$^{5}$4p$^{7}$P$_{J}$ -- 3d$^{5}$4s$^{7}$S$_{3}$ & 425.4 & 427.5, 429 & \cite{sdorra1989basic} \\
    \hline
    \rownumber & Cu & 4s$^{2}$S$_{1/2}$ -- 4p$^{2}$P$^{0}_{3/2}$ & 324.8 & 510.6, 570 & \cite{gellert1988measurement} \\
    \hline
    \rownumber &  F & 2p$^5$ $^2$P$^0$ -- 3p $^2$D$^0$ & 170 & 776 & \cite{herring1988two} \\
    \hline
    \multirow{2}{*}{\rownumber} &  \multirow{2}{*}{Fe} & 3d$^{6}$4s$^{2}$a $^{5}$D$_{4}$ --  3d$^{7}$($^{4}$F)4p y $^{5}$F$^{0}_{5}$ & 296.7 & 373.5 & \cite{bolshov1981atomic} \\
    \cline{3-6}
     &  & 3d$^{6}$4s$^{2}$a $^{5}$D$_{4}$ -- 3d$^{7}$($^{4}$F)4p y $^{5}$D$^{0}_{4}$ & 302.06 & 382.04 & \cite{falk1988analytical} \\
    \hline
    \rownumber & Ga & 4$^2$D$_{3/2}$ -- 4$^2$P$_{1/2}$, 5$^2$S$_{1/2}$ -- 4$^2$P$_{3/2}$ & 287.4, 403.3 &  294.4, 417.2 & \cite{ma2002determination} \\
    \hline
    \rownumber & H & 1s$^2$S -- 3d $^2$D & 2 $\times$ 205 &  656.3 & \cite{niemi2001absolute} \\
    \hline
    \rownumber &  I & 5p$^5$ $^2$P$^0$ -- $^2$D$^0$ & 2 $\times$ 304.7  & 178.3 & \cite{brewer1983measurement}\\
    \hline
    \rownumber & Kr & 4p$^6$ $^1$S$_0$ -- 5p$^{\prime}[3/2]_2$ & 2 $\times$ 204.1  & 587, 826.3 & \cite{niemi2001absolute}\\
    \hline
    \rownumber & N & 2p$^3$ $^4$S$^0$ -- 3p $^4$D$^0$ & 2 $\times$ 211 & 870, 822, 744 & \cite{niemi2001absolute} \\
    \hline
    \multirow{2}{*}{\rownumber} & \multirow{2}{*}{Na} & $NR$ & 2$\times$ 685 & 818 & \cite{zhu2020two} \\
    \cline{3-6}
     &  & $^2$P$_{1/2}$, $^{2}$P$_{3/2}$ -- $^2$S$_{1/2}$ & 589 & $NR$ & \cite{daily1978laser,erdmann1972lifetime} \\
    \hline
    \rownumber & O & 2p$^4$ $^3$P -- 3p $^3$P & 2 $\times$ 225.7 & 844.9, 777 & \cite{niemi2001absolute} \\
    \hline
    \rownumber & Pb & 6s$^2$6p$^2$ $^3$P$_0$ -- 6s$^2$6p(2P$^0_{0.5}$)7s 3P$^0_1$ & 283.31 & 405.78 & \cite{laville2009laser} \\
      \hline
    \rownumber & S & 3p$^4$ $^3$P -- 4p $^3$P & 2 $\times$ 288 & VUV & \cite{brewer1982two}\\
    \hline
    \rownumber & Si & 3p$^2$ $^3$P$_0$ -- 4s $^3$P$^0_1$& 251.4 & 252.8 & \cite{roth1984spatial} \\
    \hline
    \multirow{2}{*}{\rownumber} & \multirow{2}{*}{Ti} &  3d$_2$ 4s4p -- 3d$_2$ 4s$_2$($^3$F$_3$) & 322.3 & 511.3 & \cite{britun2008laser} \\
    \cline{3-6}
     & & $^3$D$_{1,2,3}$, $^1$F$_3$, $^3$F$_3$, $^3$G$_3$ -- $^3$F$_{2,3,4}$ & 221 -- 296 & 250 -- 341 & \cite{ljung1997detection} \\
     \hline
    \rownumber & W & 5d$^{4}$6s$^{2}$ $^{5}$D$_{0}$ -- 5d$^{5}$ $^{4}$D6s$^{0}_{1}$   & 287.9 & 302.5 & \cite{hadrath2005determination,georgiev2019detection} \\
    \hline
    \rownumber & Xe & 5p$^6$ $^1$S$_0$ -- 7p$[3/2]_2$ & 2 $\times$ 225.5  & 462.6 & \cite{niemi2001absolute}\\
    \hline
    
    \end{tabular}
    
    \caption{Laser-induced fluorescence transitions of neutral atomic species. $NR$: not reported.}
    \label{Table: Neutral atoms LIF}
\end{table}

\setcounter{magicrownumbers}{0}
\begin{table}[h]
    \centering
    \begin{tabular}{|c|c|c|p{3 cm}|p{3 cm}|c|}
    \hline
    Sl. No. & Species & Transition & Excitation (nm)  & Fluorescence (nm) & Reference\\
    \hline
    \multirow{4}{*}{\rownumber} & \multirow{4}{*}{Ar$^{+*}$} & 3d$^4$ F$_{9/2}$ -- 4p$^4$ F$_{7/2}$ & 664.4 & 434.8 & \cite{lunt2008experimental} \\
    \cline{3-6}
    & & 3d$^4$ F$_{7/2}$ -- 4p$^4$ D$_{5/2}$ & 668.6 & 442.7 & \cite{tanida2016spatial,thakur2016development,biloiu2005evolution,boivin2003laser,severn2006ion,severn2006ion,bieber2011measurements,keesee2004laser,kuwahara2015development} \\
    \cline{3-6}
    & & 4p$^{2}$D$_{5/2}$ -- 4s$^{2}$P$_{3/2}$ & 488 & 422.8 & \cite{stern1975plasma} \\
    \cline{3-6}
    & & 4p$^{'}$ $^{2}$F$^{0}_{7/2}$ -- 3d$^{'}$ $^{2}$G$_{9/2}$ & 611.7 & 461.1 & \cite{severn1998argon} \\
    \hline
    \rownumber & Br$_2^+$ &  $^{2}\Pi_{u}$ -- X$^{2}\Pi_{g}$ & 472.7 & 480 -- 640 & \cite{harris1983ax}\\
    \hline
    \rownumber & Cl$^{+*}$ & $^5$P$_3$ -- $^5$D$^0_4$ & 542.3 & 479.5 & \cite{kumagai1999detection}\\
    \hline
    \rownumber & Cl$_{2}^{+}$ & X$^2\Pi_i$ -- A$^2\Pi_i$ & 387.6 & 396 & \cite{malyshev1999laser}\\
    \hline
    \multirow{2}{*}{\rownumber} & \multirow{2}{*}{CS$_2^+$} & X$^{2}\Pi_{3/2}$ -- A$^{2}\Pi_{3/2}$  & 427 -- 476  & 460 -- 550 & \cite{bondybey1979laser} \\
    \cline{3-6}
    & & A$^{2}\Pi_{u}$--X$^{2}\Pi_{g}$ & 410 -- 480 & 507.7 & \cite{zen1995laser} \\
    \hline
     \rownumber & C$_{2}$S$_{2}^{-}$ & $^{2}\Pi_{3/2}$ -- $^{2}\Pi_{3/2}$ & 542.5 -- 543 & 555 -- 695 & \cite{nakajima2003laser} \\
    \hline
    \rownumber & I$^{+*}$ & $^{5}$D$^{0}_{4}$ -- $^{5}$P$_{3}$ & 696.1 & 516.3 &  \cite{steinberger2018laser} \\ 
    \hline
    \multirow{2}{*}{\rownumber} & \multirow{2}{*}{Kr$^{+*}$} & 4d$^4$F$_{7/2}$ --  & 820.3 & 462.9 & \cite{lejeune2012kr} \\
    \cline{3-6}
    & & 5d$^4$F$_{7/2}$ --  & 729 & 473.9 & \cite{hargus2011demonstration} \\
    \hline
    \rownumber & La$^{+*}$  & 5d6s a$^3$D$_{1}$ -- 5d6p y$^3$D$^{0}_{2}$ & 403.2 & 379.1 & \cite{tremblay1987laser} \\
    \hline
    \multirow{2}{*}{\rownumber} & \multirow{2}{*}{N$_{2}^{+}$} & \multirow{2}{*}{B$^2$ $\Sigma^+_u$ – X$^2\Sigma^+_g$} & 391 & 428 & \cite{konthasinghe2015laser}\\
    \cline{4-6}
    & & & 330 & 458.8 &\cite{woodcock1997doppler}\\
    \hline
    \rownumber & N$_{2}$O$^{+}$ & $\tilde{A}^2\Sigma^{+}$ -- $\tilde{X}^{2}\Pi_{i}$ & 337 -- 356 & 312 -- 345 & \cite{gharaibeh2012laser}\\
    \hline
    \rownumber & O$_2^+$ & X$^2\Pi_g$ -- A$^2\Pi_u$ & 225 -- 227 &  245 & \cite{li2000laser}\\
    \hline
    \rownumber & SiO$^{+}$ & B$^2\Sigma^+$ -- X$^2\Sigma^{+}$ & 385 & 383 -- 385 & \cite{matsuo1997formation}\\
    \hline
     \rownumber & Ti$^+$ &  3d$_2$ 4p(z$^4$G$^0_{5/2}$) -- 3d$_2$ 4s(a$^4$F$_{3/2}$) & 338.4 & 486.6 & \cite{britun2008laser} \\
    \hline
    \multirow{3}{*}{\rownumber} & \multirow{3}{*}{Xe$^{+*}$} & 5d$^4$F$_{5/2}$ -- 6p$^{4}$D$_{5/2}$ & 834.7 & 541.9 &  \cite{svarnas2018laser}\\
    \cline{3-6}
    & & 6p$^{4}$P$_{5/2}^0$ -- 5d$^4$D$_{7/2}$ & 605.1 & 529.2 & \cite{smith2005diode,smith2004laser}\\
    \cline{3-6}
    & & 6p$^{4}$D$^{0}_{5/2}$ -- 5d$^{4}$F$_{7/2}$ & 680.6 & 492.2 & \cite{lee2007measurements}\\
    \hline
    
    \end{tabular}
    \caption{Laser-induced fluorescence transitions of metastable and ground states of ionic species.}
    \label{Table: Ions LIF}
\end{table}

\setcounter{magicrownumbers}{0}
\begin{table}[H]
    \centering
    \begin{tabular}{|c|c|c|c|c|c|}
    \hline
    Sl. No. & Species & Transition & Excitation (nm)  & Fluorescence (nm) & Reference\\
     \hline
    \multirow{6}{*}{\rownumber} & \multirow{6}{*}{Ar$^{*}$}  & 3p$^5$P$_{1/2}$ -- 4p$^5$D$_{3/2}$ & 696.7 & 826.7 & \cite{aramaki2009high} \\
    \cline{3-6}
    & & 3p$^5$S$_{1/2}$ -- 4p$^5$P$_{1/2}$ & 772.6 & 810.6, 826.7 & \cite{sadeghi1997transport} \\
    \cline{3-6}
    & & 1s$_5$ -- 2p$_9$ & 811.5, 810.4 & 811.5, 772.6 & \cite{engeln2001flow} \\
    \cline{3-6}
    & & 1s$_2$ -- 2p$_3$ & 841.1 & 706.9 & \cite{short2016novel} \\
    \cline{3-6}
    & & 1s$_2$ -- 2p$_3$ & 842.5 & 842.5, 801.7 & \cite{bergert2021quantitative} \\
    \cline{3-6}
    & & 3d$^4$F$_{(7/2)}$ -- 4p$^4$D$_{(3/2)}$ & 667.9 & 750.6 & \cite{keesee2004laser,kuwahara2015development} \\
    \hline
    \rownumber &  F$^{*}$ & $^4$D$^0_{5/2}$ -- $^4$P$_{3/2}$ & 690.3 & 677.4 & \cite{hansen1990formation}\\
    \hline
    \multirow{2}{*}{\rownumber} & \multirow{2}{*}{He$^{*}$} & 3$^{3}$P -- 2$^{3}$S  & 388.9 &  706.5, 587.6 & \cite{frost2001laser}\\
    \cline{3-6}
     & & 2$^{1}$S -- 3$^{1}$P & 501.6 & 667.8  & \cite{frost2001laser}\\
    \hline
    \rownumber & He$^{*}_{2}$ & a$^{3}\Sigma^{+}_{u}$ -- d$^{3}\Sigma^{+}_{u}$ & 2 $\times$ 940 & 640  & \cite{rellergert2008detection}\\
    \hline    
    \multirow{2}{*}{\rownumber} & \multirow{2}{*}{Kr$^{*}$} & 5s[3/2]$^0_2$ -- 5p[3/2]$_2$ & 760.2 & 819  & \cite{hargus2010preliminary,mustafa2017krypton} \\
    \cline{3-6}
    & & 5s[3/2]$_2$ -- 5p[5/2]$_2$ & 810.44 & 877.6 & \cite{hargus2010preliminary} \\
    
    \hline
    \rownumber & N$_{2}^{*}$ & A$^3\Sigma^{+}_{u}$ -- B$^3$$\Pi_{g}$ & 687.44, 618 & 762, 676 & \cite{nemschokmichal2011laser,ono2009laser}\\
    \hline
    \rownumber & P$^{*}$ & 3s$^{2}$3p$^{3}$ $^{2}$P$^{0}_{1/2, 3/2}$ -- 3s$^{2}$3p$^{2}$($^{3}$P)4s $^{2}$P$_{3/2}$ & 253.4, 253.6 & 213.6, 213.6 & \cite{shen2009detection,kondo2009determination} \\
    \hline
   \multirow{2}{*}{\rownumber} & \multirow{2}{*}{Xe$^{*}$} & 6s$^{2}$[1/2]$^0_1$ -- 6p$^{2}$[3/2]$^0_1$ & 834.7 & 473.4 &  \cite{svarnas2018laser,mazouffre2011comprehensive}\\
   \cline{3-6}
    &  & 6s[3/2]$^{0}_{2}$ -- 6p[3/2]$_{2}$ & 823.2 & 823.2 & \cite{cedolin1997laser,mazouffre2011comprehensive}\\

    \hline
    
    \end{tabular}
    \caption{Laser-induced fluorescence transitions of neutral metastable species.}
    \label{Table: Neutral metastables LIF}
\end{table}


\section{Summary}
\label{Summary}
Laser-induced fluorescence provides both spatial and temporally resolved characteristics of the neutral species, the plasma-produced radicals, ions, and metastables. It is a useful tool that enables direct measurement of densities, ion velocities, and temperatures. LIF can help in the validation of simulation models and help in the development of the predictive capabilities of plasma-material interactions. In this report, we have reviewed the basic principles of LIF, the widely used types of LIF setups and compiled a list of the available transitions to generate a reference data for laser-induced fluorescence transitions relevant in the microelectronics industry and the sustainability applications. We have identified the groups of species with overlapping LIF/TALIF excitation and fluorescence wavelengths. This compilation intends to assist in the identification of the possible species that can be detected by LIF/TALIF an interrogation volume. 
 
\section{Acknowledgements}
This work was performed under the U.S. Department of Energy through contract DE-AC02-09CH11466. The authors would like to thank Shahid Rauf, Alexandre Likhanskii, Prashanth Kothnur, Guus Reefman, and Mu-Chien Wu for their feedback and input in providing gases relevant to processing applications. 

\section{Data Availability}
Data sharing is not applicable to this article as no new data were created or analyzed in this study.

\section{Conflict of Interest}
The authors have no conflicts to disclose.
\section{References}
\bibliography{References}

\begin{thebibliography}{395}%
\makeatletter
\providecommand \@ifxundefined [1]{%
 \@ifx{#1\undefined}
}%
\providecommand \@ifnum [1]{%
 \ifnum #1\expandafter \@firstoftwo
 \else \expandafter \@secondoftwo
 \fi
}%
\providecommand \@ifx [1]{%
 \ifx #1\expandafter \@firstoftwo
 \else \expandafter \@secondoftwo
 \fi
}%
\providecommand \natexlab [1]{#1}%
\providecommand \enquote  [1]{``#1''}%
\providecommand \bibnamefont  [1]{#1}%
\providecommand \bibfnamefont [1]{#1}%
\providecommand \citenamefont [1]{#1}%
\providecommand \href@noop [0]{\@secondoftwo}%
\providecommand \href [0]{\begingroup \@sanitize@url \@href}%
\providecommand \@href[1]{\@@startlink{#1}\@@href}%
\providecommand \@@href[1]{\endgroup#1\@@endlink}%
\providecommand \@sanitize@url [0]{\catcode `\\12\catcode `\$12\catcode `\&12\catcode `\#12\catcode `\^12\catcode `\_12\catcode `\%12\relax}%
\providecommand \@@startlink[1]{}%
\providecommand \@@endlink[0]{}%
\providecommand \url  [0]{\begingroup\@sanitize@url \@url }%
\providecommand \@url [1]{\endgroup\@href {#1}{\urlprefix }}%
\providecommand \urlprefix  [0]{URL }%
\providecommand \Eprint [0]{\href }%
\providecommand \doibase [0]{http://dx.doi.org/}%
\providecommand \selectlanguage [0]{\@gobble}%
\providecommand \bibinfo  [0]{\@secondoftwo}%
\providecommand \bibfield  [0]{\@secondoftwo}%
\providecommand \translation [1]{[#1]}%
\providecommand \BibitemOpen [0]{}%
\providecommand \bibitemStop [0]{}%
\providecommand \bibitemNoStop [0]{.\EOS\space}%
\providecommand \EOS [0]{\spacefactor3000\relax}%
\providecommand \BibitemShut  [1]{\csname bibitem#1\endcsname}%
\let\auto@bib@innerbib\@empty
\bibitem [{\citenamefont {Waldrop}(2016)}]{waldrop2016chips}%
  \BibitemOpen
  \bibfield  {author} {\bibinfo {author} {\bibfnamefont {M.~M.}\ \bibnamefont {Waldrop}},\ }\bibfield  {title} {\enquote {\bibinfo {title} {The chips are down for moore’s law},}\ }\href@noop {} {\bibfield  {journal} {\bibinfo  {journal} {Nature News}\ }\textbf {\bibinfo {volume} {530}},\ \bibinfo {pages} {144} (\bibinfo {year} {2016})}\BibitemShut {NoStop}%
\bibitem [{\citenamefont {Lundstrom}\ and\ \citenamefont {Alam}(2022)}]{lundstrom2022moore}%
  \BibitemOpen
  \bibfield  {author} {\bibinfo {author} {\bibfnamefont {M.~S.}\ \bibnamefont {Lundstrom}}\ and\ \bibinfo {author} {\bibfnamefont {M.~A.}\ \bibnamefont {Alam}},\ }\bibfield  {title} {\enquote {\bibinfo {title} {Moore’s law: The journey ahead},}\ }\href@noop {} {\bibfield  {journal} {\bibinfo  {journal} {Science}\ }\textbf {\bibinfo {volume} {378}},\ \bibinfo {pages} {722--723} (\bibinfo {year} {2022})}\BibitemShut {NoStop}%
\bibitem [{\citenamefont {Shalf}(2020)}]{shalf2020future}%
  \BibitemOpen
  \bibfield  {author} {\bibinfo {author} {\bibfnamefont {J.}~\bibnamefont {Shalf}},\ }\bibfield  {title} {\enquote {\bibinfo {title} {{The future of computing beyond Moore’s Law}},}\ }\href@noop {} {\bibfield  {journal} {\bibinfo  {journal} {Philosophical Transactions of the Royal Society A}\ }\textbf {\bibinfo {volume} {378}},\ \bibinfo {pages} {20190061} (\bibinfo {year} {2020})}\BibitemShut {NoStop}%
\bibitem [{\citenamefont {Theis}\ and\ \citenamefont {Wong}(2017)}]{theis2017end}%
  \BibitemOpen
  \bibfield  {author} {\bibinfo {author} {\bibfnamefont {T.~N.}\ \bibnamefont {Theis}}\ and\ \bibinfo {author} {\bibfnamefont {H.-S.~P.}\ \bibnamefont {Wong}},\ }\bibfield  {title} {\enquote {\bibinfo {title} {{The end of moore's law: A new beginning for information technology}},}\ }\href@noop {} {\bibfield  {journal} {\bibinfo  {journal} {Computing in science \& engineering}\ }\textbf {\bibinfo {volume} {19}},\ \bibinfo {pages} {41--50} (\bibinfo {year} {2017})}\BibitemShut {NoStop}%
\bibitem [{\citenamefont {Graves}(1994)}]{graves1994plasma}%
  \BibitemOpen
  \bibfield  {author} {\bibinfo {author} {\bibfnamefont {D.~B.}\ \bibnamefont {Graves}},\ }\bibfield  {title} {\enquote {\bibinfo {title} {Plasma processing},}\ }\href@noop {} {\bibfield  {journal} {\bibinfo  {journal} {IEEE transactions on Plasma Science}\ }\textbf {\bibinfo {volume} {22}},\ \bibinfo {pages} {31--42} (\bibinfo {year} {1994})}\BibitemShut {NoStop}%
\bibitem [{\citenamefont {Adamovich}\ \emph {et~al.}(2022)\citenamefont {Adamovich}, \citenamefont {Agarwal}, \citenamefont {Ahedo}, \citenamefont {Alves}, \citenamefont {Baalrud}, \citenamefont {Babaeva}, \citenamefont {Bogaerts}, \citenamefont {Bourdon}, \citenamefont {Bruggeman}, \citenamefont {Canal} \emph {et~al.}}]{adamovich20222022}%
  \BibitemOpen
  \bibfield  {author} {\bibinfo {author} {\bibfnamefont {I.}~\bibnamefont {Adamovich}}, \bibinfo {author} {\bibfnamefont {S.}~\bibnamefont {Agarwal}}, \bibinfo {author} {\bibfnamefont {E.}~\bibnamefont {Ahedo}}, \bibinfo {author} {\bibfnamefont {L.~L.}\ \bibnamefont {Alves}}, \bibinfo {author} {\bibfnamefont {S.}~\bibnamefont {Baalrud}}, \bibinfo {author} {\bibfnamefont {N.}~\bibnamefont {Babaeva}}, \bibinfo {author} {\bibfnamefont {A.}~\bibnamefont {Bogaerts}}, \bibinfo {author} {\bibfnamefont {A.}~\bibnamefont {Bourdon}}, \bibinfo {author} {\bibfnamefont {P.}~\bibnamefont {Bruggeman}}, \bibinfo {author} {\bibfnamefont {C.}~\bibnamefont {Canal}},  \emph {et~al.},\ }\bibfield  {title} {\enquote {\bibinfo {title} {The 2022 plasma roadmap: low temperature plasma science and technology},}\ }\href@noop {} {\bibfield  {journal} {\bibinfo  {journal} {Journal of Physics D: Applied Physics}\ }\textbf {\bibinfo {volume} {55}},\ \bibinfo {pages} {373001} (\bibinfo {year} {2022})}\BibitemShut {NoStop}%
\bibitem [{\citenamefont {Parsons}\ and\ \citenamefont {Clark}(2020)}]{parsons2020area}%
  \BibitemOpen
  \bibfield  {author} {\bibinfo {author} {\bibfnamefont {G.~N.}\ \bibnamefont {Parsons}}\ and\ \bibinfo {author} {\bibfnamefont {R.~D.}\ \bibnamefont {Clark}},\ }\bibfield  {title} {\enquote {\bibinfo {title} {Area-selective deposition: Fundamentals, applications, and future outlook},}\ }\href@noop {} {\bibfield  {journal} {\bibinfo  {journal} {Chemistry of Materials}\ }\textbf {\bibinfo {volume} {32}},\ \bibinfo {pages} {4920--4953} (\bibinfo {year} {2020})}\BibitemShut {NoStop}%
\bibitem [{\citenamefont {Knoops}\ \emph {et~al.}(2019)\citenamefont {Knoops}, \citenamefont {Faraz}, \citenamefont {Arts},\ and\ \citenamefont {Kessels}}]{knoops2019status}%
  \BibitemOpen
  \bibfield  {author} {\bibinfo {author} {\bibfnamefont {H.~C.}\ \bibnamefont {Knoops}}, \bibinfo {author} {\bibfnamefont {T.}~\bibnamefont {Faraz}}, \bibinfo {author} {\bibfnamefont {K.}~\bibnamefont {Arts}}, \ and\ \bibinfo {author} {\bibfnamefont {W.~M.}\ \bibnamefont {Kessels}},\ }\bibfield  {title} {\enquote {\bibinfo {title} {Status and prospects of plasma-assisted atomic layer deposition},}\ }\href@noop {} {\bibfield  {journal} {\bibinfo  {journal} {Journal of Vacuum Science \& Technology A: Vacuum, Surfaces, and Films}\ }\textbf {\bibinfo {volume} {37}},\ \bibinfo {pages} {030902} (\bibinfo {year} {2019})}\BibitemShut {NoStop}%
\bibitem [{\citenamefont {Oehrlein}, \citenamefont {Metzler},\ and\ \citenamefont {Li}(2015)}]{oehrlein2015atomic}%
  \BibitemOpen
  \bibfield  {author} {\bibinfo {author} {\bibfnamefont {G.}~\bibnamefont {Oehrlein}}, \bibinfo {author} {\bibfnamefont {D.}~\bibnamefont {Metzler}}, \ and\ \bibinfo {author} {\bibfnamefont {C.}~\bibnamefont {Li}},\ }\bibfield  {title} {\enquote {\bibinfo {title} {Atomic layer etching at the tipping point: an overview},}\ }\href@noop {} {\bibfield  {journal} {\bibinfo  {journal} {ECS Journal of Solid State Science and Technology}\ }\textbf {\bibinfo {volume} {4}},\ \bibinfo {pages} {N5041} (\bibinfo {year} {2015})}\BibitemShut {NoStop}%
\bibitem [{\citenamefont {Dorf}\ \emph {et~al.}(2016)\citenamefont {Dorf}, \citenamefont {Wang}, \citenamefont {Rauf}, \citenamefont {Zhang}, \citenamefont {Agarwal}, \citenamefont {Kenney}, \citenamefont {Ramaswamy},\ and\ \citenamefont {Collins}}]{dorf2016atomic}%
  \BibitemOpen
  \bibfield  {author} {\bibinfo {author} {\bibfnamefont {L.}~\bibnamefont {Dorf}}, \bibinfo {author} {\bibfnamefont {J.-C.}\ \bibnamefont {Wang}}, \bibinfo {author} {\bibfnamefont {S.}~\bibnamefont {Rauf}}, \bibinfo {author} {\bibfnamefont {Y.}~\bibnamefont {Zhang}}, \bibinfo {author} {\bibfnamefont {A.}~\bibnamefont {Agarwal}}, \bibinfo {author} {\bibfnamefont {J.}~\bibnamefont {Kenney}}, \bibinfo {author} {\bibfnamefont {K.}~\bibnamefont {Ramaswamy}}, \ and\ \bibinfo {author} {\bibfnamefont {K.}~\bibnamefont {Collins}},\ }\bibfield  {title} {\enquote {\bibinfo {title} {Atomic precision etch using a low-electron temperature plasma},}\ }in\ \href@noop {} {\emph {\bibinfo {booktitle} {Advanced Etch Technology for Nanopatterning V}}},\ Vol.\ \bibinfo {volume} {9782}\ (\bibinfo {organization} {SPIE},\ \bibinfo {year} {2016})\ pp.\ \bibinfo {pages} {30--37}\BibitemShut {NoStop}%
\bibitem [{\citenamefont {Jagtiani}\ \emph {et~al.}(2016)\citenamefont {Jagtiani}, \citenamefont {Miyazoe}, \citenamefont {Chang}, \citenamefont {Farmer}, \citenamefont {Engel}, \citenamefont {Neumayer}, \citenamefont {Han}, \citenamefont {Engelmann}, \citenamefont {Boris}, \citenamefont {Hern{\'a}ndez} \emph {et~al.}}]{jagtiani2016initial}%
  \BibitemOpen
  \bibfield  {author} {\bibinfo {author} {\bibfnamefont {A.~V.}\ \bibnamefont {Jagtiani}}, \bibinfo {author} {\bibfnamefont {H.}~\bibnamefont {Miyazoe}}, \bibinfo {author} {\bibfnamefont {J.}~\bibnamefont {Chang}}, \bibinfo {author} {\bibfnamefont {D.~B.}\ \bibnamefont {Farmer}}, \bibinfo {author} {\bibfnamefont {M.}~\bibnamefont {Engel}}, \bibinfo {author} {\bibfnamefont {D.}~\bibnamefont {Neumayer}}, \bibinfo {author} {\bibfnamefont {S.-J.}\ \bibnamefont {Han}}, \bibinfo {author} {\bibfnamefont {S.~U.}\ \bibnamefont {Engelmann}}, \bibinfo {author} {\bibfnamefont {D.~R.}\ \bibnamefont {Boris}}, \bibinfo {author} {\bibfnamefont {S.~C.}\ \bibnamefont {Hern{\'a}ndez}},  \emph {et~al.},\ }\bibfield  {title} {\enquote {\bibinfo {title} {Initial evaluation and comparison of plasma damage to atomic layer carbon materials using conventional and low t$_e$ plasma sources},}\ }\href@noop {} {\bibfield  {journal} {\bibinfo  {journal} {Journal of Vacuum Science \& Technology A}\ }\textbf {\bibinfo {volume} {34}}
  (\bibinfo {year} {2016})}\BibitemShut {NoStop}%
\bibitem [{\citenamefont {Rauf}\ \emph {et~al.}(2017)\citenamefont {Rauf}, \citenamefont {Balakrishna}, \citenamefont {Agarwal}, \citenamefont {Dorf}, \citenamefont {Collins}, \citenamefont {Boris},\ and\ \citenamefont {Walton}}]{rauf2017three}%
  \BibitemOpen
  \bibfield  {author} {\bibinfo {author} {\bibfnamefont {S.}~\bibnamefont {Rauf}}, \bibinfo {author} {\bibfnamefont {A.}~\bibnamefont {Balakrishna}}, \bibinfo {author} {\bibfnamefont {A.}~\bibnamefont {Agarwal}}, \bibinfo {author} {\bibfnamefont {L.}~\bibnamefont {Dorf}}, \bibinfo {author} {\bibfnamefont {K.}~\bibnamefont {Collins}}, \bibinfo {author} {\bibfnamefont {D.~R.}\ \bibnamefont {Boris}}, \ and\ \bibinfo {author} {\bibfnamefont {S.~G.}\ \bibnamefont {Walton}},\ }\bibfield  {title} {\enquote {\bibinfo {title} {Three-dimensional model of electron beam generated plasma},}\ }\href@noop {} {\bibfield  {journal} {\bibinfo  {journal} {Plasma Sources Science and Technology}\ }\textbf {\bibinfo {volume} {26}},\ \bibinfo {pages} {065006} (\bibinfo {year} {2017})}\BibitemShut {NoStop}%
\bibitem [{\citenamefont {Dorf}\ \emph {et~al.}(2022{\natexlab{a}})\citenamefont {Dorf}, \citenamefont {Dhindsa}, \citenamefont {Rogers}, \citenamefont {Byun}, \citenamefont {Kamenetskiy}, \citenamefont {Guo}, \citenamefont {Ramaswamy}, \citenamefont {Todorow}, \citenamefont {Luere}, \citenamefont {Linying} \emph {et~al.}}]{dorf2022plasma1}%
  \BibitemOpen
  \bibfield  {author} {\bibinfo {author} {\bibfnamefont {L.}~\bibnamefont {Dorf}}, \bibinfo {author} {\bibfnamefont {R.}~\bibnamefont {Dhindsa}}, \bibinfo {author} {\bibfnamefont {J.}~\bibnamefont {Rogers}}, \bibinfo {author} {\bibfnamefont {D.~S.}\ \bibnamefont {Byun}}, \bibinfo {author} {\bibfnamefont {E.}~\bibnamefont {Kamenetskiy}}, \bibinfo {author} {\bibfnamefont {Y.}~\bibnamefont {Guo}}, \bibinfo {author} {\bibfnamefont {K.}~\bibnamefont {Ramaswamy}}, \bibinfo {author} {\bibfnamefont {V.~N.}\ \bibnamefont {Todorow}}, \bibinfo {author} {\bibfnamefont {O.}~\bibnamefont {Luere}}, \bibinfo {author} {\bibfnamefont {C.}~\bibnamefont {Linying}},  \emph {et~al.},\ }\href@noop {} {\enquote {\bibinfo {title} {Plasma processing assembly using pulsed-voltage and radio-frequency power},}\ } (\bibinfo {year} {2022}{\natexlab{a}}),\ \bibinfo {note} {uS Patent 11,462,388}\BibitemShut {NoStop}%
\bibitem [{\citenamefont {Dorf}\ \emph {et~al.}(2022{\natexlab{b}})\citenamefont {Dorf}, \citenamefont {Dhindsa}, \citenamefont {Rogers}, \citenamefont {Byun}, \citenamefont {Kamenetskiy}, \citenamefont {Guo}, \citenamefont {Ramaswamy}, \citenamefont {Todorow},\ and\ \citenamefont {Luere}}]{dorf2022plasma2}%
  \BibitemOpen
  \bibfield  {author} {\bibinfo {author} {\bibfnamefont {L.}~\bibnamefont {Dorf}}, \bibinfo {author} {\bibfnamefont {R.}~\bibnamefont {Dhindsa}}, \bibinfo {author} {\bibfnamefont {J.}~\bibnamefont {Rogers}}, \bibinfo {author} {\bibfnamefont {D.~S.}\ \bibnamefont {Byun}}, \bibinfo {author} {\bibfnamefont {E.}~\bibnamefont {Kamenetskiy}}, \bibinfo {author} {\bibfnamefont {Y.}~\bibnamefont {Guo}}, \bibinfo {author} {\bibfnamefont {K.}~\bibnamefont {Ramaswamy}}, \bibinfo {author} {\bibfnamefont {V.~N.}\ \bibnamefont {Todorow}}, \ and\ \bibinfo {author} {\bibfnamefont {O.}~\bibnamefont {Luere}},\ }\href@noop {} {\enquote {\bibinfo {title} {Plasma processing using pulsed-voltage and radio-frequency power},}\ } (\bibinfo {year} {2022}{\natexlab{b}}),\ \bibinfo {note} {uS Patent App. 17/315,234}\BibitemShut {NoStop}%
\bibitem [{\citenamefont {Zhang}(2014)}]{zhang2014silicon}%
  \BibitemOpen
  \bibfield  {author} {\bibinfo {author} {\bibfnamefont {L.}~\bibnamefont {Zhang}},\ }\bibfield  {title} {\enquote {\bibinfo {title} {Silicon process and manufacturing technology evolution: An overview of advancements in chip making},}\ }\href@noop {} {\bibfield  {journal} {\bibinfo  {journal} {IEEE Consumer Electronics Magazine}\ }\textbf {\bibinfo {volume} {3}},\ \bibinfo {pages} {44--48} (\bibinfo {year} {2014})}\BibitemShut {NoStop}%
\bibitem [{\citenamefont {Bera}\ \emph {et~al.}(2021)\citenamefont {Bera}, \citenamefont {Rauf}, \citenamefont {Forster},\ and\ \citenamefont {Collins}}]{bera2021self}%
  \BibitemOpen
  \bibfield  {author} {\bibinfo {author} {\bibfnamefont {K.}~\bibnamefont {Bera}}, \bibinfo {author} {\bibfnamefont {S.}~\bibnamefont {Rauf}}, \bibinfo {author} {\bibfnamefont {J.}~\bibnamefont {Forster}}, \ and\ \bibinfo {author} {\bibfnamefont {K.}~\bibnamefont {Collins}},\ }\bibfield  {title} {\enquote {\bibinfo {title} {Self-organized pattern formation in radio frequency capacitively coupled discharges},}\ }\href@noop {} {\bibfield  {journal} {\bibinfo  {journal} {Journal of Applied Physics}\ }\textbf {\bibinfo {volume} {129}} (\bibinfo {year} {2021})}\BibitemShut {NoStop}%
\bibitem [{\citenamefont {Cunge}\ \emph {et~al.}(2005)\citenamefont {Cunge}, \citenamefont {Kogelschatz}, \citenamefont {Joubert},\ and\ \citenamefont {Sadeghi}}]{cunge2005plasma}%
  \BibitemOpen
  \bibfield  {author} {\bibinfo {author} {\bibfnamefont {G.}~\bibnamefont {Cunge}}, \bibinfo {author} {\bibfnamefont {M.}~\bibnamefont {Kogelschatz}}, \bibinfo {author} {\bibfnamefont {O.}~\bibnamefont {Joubert}}, \ and\ \bibinfo {author} {\bibfnamefont {N.}~\bibnamefont {Sadeghi}},\ }\bibfield  {title} {\enquote {\bibinfo {title} {Plasma--wall interactions during silicon etching processes in high-density {HBr}/{Cl}$_2$/{O}$_2$ plasmas},}\ }\href@noop {} {\bibfield  {journal} {\bibinfo  {journal} {Plasma Sources Science and Technology}\ }\textbf {\bibinfo {volume} {14}},\ \bibinfo {pages} {S42} (\bibinfo {year} {2005})}\BibitemShut {NoStop}%
\bibitem [{\citenamefont {Badaroglu}(2017)}]{badaroglu2017international}%
  \BibitemOpen
  \bibfield  {author} {\bibinfo {author} {\bibfnamefont {M.}~\bibnamefont {Badaroglu}},\ }\bibfield  {title} {\enquote {\bibinfo {title} {International roadmap for devices and systems},}\ }\href@noop {} {\bibfield  {journal} {\bibinfo  {journal} {More Moore: https://irds. ieee. org/roadmap-2017}\ } (\bibinfo {year} {2017})}\BibitemShut {NoStop}%
\bibitem [{\citenamefont {Atature}\ \emph {et~al.}(2018)\citenamefont {Atature}, \citenamefont {Englund}, \citenamefont {Vamivakas}, \citenamefont {Lee},\ and\ \citenamefont {Wrachtrup}}]{atature2018material}%
  \BibitemOpen
  \bibfield  {author} {\bibinfo {author} {\bibfnamefont {M.}~\bibnamefont {Atature}}, \bibinfo {author} {\bibfnamefont {D.}~\bibnamefont {Englund}}, \bibinfo {author} {\bibfnamefont {N.}~\bibnamefont {Vamivakas}}, \bibinfo {author} {\bibfnamefont {S.-Y.}\ \bibnamefont {Lee}}, \ and\ \bibinfo {author} {\bibfnamefont {J.}~\bibnamefont {Wrachtrup}},\ }\bibfield  {title} {\enquote {\bibinfo {title} {Material platforms for spin-based photonic quantum technologies},}\ }\href@noop {} {\bibfield  {journal} {\bibinfo  {journal} {Nature Reviews Materials}\ }\textbf {\bibinfo {volume} {3}},\ \bibinfo {pages} {38--51} (\bibinfo {year} {2018})}\BibitemShut {NoStop}%
\bibitem [{\citenamefont {Zwanenburg}\ \emph {et~al.}(2013)\citenamefont {Zwanenburg}, \citenamefont {Dzurak}, \citenamefont {Morello}, \citenamefont {Simmons}, \citenamefont {Hollenberg}, \citenamefont {Klimeck}, \citenamefont {Rogge}, \citenamefont {Coppersmith},\ and\ \citenamefont {Eriksson}}]{zwanenburg2013silicon}%
  \BibitemOpen
  \bibfield  {author} {\bibinfo {author} {\bibfnamefont {F.~A.}\ \bibnamefont {Zwanenburg}}, \bibinfo {author} {\bibfnamefont {A.~S.}\ \bibnamefont {Dzurak}}, \bibinfo {author} {\bibfnamefont {A.}~\bibnamefont {Morello}}, \bibinfo {author} {\bibfnamefont {M.~Y.}\ \bibnamefont {Simmons}}, \bibinfo {author} {\bibfnamefont {L.~C.}\ \bibnamefont {Hollenberg}}, \bibinfo {author} {\bibfnamefont {G.}~\bibnamefont {Klimeck}}, \bibinfo {author} {\bibfnamefont {S.}~\bibnamefont {Rogge}}, \bibinfo {author} {\bibfnamefont {S.~N.}\ \bibnamefont {Coppersmith}}, \ and\ \bibinfo {author} {\bibfnamefont {M.~A.}\ \bibnamefont {Eriksson}},\ }\bibfield  {title} {\enquote {\bibinfo {title} {Silicon quantum electronics},}\ }\href@noop {} {\bibfield  {journal} {\bibinfo  {journal} {Reviews of modern physics}\ }\textbf {\bibinfo {volume} {85}},\ \bibinfo {pages} {961} (\bibinfo {year} {2013})}\BibitemShut {NoStop}%
\bibitem [{\citenamefont {Ladd}\ \emph {et~al.}(2010)\citenamefont {Ladd}, \citenamefont {Jelezko}, \citenamefont {Laflamme}, \citenamefont {Nakamura}, \citenamefont {Monroe},\ and\ \citenamefont {O’Brien}}]{ladd2010quantum}%
  \BibitemOpen
  \bibfield  {author} {\bibinfo {author} {\bibfnamefont {T.~D.}\ \bibnamefont {Ladd}}, \bibinfo {author} {\bibfnamefont {F.}~\bibnamefont {Jelezko}}, \bibinfo {author} {\bibfnamefont {R.}~\bibnamefont {Laflamme}}, \bibinfo {author} {\bibfnamefont {Y.}~\bibnamefont {Nakamura}}, \bibinfo {author} {\bibfnamefont {C.}~\bibnamefont {Monroe}}, \ and\ \bibinfo {author} {\bibfnamefont {J.~L.}\ \bibnamefont {O’Brien}},\ }\bibfield  {title} {\enquote {\bibinfo {title} {Quantum computers},}\ }\href@noop {} {\bibfield  {journal} {\bibinfo  {journal} {Nature}\ }\textbf {\bibinfo {volume} {464}},\ \bibinfo {pages} {45--53} (\bibinfo {year} {2010})}\BibitemShut {NoStop}%
\bibitem [{\citenamefont {Samukawa}\ \emph {et~al.}(2012)\citenamefont {Samukawa}, \citenamefont {Hori}, \citenamefont {Rauf}, \citenamefont {Tachibana}, \citenamefont {Bruggeman}, \citenamefont {Kroesen}, \citenamefont {Whitehead}, \citenamefont {Murphy}, \citenamefont {Gutsol}, \citenamefont {Starikovskaia} \emph {et~al.}}]{samukawa20122012}%
  \BibitemOpen
  \bibfield  {author} {\bibinfo {author} {\bibfnamefont {S.}~\bibnamefont {Samukawa}}, \bibinfo {author} {\bibfnamefont {M.}~\bibnamefont {Hori}}, \bibinfo {author} {\bibfnamefont {S.}~\bibnamefont {Rauf}}, \bibinfo {author} {\bibfnamefont {K.}~\bibnamefont {Tachibana}}, \bibinfo {author} {\bibfnamefont {P.}~\bibnamefont {Bruggeman}}, \bibinfo {author} {\bibfnamefont {G.}~\bibnamefont {Kroesen}}, \bibinfo {author} {\bibfnamefont {J.~C.}\ \bibnamefont {Whitehead}}, \bibinfo {author} {\bibfnamefont {A.~B.}\ \bibnamefont {Murphy}}, \bibinfo {author} {\bibfnamefont {A.~F.}\ \bibnamefont {Gutsol}}, \bibinfo {author} {\bibfnamefont {S.}~\bibnamefont {Starikovskaia}},  \emph {et~al.},\ }\bibfield  {title} {\enquote {\bibinfo {title} {The 2012 plasma roadmap},}\ }\href@noop {} {\bibfield  {journal} {\bibinfo  {journal} {Journal of Physics D: Applied Physics}\ }\textbf {\bibinfo {volume} {45}},\ \bibinfo {pages} {253001} (\bibinfo {year} {2012})}\BibitemShut {NoStop}%
\bibitem [{\citenamefont {Adamovich}\ \emph {et~al.}(2017)\citenamefont {Adamovich}, \citenamefont {Baalrud}, \citenamefont {Bogaerts}, \citenamefont {Bruggeman}, \citenamefont {Cappelli}, \citenamefont {Colombo}, \citenamefont {Czarnetzki}, \citenamefont {Ebert}, \citenamefont {Eden}, \citenamefont {Favia} \emph {et~al.}}]{adamovich20172017}%
  \BibitemOpen
  \bibfield  {author} {\bibinfo {author} {\bibfnamefont {I.}~\bibnamefont {Adamovich}}, \bibinfo {author} {\bibfnamefont {S.}~\bibnamefont {Baalrud}}, \bibinfo {author} {\bibfnamefont {A.}~\bibnamefont {Bogaerts}}, \bibinfo {author} {\bibfnamefont {P.}~\bibnamefont {Bruggeman}}, \bibinfo {author} {\bibfnamefont {M.}~\bibnamefont {Cappelli}}, \bibinfo {author} {\bibfnamefont {V.}~\bibnamefont {Colombo}}, \bibinfo {author} {\bibfnamefont {U.}~\bibnamefont {Czarnetzki}}, \bibinfo {author} {\bibfnamefont {U.}~\bibnamefont {Ebert}}, \bibinfo {author} {\bibfnamefont {J.~G.}\ \bibnamefont {Eden}}, \bibinfo {author} {\bibfnamefont {P.}~\bibnamefont {Favia}},  \emph {et~al.},\ }\bibfield  {title} {\enquote {\bibinfo {title} {{The 2017 Plasma Roadmap: Low temperature plasma science and technology}},}\ }\href@noop {} {\bibfield  {journal} {\bibinfo  {journal} {Journal of Physics D: Applied Physics}\ }\textbf {\bibinfo {volume} {50}},\ \bibinfo {pages} {323001} (\bibinfo {year} {2017})}\BibitemShut {NoStop}%
\bibitem [{\citenamefont {Kong}\ \emph {et~al.}(2009)\citenamefont {Kong}, \citenamefont {Kroesen}, \citenamefont {Morfill}, \citenamefont {Nosenko}, \citenamefont {Shimizu}, \citenamefont {Van~Dijk},\ and\ \citenamefont {Zimmermann}}]{kong2009plasma}%
  \BibitemOpen
  \bibfield  {author} {\bibinfo {author} {\bibfnamefont {M.~G.}\ \bibnamefont {Kong}}, \bibinfo {author} {\bibfnamefont {G.}~\bibnamefont {Kroesen}}, \bibinfo {author} {\bibfnamefont {G.}~\bibnamefont {Morfill}}, \bibinfo {author} {\bibfnamefont {T.}~\bibnamefont {Nosenko}}, \bibinfo {author} {\bibfnamefont {T.}~\bibnamefont {Shimizu}}, \bibinfo {author} {\bibfnamefont {J.}~\bibnamefont {Van~Dijk}}, \ and\ \bibinfo {author} {\bibfnamefont {J.}~\bibnamefont {Zimmermann}},\ }\bibfield  {title} {\enquote {\bibinfo {title} {Plasma medicine: an introductory review},}\ }\href@noop {} {\bibfield  {journal} {\bibinfo  {journal} {new Journal of Physics}\ }\textbf {\bibinfo {volume} {11}},\ \bibinfo {pages} {115012} (\bibinfo {year} {2009})}\BibitemShut {NoStop}%
\bibitem [{\citenamefont {Ranieri}\ \emph {et~al.}(2021)\citenamefont {Ranieri}, \citenamefont {Sponsel}, \citenamefont {Kizer}, \citenamefont {Rojas-Pierce}, \citenamefont {Hern{\'a}ndez}, \citenamefont {Gatiboni}, \citenamefont {Grunden},\ and\ \citenamefont {Stapelmann}}]{ranieri2021plasma}%
  \BibitemOpen
  \bibfield  {author} {\bibinfo {author} {\bibfnamefont {P.}~\bibnamefont {Ranieri}}, \bibinfo {author} {\bibfnamefont {N.}~\bibnamefont {Sponsel}}, \bibinfo {author} {\bibfnamefont {J.}~\bibnamefont {Kizer}}, \bibinfo {author} {\bibfnamefont {M.}~\bibnamefont {Rojas-Pierce}}, \bibinfo {author} {\bibfnamefont {R.}~\bibnamefont {Hern{\'a}ndez}}, \bibinfo {author} {\bibfnamefont {L.}~\bibnamefont {Gatiboni}}, \bibinfo {author} {\bibfnamefont {A.}~\bibnamefont {Grunden}}, \ and\ \bibinfo {author} {\bibfnamefont {K.}~\bibnamefont {Stapelmann}},\ }\bibfield  {title} {\enquote {\bibinfo {title} {Plasma agriculture: Review from the perspective of the plant and its ecosystem},}\ }\href@noop {} {\bibfield  {journal} {\bibinfo  {journal} {Plasma Processes and Polymers}\ }\textbf {\bibinfo {volume} {18}},\ \bibinfo {pages} {2000162} (\bibinfo {year} {2021})}\BibitemShut {NoStop}%
\bibitem [{\citenamefont {Chen}, \citenamefont {Li},\ and\ \citenamefont {Li}(2015)}]{chen2015review}%
  \BibitemOpen
  \bibfield  {author} {\bibinfo {author} {\bibfnamefont {Q.}~\bibnamefont {Chen}}, \bibinfo {author} {\bibfnamefont {J.}~\bibnamefont {Li}}, \ and\ \bibinfo {author} {\bibfnamefont {Y.}~\bibnamefont {Li}},\ }\bibfield  {title} {\enquote {\bibinfo {title} {A review of plasma--liquid interactions for nanomaterial synthesis},}\ }\href@noop {} {\bibfield  {journal} {\bibinfo  {journal} {Journal of Physics D: Applied Physics}\ }\textbf {\bibinfo {volume} {48}},\ \bibinfo {pages} {424005} (\bibinfo {year} {2015})}\BibitemShut {NoStop}%
\bibitem [{\citenamefont {Bogaerts}\ \emph {et~al.}(2020)\citenamefont {Bogaerts}, \citenamefont {Tu}, \citenamefont {Whitehead}, \citenamefont {Centi}, \citenamefont {Lefferts}, \citenamefont {Guaitella}, \citenamefont {Azzolina-Jury}, \citenamefont {Kim}, \citenamefont {Murphy}, \citenamefont {Schneider} \emph {et~al.}}]{bogaerts20202020}%
  \BibitemOpen
  \bibfield  {author} {\bibinfo {author} {\bibfnamefont {A.}~\bibnamefont {Bogaerts}}, \bibinfo {author} {\bibfnamefont {X.}~\bibnamefont {Tu}}, \bibinfo {author} {\bibfnamefont {J.~C.}\ \bibnamefont {Whitehead}}, \bibinfo {author} {\bibfnamefont {G.}~\bibnamefont {Centi}}, \bibinfo {author} {\bibfnamefont {L.}~\bibnamefont {Lefferts}}, \bibinfo {author} {\bibfnamefont {O.}~\bibnamefont {Guaitella}}, \bibinfo {author} {\bibfnamefont {F.}~\bibnamefont {Azzolina-Jury}}, \bibinfo {author} {\bibfnamefont {H.-H.}\ \bibnamefont {Kim}}, \bibinfo {author} {\bibfnamefont {A.~B.}\ \bibnamefont {Murphy}}, \bibinfo {author} {\bibfnamefont {W.~F.}\ \bibnamefont {Schneider}},  \emph {et~al.},\ }\bibfield  {title} {\enquote {\bibinfo {title} {The 2020 plasma catalysis roadmap},}\ }\href@noop {} {\bibfield  {journal} {\bibinfo  {journal} {Journal of Physics D: Applied Physics}\ }\textbf {\bibinfo {volume} {53}},\ \bibinfo {pages} {443001} (\bibinfo {year} {2020})}\BibitemShut {NoStop}%
\bibitem [{\citenamefont {Whitehead}(2016)}]{whitehead2016plasma}%
  \BibitemOpen
  \bibfield  {author} {\bibinfo {author} {\bibfnamefont {J.~C.}\ \bibnamefont {Whitehead}},\ }\bibfield  {title} {\enquote {\bibinfo {title} {Plasma--catalysis: the known knowns, the known unknowns and the unknown unknowns},}\ }\href@noop {} {\bibfield  {journal} {\bibinfo  {journal} {Journal of Physics D: Applied Physics}\ }\textbf {\bibinfo {volume} {49}},\ \bibinfo {pages} {243001} (\bibinfo {year} {2016})}\BibitemShut {NoStop}%
\bibitem [{\citenamefont {Vardelle}\ \emph {et~al.}(2015)\citenamefont {Vardelle}, \citenamefont {Moreau}, \citenamefont {Themelis},\ and\ \citenamefont {Chazelas}}]{vardelle2015perspective}%
  \BibitemOpen
  \bibfield  {author} {\bibinfo {author} {\bibfnamefont {A.}~\bibnamefont {Vardelle}}, \bibinfo {author} {\bibfnamefont {C.}~\bibnamefont {Moreau}}, \bibinfo {author} {\bibfnamefont {N.~J.}\ \bibnamefont {Themelis}}, \ and\ \bibinfo {author} {\bibfnamefont {C.}~\bibnamefont {Chazelas}},\ }\bibfield  {title} {\enquote {\bibinfo {title} {A perspective on plasma spray technology},}\ }\href@noop {} {\bibfield  {journal} {\bibinfo  {journal} {Plasma Chemistry and Plasma Processing}\ }\textbf {\bibinfo {volume} {35}},\ \bibinfo {pages} {491--509} (\bibinfo {year} {2015})}\BibitemShut {NoStop}%
\bibitem [{\citenamefont {Starikovskaia}(2006)}]{starikovskaia2006plasma}%
  \BibitemOpen
  \bibfield  {author} {\bibinfo {author} {\bibfnamefont {S.~M.}\ \bibnamefont {Starikovskaia}},\ }\bibfield  {title} {\enquote {\bibinfo {title} {Plasma assisted ignition and combustion},}\ }\href@noop {} {\bibfield  {journal} {\bibinfo  {journal} {Journal of Physics D: Applied Physics}\ }\textbf {\bibinfo {volume} {39}},\ \bibinfo {pages} {R265} (\bibinfo {year} {2006})}\BibitemShut {NoStop}%
\bibitem [{\citenamefont {Huang}\ \emph {et~al.}(2015)\citenamefont {Huang}, \citenamefont {Li}, \citenamefont {Zheng}, \citenamefont {Shen}, \citenamefont {Wang}, \citenamefont {Han}, \citenamefont {Liu},\ and\ \citenamefont {Yan}}]{huang2015recent}%
  \BibitemOpen
  \bibfield  {author} {\bibinfo {author} {\bibfnamefont {Y.}~\bibnamefont {Huang}}, \bibinfo {author} {\bibfnamefont {S.}~\bibnamefont {Li}}, \bibinfo {author} {\bibfnamefont {Q.}~\bibnamefont {Zheng}}, \bibinfo {author} {\bibfnamefont {X.}~\bibnamefont {Shen}}, \bibinfo {author} {\bibfnamefont {S.}~\bibnamefont {Wang}}, \bibinfo {author} {\bibfnamefont {P.}~\bibnamefont {Han}}, \bibinfo {author} {\bibfnamefont {Z.}~\bibnamefont {Liu}}, \ and\ \bibinfo {author} {\bibfnamefont {K.}~\bibnamefont {Yan}},\ }\bibfield  {title} {\enquote {\bibinfo {title} {{Recent progress of dry electrostatic precipitation for PM2.5 emission control from coal-fired boilers}},}\ }\href@noop {} {\bibfield  {journal} {\bibinfo  {journal} {Int. J. Plasma Environ. Sci. Technol}\ }\textbf {\bibinfo {volume} {9}},\ \bibinfo {pages} {69--95} (\bibinfo {year} {2015})}\BibitemShut {NoStop}%
\bibitem [{\citenamefont {Nijdam}\ \emph {et~al.}(2012)\citenamefont {Nijdam}, \citenamefont {Van~Veldhuizen}, \citenamefont {Bruggeman},\ and\ \citenamefont {Ebert}}]{nijdam2012introduction}%
  \BibitemOpen
  \bibfield  {author} {\bibinfo {author} {\bibfnamefont {S.}~\bibnamefont {Nijdam}}, \bibinfo {author} {\bibfnamefont {E.}~\bibnamefont {Van~Veldhuizen}}, \bibinfo {author} {\bibfnamefont {P.}~\bibnamefont {Bruggeman}}, \ and\ \bibinfo {author} {\bibfnamefont {U.}~\bibnamefont {Ebert}},\ }\bibfield  {title} {\enquote {\bibinfo {title} {An introduction to nonequilibrium plasmas at atmospheric pressure},}\ }\href@noop {} {\bibfield  {journal} {\bibinfo  {journal} {Plasma chemistry and catalysis in gases and liquids}\ ,\ \bibinfo {pages} {1--44}} (\bibinfo {year} {2012})}\BibitemShut {NoStop}%
\bibitem [{\citenamefont {Byeon}\ \emph {et~al.}(2017)\citenamefont {Byeon}, \citenamefont {Hong}, \citenamefont {Yoo}, \citenamefont {Lho}, \citenamefont {Yoon}, \citenamefont {Kim}, \citenamefont {Yoo},\ and\ \citenamefont {Ryu}}]{byeon2017ballast}%
  \BibitemOpen
  \bibfield  {author} {\bibinfo {author} {\bibfnamefont {Y.-S.}\ \bibnamefont {Byeon}}, \bibinfo {author} {\bibfnamefont {E.~J.}\ \bibnamefont {Hong}}, \bibinfo {author} {\bibfnamefont {S.}~\bibnamefont {Yoo}}, \bibinfo {author} {\bibfnamefont {T.}~\bibnamefont {Lho}}, \bibinfo {author} {\bibfnamefont {S.-Y.}\ \bibnamefont {Yoon}}, \bibinfo {author} {\bibfnamefont {S.~B.}\ \bibnamefont {Kim}}, \bibinfo {author} {\bibfnamefont {S.~J.}\ \bibnamefont {Yoo}}, \ and\ \bibinfo {author} {\bibfnamefont {S.}~\bibnamefont {Ryu}},\ }\bibfield  {title} {\enquote {\bibinfo {title} {Ballast water treatment test at pilot-scale using an underwater capillary discharge device},}\ }\href@noop {} {\bibfield  {journal} {\bibinfo  {journal} {Plasma Chemistry and Plasma Processing}\ }\textbf {\bibinfo {volume} {37}},\ \bibinfo {pages} {1405--1416} (\bibinfo {year} {2017})}\BibitemShut {NoStop}%
\bibitem [{\citenamefont {Trenchev}\ and\ \citenamefont {Bogaerts}(2020)}]{trenchev2020dual}%
  \BibitemOpen
  \bibfield  {author} {\bibinfo {author} {\bibfnamefont {G.}~\bibnamefont {Trenchev}}\ and\ \bibinfo {author} {\bibfnamefont {A.}~\bibnamefont {Bogaerts}},\ }\bibfield  {title} {\enquote {\bibinfo {title} {Dual-vortex plasmatron: A novel plasma source for co$_2$ conversion},}\ }\href@noop {} {\bibfield  {journal} {\bibinfo  {journal} {{Journal of CO$_2$ Utilization}}\ }\textbf {\bibinfo {volume} {39}},\ \bibinfo {pages} {101152} (\bibinfo {year} {2020})}\BibitemShut {NoStop}%
\bibitem [{\citenamefont {Sui}, \citenamefont {Zorman},\ and\ \citenamefont {Sankaran}(2020)}]{sui2020plasmas}%
  \BibitemOpen
  \bibfield  {author} {\bibinfo {author} {\bibfnamefont {Y.}~\bibnamefont {Sui}}, \bibinfo {author} {\bibfnamefont {C.~A.}\ \bibnamefont {Zorman}}, \ and\ \bibinfo {author} {\bibfnamefont {R.~M.}\ \bibnamefont {Sankaran}},\ }\bibfield  {title} {\enquote {\bibinfo {title} {Plasmas for additive manufacturing},}\ }\href@noop {} {\bibfield  {journal} {\bibinfo  {journal} {Plasma Processes and Polymers}\ }\textbf {\bibinfo {volume} {17}},\ \bibinfo {pages} {2000009} (\bibinfo {year} {2020})}\BibitemShut {NoStop}%
\bibitem [{\citenamefont {Bruggeman}\ \emph {et~al.}(2016)\citenamefont {Bruggeman}, \citenamefont {Kushner}, \citenamefont {Locke}, \citenamefont {Gardeniers}, \citenamefont {Graham}, \citenamefont {Graves}, \citenamefont {Hofman-Caris}, \citenamefont {Maric}, \citenamefont {Reid}, \citenamefont {Ceriani} \emph {et~al.}}]{bruggeman2016plasma}%
  \BibitemOpen
  \bibfield  {author} {\bibinfo {author} {\bibfnamefont {P.~J.}\ \bibnamefont {Bruggeman}}, \bibinfo {author} {\bibfnamefont {M.~J.}\ \bibnamefont {Kushner}}, \bibinfo {author} {\bibfnamefont {B.~R.}\ \bibnamefont {Locke}}, \bibinfo {author} {\bibfnamefont {J.~G.}\ \bibnamefont {Gardeniers}}, \bibinfo {author} {\bibfnamefont {W.}~\bibnamefont {Graham}}, \bibinfo {author} {\bibfnamefont {D.~B.}\ \bibnamefont {Graves}}, \bibinfo {author} {\bibfnamefont {R.}~\bibnamefont {Hofman-Caris}}, \bibinfo {author} {\bibfnamefont {D.}~\bibnamefont {Maric}}, \bibinfo {author} {\bibfnamefont {J.~P.}\ \bibnamefont {Reid}}, \bibinfo {author} {\bibfnamefont {E.}~\bibnamefont {Ceriani}},  \emph {et~al.},\ }\bibfield  {title} {\enquote {\bibinfo {title} {Plasma--liquid interactions: a review and roadmap},}\ }\href@noop {} {\bibfield  {journal} {\bibinfo  {journal} {Plasma sources science and technology}\ }\textbf {\bibinfo {volume} {25}},\ \bibinfo {pages} {053002} (\bibinfo {year} {2016})}\BibitemShut {NoStop}%
\bibitem [{\citenamefont {Oehrlein}, \citenamefont {Phaneuf},\ and\ \citenamefont {Graves}(2011)}]{oehrlein2011plasma}%
  \BibitemOpen
  \bibfield  {author} {\bibinfo {author} {\bibfnamefont {G.~S.}\ \bibnamefont {Oehrlein}}, \bibinfo {author} {\bibfnamefont {R.~J.}\ \bibnamefont {Phaneuf}}, \ and\ \bibinfo {author} {\bibfnamefont {D.~B.}\ \bibnamefont {Graves}},\ }\bibfield  {title} {\enquote {\bibinfo {title} {Plasma-polymer interactions: A review of progress in understanding polymer resist mask durability during plasma etching for nanoscale fabrication},}\ }\href@noop {} {\bibfield  {journal} {\bibinfo  {journal} {Journal of Vacuum Science \& Technology B}\ }\textbf {\bibinfo {volume} {29}} (\bibinfo {year} {2011})}\BibitemShut {NoStop}%
\bibitem [{\citenamefont {Becker}, \citenamefont {Schoenbach},\ and\ \citenamefont {Eden}(2006)}]{becker2006microplasmas}%
  \BibitemOpen
  \bibfield  {author} {\bibinfo {author} {\bibfnamefont {K.}~\bibnamefont {Becker}}, \bibinfo {author} {\bibfnamefont {K.}~\bibnamefont {Schoenbach}}, \ and\ \bibinfo {author} {\bibfnamefont {J.}~\bibnamefont {Eden}},\ }\bibfield  {title} {\enquote {\bibinfo {title} {Microplasmas and applications},}\ }\href@noop {} {\bibfield  {journal} {\bibinfo  {journal} {Journal of Physics D: Applied Physics}\ }\textbf {\bibinfo {volume} {39}},\ \bibinfo {pages} {R55} (\bibinfo {year} {2006})}\BibitemShut {NoStop}%
\bibitem [{\citenamefont {Kusano}(2014)}]{kusano2014atmospheric}%
  \BibitemOpen
  \bibfield  {author} {\bibinfo {author} {\bibfnamefont {Y.}~\bibnamefont {Kusano}},\ }\bibfield  {title} {\enquote {\bibinfo {title} {Atmospheric pressure plasma processing for polymer adhesion: A review},}\ }\href@noop {} {\bibfield  {journal} {\bibinfo  {journal} {The Journal of Adhesion}\ }\textbf {\bibinfo {volume} {90}},\ \bibinfo {pages} {755--777} (\bibinfo {year} {2014})}\BibitemShut {NoStop}%
\bibitem [{\citenamefont {Amorim}, \citenamefont {Baravian},\ and\ \citenamefont {Jolly}(2000)}]{amorim2000laser}%
  \BibitemOpen
  \bibfield  {author} {\bibinfo {author} {\bibfnamefont {J.}~\bibnamefont {Amorim}}, \bibinfo {author} {\bibfnamefont {G.}~\bibnamefont {Baravian}}, \ and\ \bibinfo {author} {\bibfnamefont {J.}~\bibnamefont {Jolly}},\ }\bibfield  {title} {\enquote {\bibinfo {title} {Laser-induced resonance fluorescence as a diagnostic technique in non-thermal equilibrium plasmas},}\ }\href@noop {} {\bibfield  {journal} {\bibinfo  {journal} {Journal of Physics D: Applied Physics}\ }\textbf {\bibinfo {volume} {33}},\ \bibinfo {pages} {R51} (\bibinfo {year} {2000})}\BibitemShut {NoStop}%
\bibitem [{\citenamefont {D{\"o}bele}\ \emph {et~al.}(2005)\citenamefont {D{\"o}bele}, \citenamefont {Mosbach}, \citenamefont {Niemi},\ and\ \citenamefont {Schulz-Von Der~Gathen}}]{dobele2005laser}%
  \BibitemOpen
  \bibfield  {author} {\bibinfo {author} {\bibfnamefont {H.}~\bibnamefont {D{\"o}bele}}, \bibinfo {author} {\bibfnamefont {T.}~\bibnamefont {Mosbach}}, \bibinfo {author} {\bibfnamefont {K.}~\bibnamefont {Niemi}}, \ and\ \bibinfo {author} {\bibfnamefont {V.}~\bibnamefont {Schulz-Von Der~Gathen}},\ }\bibfield  {title} {\enquote {\bibinfo {title} {Laser-induced fluorescence measurements of absolute atomic densities: concepts and limitations},}\ }\href@noop {} {\bibfield  {journal} {\bibinfo  {journal} {Plasma Sources Science and Technology}\ }\textbf {\bibinfo {volume} {14}},\ \bibinfo {pages} {S31} (\bibinfo {year} {2005})}\BibitemShut {NoStop}%
\bibitem [{\citenamefont {Donnelly}, \citenamefont {Flamm},\ and\ \citenamefont {Collins}(1982)}]{donnelly1982laser}%
  \BibitemOpen
  \bibfield  {author} {\bibinfo {author} {\bibfnamefont {V.}~\bibnamefont {Donnelly}}, \bibinfo {author} {\bibfnamefont {D.}~\bibnamefont {Flamm}}, \ and\ \bibinfo {author} {\bibfnamefont {G.}~\bibnamefont {Collins}},\ }\bibfield  {title} {\enquote {\bibinfo {title} {{Laser diagnostics of plasma etching: Measurement of Cl$_{2}^{+}$ in a chlorine discharge}},}\ }\href@noop {} {\bibfield  {journal} {\bibinfo  {journal} {Journal of Vacuum Science and Technology}\ }\textbf {\bibinfo {volume} {21}},\ \bibinfo {pages} {817--823} (\bibinfo {year} {1982})}\BibitemShut {NoStop}%
\bibitem [{\citenamefont {Konthasinghe}\ \emph {et~al.}(2015)\citenamefont {Konthasinghe}, \citenamefont {Fitzmorris}, \citenamefont {Peiris}, \citenamefont {Hopkins}, \citenamefont {Petrak}, \citenamefont {Killinger},\ and\ \citenamefont {Muller}}]{konthasinghe2015laser}%
  \BibitemOpen
  \bibfield  {author} {\bibinfo {author} {\bibfnamefont {K.}~\bibnamefont {Konthasinghe}}, \bibinfo {author} {\bibfnamefont {K.}~\bibnamefont {Fitzmorris}}, \bibinfo {author} {\bibfnamefont {M.}~\bibnamefont {Peiris}}, \bibinfo {author} {\bibfnamefont {A.~J.}\ \bibnamefont {Hopkins}}, \bibinfo {author} {\bibfnamefont {B.}~\bibnamefont {Petrak}}, \bibinfo {author} {\bibfnamefont {D.~K.}\ \bibnamefont {Killinger}}, \ and\ \bibinfo {author} {\bibfnamefont {A.}~\bibnamefont {Muller}},\ }\bibfield  {title} {\enquote {\bibinfo {title} {Laser-induced fluorescence from {N$_2^+$} ions generated by a corona discharge in ambient air},}\ }\href@noop {} {\bibfield  {journal} {\bibinfo  {journal} {Applied Spectroscopy}\ }\textbf {\bibinfo {volume} {69}},\ \bibinfo {pages} {1042--1046} (\bibinfo {year} {2015})}\BibitemShut {NoStop}%
\bibitem [{\citenamefont {Lee}, \citenamefont {Hershkowitz},\ and\ \citenamefont {Severn}(2007)}]{lee2007measurements}%
  \BibitemOpen
  \bibfield  {author} {\bibinfo {author} {\bibfnamefont {D.}~\bibnamefont {Lee}}, \bibinfo {author} {\bibfnamefont {N.}~\bibnamefont {Hershkowitz}}, \ and\ \bibinfo {author} {\bibfnamefont {G.~D.}\ \bibnamefont {Severn}},\ }\bibfield  {title} {\enquote {\bibinfo {title} {{Measurements of Ar$^{+}$ and Xe$^{+}$ velocities near the sheath boundary of Ar--Xe plasma using two diode lasers}},}\ }\href@noop {} {\bibfield  {journal} {\bibinfo  {journal} {Applied Physics Letters}\ }\textbf {\bibinfo {volume} {91}} (\bibinfo {year} {2007})}\BibitemShut {NoStop}%
\bibitem [{\citenamefont {Tachibana}, \citenamefont {Mukai},\ and\ \citenamefont {Harima}(1991)}]{tachibana1991measurement}%
  \BibitemOpen
  \bibfield  {author} {\bibinfo {author} {\bibfnamefont {K.}~\bibnamefont {Tachibana}}, \bibinfo {author} {\bibfnamefont {T.~M.~T.}\ \bibnamefont {Mukai}}, \ and\ \bibinfo {author} {\bibfnamefont {H.~H.~H.}\ \bibnamefont {Harima}},\ }\bibfield  {title} {\enquote {\bibinfo {title} {Measurement of absolute densities and spatial distributions of {Si} and {SiH} in an {RF}-discharge silane plasma for the chemical vapor deposition of a-{Si}:{H} films},}\ }\href@noop {} {\bibfield  {journal} {\bibinfo  {journal} {Japanese journal of applied physics}\ }\textbf {\bibinfo {volume} {30}},\ \bibinfo {pages} {L1208} (\bibinfo {year} {1991})}\BibitemShut {NoStop}%
\bibitem [{\citenamefont {Herring}\ \emph {et~al.}(1988)\citenamefont {Herring}, \citenamefont {Dyer}, \citenamefont {Jusinski},\ and\ \citenamefont {Bischel}}]{herring1988two}%
  \BibitemOpen
  \bibfield  {author} {\bibinfo {author} {\bibfnamefont {G.}~\bibnamefont {Herring}}, \bibinfo {author} {\bibfnamefont {M.~J.}\ \bibnamefont {Dyer}}, \bibinfo {author} {\bibfnamefont {L.~E.}\ \bibnamefont {Jusinski}}, \ and\ \bibinfo {author} {\bibfnamefont {W.~K.}\ \bibnamefont {Bischel}},\ }\bibfield  {title} {\enquote {\bibinfo {title} {Two-photon-excited fluorescence spectroscopy of atomic fluorine at 170 nm},}\ }\href@noop {} {\bibfield  {journal} {\bibinfo  {journal} {Optics letters}\ }\textbf {\bibinfo {volume} {13}},\ \bibinfo {pages} {360--362} (\bibinfo {year} {1988})}\BibitemShut {NoStop}%
\bibitem [{\citenamefont {Selwyn}, \citenamefont {Baston},\ and\ \citenamefont {Sawin}(1987)}]{selwyn1987detection}%
  \BibitemOpen
  \bibfield  {author} {\bibinfo {author} {\bibfnamefont {G.~S.}\ \bibnamefont {Selwyn}}, \bibinfo {author} {\bibfnamefont {L.}~\bibnamefont {Baston}}, \ and\ \bibinfo {author} {\bibfnamefont {H.}~\bibnamefont {Sawin}},\ }\bibfield  {title} {\enquote {\bibinfo {title} {Detection of {Cl} and chlorine-containing negative ions in {RF} plasmas by two-photon laser-induced fluorescence},}\ }\href@noop {} {\bibfield  {journal} {\bibinfo  {journal} {Applied physics letters}\ }\textbf {\bibinfo {volume} {51}},\ \bibinfo {pages} {898--900} (\bibinfo {year} {1987})}\BibitemShut {NoStop}%
\bibitem [{\citenamefont {van~de Weijer}\ and\ \citenamefont {Zwerver}(1989)}]{van1989laser}%
  \BibitemOpen
  \bibfield  {author} {\bibinfo {author} {\bibfnamefont {P.}~\bibnamefont {van~de Weijer}}\ and\ \bibinfo {author} {\bibfnamefont {B.~H.}\ \bibnamefont {Zwerver}},\ }\bibfield  {title} {\enquote {\bibinfo {title} {Laser-induced fluorescence of {OH} and {SiO} molecules during thermal chemical vapour deposition of {SiO}$_{2}$ from silane-oxygen mixtures},}\ }\href@noop {} {\bibfield  {journal} {\bibinfo  {journal} {Chemical physics letters}\ }\textbf {\bibinfo {volume} {163}},\ \bibinfo {pages} {48--54} (\bibinfo {year} {1989})}\BibitemShut {NoStop}%
\bibitem [{\citenamefont {Diallo}\ \emph {et~al.}(2015)\citenamefont {Diallo}, \citenamefont {Keller}, \citenamefont {Shi}, \citenamefont {Raitses},\ and\ \citenamefont {Mazouffre}}]{diallo2015time}%
  \BibitemOpen
  \bibfield  {author} {\bibinfo {author} {\bibfnamefont {A.}~\bibnamefont {Diallo}}, \bibinfo {author} {\bibfnamefont {S.}~\bibnamefont {Keller}}, \bibinfo {author} {\bibfnamefont {Y.}~\bibnamefont {Shi}}, \bibinfo {author} {\bibfnamefont {Y.}~\bibnamefont {Raitses}}, \ and\ \bibinfo {author} {\bibfnamefont {S.}~\bibnamefont {Mazouffre}},\ }\bibfield  {title} {\enquote {\bibinfo {title} {Time-resolved ion velocity distribution in a cylindrical hall thruster: Heterodyne-based experiment and modeling},}\ }\href@noop {} {\bibfield  {journal} {\bibinfo  {journal} {Review of Scientific Instruments}\ }\textbf {\bibinfo {volume} {86}} (\bibinfo {year} {2015})}\BibitemShut {NoStop}%
\bibitem [{\citenamefont {Bruggeman}\ \emph {et~al.}(2014)\citenamefont {Bruggeman}, \citenamefont {Sadeghi}, \citenamefont {Schram},\ and\ \citenamefont {Linss}}]{bruggeman2014gas}%
  \BibitemOpen
  \bibfield  {author} {\bibinfo {author} {\bibfnamefont {P.~J.}\ \bibnamefont {Bruggeman}}, \bibinfo {author} {\bibfnamefont {N.}~\bibnamefont {Sadeghi}}, \bibinfo {author} {\bibfnamefont {D.}~\bibnamefont {Schram}}, \ and\ \bibinfo {author} {\bibfnamefont {V.}~\bibnamefont {Linss}},\ }\bibfield  {title} {\enquote {\bibinfo {title} {Gas temperature determination from rotational lines in non-equilibrium plasmas: a review},}\ }\href@noop {} {\bibfield  {journal} {\bibinfo  {journal} {Plasma Sources Science and Technology}\ }\textbf {\bibinfo {volume} {23}},\ \bibinfo {pages} {023001} (\bibinfo {year} {2014})}\BibitemShut {NoStop}%
\bibitem [{\citenamefont {Yatom}\ \emph {et~al.}(2023)\citenamefont {Yatom}, \citenamefont {Chopra}, \citenamefont {Kondeti}, \citenamefont {Petrova}, \citenamefont {Raitses}, \citenamefont {Boris}, \citenamefont {Johnson},\ and\ \citenamefont {Walton}}]{yatom2023measurement}%
  \BibitemOpen
  \bibfield  {author} {\bibinfo {author} {\bibfnamefont {S.}~\bibnamefont {Yatom}}, \bibinfo {author} {\bibfnamefont {N.}~\bibnamefont {Chopra}}, \bibinfo {author} {\bibfnamefont {S.}~\bibnamefont {Kondeti}}, \bibinfo {author} {\bibfnamefont {T.~B.}\ \bibnamefont {Petrova}}, \bibinfo {author} {\bibfnamefont {Y.}~\bibnamefont {Raitses}}, \bibinfo {author} {\bibfnamefont {D.~R.}\ \bibnamefont {Boris}}, \bibinfo {author} {\bibfnamefont {M.~J.}\ \bibnamefont {Johnson}}, \ and\ \bibinfo {author} {\bibfnamefont {S.~G.}\ \bibnamefont {Walton}},\ }\bibfield  {title} {\enquote {\bibinfo {title} {{Measurement and reduction of Ar metastable densities by nitrogen admixing in electron beam-generated plasmas}},}\ }\href@noop {} {\bibfield  {journal} {\bibinfo  {journal} {Plasma Sources Science and Technology}\ }\textbf {\bibinfo {volume} {32}},\ \bibinfo {pages} {115005} (\bibinfo {year} {2023})}\BibitemShut {NoStop}%
\bibitem [{\citenamefont {Forster}\ \emph {et~al.}(1995)\citenamefont {Forster}, \citenamefont {Frost}, \citenamefont {Fulle}, \citenamefont {Hamann}, \citenamefont {Hippler}, \citenamefont {Schlepegrell},\ and\ \citenamefont {Troe}}]{forster1995high}%
  \BibitemOpen
  \bibfield  {author} {\bibinfo {author} {\bibfnamefont {R.}~\bibnamefont {Forster}}, \bibinfo {author} {\bibfnamefont {M.}~\bibnamefont {Frost}}, \bibinfo {author} {\bibfnamefont {D.}~\bibnamefont {Fulle}}, \bibinfo {author} {\bibfnamefont {H.}~\bibnamefont {Hamann}}, \bibinfo {author} {\bibfnamefont {H.}~\bibnamefont {Hippler}}, \bibinfo {author} {\bibfnamefont {A.}~\bibnamefont {Schlepegrell}}, \ and\ \bibinfo {author} {\bibfnamefont {J.}~\bibnamefont {Troe}},\ }\bibfield  {title} {\enquote {\bibinfo {title} {{High pressure range of the addition of HO to HO, NO, NO2, and CO. I. Saturated laser induced fluorescence measurements at 298 K}},}\ }\href@noop {} {\bibfield  {journal} {\bibinfo  {journal} {The Journal of chemical physics}\ }\textbf {\bibinfo {volume} {103}},\ \bibinfo {pages} {2949--2958} (\bibinfo {year} {1995})}\BibitemShut {NoStop}%
\bibitem [{\citenamefont {Mrkvi{\v{c}}kov{\'a}}\ \emph {et~al.}(2022)\citenamefont {Mrkvi{\v{c}}kov{\'a}}, \citenamefont {Dvo{\v{r}}{\'a}k}, \citenamefont {Svoboda}, \citenamefont {Kratzer}, \citenamefont {Vor{\'a}{\v{c}}},\ and\ \citenamefont {D{\v{e}}dina}}]{mrkvivckova2022dealing}%
  \BibitemOpen
  \bibfield  {author} {\bibinfo {author} {\bibfnamefont {M.}~\bibnamefont {Mrkvi{\v{c}}kov{\'a}}}, \bibinfo {author} {\bibfnamefont {P.}~\bibnamefont {Dvo{\v{r}}{\'a}k}}, \bibinfo {author} {\bibfnamefont {M.}~\bibnamefont {Svoboda}}, \bibinfo {author} {\bibfnamefont {J.}~\bibnamefont {Kratzer}}, \bibinfo {author} {\bibfnamefont {J.}~\bibnamefont {Vor{\'a}{\v{c}}}}, \ and\ \bibinfo {author} {\bibfnamefont {J.}~\bibnamefont {D{\v{e}}dina}},\ }\bibfield  {title} {\enquote {\bibinfo {title} {{Dealing with saturation of the laser-induced fluorescence signal: An application to lead atoms}},}\ }\href@noop {} {\bibfield  {journal} {\bibinfo  {journal} {Combustion and Flame}\ }\textbf {\bibinfo {volume} {241}},\ \bibinfo {pages} {112100} (\bibinfo {year} {2022})}\BibitemShut {NoStop}%
\bibitem [{\citenamefont {Goeckner}, \citenamefont {Goree},\ and\ \citenamefont {Sheridan}(1993)}]{goeckner1993saturation}%
  \BibitemOpen
  \bibfield  {author} {\bibinfo {author} {\bibfnamefont {M.}~\bibnamefont {Goeckner}}, \bibinfo {author} {\bibfnamefont {J.}~\bibnamefont {Goree}}, \ and\ \bibinfo {author} {\bibfnamefont {T.}~\bibnamefont {Sheridan}},\ }\bibfield  {title} {\enquote {\bibinfo {title} {Saturation broadening of laser-induced fluorescence from plasma ions},}\ }\href@noop {} {\bibfield  {journal} {\bibinfo  {journal} {Review of scientific instruments}\ }\textbf {\bibinfo {volume} {64}},\ \bibinfo {pages} {996--1000} (\bibinfo {year} {1993})}\BibitemShut {NoStop}%
\bibitem [{\citenamefont {Mazouffre}(2016)}]{mazouffre2016laser}%
  \BibitemOpen
  \bibfield  {author} {\bibinfo {author} {\bibfnamefont {S.}~\bibnamefont {Mazouffre}},\ }\bibfield  {title} {\enquote {\bibinfo {title} {Laser-induced fluorescence spectroscopy applied to electric thrusters},}\ }\href@noop {} {\bibfield  {journal} {\bibinfo  {journal} {Electric Propulsion Systems: from Recent Research Developments to Industrial Space Applications—STO-AVT}\ }\textbf {\bibinfo {volume} {263}},\ \bibinfo {pages} {1--10} (\bibinfo {year} {2016})}\BibitemShut {NoStop}%
\bibitem [{\citenamefont {Kondeti}\ \emph {et~al.}(2020)\citenamefont {Kondeti}, \citenamefont {Zheng}, \citenamefont {Luan}, \citenamefont {Oehrlein},\ and\ \citenamefont {Bruggeman}}]{kondeti2020h}%
  \BibitemOpen
  \bibfield  {author} {\bibinfo {author} {\bibfnamefont {V.}~\bibnamefont {Kondeti}}, \bibinfo {author} {\bibfnamefont {Y.}~\bibnamefont {Zheng}}, \bibinfo {author} {\bibfnamefont {P.}~\bibnamefont {Luan}}, \bibinfo {author} {\bibfnamefont {G.~S.}\ \bibnamefont {Oehrlein}}, \ and\ \bibinfo {author} {\bibfnamefont {P.~J.}\ \bibnamefont {Bruggeman}},\ }\bibfield  {title} {\enquote {\bibinfo {title} {{O}, {H}, and {OH} radical etching probability of polystyrene obtained for a radio frequency driven atmospheric pressure plasma jet},}\ }\href@noop {} {\bibfield  {journal} {\bibinfo  {journal} {Journal of Vacuum Science \& Technology A}\ }\textbf {\bibinfo {volume} {38}} (\bibinfo {year} {2020})}\BibitemShut {NoStop}%
\bibitem [{\citenamefont {Yatom}\ \emph {et~al.}(2017)\citenamefont {Yatom}, \citenamefont {Luo}, \citenamefont {Xiong},\ and\ \citenamefont {Bruggeman}}]{yatom2017nanosecond}%
  \BibitemOpen
  \bibfield  {author} {\bibinfo {author} {\bibfnamefont {S.}~\bibnamefont {Yatom}}, \bibinfo {author} {\bibfnamefont {Y.}~\bibnamefont {Luo}}, \bibinfo {author} {\bibfnamefont {Q.}~\bibnamefont {Xiong}}, \ and\ \bibinfo {author} {\bibfnamefont {P.~J.}\ \bibnamefont {Bruggeman}},\ }\bibfield  {title} {\enquote {\bibinfo {title} {Nanosecond pulsed humid {Ar} plasma jet in air: shielding, discharge characteristics and atomic hydrogen production},}\ }\href@noop {} {\bibfield  {journal} {\bibinfo  {journal} {Journal of Physics D: Applied Physics}\ }\textbf {\bibinfo {volume} {50}},\ \bibinfo {pages} {415204} (\bibinfo {year} {2017})}\BibitemShut {NoStop}%
\bibitem [{\citenamefont {Kondeti}\ \emph {et~al.}(2017)\citenamefont {Kondeti}, \citenamefont {Gangal}, \citenamefont {Yatom},\ and\ \citenamefont {Bruggeman}}]{kondeti2017ag+}%
  \BibitemOpen
  \bibfield  {author} {\bibinfo {author} {\bibfnamefont {V.}~\bibnamefont {Kondeti}}, \bibinfo {author} {\bibfnamefont {U.}~\bibnamefont {Gangal}}, \bibinfo {author} {\bibfnamefont {S.}~\bibnamefont {Yatom}}, \ and\ \bibinfo {author} {\bibfnamefont {P.~J.}\ \bibnamefont {Bruggeman}},\ }\bibfield  {title} {\enquote {\bibinfo {title} {{Ag$^{+}$ reduction and silver nanoparticle synthesis at the plasma--liquid interface by an RF driven atmospheric pressure plasma jet: Mechanisms and the effect of surfactant}},}\ }\href@noop {} {\bibfield  {journal} {\bibinfo  {journal} {Journal of Vacuum Science \& Technology A}\ }\textbf {\bibinfo {volume} {35}} (\bibinfo {year} {2017})}\BibitemShut {NoStop}%
\bibitem [{\citenamefont {Svarnas}\ \emph {et~al.}(2018)\citenamefont {Svarnas}, \citenamefont {Romadanov}, \citenamefont {Diallo},\ and\ \citenamefont {Raitses}}]{svarnas2018laser}%
  \BibitemOpen
  \bibfield  {author} {\bibinfo {author} {\bibfnamefont {P.}~\bibnamefont {Svarnas}}, \bibinfo {author} {\bibfnamefont {I.}~\bibnamefont {Romadanov}}, \bibinfo {author} {\bibfnamefont {A.}~\bibnamefont {Diallo}}, \ and\ \bibinfo {author} {\bibfnamefont {Y.}~\bibnamefont {Raitses}},\ }\bibfield  {title} {\enquote {\bibinfo {title} {{Laser-induced fluorescence of Xe I and Xe II in ambipolar plasma flow}},}\ }\href@noop {} {\bibfield  {journal} {\bibinfo  {journal} {IEEE Transactions on Plasma Science}\ }\textbf {\bibinfo {volume} {46}},\ \bibinfo {pages} {3998--4009} (\bibinfo {year} {2018})}\BibitemShut {NoStop}%
\bibitem [{\citenamefont {Pietzonka}\ \emph {et~al.}(2023)\citenamefont {Pietzonka}, \citenamefont {Eichhorn}, \citenamefont {Scholze},\ and\ \citenamefont {Spemann}}]{pietzonka2023laser}%
  \BibitemOpen
  \bibfield  {author} {\bibinfo {author} {\bibfnamefont {L.}~\bibnamefont {Pietzonka}}, \bibinfo {author} {\bibfnamefont {C.}~\bibnamefont {Eichhorn}}, \bibinfo {author} {\bibfnamefont {F.}~\bibnamefont {Scholze}}, \ and\ \bibinfo {author} {\bibfnamefont {D.}~\bibnamefont {Spemann}},\ }\bibfield  {title} {\enquote {\bibinfo {title} {Laser-induced fluorescence spectroscopy for kinetic temperature measurement of xenon neutrals and ions in the discharge chamber of a radiofrequency ion source},}\ }\href@noop {} {\bibfield  {journal} {\bibinfo  {journal} {Journal of Electric Propulsion}\ }\textbf {\bibinfo {volume} {2}},\ \bibinfo {pages} {4} (\bibinfo {year} {2023})}\BibitemShut {NoStop}%
\bibitem [{\citenamefont {Vinci}\ \emph {et~al.}(2022)\citenamefont {Vinci}, \citenamefont {Mazouffre}, \citenamefont {G{\'o}mez}, \citenamefont {Fajardo},\ and\ \citenamefont {Navarro-Cavall{\'e}}}]{vinci2022laser}%
  \BibitemOpen
  \bibfield  {author} {\bibinfo {author} {\bibfnamefont {A.~E.}\ \bibnamefont {Vinci}}, \bibinfo {author} {\bibfnamefont {S.}~\bibnamefont {Mazouffre}}, \bibinfo {author} {\bibfnamefont {V.}~\bibnamefont {G{\'o}mez}}, \bibinfo {author} {\bibfnamefont {P.}~\bibnamefont {Fajardo}}, \ and\ \bibinfo {author} {\bibfnamefont {J.}~\bibnamefont {Navarro-Cavall{\'e}}},\ }\bibfield  {title} {\enquote {\bibinfo {title} {Laser-induced fluorescence spectroscopy on xenon atoms and ions in the magnetic nozzle of a helicon plasma thruster},}\ }\href@noop {} {\bibfield  {journal} {\bibinfo  {journal} {Plasma Sources Science and Technology}\ }\textbf {\bibinfo {volume} {31}},\ \bibinfo {pages} {095007} (\bibinfo {year} {2022})}\BibitemShut {NoStop}%
\bibitem [{\citenamefont {Brackmann}\ \emph {et~al.}(2016)\citenamefont {Brackmann}, \citenamefont {Zhou}, \citenamefont {Li},\ and\ \citenamefont {Alden}}]{brackmann2016strategies}%
  \BibitemOpen
  \bibfield  {author} {\bibinfo {author} {\bibfnamefont {C.}~\bibnamefont {Brackmann}}, \bibinfo {author} {\bibfnamefont {B.}~\bibnamefont {Zhou}}, \bibinfo {author} {\bibfnamefont {Z.}~\bibnamefont {Li}}, \ and\ \bibinfo {author} {\bibfnamefont {M.}~\bibnamefont {Alden}},\ }\bibfield  {title} {\enquote {\bibinfo {title} {Strategies for quantitative planar laser-induced fluorescence of {NH} radicals in flames},}\ }\href@noop {} {\bibfield  {journal} {\bibinfo  {journal} {Combustion Science and Technology}\ }\textbf {\bibinfo {volume} {188}},\ \bibinfo {pages} {529--541} (\bibinfo {year} {2016})}\BibitemShut {NoStop}%
\bibitem [{\citenamefont {Hirano}\ and\ \citenamefont {Tsujishita}(1994)}]{hirano1994visualization}%
  \BibitemOpen
  \bibfield  {author} {\bibinfo {author} {\bibfnamefont {A.}~\bibnamefont {Hirano}}\ and\ \bibinfo {author} {\bibfnamefont {M.}~\bibnamefont {Tsujishita}},\ }\bibfield  {title} {\enquote {\bibinfo {title} {Visualization of {CN} by the use of planar laser-induced fluorescence in a cross section of an unseeded turbulent {CH}$_{4}$--air flame},}\ }\href@noop {} {\bibfield  {journal} {\bibinfo  {journal} {Applied optics}\ }\textbf {\bibinfo {volume} {33}},\ \bibinfo {pages} {7777--7780} (\bibinfo {year} {1994})}\BibitemShut {NoStop}%
\bibitem [{\citenamefont {Kirby}\ and\ \citenamefont {Hanson}(2000)}]{kirby2000imaging}%
  \BibitemOpen
  \bibfield  {author} {\bibinfo {author} {\bibfnamefont {B.~J.}\ \bibnamefont {Kirby}}\ and\ \bibinfo {author} {\bibfnamefont {B.~K.}\ \bibnamefont {Hanson}},\ }\bibfield  {title} {\enquote {\bibinfo {title} {Imaging of {CO} and {CO}$_2$ using infrared planar laser-induced fluorescence},}\ }\href@noop {} {\bibfield  {journal} {\bibinfo  {journal} {Proceedings of the Combustion Institute}\ }\textbf {\bibinfo {volume} {28}},\ \bibinfo {pages} {253--259} (\bibinfo {year} {2000})}\BibitemShut {NoStop}%
\bibitem [{\citenamefont {Sappey}\ and\ \citenamefont {Gamble}(1992)}]{sappey1992planar}%
  \BibitemOpen
  \bibfield  {author} {\bibinfo {author} {\bibfnamefont {A.~D.}\ \bibnamefont {Sappey}}\ and\ \bibinfo {author} {\bibfnamefont {T.~K.}\ \bibnamefont {Gamble}},\ }\bibfield  {title} {\enquote {\bibinfo {title} {Planar laser-induced fluorescence imaging of {Cu} atom and {Cu}$_2$ in a condensing laser-ablated copper plasma plume},}\ }\href@noop {} {\bibfield  {journal} {\bibinfo  {journal} {Journal of applied physics}\ }\textbf {\bibinfo {volume} {72}},\ \bibinfo {pages} {5095--5107} (\bibinfo {year} {1992})}\BibitemShut {NoStop}%
\bibitem [{\citenamefont {Niemi}, \citenamefont {Schulz-Von Der~Gathen},\ and\ \citenamefont {D{\"o}bele}(2001)}]{niemi2001absolute}%
  \BibitemOpen
  \bibfield  {author} {\bibinfo {author} {\bibfnamefont {K.}~\bibnamefont {Niemi}}, \bibinfo {author} {\bibfnamefont {V.}~\bibnamefont {Schulz-Von Der~Gathen}}, \ and\ \bibinfo {author} {\bibfnamefont {H.}~\bibnamefont {D{\"o}bele}},\ }\bibfield  {title} {\enquote {\bibinfo {title} {Absolute calibration of atomic density measurements by laser-induced fluorescence spectroscopy with two-photon excitation},}\ }\href@noop {} {\bibfield  {journal} {\bibinfo  {journal} {Journal of Physics D: Applied Physics}\ }\textbf {\bibinfo {volume} {34}},\ \bibinfo {pages} {2330} (\bibinfo {year} {2001})}\BibitemShut {NoStop}%
\bibitem [{\citenamefont {Kulatilaka}, \citenamefont {Frank},\ and\ \citenamefont {Settersten}(2009)}]{kulatilaka2009interference}%
  \BibitemOpen
  \bibfield  {author} {\bibinfo {author} {\bibfnamefont {W.~D.}\ \bibnamefont {Kulatilaka}}, \bibinfo {author} {\bibfnamefont {J.~H.}\ \bibnamefont {Frank}}, \ and\ \bibinfo {author} {\bibfnamefont {T.~B.}\ \bibnamefont {Settersten}},\ }\bibfield  {title} {\enquote {\bibinfo {title} {Interference-free two-photon {LIF} imaging of atomic hydrogen in flames using picosecond excitation},}\ }\href@noop {} {\bibfield  {journal} {\bibinfo  {journal} {Proceedings of the Combustion Institute}\ }\textbf {\bibinfo {volume} {32}},\ \bibinfo {pages} {955--962} (\bibinfo {year} {2009})}\BibitemShut {NoStop}%
\bibitem [{\citenamefont {Dogariu}\ \emph {et~al.}(2021)\citenamefont {Dogariu}, \citenamefont {Evans}, \citenamefont {Vinoth},\ and\ \citenamefont {Cohen}}]{dogariu2021non}%
  \BibitemOpen
  \bibfield  {author} {\bibinfo {author} {\bibfnamefont {A.}~\bibnamefont {Dogariu}}, \bibinfo {author} {\bibfnamefont {E.}~\bibnamefont {Evans}}, \bibinfo {author} {\bibfnamefont {S.~P.}\ \bibnamefont {Vinoth}}, \ and\ \bibinfo {author} {\bibfnamefont {S.~A.}\ \bibnamefont {Cohen}},\ }\bibfield  {title} {\enquote {\bibinfo {title} {Non-invasive neutral atom density measurements using fs-{TALIF} in a magnetic linear plasma device},}\ }in\ \href@noop {} {\emph {\bibinfo {booktitle} {CLEO: Science and Innovations}}}\ (\bibinfo {organization} {Optica Publishing Group},\ \bibinfo {year} {2021})\ pp.\ \bibinfo {pages} {SM1E--1}\BibitemShut {NoStop}%
\bibitem [{\citenamefont {Yatom}\ and\ \citenamefont {Dobrynin}(2022)}]{yatom2022examination}%
  \BibitemOpen
  \bibfield  {author} {\bibinfo {author} {\bibfnamefont {S.}~\bibnamefont {Yatom}}\ and\ \bibinfo {author} {\bibfnamefont {D.}~\bibnamefont {Dobrynin}},\ }\bibfield  {title} {\enquote {\bibinfo {title} {{Examination of OH and H$_2$O$_2$ production by uniform and non-uniform modes of dielectric barrier discharge in He/air mixture}},}\ }\href@noop {} {\bibfield  {journal} {\bibinfo  {journal} {Journal of Physics D: Applied Physics}\ }\textbf {\bibinfo {volume} {55}},\ \bibinfo {pages} {485203} (\bibinfo {year} {2022})}\BibitemShut {NoStop}%
\bibitem [{\citenamefont {Tr{\"a}ger}(2012)}]{trager2012springer}%
  \BibitemOpen
  \bibfield  {author} {\bibinfo {author} {\bibfnamefont {F.}~\bibnamefont {Tr{\"a}ger}},\ }\href@noop {} {\emph {\bibinfo {title} {Springer handbook of lasers and optics}}},\ Vol.~\bibinfo {volume} {2}\ (\bibinfo  {publisher} {Springer},\ \bibinfo {year} {2012})\BibitemShut {NoStop}%
\bibitem [{\citenamefont {Renk}(2012)}]{renk2012basics}%
  \BibitemOpen
  \bibfield  {author} {\bibinfo {author} {\bibfnamefont {K.~F.}\ \bibnamefont {Renk}},\ }\href@noop {} {\emph {\bibinfo {title} {Basics of laser physics}}}\ (\bibinfo  {publisher} {Springer},\ \bibinfo {year} {2012})\BibitemShut {NoStop}%
\bibitem [{\citenamefont {Vekselman}\ \emph {et~al.}(2010)\citenamefont {Vekselman}, \citenamefont {Gleizer}, \citenamefont {Yatom}, \citenamefont {Gurovich},\ and\ \citenamefont {Krasik}}]{vekselman2010high}%
  \BibitemOpen
  \bibfield  {author} {\bibinfo {author} {\bibfnamefont {V.}~\bibnamefont {Vekselman}}, \bibinfo {author} {\bibfnamefont {J.}~\bibnamefont {Gleizer}}, \bibinfo {author} {\bibfnamefont {S.}~\bibnamefont {Yatom}}, \bibinfo {author} {\bibfnamefont {V.~T.}\ \bibnamefont {Gurovich}}, \ and\ \bibinfo {author} {\bibfnamefont {Y.~E.}\ \bibnamefont {Krasik}},\ }\bibfield  {title} {\enquote {\bibinfo {title} {High-current diode with ferroelectric plasma source-assisted hollow anode},}\ }\href@noop {} {\bibfield  {journal} {\bibinfo  {journal} {Journal of Applied Physics}\ }\textbf {\bibinfo {volume} {108}} (\bibinfo {year} {2010})}\BibitemShut {NoStop}%
\bibitem [{\citenamefont {Verreycken}\ \emph {et~al.}(2013)\citenamefont {Verreycken}, \citenamefont {Mensink}, \citenamefont {Van Der~Horst}, \citenamefont {Sadeghi},\ and\ \citenamefont {Bruggeman}}]{verreycken2013absolute}%
  \BibitemOpen
  \bibfield  {author} {\bibinfo {author} {\bibfnamefont {T.}~\bibnamefont {Verreycken}}, \bibinfo {author} {\bibfnamefont {R.}~\bibnamefont {Mensink}}, \bibinfo {author} {\bibfnamefont {R.}~\bibnamefont {Van Der~Horst}}, \bibinfo {author} {\bibfnamefont {N.}~\bibnamefont {Sadeghi}}, \ and\ \bibinfo {author} {\bibfnamefont {P.~J.}\ \bibnamefont {Bruggeman}},\ }\bibfield  {title} {\enquote {\bibinfo {title} {Absolute {OH} density measurements in the effluent of a cold atmospheric-pressure {Ar}--{H}$_2${O} {RF} plasma jet in air},}\ }\href@noop {} {\bibfield  {journal} {\bibinfo  {journal} {Plasma Sources Science and Technology}\ }\textbf {\bibinfo {volume} {22}},\ \bibinfo {pages} {055014} (\bibinfo {year} {2013})}\BibitemShut {NoStop}%
\bibitem [{\citenamefont {Miles}, \citenamefont {Lempert},\ and\ \citenamefont {Forkey}(2001)}]{miles2001laser}%
  \BibitemOpen
  \bibfield  {author} {\bibinfo {author} {\bibfnamefont {R.~B.}\ \bibnamefont {Miles}}, \bibinfo {author} {\bibfnamefont {W.~R.}\ \bibnamefont {Lempert}}, \ and\ \bibinfo {author} {\bibfnamefont {J.~N.}\ \bibnamefont {Forkey}},\ }\bibfield  {title} {\enquote {\bibinfo {title} {{Laser Rayleigh scattering}},}\ }\href@noop {} {\bibfield  {journal} {\bibinfo  {journal} {Measurement Science and Technology}\ }\textbf {\bibinfo {volume} {12}},\ \bibinfo {pages} {R33} (\bibinfo {year} {2001})}\BibitemShut {NoStop}%
\bibitem [{\citenamefont {Kauzmann}(2012)}]{kauzmann2012kinetic}%
  \BibitemOpen
  \bibfield  {author} {\bibinfo {author} {\bibfnamefont {W.}~\bibnamefont {Kauzmann}},\ }\href@noop {} {\emph {\bibinfo {title} {Kinetic theory of gases}}}\ (\bibinfo  {publisher} {Courier Corporation},\ \bibinfo {year} {2012})\BibitemShut {NoStop}%
\bibitem [{\citenamefont {Fiechtner}\ and\ \citenamefont {Gord}(2001)}]{fiechtner2001absorption}%
  \BibitemOpen
  \bibfield  {author} {\bibinfo {author} {\bibfnamefont {G.~J.}\ \bibnamefont {Fiechtner}}\ and\ \bibinfo {author} {\bibfnamefont {J.~R.}\ \bibnamefont {Gord}},\ }\bibfield  {title} {\enquote {\bibinfo {title} {Absorption and the dimensionless overlap integral for two-photon excitation},}\ }\href@noop {} {\bibfield  {journal} {\bibinfo  {journal} {Journal of Quantitative Spectroscopy and Radiative Transfer}\ }\textbf {\bibinfo {volume} {68}},\ \bibinfo {pages} {543--557} (\bibinfo {year} {2001})}\BibitemShut {NoStop}%
\bibitem [{\citenamefont {Luque}\ and\ \citenamefont {Crosley}(1996)}]{luque1996absolute}%
  \BibitemOpen
  \bibfield  {author} {\bibinfo {author} {\bibfnamefont {J.}~\bibnamefont {Luque}}\ and\ \bibinfo {author} {\bibfnamefont {D.}~\bibnamefont {Crosley}},\ }\bibfield  {title} {\enquote {\bibinfo {title} {{Absolute CH concentrations in low-pressure flames measured with laser-induced fluorescence}},}\ }\href@noop {} {\bibfield  {journal} {\bibinfo  {journal} {Applied Physics B}\ }\textbf {\bibinfo {volume} {63}},\ \bibinfo {pages} {91--98} (\bibinfo {year} {1996})}\BibitemShut {NoStop}%
\bibitem [{\citenamefont {Nemschokmichal}\ and\ \citenamefont {Meichsner}(2012)}]{nemschokmichal2012metastable}%
  \BibitemOpen
  \bibfield  {author} {\bibinfo {author} {\bibfnamefont {S.}~\bibnamefont {Nemschokmichal}}\ and\ \bibinfo {author} {\bibfnamefont {J.}~\bibnamefont {Meichsner}},\ }\bibfield  {title} {\enquote {\bibinfo {title} {{N$_2$(A$^{3}\Sigma^{+}_{u}$) metastable density in nitrogen barrier discharges: I. LIF diagnostics and absolute calibration by Rayleigh scattering}},}\ }\href@noop {} {\bibfield  {journal} {\bibinfo  {journal} {Plasma Sources Science and Technology}\ }\textbf {\bibinfo {volume} {22}},\ \bibinfo {pages} {015005} (\bibinfo {year} {2012})}\BibitemShut {NoStop}%
\bibitem [{\citenamefont {Kramida}\ \emph {et~al.}(2024)\citenamefont {Kramida}, \citenamefont {{Yu.~Ralchenko}}, \citenamefont {Reader},\ and\ \citenamefont {{and NIST ASD Team}}}]{NIST_ASD}%
  \BibitemOpen
  \bibfield  {author} {\bibinfo {author} {\bibfnamefont {A.}~\bibnamefont {Kramida}}, \bibinfo {author} {\bibnamefont {{Yu.~Ralchenko}}}, \bibinfo {author} {\bibfnamefont {J.}~\bibnamefont {Reader}}, \ and\ \bibinfo {author} {\bibnamefont {{and NIST ASD Team}}},\ }\href@noop {} {}\bibinfo {howpublished} {{NIST Atomic Spectra Database (ver. 5.11), [Online]. Available: {\tt{https://physics.nist.gov/asd}} [2024, June 1]. National Institute of Standards and Technology, Gaithersburg, MD.}} (\bibinfo {year} {2024})\BibitemShut {NoStop}%
\bibitem [{\citenamefont {Huber}(2013)}]{huber2013molecular}%
  \BibitemOpen
  \bibfield  {author} {\bibinfo {author} {\bibfnamefont {K.}~\bibnamefont {Huber}},\ }\href@noop {} {\emph {\bibinfo {title} {{Molecular spectra and molecular structure: IV. Constants of diatomic molecules}}}}\ (\bibinfo  {publisher} {Springer Science \& Business Media},\ \bibinfo {year} {2013})\BibitemShut {NoStop}%
\bibitem [{\citenamefont {Kaminski}\ and\ \citenamefont {Ewart}(1995)}]{kaminski1995absolute}%
  \BibitemOpen
  \bibfield  {author} {\bibinfo {author} {\bibfnamefont {C.}~\bibnamefont {Kaminski}}\ and\ \bibinfo {author} {\bibfnamefont {P.}~\bibnamefont {Ewart}},\ }\bibfield  {title} {\enquote {\bibinfo {title} {{Absolute concentration measurements of {C}$_2$ in a diamond CVD reactor by laser-induced fluorescence}},}\ }\href@noop {} {\bibfield  {journal} {\bibinfo  {journal} {Applied Physics B}\ }\textbf {\bibinfo {volume} {61}},\ \bibinfo {pages} {585--592} (\bibinfo {year} {1995})}\BibitemShut {NoStop}%
\bibitem [{\citenamefont {Luque}, \citenamefont {Juchmann},\ and\ \citenamefont {Jeffries}(1997{\natexlab{a}})}]{luque1997spatial}%
  \BibitemOpen
  \bibfield  {author} {\bibinfo {author} {\bibfnamefont {J.}~\bibnamefont {Luque}}, \bibinfo {author} {\bibfnamefont {W.}~\bibnamefont {Juchmann}}, \ and\ \bibinfo {author} {\bibfnamefont {J.}~\bibnamefont {Jeffries}},\ }\bibfield  {title} {\enquote {\bibinfo {title} {{Spatial density distributions of C$_2$, C$_3$, and CH radicals by laser-induced fluorescence in a diamond depositing DC-arcjet}},}\ }\href@noop {} {\bibfield  {journal} {\bibinfo  {journal} {Journal of applied physics}\ }\textbf {\bibinfo {volume} {82}},\ \bibinfo {pages} {2072--2081} (\bibinfo {year} {1997}{\natexlab{a}})}\BibitemShut {NoStop}%
\bibitem [{\citenamefont {Schmidt}(2015)}]{schmidt2015ultrashort}%
  \BibitemOpen
  \bibfield  {author} {\bibinfo {author} {\bibfnamefont {J.~B.}\ \bibnamefont {Schmidt}},\ }\emph {\bibinfo {title} {Ultrashort Two-Photon-Absorption Laser-Induced Fluorescence in Nanosecond-Duration, Repetitively Pulsed Discharges}},\ \href@noop {} {Ph.D. thesis},\ \bibinfo  {school} {The Ohio State University} (\bibinfo {year} {2015})\BibitemShut {NoStop}%
\bibitem [{\citenamefont {Stern}\ and\ \citenamefont {Johnson~III}(1975)}]{stern1975plasma}%
  \BibitemOpen
  \bibfield  {author} {\bibinfo {author} {\bibfnamefont {R.}~\bibnamefont {Stern}}\ and\ \bibinfo {author} {\bibfnamefont {J.}~\bibnamefont {Johnson~III}},\ }\bibfield  {title} {\enquote {\bibinfo {title} {Plasma ion diagnostics using resonant fluorescence},}\ }\href@noop {} {\bibfield  {journal} {\bibinfo  {journal} {Physical Review Letters}\ }\textbf {\bibinfo {volume} {34}},\ \bibinfo {pages} {1548} (\bibinfo {year} {1975})}\BibitemShut {NoStop}%
\bibitem [{\citenamefont {Vekselman}\ \emph {et~al.}(2009)\citenamefont {Vekselman}, \citenamefont {Gleizer}, \citenamefont {Yatom}, \citenamefont {Yarmolich}, \citenamefont {Gurovich}, \citenamefont {Bazalitski}, \citenamefont {Krasik},\ and\ \citenamefont {Bernshtam}}]{vekselman2009laser}%
  \BibitemOpen
  \bibfield  {author} {\bibinfo {author} {\bibfnamefont {V.}~\bibnamefont {Vekselman}}, \bibinfo {author} {\bibfnamefont {J.}~\bibnamefont {Gleizer}}, \bibinfo {author} {\bibfnamefont {S.}~\bibnamefont {Yatom}}, \bibinfo {author} {\bibfnamefont {D.}~\bibnamefont {Yarmolich}}, \bibinfo {author} {\bibfnamefont {V.~T.}\ \bibnamefont {Gurovich}}, \bibinfo {author} {\bibfnamefont {G.}~\bibnamefont {Bazalitski}}, \bibinfo {author} {\bibfnamefont {Y.~E.}\ \bibnamefont {Krasik}}, \ and\ \bibinfo {author} {\bibfnamefont {V.}~\bibnamefont {Bernshtam}},\ }\bibfield  {title} {\enquote {\bibinfo {title} {Laser induced fluorescence of the ferroelectric plasma source assisted hollow anode discharge},}\ }\href@noop {} {\bibfield  {journal} {\bibinfo  {journal} {Physics of Plasmas}\ }\textbf {\bibinfo {volume} {16}} (\bibinfo {year} {2009})}\BibitemShut {NoStop}%
\bibitem [{\citenamefont {Gavrilenko}\ \emph {et~al.}(2000)\citenamefont {Gavrilenko}, \citenamefont {Kim}, \citenamefont {Ikutake}, \citenamefont {Kim}, \citenamefont {Choi}, \citenamefont {Bowden},\ and\ \citenamefont {Muraoka}}]{gavrilenko2000measurement}%
  \BibitemOpen
  \bibfield  {author} {\bibinfo {author} {\bibfnamefont {V.}~\bibnamefont {Gavrilenko}}, \bibinfo {author} {\bibfnamefont {H.}~\bibnamefont {Kim}}, \bibinfo {author} {\bibfnamefont {T.}~\bibnamefont {Ikutake}}, \bibinfo {author} {\bibfnamefont {J.}~\bibnamefont {Kim}}, \bibinfo {author} {\bibfnamefont {Y.}~\bibnamefont {Choi}}, \bibinfo {author} {\bibfnamefont {M.}~\bibnamefont {Bowden}}, \ and\ \bibinfo {author} {\bibfnamefont {K.}~\bibnamefont {Muraoka}},\ }\bibfield  {title} {\enquote {\bibinfo {title} {Measurement method for electric fields based on stark spectroscopy of argon atoms},}\ }\href@noop {} {\bibfield  {journal} {\bibinfo  {journal} {Physical Review E}\ }\textbf {\bibinfo {volume} {62}},\ \bibinfo {pages} {7201} (\bibinfo {year} {2000})}\BibitemShut {NoStop}%
\bibitem [{\citenamefont {Ngom}\ \emph {et~al.}(2008)\citenamefont {Ngom}, \citenamefont {Smith}, \citenamefont {Huang},\ and\ \citenamefont {Gallimore}}]{ngom2008numerical}%
  \BibitemOpen
  \bibfield  {author} {\bibinfo {author} {\bibfnamefont {B.~B.}\ \bibnamefont {Ngom}}, \bibinfo {author} {\bibfnamefont {T.~B.}\ \bibnamefont {Smith}}, \bibinfo {author} {\bibfnamefont {W.}~\bibnamefont {Huang}}, \ and\ \bibinfo {author} {\bibfnamefont {A.~D.}\ \bibnamefont {Gallimore}},\ }\bibfield  {title} {\enquote {\bibinfo {title} {Numerical simulation of the zeeman effect in neutral xenon from nir diode-laser spectroscopy},}\ }\href@noop {} {\bibfield  {journal} {\bibinfo  {journal} {Journal of Applied Physics}\ }\textbf {\bibinfo {volume} {104}} (\bibinfo {year} {2008})}\BibitemShut {NoStop}%
\bibitem [{\citenamefont {Pérez-Luna}\ \emph {et~al.}(2009)\citenamefont {Pérez-Luna}, \citenamefont {Hagelaar}, \citenamefont {Garrigues},\ and\ \citenamefont {Boeuf}}]{Pérez-Luna_2009}%
  \BibitemOpen
  \bibfield  {author} {\bibinfo {author} {\bibfnamefont {J.}~\bibnamefont {Pérez-Luna}}, \bibinfo {author} {\bibfnamefont {G.~J.~M.}\ \bibnamefont {Hagelaar}}, \bibinfo {author} {\bibfnamefont {L.}~\bibnamefont {Garrigues}}, \ and\ \bibinfo {author} {\bibfnamefont {J.~P.}\ \bibnamefont {Boeuf}},\ }\bibfield  {title} {\enquote {\bibinfo {title} {Method to obtain the electric field and the ionization frequency from laser induced fluorescence measurements},}\ }\href {\doibase 10.1088/0963-0252/18/3/034008} {\bibfield  {journal} {\bibinfo  {journal} {Plasma Sources Science and Technology}\ }\textbf {\bibinfo {volume} {18}},\ \bibinfo {pages} {034008} (\bibinfo {year} {2009})}\BibitemShut {NoStop}%
\bibitem [{\citenamefont {Cohen-Tannoudji}, \citenamefont {Diu},\ and\ \citenamefont {Lalo{\"e}}(1977)}]{cohen1977quantum}%
  \BibitemOpen
  \bibfield  {author} {\bibinfo {author} {\bibfnamefont {C.}~\bibnamefont {Cohen-Tannoudji}}, \bibinfo {author} {\bibfnamefont {B.}~\bibnamefont {Diu}}, \ and\ \bibinfo {author} {\bibfnamefont {F.}~\bibnamefont {Lalo{\"e}}},\ }\href@noop {} {\emph {\bibinfo {title} {Quantum Mechanics Volume 1}}}\ (\bibinfo  {publisher} {Hermann},\ \bibinfo {year} {1977})\BibitemShut {NoStop}%
\bibitem [{\citenamefont {Romadanov}\ \emph {et~al.}(2018)\citenamefont {Romadanov}, \citenamefont {Raitses}, \citenamefont {Diallo}, \citenamefont {Hara}, \citenamefont {Kaganovich},\ and\ \citenamefont {Smolyakov}}]{romadanov2018limitations}%
  \BibitemOpen
  \bibfield  {author} {\bibinfo {author} {\bibfnamefont {I.}~\bibnamefont {Romadanov}}, \bibinfo {author} {\bibfnamefont {Y.}~\bibnamefont {Raitses}}, \bibinfo {author} {\bibfnamefont {A.}~\bibnamefont {Diallo}}, \bibinfo {author} {\bibfnamefont {K.}~\bibnamefont {Hara}}, \bibinfo {author} {\bibfnamefont {I.}~\bibnamefont {Kaganovich}}, \ and\ \bibinfo {author} {\bibfnamefont {A.}~\bibnamefont {Smolyakov}},\ }\bibfield  {title} {\enquote {\bibinfo {title} {On limitations of laser-induced fluorescence diagnostics for xenon ion velocity distribution function measurements in hall thrusters},}\ }\href@noop {} {\bibfield  {journal} {\bibinfo  {journal} {Physics of Plasmas}\ }\textbf {\bibinfo {volume} {25}} (\bibinfo {year} {2018})}\BibitemShut {NoStop}%
\bibitem [{\citenamefont {Mazouffre}(2012)}]{Mazouffre_2013}%
  \BibitemOpen
  \bibfield  {author} {\bibinfo {author} {\bibfnamefont {S.}~\bibnamefont {Mazouffre}},\ }\bibfield  {title} {\enquote {\bibinfo {title} {Laser-induced fluorescence diagnostics of the cross-field discharge of hall thrusters},}\ }\href {\doibase 10.1088/0963-0252/22/1/013001} {\bibfield  {journal} {\bibinfo  {journal} {Plasma Sources Science and Technology}\ }\textbf {\bibinfo {volume} {22}},\ \bibinfo {pages} {013001} (\bibinfo {year} {2012})}\BibitemShut {NoStop}%
\bibitem [{\citenamefont {Chaplin}\ \emph {et~al.}(2020)\citenamefont {Chaplin}, \citenamefont {Lobbia}, \citenamefont {Lopez~Ortega}, \citenamefont {Mikellides}, \citenamefont {Hofer}, \citenamefont {Polk},\ and\ \citenamefont {Friss}}]{chaplin2020time}%
  \BibitemOpen
  \bibfield  {author} {\bibinfo {author} {\bibfnamefont {V.}~\bibnamefont {Chaplin}}, \bibinfo {author} {\bibfnamefont {R.}~\bibnamefont {Lobbia}}, \bibinfo {author} {\bibfnamefont {A.}~\bibnamefont {Lopez~Ortega}}, \bibinfo {author} {\bibfnamefont {I.}~\bibnamefont {Mikellides}}, \bibinfo {author} {\bibfnamefont {R.}~\bibnamefont {Hofer}}, \bibinfo {author} {\bibfnamefont {J.}~\bibnamefont {Polk}}, \ and\ \bibinfo {author} {\bibfnamefont {A.}~\bibnamefont {Friss}},\ }\bibfield  {title} {\enquote {\bibinfo {title} {Time-resolved ion velocity measurements in a high-power hall thruster using laser-induced fluorescence with transfer function averaging},}\ }\href@noop {} {\bibfield  {journal} {\bibinfo  {journal} {Applied Physics Letters}\ }\textbf {\bibinfo {volume} {116}} (\bibinfo {year} {2020})}\BibitemShut {NoStop}%
\bibitem [{\citenamefont {Vaudolon}, \citenamefont {Khiar},\ and\ \citenamefont {Mazouffre}(2014)}]{vaudolon2014time}%
  \BibitemOpen
  \bibfield  {author} {\bibinfo {author} {\bibfnamefont {J.}~\bibnamefont {Vaudolon}}, \bibinfo {author} {\bibfnamefont {B.}~\bibnamefont {Khiar}}, \ and\ \bibinfo {author} {\bibfnamefont {S.}~\bibnamefont {Mazouffre}},\ }\bibfield  {title} {\enquote {\bibinfo {title} {Time evolution of the electric field in a hall thruster},}\ }\href@noop {} {\bibfield  {journal} {\bibinfo  {journal} {Plasma Sources Science and Technology}\ }\textbf {\bibinfo {volume} {23}},\ \bibinfo {pages} {022002} (\bibinfo {year} {2014})}\BibitemShut {NoStop}%
\bibitem [{\citenamefont {Young}\ \emph {et~al.}(2018)\citenamefont {Young}, \citenamefont {Fabris}, \citenamefont {MacDonald-Tenenbaum}, \citenamefont {Hargus},\ and\ \citenamefont {Cappelli}}]{Young_2018}%
  \BibitemOpen
  \bibfield  {author} {\bibinfo {author} {\bibfnamefont {C.~V.}\ \bibnamefont {Young}}, \bibinfo {author} {\bibfnamefont {A.~L.}\ \bibnamefont {Fabris}}, \bibinfo {author} {\bibfnamefont {N.~A.}\ \bibnamefont {MacDonald-Tenenbaum}}, \bibinfo {author} {\bibfnamefont {W.~A.}\ \bibnamefont {Hargus}}, \ and\ \bibinfo {author} {\bibfnamefont {M.~A.}\ \bibnamefont {Cappelli}},\ }\bibfield  {title} {\enquote {\bibinfo {title} {Time-resolved laser-induced fluorescence diagnostics for electric propulsion and their application to breathing mode dynamics},}\ }\href {\doibase 10.1088/1361-6595/aade42} {\bibfield  {journal} {\bibinfo  {journal} {Plasma Sources Science and Technology}\ }\textbf {\bibinfo {volume} {27}},\ \bibinfo {pages} {094004} (\bibinfo {year} {2018})}\BibitemShut {NoStop}%
\bibitem [{\citenamefont {Aramaki}\ \emph {et~al.}(2009)\citenamefont {Aramaki}, \citenamefont {Ogiwara}, \citenamefont {Etoh}, \citenamefont {Yoshimura},\ and\ \citenamefont {Tanaka}}]{aramaki2009high}%
  \BibitemOpen
  \bibfield  {author} {\bibinfo {author} {\bibfnamefont {M.}~\bibnamefont {Aramaki}}, \bibinfo {author} {\bibfnamefont {K.}~\bibnamefont {Ogiwara}}, \bibinfo {author} {\bibfnamefont {S.}~\bibnamefont {Etoh}}, \bibinfo {author} {\bibfnamefont {S.}~\bibnamefont {Yoshimura}}, \ and\ \bibinfo {author} {\bibfnamefont {M.~Y.}\ \bibnamefont {Tanaka}},\ }\bibfield  {title} {\enquote {\bibinfo {title} {{High resolution laser induced fluorescence Doppler velocimetry utilizing saturated absorption spectroscopy}},}\ }\href@noop {} {\bibfield  {journal} {\bibinfo  {journal} {Review of Scientific Instruments}\ }\textbf {\bibinfo {volume} {80}} (\bibinfo {year} {2009})}\BibitemShut {NoStop}%
\bibitem [{\citenamefont {Boivin}\ and\ \citenamefont {Scime}(2003)}]{boivin2003laser}%
  \BibitemOpen
  \bibfield  {author} {\bibinfo {author} {\bibfnamefont {R.}~\bibnamefont {Boivin}}\ and\ \bibinfo {author} {\bibfnamefont {E.}~\bibnamefont {Scime}},\ }\bibfield  {title} {\enquote {\bibinfo {title} {Laser induced fluorescence in {Ar} and {He} plasmas with a tunable diode laser},}\ }\href@noop {} {\bibfield  {journal} {\bibinfo  {journal} {Review of scientific instruments}\ }\textbf {\bibinfo {volume} {74}},\ \bibinfo {pages} {4352--4360} (\bibinfo {year} {2003})}\BibitemShut {NoStop}%
\bibitem [{\citenamefont {Huang}, \citenamefont {Smith},\ and\ \citenamefont {Gallimore}(2009)}]{huang2009obtaining}%
  \BibitemOpen
  \bibfield  {author} {\bibinfo {author} {\bibfnamefont {W.}~\bibnamefont {Huang}}, \bibinfo {author} {\bibfnamefont {T.}~\bibnamefont {Smith}}, \ and\ \bibinfo {author} {\bibfnamefont {A.}~\bibnamefont {Gallimore}},\ }\bibfield  {title} {\enquote {\bibinfo {title} {Obtaining velocity distribution using a xenon ion line with unknown hyperfine constants},}\ }in\ \href@noop {} {\emph {\bibinfo {booktitle} {40th AIAA Plasmadynamics and Lasers Conference}}}\ (\bibinfo {year} {2009})\ p.\ \bibinfo {pages} {4226}\BibitemShut {NoStop}%
\bibitem [{\citenamefont {Kunze}(2009)}]{kunze2009introduction}%
  \BibitemOpen
  \bibfield  {author} {\bibinfo {author} {\bibfnamefont {H.-J.}\ \bibnamefont {Kunze}},\ }\href@noop {} {\emph {\bibinfo {title} {Introduction to plasma spectroscopy}}},\ Vol.~\bibinfo {volume} {56}\ (\bibinfo  {publisher} {Springer Science \& Business Media},\ \bibinfo {year} {2009})\BibitemShut {NoStop}%
\bibitem [{\citenamefont {H{\"u}bner}\ \emph {et~al.}(2013)\citenamefont {H{\"u}bner}, \citenamefont {Sadeghi}, \citenamefont {Carbone},\ and\ \citenamefont {Van Der~Mullen}}]{hubner2013density}%
  \BibitemOpen
  \bibfield  {author} {\bibinfo {author} {\bibfnamefont {S.}~\bibnamefont {H{\"u}bner}}, \bibinfo {author} {\bibfnamefont {N.}~\bibnamefont {Sadeghi}}, \bibinfo {author} {\bibfnamefont {E.}~\bibnamefont {Carbone}}, \ and\ \bibinfo {author} {\bibfnamefont {J.}~\bibnamefont {Van Der~Mullen}},\ }\bibfield  {title} {\enquote {\bibinfo {title} {Density of atoms in ar*(3p54s) states and gas temperatures in an argon surfatron plasma measured by tunable laser spectroscopy},}\ }\href@noop {} {\bibfield  {journal} {\bibinfo  {journal} {Journal of Applied Physics}\ }\textbf {\bibinfo {volume} {113}} (\bibinfo {year} {2013})}\BibitemShut {NoStop}%
\bibitem [{\citenamefont {Scime}\ \emph {et~al.}(2005)\citenamefont {Scime}, \citenamefont {Biloiu}, \citenamefont {Compton}, \citenamefont {Doss}, \citenamefont {Venture}, \citenamefont {Heard}, \citenamefont {Choueiri},\ and\ \citenamefont {Spektor}}]{scime2005laser}%
  \BibitemOpen
  \bibfield  {author} {\bibinfo {author} {\bibfnamefont {E.}~\bibnamefont {Scime}}, \bibinfo {author} {\bibfnamefont {C.}~\bibnamefont {Biloiu}}, \bibinfo {author} {\bibfnamefont {C.}~\bibnamefont {Compton}}, \bibinfo {author} {\bibfnamefont {F.}~\bibnamefont {Doss}}, \bibinfo {author} {\bibfnamefont {D.}~\bibnamefont {Venture}}, \bibinfo {author} {\bibfnamefont {J.}~\bibnamefont {Heard}}, \bibinfo {author} {\bibfnamefont {E.}~\bibnamefont {Choueiri}}, \ and\ \bibinfo {author} {\bibfnamefont {R.}~\bibnamefont {Spektor}},\ }\bibfield  {title} {\enquote {\bibinfo {title} {Laser induced fluorescence in a pulsed argon plasma},}\ }\href@noop {} {\bibfield  {journal} {\bibinfo  {journal} {Review of Scientific Instruments}\ }\textbf {\bibinfo {volume} {76}} (\bibinfo {year} {2005})}\BibitemShut {NoStop}%
\bibitem [{\citenamefont {Pelissier}\ and\ \citenamefont {Sadeghi}(1996)}]{pelissier1996time}%
  \BibitemOpen
  \bibfield  {author} {\bibinfo {author} {\bibfnamefont {B.}~\bibnamefont {Pelissier}}\ and\ \bibinfo {author} {\bibfnamefont {N.}~\bibnamefont {Sadeghi}},\ }\bibfield  {title} {\enquote {\bibinfo {title} {Time-resolved pulse-counting lock-in detection of laser induced fluorescence in the presence of a strong background emission},}\ }\href@noop {} {\bibfield  {journal} {\bibinfo  {journal} {Review of scientific instruments}\ }\textbf {\bibinfo {volume} {67}},\ \bibinfo {pages} {3405--3410} (\bibinfo {year} {1996})}\BibitemShut {NoStop}%
\bibitem [{\citenamefont {Vaudolon}, \citenamefont {Balika},\ and\ \citenamefont {Mazouffre}(2013)}]{vaudolon2013photon}%
  \BibitemOpen
  \bibfield  {author} {\bibinfo {author} {\bibfnamefont {J.}~\bibnamefont {Vaudolon}}, \bibinfo {author} {\bibfnamefont {L.}~\bibnamefont {Balika}}, \ and\ \bibinfo {author} {\bibfnamefont {S.}~\bibnamefont {Mazouffre}},\ }\bibfield  {title} {\enquote {\bibinfo {title} {Photon counting technique applied to time-resolved laser-induced fluorescence measurements on a stabilized discharge},}\ }\href@noop {} {\bibfield  {journal} {\bibinfo  {journal} {Review of Scientific Instruments}\ }\textbf {\bibinfo {volume} {84}} (\bibinfo {year} {2013})}\BibitemShut {NoStop}%
\bibitem [{\citenamefont {Jacobs}\ \emph {et~al.}(2007)\citenamefont {Jacobs}, \citenamefont {Gekelman}, \citenamefont {Pribyl}, \citenamefont {Barnes},\ and\ \citenamefont {Kilgore}}]{jacobs2007laser}%
  \BibitemOpen
  \bibfield  {author} {\bibinfo {author} {\bibfnamefont {B.}~\bibnamefont {Jacobs}}, \bibinfo {author} {\bibfnamefont {W.}~\bibnamefont {Gekelman}}, \bibinfo {author} {\bibfnamefont {P.}~\bibnamefont {Pribyl}}, \bibinfo {author} {\bibfnamefont {M.}~\bibnamefont {Barnes}}, \ and\ \bibinfo {author} {\bibfnamefont {M.}~\bibnamefont {Kilgore}},\ }\bibfield  {title} {\enquote {\bibinfo {title} {Laser-induced fluorescence measurements in an inductively coupled plasma reactor},}\ }\href@noop {} {\bibfield  {journal} {\bibinfo  {journal} {Applied Physics Letters}\ }\textbf {\bibinfo {volume} {91}} (\bibinfo {year} {2007})}\BibitemShut {NoStop}%
\bibitem [{\citenamefont {Romadanov}, \citenamefont {Raitses},\ and\ \citenamefont {Smolyakov}(2024)}]{romadanov2024wavelength}%
  \BibitemOpen
  \bibfield  {author} {\bibinfo {author} {\bibfnamefont {I.}~\bibnamefont {Romadanov}}, \bibinfo {author} {\bibfnamefont {Y.}~\bibnamefont {Raitses}}, \ and\ \bibinfo {author} {\bibfnamefont {A.}~\bibnamefont {Smolyakov}},\ }\bibfield  {title} {\enquote {\bibinfo {title} {Wavelength modulation laser-induced fluorescence for plasma characterization},}\ }\href@noop {} {\bibfield  {journal} {\bibinfo  {journal} {arXiv preprint arXiv:2403.11045}\ } (\bibinfo {year} {2024})}\BibitemShut {NoStop}%
\bibitem [{\citenamefont {Dixit}\ and\ \citenamefont {Ram}(1985)}]{dixit1985quantitative}%
  \BibitemOpen
  \bibfield  {author} {\bibinfo {author} {\bibfnamefont {L.}~\bibnamefont {Dixit}}\ and\ \bibinfo {author} {\bibfnamefont {S.}~\bibnamefont {Ram}},\ }\bibfield  {title} {\enquote {\bibinfo {title} {Quantitative analysis by derivative electronic spectroscopy},}\ }\href@noop {} {\bibfield  {journal} {\bibinfo  {journal} {Applied Spectroscopy Reviews}\ }\textbf {\bibinfo {volume} {21}},\ \bibinfo {pages} {311--418} (\bibinfo {year} {1985})}\BibitemShut {NoStop}%
\bibitem [{\citenamefont {Amos}\ and\ \citenamefont {White}(2003)}]{amos2003confocal}%
  \BibitemOpen
  \bibfield  {author} {\bibinfo {author} {\bibfnamefont {W.}~\bibnamefont {Amos}}\ and\ \bibinfo {author} {\bibfnamefont {J.}~\bibnamefont {White}},\ }\bibfield  {title} {\enquote {\bibinfo {title} {How the confocal laser scanning microscope entered biological research},}\ }\href@noop {} {\bibfield  {journal} {\bibinfo  {journal} {Biology of the Cell}\ }\textbf {\bibinfo {volume} {95}},\ \bibinfo {pages} {335--342} (\bibinfo {year} {2003})}\BibitemShut {NoStop}%
\bibitem [{\citenamefont {VanDervort}\ \emph {et~al.}(2014)\citenamefont {VanDervort}, \citenamefont {Elliott}, \citenamefont {McCarren}, \citenamefont {McKee}, \citenamefont {Soderholm}, \citenamefont {Sears},\ and\ \citenamefont {Scime}}]{vandervort2014optimization}%
  \BibitemOpen
  \bibfield  {author} {\bibinfo {author} {\bibfnamefont {R.}~\bibnamefont {VanDervort}}, \bibinfo {author} {\bibfnamefont {D.}~\bibnamefont {Elliott}}, \bibinfo {author} {\bibfnamefont {D.}~\bibnamefont {McCarren}}, \bibinfo {author} {\bibfnamefont {J.}~\bibnamefont {McKee}}, \bibinfo {author} {\bibfnamefont {M.}~\bibnamefont {Soderholm}}, \bibinfo {author} {\bibfnamefont {S.}~\bibnamefont {Sears}}, \ and\ \bibinfo {author} {\bibfnamefont {E.}~\bibnamefont {Scime}},\ }\bibfield  {title} {\enquote {\bibinfo {title} {Optimization of confocal laser induced fluorescence in a plasma},}\ }\href@noop {} {\bibfield  {journal} {\bibinfo  {journal} {Review of Scientific Instruments}\ }\textbf {\bibinfo {volume} {85}} (\bibinfo {year} {2014})}\BibitemShut {NoStop}%
\bibitem [{\citenamefont {Thompson}\ \emph {et~al.}(2017)\citenamefont {Thompson}, \citenamefont {Henriquez}, \citenamefont {Scime},\ and\ \citenamefont {Good}}]{thompson2017confocal}%
  \BibitemOpen
  \bibfield  {author} {\bibinfo {author} {\bibfnamefont {D.~S.}\ \bibnamefont {Thompson}}, \bibinfo {author} {\bibfnamefont {M.~F.}\ \bibnamefont {Henriquez}}, \bibinfo {author} {\bibfnamefont {E.~E.}\ \bibnamefont {Scime}}, \ and\ \bibinfo {author} {\bibfnamefont {T.~N.}\ \bibnamefont {Good}},\ }\bibfield  {title} {\enquote {\bibinfo {title} {Confocal laser induced fluorescence with comparable spatial localization to the conventional method},}\ }\href@noop {} {\bibfield  {journal} {\bibinfo  {journal} {Review of Scientific Instruments}\ }\textbf {\bibinfo {volume} {88}} (\bibinfo {year} {2017})}\BibitemShut {NoStop}%
\bibitem [{\citenamefont {Kajiwara}\ \emph {et~al.}(1990)\citenamefont {Kajiwara}, \citenamefont {Takeda}, \citenamefont {Muraoka}, \citenamefont {Okada}, \citenamefont {Maeda},\ and\ \citenamefont {Akazaki}}]{Kajiwara_1990}%
  \BibitemOpen
  \bibfield  {author} {\bibinfo {author} {\bibfnamefont {T.}~\bibnamefont {Kajiwara}}, \bibinfo {author} {\bibfnamefont {K.}~\bibnamefont {Takeda}}, \bibinfo {author} {\bibfnamefont {K.}~\bibnamefont {Muraoka}}, \bibinfo {author} {\bibfnamefont {T.}~\bibnamefont {Okada}}, \bibinfo {author} {\bibfnamefont {M.}~\bibnamefont {Maeda}}, \ and\ \bibinfo {author} {\bibfnamefont {M.}~\bibnamefont {Akazaki}},\ }\bibfield  {title} {\enquote {\bibinfo {title} {Coaxial laser fluorescence system by two-photon excitation for atomic hydrogen detection in high-temperature plasmas},}\ }\href {\doibase 10.1143/JJAP.29.L826} {\bibfield  {journal} {\bibinfo  {journal} {Japanese Journal of Applied Physics}\ }\textbf {\bibinfo {volume} {29}},\ \bibinfo {pages} {L826} (\bibinfo {year} {1990})}\BibitemShut {NoStop}%
\bibitem [{\citenamefont {Caron}\ \emph {et~al.}(2023)\citenamefont {Caron}, \citenamefont {John}, \citenamefont {Scime},\ and\ \citenamefont {Steinberger}}]{caron2023ion}%
  \BibitemOpen
  \bibfield  {author} {\bibinfo {author} {\bibfnamefont {D.}~\bibnamefont {Caron}}, \bibinfo {author} {\bibfnamefont {R.}~\bibnamefont {John}}, \bibinfo {author} {\bibfnamefont {E.}~\bibnamefont {Scime}}, \ and\ \bibinfo {author} {\bibfnamefont {T.}~\bibnamefont {Steinberger}},\ }\bibfield  {title} {\enquote {\bibinfo {title} {Ion velocity distribution functions across a plasma meniscus},}\ }\href@noop {} {\bibfield  {journal} {\bibinfo  {journal} {Journal of Vacuum Science \& Technology A}\ }\textbf {\bibinfo {volume} {41}} (\bibinfo {year} {2023})}\BibitemShut {NoStop}%
\bibitem [{\citenamefont {Romadanov}\ and\ \citenamefont {Raitses}(2023)}]{romadanov2023confocal}%
  \BibitemOpen
  \bibfield  {author} {\bibinfo {author} {\bibfnamefont {I.}~\bibnamefont {Romadanov}}\ and\ \bibinfo {author} {\bibfnamefont {Y.}~\bibnamefont {Raitses}},\ }\bibfield  {title} {\enquote {\bibinfo {title} {A confocal laser-induced fluorescence diagnostic with an annular laser beam},}\ }\href@noop {} {\bibfield  {journal} {\bibinfo  {journal} {Review of Scientific Instruments}\ }\textbf {\bibinfo {volume} {94}} (\bibinfo {year} {2023})}\BibitemShut {NoStop}%
\bibitem [{\citenamefont {Lee}, \citenamefont {Efremov},\ and\ \citenamefont {Kwon}(2018)}]{lee2018relationships}%
  \BibitemOpen
  \bibfield  {author} {\bibinfo {author} {\bibfnamefont {J.}~\bibnamefont {Lee}}, \bibinfo {author} {\bibfnamefont {A.}~\bibnamefont {Efremov}}, \ and\ \bibinfo {author} {\bibfnamefont {K.-H.}\ \bibnamefont {Kwon}},\ }\bibfield  {title} {\enquote {\bibinfo {title} {On the relationships between plasma chemistry, etching kinetics and etching residues in {CF}$_4$+ {C}$_4${F}$_8$+ {Ar} and {CF}$_4$+ {CH}$_2${F}$_2$ + {Ar} plasmas with various {CF}$_4$/{C}$_4${F}$_8$ and {CF}$_4$/{CH}$_2${F}$_2$ mixing ratios},}\ }\href@noop {} {\bibfield  {journal} {\bibinfo  {journal} {Vacuum}\ }\textbf {\bibinfo {volume} {148}},\ \bibinfo {pages} {214--223} (\bibinfo {year} {2018})}\BibitemShut {NoStop}%
\bibitem [{\citenamefont {Foad}\ \emph {et~al.}(1992)\citenamefont {Foad}, \citenamefont {Wilkinson}, \citenamefont {Dunscomb},\ and\ \citenamefont {Williams}}]{foad1992ch4}%
  \BibitemOpen
  \bibfield  {author} {\bibinfo {author} {\bibfnamefont {M.}~\bibnamefont {Foad}}, \bibinfo {author} {\bibfnamefont {C.}~\bibnamefont {Wilkinson}}, \bibinfo {author} {\bibfnamefont {C.}~\bibnamefont {Dunscomb}}, \ and\ \bibinfo {author} {\bibfnamefont {R.}~\bibnamefont {Williams}},\ }\bibfield  {title} {\enquote {\bibinfo {title} {{CH}$_4$/{H}$_2$: A universal reactive ion etch for {II}-{VI} semiconductors?}}\ }\href@noop {} {\bibfield  {journal} {\bibinfo  {journal} {Applied physics letters}\ }\textbf {\bibinfo {volume} {60}},\ \bibinfo {pages} {2531--2533} (\bibinfo {year} {1992})}\BibitemShut {NoStop}%
\bibitem [{\citenamefont {Bell}\ \emph {et~al.}(1994)\citenamefont {Bell}, \citenamefont {Joubert}, \citenamefont {Oehrlein}, \citenamefont {Zhang},\ and\ \citenamefont {Vender}}]{bell1994investigation}%
  \BibitemOpen
  \bibfield  {author} {\bibinfo {author} {\bibfnamefont {F.}~\bibnamefont {Bell}}, \bibinfo {author} {\bibfnamefont {O.}~\bibnamefont {Joubert}}, \bibinfo {author} {\bibfnamefont {G.}~\bibnamefont {Oehrlein}}, \bibinfo {author} {\bibfnamefont {Y.}~\bibnamefont {Zhang}}, \ and\ \bibinfo {author} {\bibfnamefont {D.}~\bibnamefont {Vender}},\ }\bibfield  {title} {\enquote {\bibinfo {title} {Investigation of selective {SiO}$_2$-to-{Si} etching in an inductively coupled high-density plasma using fluorocarbon gases},}\ }\href@noop {} {\bibfield  {journal} {\bibinfo  {journal} {Journal of Vacuum Science \& Technology A: Vacuum, Surfaces, and Films}\ }\textbf {\bibinfo {volume} {12}},\ \bibinfo {pages} {3095--3101} (\bibinfo {year} {1994})}\BibitemShut {NoStop}%
\bibitem [{\citenamefont {Li}\ \emph {et~al.}(2002)\citenamefont {Li}, \citenamefont {Hua}, \citenamefont {Ling}, \citenamefont {Oehrlein}, \citenamefont {Barela},\ and\ \citenamefont {Anderson}}]{li2002fluorocarbon}%
  \BibitemOpen
  \bibfield  {author} {\bibinfo {author} {\bibfnamefont {X.}~\bibnamefont {Li}}, \bibinfo {author} {\bibfnamefont {X.}~\bibnamefont {Hua}}, \bibinfo {author} {\bibfnamefont {L.}~\bibnamefont {Ling}}, \bibinfo {author} {\bibfnamefont {G.~S.}\ \bibnamefont {Oehrlein}}, \bibinfo {author} {\bibfnamefont {M.}~\bibnamefont {Barela}}, \ and\ \bibinfo {author} {\bibfnamefont {H.~M.}\ \bibnamefont {Anderson}},\ }\bibfield  {title} {\enquote {\bibinfo {title} {Fluorocarbon-based plasma etching of {SiO}$_2$: Comparison of {C}$_4${F}$_6$/{Ar} and {C}$_4${F}$_8$/{Ar} discharges},}\ }\href@noop {} {\bibfield  {journal} {\bibinfo  {journal} {Journal of Vacuum Science \& Technology A: Vacuum, Surfaces, and Films}\ }\textbf {\bibinfo {volume} {20}},\ \bibinfo {pages} {2052--2061} (\bibinfo {year} {2002})}\BibitemShut {NoStop}%
\bibitem [{\citenamefont {Cho}\ \emph {et~al.}(2012)\citenamefont {Cho}, \citenamefont {Kim}, \citenamefont {Lee}, \citenamefont {Moon},\ and\ \citenamefont {Chae}}]{cho2012angular}%
  \BibitemOpen
  \bibfield  {author} {\bibinfo {author} {\bibfnamefont {S.-W.}\ \bibnamefont {Cho}}, \bibinfo {author} {\bibfnamefont {C.-K.}\ \bibnamefont {Kim}}, \bibinfo {author} {\bibfnamefont {J.-K.}\ \bibnamefont {Lee}}, \bibinfo {author} {\bibfnamefont {S.~H.}\ \bibnamefont {Moon}}, \ and\ \bibinfo {author} {\bibfnamefont {H.}~\bibnamefont {Chae}},\ }\bibfield  {title} {\enquote {\bibinfo {title} {Angular dependences of {SiO}$_2$ etch rates in {C}$_4${F}$_6$/{O}$_2$/{Ar} and {C}$_4${F}$_6$/{CH}$_2${F}$_2$/{O}$_2$/{Ar} plasmas},}\ }\href@noop {} {\bibfield  {journal} {\bibinfo  {journal} {Journal of Vacuum Science \& Technology A}\ }\textbf {\bibinfo {volume} {30}} (\bibinfo {year} {2012})}\BibitemShut {NoStop}%
\bibitem [{\citenamefont {Rhee}\ \emph {et~al.}(2008)\citenamefont {Rhee}, \citenamefont {Kwon}, \citenamefont {Kim}, \citenamefont {Kim}, \citenamefont {Yoo},\ and\ \citenamefont {Kim}}]{rhee2008comparison}%
  \BibitemOpen
  \bibfield  {author} {\bibinfo {author} {\bibfnamefont {H.}~\bibnamefont {Rhee}}, \bibinfo {author} {\bibfnamefont {H.}~\bibnamefont {Kwon}}, \bibinfo {author} {\bibfnamefont {C.-K.}\ \bibnamefont {Kim}}, \bibinfo {author} {\bibfnamefont {H.}~\bibnamefont {Kim}}, \bibinfo {author} {\bibfnamefont {J.}~\bibnamefont {Yoo}}, \ and\ \bibinfo {author} {\bibfnamefont {Y.~W.}\ \bibnamefont {Kim}},\ }\bibfield  {title} {\enquote {\bibinfo {title} {Comparison of deep silicon etching using {SF}$_6$/{C}$_4${F}$_8$ and {SF}$_6$/{C}$_4${F}$_6$ plasmas in the bosch process},}\ }\href@noop {} {\bibfield  {journal} {\bibinfo  {journal} {Journal of Vacuum Science \& Technology B: Microelectronics and Nanometer Structures Processing, Measurement, and Phenomena}\ }\textbf {\bibinfo {volume} {26}},\ \bibinfo {pages} {576--581} (\bibinfo {year} {2008})}\BibitemShut {NoStop}%
\bibitem [{\citenamefont {d’Agostino}\ and\ \citenamefont {Flamm}(1981)}]{d1981plasma}%
  \BibitemOpen
  \bibfield  {author} {\bibinfo {author} {\bibfnamefont {R.}~\bibnamefont {d’Agostino}}\ and\ \bibinfo {author} {\bibfnamefont {D.~L.}\ \bibnamefont {Flamm}},\ }\bibfield  {title} {\enquote {\bibinfo {title} {Plasma etching of {Si} and {SiO}$_2$ in {SF}$_6$--{O}$_2$ mixtures},}\ }\href@noop {} {\bibfield  {journal} {\bibinfo  {journal} {Journal of Applied Physics}\ }\textbf {\bibinfo {volume} {52}},\ \bibinfo {pages} {162--167} (\bibinfo {year} {1981})}\BibitemShut {NoStop}%
\bibitem [{\citenamefont {Efremov}, \citenamefont {Betelin},\ and\ \citenamefont {Kwon}(2020)}]{efremov2020kinetics}%
  \BibitemOpen
  \bibfield  {author} {\bibinfo {author} {\bibfnamefont {A.}~\bibnamefont {Efremov}}, \bibinfo {author} {\bibfnamefont {V.}~\bibnamefont {Betelin}}, \ and\ \bibinfo {author} {\bibfnamefont {K.-H.}\ \bibnamefont {Kwon}},\ }\bibfield  {title} {\enquote {\bibinfo {title} {Kinetics and mechanisms of reactive-ion etching of {Si} and {SiO}$_2$ in a plasma of a mixture of {HBr} + {O}$_2$},}\ }\href@noop {} {\bibfield  {journal} {\bibinfo  {journal} {Russian Microelectronics}\ }\textbf {\bibinfo {volume} {49}},\ \bibinfo {pages} {379--384} (\bibinfo {year} {2020})}\BibitemShut {NoStop}%
\bibitem [{\citenamefont {Kim}\ \emph {et~al.}(2013)\citenamefont {Kim}, \citenamefont {Cho}, \citenamefont {Kim}, \citenamefont {Jhon}, \citenamefont {Min}, \citenamefont {Kim},\ and\ \citenamefont {Yeom}}]{kim2013study}%
  \BibitemOpen
  \bibfield  {author} {\bibinfo {author} {\bibfnamefont {J.~K.}\ \bibnamefont {Kim}}, \bibinfo {author} {\bibfnamefont {S.~I.}\ \bibnamefont {Cho}}, \bibinfo {author} {\bibfnamefont {N.~G.}\ \bibnamefont {Kim}}, \bibinfo {author} {\bibfnamefont {M.~S.}\ \bibnamefont {Jhon}}, \bibinfo {author} {\bibfnamefont {K.~S.}\ \bibnamefont {Min}}, \bibinfo {author} {\bibfnamefont {C.~K.}\ \bibnamefont {Kim}}, \ and\ \bibinfo {author} {\bibfnamefont {G.~Y.}\ \bibnamefont {Yeom}},\ }\bibfield  {title} {\enquote {\bibinfo {title} {Study on the etching characteristics of amorphous carbon layer in oxygen plasma with carbonyl sulfide},}\ }\href@noop {} {\bibfield  {journal} {\bibinfo  {journal} {Journal of Vacuum Science \& Technology A}\ }\textbf {\bibinfo {volume} {31}} (\bibinfo {year} {2013})}\BibitemShut {NoStop}%
\bibitem [{\citenamefont {Jiang}\ \emph {et~al.}(2004)\citenamefont {Jiang}, \citenamefont {Plank}, \citenamefont {Blauw}, \citenamefont {Cheung},\ and\ \citenamefont {van~der Drift}}]{jiang2004dry}%
  \BibitemOpen
  \bibfield  {author} {\bibinfo {author} {\bibfnamefont {L.}~\bibnamefont {Jiang}}, \bibinfo {author} {\bibfnamefont {N.}~\bibnamefont {Plank}}, \bibinfo {author} {\bibfnamefont {M.}~\bibnamefont {Blauw}}, \bibinfo {author} {\bibfnamefont {R.}~\bibnamefont {Cheung}}, \ and\ \bibinfo {author} {\bibfnamefont {E.}~\bibnamefont {van~der Drift}},\ }\bibfield  {title} {\enquote {\bibinfo {title} {Dry etching of {SiC} in inductively coupled {Cl}$_2$/{Ar} plasma},}\ }\href@noop {} {\bibfield  {journal} {\bibinfo  {journal} {Journal of Physics D: Applied Physics}\ }\textbf {\bibinfo {volume} {37}},\ \bibinfo {pages} {1809} (\bibinfo {year} {2004})}\BibitemShut {NoStop}%
\bibitem [{\citenamefont {Donnelly}\ \emph {et~al.}(1984)\citenamefont {Donnelly}, \citenamefont {Flamm}, \citenamefont {Dautremont-Smith},\ and\ \citenamefont {Werder}}]{donnelly1984anisotropic}%
  \BibitemOpen
  \bibfield  {author} {\bibinfo {author} {\bibfnamefont {V.~M.}\ \bibnamefont {Donnelly}}, \bibinfo {author} {\bibfnamefont {D.~L.}\ \bibnamefont {Flamm}}, \bibinfo {author} {\bibfnamefont {W.}~\bibnamefont {Dautremont-Smith}}, \ and\ \bibinfo {author} {\bibfnamefont {D.}~\bibnamefont {Werder}},\ }\bibfield  {title} {\enquote {\bibinfo {title} {Anisotropic etching of {SiO}$_2$ in low-frequency {CF}$_4$/{O}$_2$ and {NF}$_3$/{Ar} plasmas},}\ }\href@noop {} {\bibfield  {journal} {\bibinfo  {journal} {Journal of applied physics}\ }\textbf {\bibinfo {volume} {55}},\ \bibinfo {pages} {242--252} (\bibinfo {year} {1984})}\BibitemShut {NoStop}%
\bibitem [{\citenamefont {Nishino}, \citenamefont {Hayasaka},\ and\ \citenamefont {Okano}(1993)}]{nishino1993damage}%
  \BibitemOpen
  \bibfield  {author} {\bibinfo {author} {\bibfnamefont {H.}~\bibnamefont {Nishino}}, \bibinfo {author} {\bibfnamefont {N.}~\bibnamefont {Hayasaka}}, \ and\ \bibinfo {author} {\bibfnamefont {H.}~\bibnamefont {Okano}},\ }\bibfield  {title} {\enquote {\bibinfo {title} {Damage-free selective etching of {Si} native oxides using {NH}$_3$/{NF}$_3$ and {SF}$_6$/{H}$_2${O} down-flow etching},}\ }\href@noop {} {\bibfield  {journal} {\bibinfo  {journal} {Journal of applied physics}\ }\textbf {\bibinfo {volume} {74}},\ \bibinfo {pages} {1345--1348} (\bibinfo {year} {1993})}\BibitemShut {NoStop}%
\bibitem [{\citenamefont {Hefny}\ \emph {et~al.}(2016)\citenamefont {Hefny}, \citenamefont {Pattyn}, \citenamefont {Lukes},\ and\ \citenamefont {Benedikt}}]{hefny2016atmospheric}%
  \BibitemOpen
  \bibfield  {author} {\bibinfo {author} {\bibfnamefont {M.~M.}\ \bibnamefont {Hefny}}, \bibinfo {author} {\bibfnamefont {C.}~\bibnamefont {Pattyn}}, \bibinfo {author} {\bibfnamefont {P.}~\bibnamefont {Lukes}}, \ and\ \bibinfo {author} {\bibfnamefont {J.}~\bibnamefont {Benedikt}},\ }\bibfield  {title} {\enquote {\bibinfo {title} {Atmospheric plasma generates oxygen atoms as oxidizing species in aqueous solutions},}\ }\href@noop {} {\bibfield  {journal} {\bibinfo  {journal} {Journal of Physics D: Applied Physics}\ }\textbf {\bibinfo {volume} {49}},\ \bibinfo {pages} {404002} (\bibinfo {year} {2016})}\BibitemShut {NoStop}%
\bibitem [{\citenamefont {Yatom}(2023)}]{yatom2023diagnostics}%
  \BibitemOpen
  \bibfield  {author} {\bibinfo {author} {\bibfnamefont {S.}~\bibnamefont {Yatom}},\ }\bibfield  {title} {\enquote {\bibinfo {title} {Diagnostics of plasma--liquids systems: Challenges and their mitigation},}\ }\href@noop {} {\bibfield  {journal} {\bibinfo  {journal} {Physics of Plasmas}\ }\textbf {\bibinfo {volume} {30}} (\bibinfo {year} {2023})}\BibitemShut {NoStop}%
\bibitem [{\citenamefont {Colibaba-Evulet}, \citenamefont {Singhal},\ and\ \citenamefont {Glumac}(2000)}]{colibaba2000detection}%
  \BibitemOpen
  \bibfield  {author} {\bibinfo {author} {\bibfnamefont {A.}~\bibnamefont {Colibaba-Evulet}}, \bibinfo {author} {\bibfnamefont {A.}~\bibnamefont {Singhal}}, \ and\ \bibinfo {author} {\bibfnamefont {N.}~\bibnamefont {Glumac}},\ }\bibfield  {title} {\enquote {\bibinfo {title} {Detection of {AlO} and {TiO} by laser-induced fluorescence in powder synthesis flames},}\ }\href@noop {} {\bibfield  {journal} {\bibinfo  {journal} {Combustion science and technology}\ }\textbf {\bibinfo {volume} {157}},\ \bibinfo {pages} {129--139} (\bibinfo {year} {2000})}\BibitemShut {NoStop}%
\bibitem [{\citenamefont {Honma}(2003)}]{honma2003reaction}%
  \BibitemOpen
  \bibfield  {author} {\bibinfo {author} {\bibfnamefont {K.}~\bibnamefont {Honma}},\ }\bibfield  {title} {\enquote {\bibinfo {title} {Reaction dynamics of {Al} + {O}$_2$ $\rightarrow$ {AlO} + {O} studied by the crossed-beam laser-induced fluorescence technique},}\ }\href@noop {} {\bibfield  {journal} {\bibinfo  {journal} {The Journal of chemical physics}\ }\textbf {\bibinfo {volume} {119}},\ \bibinfo {pages} {3641--3649} (\bibinfo {year} {2003})}\BibitemShut {NoStop}%
\bibitem [{\citenamefont {Li}\ \emph {et~al.}(2019)\citenamefont {Li}, \citenamefont {Xu}, \citenamefont {Li}, \citenamefont {Ma}, \citenamefont {Zhao}, \citenamefont {Zhang}, \citenamefont {Guo},\ and\ \citenamefont {Lu}}]{li2019laser}%
  \BibitemOpen
  \bibfield  {author} {\bibinfo {author} {\bibfnamefont {J.}~\bibnamefont {Li}}, \bibinfo {author} {\bibfnamefont {M.}~\bibnamefont {Xu}}, \bibinfo {author} {\bibfnamefont {X.}~\bibnamefont {Li}}, \bibinfo {author} {\bibfnamefont {Q.}~\bibnamefont {Ma}}, \bibinfo {author} {\bibfnamefont {N.}~\bibnamefont {Zhao}}, \bibinfo {author} {\bibfnamefont {Q.}~\bibnamefont {Zhang}}, \bibinfo {author} {\bibfnamefont {L.}~\bibnamefont {Guo}}, \ and\ \bibinfo {author} {\bibfnamefont {Y.}~\bibnamefont {Lu}},\ }\bibfield  {title} {\enquote {\bibinfo {title} {Laser-induced molecular fluorescence diagnosis of aluminum monoxide evolution in laser-induced plasma},}\ }\href@noop {} {\bibfield  {journal} {\bibinfo  {journal} {Laser Physics Letters}\ }\textbf {\bibinfo {volume} {16}},\ \bibinfo {pages} {055701} (\bibinfo {year} {2019})}\BibitemShut {NoStop}%
\bibitem [{\citenamefont {He}, \citenamefont {Wang},\ and\ \citenamefont {Weiner}(1993)}]{he1993production}%
  \BibitemOpen
  \bibfield  {author} {\bibinfo {author} {\bibfnamefont {M.}~\bibnamefont {He}}, \bibinfo {author} {\bibfnamefont {H.}~\bibnamefont {Wang}}, \ and\ \bibinfo {author} {\bibfnamefont {B.~R.}\ \bibnamefont {Weiner}},\ }\bibfield  {title} {\enquote {\bibinfo {title} {Production and laser-induced fluorescence spectrum of aluminum sulfide},}\ }\href@noop {} {\bibfield  {journal} {\bibinfo  {journal} {Chemical physics letters}\ }\textbf {\bibinfo {volume} {204}},\ \bibinfo {pages} {563--566} (\bibinfo {year} {1993})}\BibitemShut {NoStop}%
\bibitem [{\citenamefont {Alstrin}\ \emph {et~al.}(1992)\citenamefont {Alstrin}, \citenamefont {Smilgys}, \citenamefont {Strupp},\ and\ \citenamefont {Leone}}]{alstrin1992vibrational}%
  \BibitemOpen
  \bibfield  {author} {\bibinfo {author} {\bibfnamefont {A.~L.}\ \bibnamefont {Alstrin}}, \bibinfo {author} {\bibfnamefont {R.~V.}\ \bibnamefont {Smilgys}}, \bibinfo {author} {\bibfnamefont {P.~G.}\ \bibnamefont {Strupp}}, \ and\ \bibinfo {author} {\bibfnamefont {S.~R.}\ \bibnamefont {Leone}},\ }\bibfield  {title} {\enquote {\bibinfo {title} {Vibrational distributions of {As}$_2$ in the cracking of {As}$_4$ on {Si} (100) and {Si} (111)},}\ }\href@noop {} {\bibfield  {journal} {\bibinfo  {journal} {The Journal of chemical physics}\ }\textbf {\bibinfo {volume} {97}},\ \bibinfo {pages} {6864--6870} (\bibinfo {year} {1992})}\BibitemShut {NoStop}%
\bibitem [{\citenamefont {Donnelly}\ and\ \citenamefont {Karlicek}(1982)}]{donnelly1982development}%
  \BibitemOpen
  \bibfield  {author} {\bibinfo {author} {\bibfnamefont {V.}~\bibnamefont {Donnelly}}\ and\ \bibinfo {author} {\bibfnamefont {R.}~\bibnamefont {Karlicek}},\ }\bibfield  {title} {\enquote {\bibinfo {title} {Development of laser diagnostic probes for chemical vapor deposition of {InP}/{InGaAsP} epitaxial layers},}\ }\href@noop {} {\bibfield  {journal} {\bibinfo  {journal} {Journal of Applied Physics}\ }\textbf {\bibinfo {volume} {53}},\ \bibinfo {pages} {6399--6407} (\bibinfo {year} {1982})}\BibitemShut {NoStop}%
\bibitem [{\citenamefont {Smilgys}\ and\ \citenamefont {Leone}(1990)}]{smilgys1990state}%
  \BibitemOpen
  \bibfield  {author} {\bibinfo {author} {\bibfnamefont {R.~V.}\ \bibnamefont {Smilgys}}\ and\ \bibinfo {author} {\bibfnamefont {S.~R.}\ \bibnamefont {Leone}},\ }\bibfield  {title} {\enquote {\bibinfo {title} {{State-resolved laser probing of As$_{2}$ in a molecular-beam epitaxy reactor}},}\ }\href@noop {} {\bibfield  {journal} {\bibinfo  {journal} {Journal of Vacuum Science \& Technology B: Microelectronics Processing and Phenomena}\ }\textbf {\bibinfo {volume} {8}},\ \bibinfo {pages} {416--421} (\bibinfo {year} {1990})}\BibitemShut {NoStop}%
\bibitem [{\citenamefont {Ng}, \citenamefont {Pang},\ and\ \citenamefont {Cheung}(2011)}]{ng2011laser}%
  \BibitemOpen
  \bibfield  {author} {\bibinfo {author} {\bibfnamefont {Y.}~\bibnamefont {Ng}}, \bibinfo {author} {\bibfnamefont {H.}~\bibnamefont {Pang}}, \ and\ \bibinfo {author} {\bibfnamefont {A.-C.}\ \bibnamefont {Cheung}},\ }\bibfield  {title} {\enquote {\bibinfo {title} {Laser induced fluorescence spectroscopy of boron carbide},}\ }\href@noop {} {\bibfield  {journal} {\bibinfo  {journal} {Chemical Physics Letters}\ }\textbf {\bibinfo {volume} {509}},\ \bibinfo {pages} {16--19} (\bibinfo {year} {2011})}\BibitemShut {NoStop}%
\bibitem [{\citenamefont {Sunahori}, \citenamefont {Nagarajan},\ and\ \citenamefont {Clouthier}(2015)}]{sunahori2015optical}%
  \BibitemOpen
  \bibfield  {author} {\bibinfo {author} {\bibfnamefont {F.~X.}\ \bibnamefont {Sunahori}}, \bibinfo {author} {\bibfnamefont {R.}~\bibnamefont {Nagarajan}}, \ and\ \bibinfo {author} {\bibfnamefont {D.~J.}\ \bibnamefont {Clouthier}},\ }\bibfield  {title} {\enquote {\bibinfo {title} {{Optical-optical double resonance, laser induced fluorescence, and revision of the signs of the spin-spin constants of the boron carbide (BC) free radical}},}\ }\href@noop {} {\bibfield  {journal} {\bibinfo  {journal} {The Journal of Chemical Physics}\ }\textbf {\bibinfo {volume} {143}} (\bibinfo {year} {2015})}\BibitemShut {NoStop}%
\bibitem [{\citenamefont {Fleddermann}\ and\ \citenamefont {Hebner}(1998)}]{fleddermann1998measurements}%
  \BibitemOpen
  \bibfield  {author} {\bibinfo {author} {\bibfnamefont {C.}~\bibnamefont {Fleddermann}}\ and\ \bibinfo {author} {\bibfnamefont {G.}~\bibnamefont {Hebner}},\ }\bibfield  {title} {\enquote {\bibinfo {title} {Measurements of relative {BCl} density in {BCl}$_3$-containing inductively coupled radio frequency plasmas},}\ }\href@noop {} {\bibfield  {journal} {\bibinfo  {journal} {Journal of applied physics}\ }\textbf {\bibinfo {volume} {83}},\ \bibinfo {pages} {4030--4036} (\bibinfo {year} {1998})}\BibitemShut {NoStop}%
\bibitem [{\citenamefont {Rice}, \citenamefont {Caldwell},\ and\ \citenamefont {Nelson}(1989)}]{rice1989gas}%
  \BibitemOpen
  \bibfield  {author} {\bibinfo {author} {\bibfnamefont {J.}~\bibnamefont {Rice}}, \bibinfo {author} {\bibfnamefont {N.}~\bibnamefont {Caldwell}}, \ and\ \bibinfo {author} {\bibfnamefont {H.}~\bibnamefont {Nelson}},\ }\bibfield  {title} {\enquote {\bibinfo {title} {Gas-phase reaction kinetics of boron monohydride},}\ }\href@noop {} {\bibfield  {journal} {\bibinfo  {journal} {The Journal of Physical Chemistry}\ }\textbf {\bibinfo {volume} {93}},\ \bibinfo {pages} {3600--3605} (\bibinfo {year} {1989})}\BibitemShut {NoStop}%
\bibitem [{\citenamefont {Harrison}, \citenamefont {Meads},\ and\ \citenamefont {Phillips}(1988)}]{harrison1988kinetics}%
  \BibitemOpen
  \bibfield  {author} {\bibinfo {author} {\bibfnamefont {J.}~\bibnamefont {Harrison}}, \bibinfo {author} {\bibfnamefont {R.}~\bibnamefont {Meads}}, \ and\ \bibinfo {author} {\bibfnamefont {L.}~\bibnamefont {Phillips}},\ }\bibfield  {title} {\enquote {\bibinfo {title} {Kinetics of reactions of {BH} with {NO} and {C}$_{2}${H}$_4$},}\ }\href@noop {} {\bibfield  {journal} {\bibinfo  {journal} {Chemical physics letters}\ }\textbf {\bibinfo {volume} {150}},\ \bibinfo {pages} {299--302} (\bibinfo {year} {1988})}\BibitemShut {NoStop}%
\bibitem [{\citenamefont {Sunahori}\ \emph {et~al.}(2015)\citenamefont {Sunahori}, \citenamefont {Gharaibeh}, \citenamefont {Clouthier},\ and\ \citenamefont {Tarroni}}]{sunahori2015bh2}%
  \BibitemOpen
  \bibfield  {author} {\bibinfo {author} {\bibfnamefont {F.~X.}\ \bibnamefont {Sunahori}}, \bibinfo {author} {\bibfnamefont {M.}~\bibnamefont {Gharaibeh}}, \bibinfo {author} {\bibfnamefont {D.~J.}\ \bibnamefont {Clouthier}}, \ and\ \bibinfo {author} {\bibfnamefont {R.}~\bibnamefont {Tarroni}},\ }\bibfield  {title} {\enquote {\bibinfo {title} {{BH}$_{2}$ revisited: New, extensive measurements of laser-induced fluorescence transitions and ab initio calculations of near-spectroscopic accuracy},}\ }\href@noop {} {\bibfield  {journal} {\bibinfo  {journal} {The Journal of Chemical Physics}\ }\textbf {\bibinfo {volume} {142}} (\bibinfo {year} {2015})}\BibitemShut {NoStop}%
\bibitem [{\citenamefont {Clyne}\ and\ \citenamefont {Heaven}(1980)}]{clyne1980laser}%
  \BibitemOpen
  \bibfield  {author} {\bibinfo {author} {\bibfnamefont {M.~A.}\ \bibnamefont {Clyne}}\ and\ \bibinfo {author} {\bibfnamefont {M.~C.}\ \bibnamefont {Heaven}},\ }\bibfield  {title} {\enquote {\bibinfo {title} {Laser-induced fluorescence of the {BO} and {BO}$_2$ free radicals},}\ }\href@noop {} {\bibfield  {journal} {\bibinfo  {journal} {Chemical Physics}\ }\textbf {\bibinfo {volume} {51}},\ \bibinfo {pages} {299--309} (\bibinfo {year} {1980})}\BibitemShut {NoStop}%
\bibitem [{\citenamefont {Maksyutenko}\ \emph {et~al.}(2011)\citenamefont {Maksyutenko}, \citenamefont {Parker}, \citenamefont {Zhang},\ and\ \citenamefont {Kaiser}}]{maksyutenko2011lif}%
  \BibitemOpen
  \bibfield  {author} {\bibinfo {author} {\bibfnamefont {P.}~\bibnamefont {Maksyutenko}}, \bibinfo {author} {\bibfnamefont {D.~S.}\ \bibnamefont {Parker}}, \bibinfo {author} {\bibfnamefont {F.}~\bibnamefont {Zhang}}, \ and\ \bibinfo {author} {\bibfnamefont {R.~I.}\ \bibnamefont {Kaiser}},\ }\bibfield  {title} {\enquote {\bibinfo {title} {{An LIF characterization of supersonic BO (X$^{2}\Sigma^{+}$) and CN (X$^{2}\Sigma^{+}$) radical sources for crossed beam studies}},}\ }\href@noop {} {\bibfield  {journal} {\bibinfo  {journal} {Review of Scientific Instruments}\ }\textbf {\bibinfo {volume} {82}} (\bibinfo {year} {2011})}\BibitemShut {NoStop}%
\bibitem [{\citenamefont {Schneider}\ and\ \citenamefont {Roh}(1988)}]{schneider1988application}%
  \BibitemOpen
  \bibfield  {author} {\bibinfo {author} {\bibfnamefont {G.~R.}\ \bibnamefont {Schneider}}\ and\ \bibinfo {author} {\bibfnamefont {W.~B.}\ \bibnamefont {Roh}},\ }\bibfield  {title} {\enquote {\bibinfo {title} {{Application of laser-induced fluorescence in an atmospheric-pressure boron-seeded flame}},}\ }\href@noop {} {\bibfield  {journal} {\bibinfo  {journal} {AIP Conference Proceedings}\ }\textbf {\bibinfo {volume} {172}},\ \bibinfo {pages} {753--755} (\bibinfo {year} {1988})}\BibitemShut {NoStop}%
\bibitem [{\citenamefont {Weyer}\ \emph {et~al.}(1980)\citenamefont {Weyer}, \citenamefont {Beaudet}, \citenamefont {Straubinger},\ and\ \citenamefont {Walther}}]{weyer1980laser}%
  \BibitemOpen
  \bibfield  {author} {\bibinfo {author} {\bibfnamefont {K.}~\bibnamefont {Weyer}}, \bibinfo {author} {\bibfnamefont {R.}~\bibnamefont {Beaudet}}, \bibinfo {author} {\bibfnamefont {R.}~\bibnamefont {Straubinger}}, \ and\ \bibinfo {author} {\bibfnamefont {H.}~\bibnamefont {Walther}},\ }\bibfield  {title} {\enquote {\bibinfo {title} {{Laser excited fluorescence of the $\Pi$ vibronic states of BO$_2$}},}\ }\href@noop {} {\bibfield  {journal} {\bibinfo  {journal} {Chemical Physics}\ }\textbf {\bibinfo {volume} {47}},\ \bibinfo {pages} {171--178} (\bibinfo {year} {1980})}\BibitemShut {NoStop}%
\bibitem [{\citenamefont {He}, \citenamefont {Evans},\ and\ \citenamefont {Clouthier}(2003)}]{he2003study}%
  \BibitemOpen
  \bibfield  {author} {\bibinfo {author} {\bibfnamefont {S.-G.}\ \bibnamefont {He}}, \bibinfo {author} {\bibfnamefont {C.~J.}\ \bibnamefont {Evans}}, \ and\ \bibinfo {author} {\bibfnamefont {D.~J.}\ \bibnamefont {Clouthier}},\ }\bibfield  {title} {\enquote {\bibinfo {title} {{A study of the molecular structure and Renner--Teller effect in the $\tilde{A}^{2}\Pi_{u}$ -- $\tilde{X}^{2}\Pi_{g}$ electronic spectrum of jet-cooled boron disulfide, BS$_2$}},}\ }\href@noop {} {\bibfield  {journal} {\bibinfo  {journal} {The Journal of chemical physics}\ }\textbf {\bibinfo {volume} {119}},\ \bibinfo {pages} {2047--2056} (\bibinfo {year} {2003})}\BibitemShut {NoStop}%
\bibitem [{\citenamefont {Focsa}, \citenamefont {Li},\ and\ \citenamefont {Bernath}(2000)}]{focsa2000characterization}%
  \BibitemOpen
  \bibfield  {author} {\bibinfo {author} {\bibfnamefont {C.}~\bibnamefont {Focsa}}, \bibinfo {author} {\bibfnamefont {H.}~\bibnamefont {Li}}, \ and\ \bibinfo {author} {\bibfnamefont {P.}~\bibnamefont {Bernath}},\ }\bibfield  {title} {\enquote {\bibinfo {title} {{Characterization of the Ground State of Br$_2$ by Laser-Induced Fluorescence Fourier Transform Spectroscopy of the B$^{3}\Pi_{0^{+}u}$ -- X$^1\Sigma^{+}_{g}$ System}},}\ }\href@noop {} {\bibfield  {journal} {\bibinfo  {journal} {Journal of Molecular Spectroscopy}\ }\textbf {\bibinfo {volume} {200}},\ \bibinfo {pages} {104--119} (\bibinfo {year} {2000})}\BibitemShut {NoStop}%
\bibitem [{\citenamefont {Bullman}, \citenamefont {Farthing},\ and\ \citenamefont {Whitehead}(1981)}]{bullman1981laser}%
  \BibitemOpen
  \bibfield  {author} {\bibinfo {author} {\bibfnamefont {S.}~\bibnamefont {Bullman}}, \bibinfo {author} {\bibfnamefont {J.}~\bibnamefont {Farthing}}, \ and\ \bibinfo {author} {\bibfnamefont {J.}~\bibnamefont {Whitehead}},\ }\bibfield  {title} {\enquote {\bibinfo {title} {Laser-induced fluorescence studies of a supersonic molecular beam of bromine: Vibrational and rotational relaxation of bromine and collision-free lifetimes for {Br}$_2$ ({B}$_{3}{\Pi}$(0$_u^+$))},}\ }\href@noop {} {\bibfield  {journal} {\bibinfo  {journal} {Molecular Physics}\ }\textbf {\bibinfo {volume} {44}},\ \bibinfo {pages} {97--109} (\bibinfo {year} {1981})}\BibitemShut {NoStop}%
\bibitem [{\citenamefont {Nakajima}\ \emph {et~al.}(2007{\natexlab{a}})\citenamefont {Nakajima}, \citenamefont {Schmidt}, \citenamefont {Sumiyoshi},\ and\ \citenamefont {Endo}}]{nakajima2007rotationally}%
  \BibitemOpen
  \bibfield  {author} {\bibinfo {author} {\bibfnamefont {M.}~\bibnamefont {Nakajima}}, \bibinfo {author} {\bibfnamefont {T.~W.}\ \bibnamefont {Schmidt}}, \bibinfo {author} {\bibfnamefont {Y.}~\bibnamefont {Sumiyoshi}}, \ and\ \bibinfo {author} {\bibfnamefont {Y.}~\bibnamefont {Endo}},\ }\bibfield  {title} {\enquote {\bibinfo {title} {Rotationally-resolved excitation spectrum of the jet-cooled cyclohexadienyl radical},}\ }\href@noop {} {\bibfield  {journal} {\bibinfo  {journal} {Chemical Physics Letters}\ }\textbf {\bibinfo {volume} {449}},\ \bibinfo {pages} {57--62} (\bibinfo {year} {2007}{\natexlab{a}})}\BibitemShut {NoStop}%
\bibitem [{\citenamefont {Zhou}\ \emph {et~al.}(1990)\citenamefont {Zhou}, \citenamefont {Zhan}, \citenamefont {Shi},\ and\ \citenamefont {Wang}}]{zhou1990gas}%
  \BibitemOpen
  \bibfield  {author} {\bibinfo {author} {\bibfnamefont {S.}~\bibnamefont {Zhou}}, \bibinfo {author} {\bibfnamefont {M.}~\bibnamefont {Zhan}}, \bibinfo {author} {\bibfnamefont {J.}~\bibnamefont {Shi}}, \ and\ \bibinfo {author} {\bibfnamefont {C.}~\bibnamefont {Wang}},\ }\bibfield  {title} {\enquote {\bibinfo {title} {Gas phase spectrum of dibromocarbene studied by laser-induced fluorescence},}\ }\href@noop {} {\bibfield  {journal} {\bibinfo  {journal} {Chemical physics letters}\ }\textbf {\bibinfo {volume} {166}},\ \bibinfo {pages} {547--550} (\bibinfo {year} {1990})}\BibitemShut {NoStop}%
\bibitem [{\citenamefont {Lee}, \citenamefont {Liu},\ and\ \citenamefont {Chang}(2003)}]{lee2003electronic}%
  \BibitemOpen
  \bibfield  {author} {\bibinfo {author} {\bibfnamefont {C.-L.}\ \bibnamefont {Lee}}, \bibinfo {author} {\bibfnamefont {M.-L.}\ \bibnamefont {Liu}}, \ and\ \bibinfo {author} {\bibfnamefont {B.-C.}\ \bibnamefont {Chang}},\ }\bibfield  {title} {\enquote {\bibinfo {title} {Electronic spectroscopy of bromomethylenes in a supersonic free jet expansion},}\ }\href@noop {} {\bibfield  {journal} {\bibinfo  {journal} {Physical Chemistry Chemical Physics}\ }\textbf {\bibinfo {volume} {5}},\ \bibinfo {pages} {3859--3863} (\bibinfo {year} {2003})}\BibitemShut {NoStop}%
\bibitem [{\citenamefont {Hack}(1985)}]{hack1985detection}%
  \BibitemOpen
  \bibfield  {author} {\bibinfo {author} {\bibfnamefont {W.}~\bibnamefont {Hack}},\ }\bibfield  {title} {\enquote {\bibinfo {title} {Detection methods for atoms and radicals in the gas phase},}\ }\href@noop {} {\bibfield  {journal} {\bibinfo  {journal} {International Reviews in Physical Chemistry}\ }\textbf {\bibinfo {volume} {4}},\ \bibinfo {pages} {165--200} (\bibinfo {year} {1985})}\BibitemShut {NoStop}%
\bibitem [{\citenamefont {Tiee}, \citenamefont {Wampler},\ and\ \citenamefont {Rice~Jr}(1980)}]{tiee1980reactions}%
  \BibitemOpen
  \bibfield  {author} {\bibinfo {author} {\bibfnamefont {J.}~\bibnamefont {Tiee}}, \bibinfo {author} {\bibfnamefont {F.}~\bibnamefont {Wampler}}, \ and\ \bibinfo {author} {\bibfnamefont {W.}~\bibnamefont {Rice~Jr}},\ }\bibfield  {title} {\enquote {\bibinfo {title} {{Reactions of CCl, CCl$_{2}$ and CClF radicals}},}\ }\href@noop {} {\bibfield  {journal} {\bibinfo  {journal} {Chemical Physics Letters}\ }\textbf {\bibinfo {volume} {73}},\ \bibinfo {pages} {519--521} (\bibinfo {year} {1980})}\BibitemShut {NoStop}%
\bibitem [{\citenamefont {Huie}, \citenamefont {Long},\ and\ \citenamefont {Thrush}(1977)}]{huie1977laser}%
  \BibitemOpen
  \bibfield  {author} {\bibinfo {author} {\bibfnamefont {R.}~\bibnamefont {Huie}}, \bibinfo {author} {\bibfnamefont {N.}~\bibnamefont {Long}}, \ and\ \bibinfo {author} {\bibfnamefont {B.}~\bibnamefont {Thrush}},\ }\bibfield  {title} {\enquote {\bibinfo {title} {Laser induced fluorescence of {CFCl} and {CCl}$_2$ in the gas phase},}\ }\href@noop {} {\bibfield  {journal} {\bibinfo  {journal} {Chemical Physics Letters}\ }\textbf {\bibinfo {volume} {51}},\ \bibinfo {pages} {197--200} (\bibinfo {year} {1977})}\BibitemShut {NoStop}%
\bibitem [{\citenamefont {G{\'o}mez}\ \emph {et~al.}(2013)\citenamefont {G{\'o}mez}, \citenamefont {D'accurso}, \citenamefont {Freytes}, \citenamefont {Manzano}, \citenamefont {Codnia},\ and\ \citenamefont {Azc{\'a}rate}}]{gomez2013kinetic}%
  \BibitemOpen
  \bibfield  {author} {\bibinfo {author} {\bibfnamefont {N.~D.}\ \bibnamefont {G{\'o}mez}}, \bibinfo {author} {\bibfnamefont {V.}~\bibnamefont {D'accurso}}, \bibinfo {author} {\bibfnamefont {V.~M.}\ \bibnamefont {Freytes}}, \bibinfo {author} {\bibfnamefont {F.~A.}\ \bibnamefont {Manzano}}, \bibinfo {author} {\bibfnamefont {J.}~\bibnamefont {Codnia}}, \ and\ \bibinfo {author} {\bibfnamefont {M.~L.}\ \bibnamefont {Azc{\'a}rate}},\ }\bibfield  {title} {\enquote {\bibinfo {title} {{Kinetic Study of the CCl$_2$ Radical Recombination Reaction by Laser-Induced Fluorescence Technique}},}\ }\href@noop {} {\bibfield  {journal} {\bibinfo  {journal} {International Journal of Chemical Kinetics}\ }\textbf {\bibinfo {volume} {45}},\ \bibinfo {pages} {306--313} (\bibinfo {year} {2013})}\BibitemShut {NoStop}%
\bibitem [{\citenamefont {Liu}\ \emph {et~al.}(2004)\citenamefont {Liu}, \citenamefont {Xin}, \citenamefont {Pei}, \citenamefont {Chen},\ and\ \citenamefont {Chen}}]{liu2004reaction}%
  \BibitemOpen
  \bibfield  {author} {\bibinfo {author} {\bibfnamefont {Y.}~\bibnamefont {Liu}}, \bibinfo {author} {\bibfnamefont {Y.}~\bibnamefont {Xin}}, \bibinfo {author} {\bibfnamefont {L.}~\bibnamefont {Pei}}, \bibinfo {author} {\bibfnamefont {Y.}~\bibnamefont {Chen}}, \ and\ \bibinfo {author} {\bibfnamefont {C.}~\bibnamefont {Chen}},\ }\bibfield  {title} {\enquote {\bibinfo {title} {{Reaction kinetic studies of CCl$_2$ (X (0, 0, 0)) with several simple molecules}},}\ }\href@noop {} {\bibfield  {journal} {\bibinfo  {journal} {Chemical physics letters}\ }\textbf {\bibinfo {volume} {385}},\ \bibinfo {pages} {314--318} (\bibinfo {year} {2004})}\BibitemShut {NoStop}%
\bibitem [{\citenamefont {Lu}\ \emph {et~al.}(1991)\citenamefont {Lu}, \citenamefont {Chen}, \citenamefont {Wang}, \citenamefont {Zhang}, \citenamefont {Yu}, \citenamefont {Chen}, \citenamefont {Koshi}, \citenamefont {Matsui}, \citenamefont {Koda},\ and\ \citenamefont {Ma}}]{lu1991laser}%
  \BibitemOpen
  \bibfield  {author} {\bibinfo {author} {\bibfnamefont {Q.}~\bibnamefont {Lu}}, \bibinfo {author} {\bibfnamefont {Y.}~\bibnamefont {Chen}}, \bibinfo {author} {\bibfnamefont {D.}~\bibnamefont {Wang}}, \bibinfo {author} {\bibfnamefont {Y.}~\bibnamefont {Zhang}}, \bibinfo {author} {\bibfnamefont {S.}~\bibnamefont {Yu}}, \bibinfo {author} {\bibfnamefont {C.}~\bibnamefont {Chen}}, \bibinfo {author} {\bibfnamefont {M.}~\bibnamefont {Koshi}}, \bibinfo {author} {\bibfnamefont {H.}~\bibnamefont {Matsui}}, \bibinfo {author} {\bibfnamefont {S.}~\bibnamefont {Koda}}, \ and\ \bibinfo {author} {\bibfnamefont {X.}~\bibnamefont {Ma}},\ }\bibfield  {title} {\enquote {\bibinfo {title} {{Laser-induced fluorescence excitation spectrum of CCl$_2$ cooled in a supersonic free jet}},}\ }\href@noop {} {\bibfield  {journal} {\bibinfo  {journal} {Chemical physics letters}\ }\textbf {\bibinfo {volume} {178}},\ \bibinfo {pages} {517--522} (\bibinfo {year} {1991})}\BibitemShut {NoStop}%
\bibitem [{\citenamefont {Tiee}, \citenamefont {Wampler},\ and\ \citenamefont {Rice}(1979)}]{tiee1979laser}%
  \BibitemOpen
  \bibfield  {author} {\bibinfo {author} {\bibfnamefont {J.}~\bibnamefont {Tiee}}, \bibinfo {author} {\bibfnamefont {F.}~\bibnamefont {Wampler}}, \ and\ \bibinfo {author} {\bibfnamefont {W.}~\bibnamefont {Rice}},\ }\bibfield  {title} {\enquote {\bibinfo {title} {{Laser-induced fluorescence excitation spectra of CCl$_2$ and CFCI radicals in the gas phase}},}\ }\href@noop {} {\bibfield  {journal} {\bibinfo  {journal} {Chemical Physics Letters}\ }\textbf {\bibinfo {volume} {65}},\ \bibinfo {pages} {425--428} (\bibinfo {year} {1979})}\BibitemShut {NoStop}%
\bibitem [{\citenamefont {Guss}\ \emph {et~al.}(2005)\citenamefont {Guss}, \citenamefont {Richmond}, \citenamefont {Nauta},\ and\ \citenamefont {Kable}}]{guss2005laser}%
  \BibitemOpen
  \bibfield  {author} {\bibinfo {author} {\bibfnamefont {J.~S.}\ \bibnamefont {Guss}}, \bibinfo {author} {\bibfnamefont {C.~A.}\ \bibnamefont {Richmond}}, \bibinfo {author} {\bibfnamefont {K.}~\bibnamefont {Nauta}}, \ and\ \bibinfo {author} {\bibfnamefont {S.~H.}\ \bibnamefont {Kable}},\ }\bibfield  {title} {\enquote {\bibinfo {title} {{Laser-induced fluorescence excitation and dispersed fluorescence spectroscopy of the $\tilde{A}$($^{1}$B$_{1}$) -- $\tilde{X}$($^{1}$A$_{1}$) transition of dichlorocarbene}},}\ }\href@noop {} {\bibfield  {journal} {\bibinfo  {journal} {Physical Chemistry Chemical Physics}\ }\textbf {\bibinfo {volume} {7}},\ \bibinfo {pages} {100--108} (\bibinfo {year} {2005})}\BibitemShut {NoStop}%
\bibitem [{\citenamefont {Ludwiczak}, \citenamefont {Latimer},\ and\ \citenamefont {Steer}(1991)}]{ludwiczak1991bx}%
  \BibitemOpen
  \bibfield  {author} {\bibinfo {author} {\bibfnamefont {M.}~\bibnamefont {Ludwiczak}}, \bibinfo {author} {\bibfnamefont {D.}~\bibnamefont {Latimer}}, \ and\ \bibinfo {author} {\bibfnamefont {R.}~\bibnamefont {Steer}},\ }\bibfield  {title} {\enquote {\bibinfo {title} {{The $\tilde{B}$ -- $\tilde{X}$ laser-induced fluorescence excitation spectrum of jet-cooled Cl$_2$CS: Origin location and partial vibronic analysis}},}\ }\href@noop {} {\bibfield  {journal} {\bibinfo  {journal} {Journal of Molecular Spectroscopy}\ }\textbf {\bibinfo {volume} {147}},\ \bibinfo {pages} {414--430} (\bibinfo {year} {1991})}\BibitemShut {NoStop}%
\bibitem [{\citenamefont {Booth}\ \emph {et~al.}(1989)\citenamefont {Booth}, \citenamefont {Hancock}, \citenamefont {Perry},\ and\ \citenamefont {Toogood}}]{booth1989spatially}%
  \BibitemOpen
  \bibfield  {author} {\bibinfo {author} {\bibfnamefont {J.}~\bibnamefont {Booth}}, \bibinfo {author} {\bibfnamefont {G.}~\bibnamefont {Hancock}}, \bibinfo {author} {\bibfnamefont {N.}~\bibnamefont {Perry}}, \ and\ \bibinfo {author} {\bibfnamefont {M.}~\bibnamefont {Toogood}},\ }\bibfield  {title} {\enquote {\bibinfo {title} {Spatially and temporally resolved laser-induced fluorescence measurements of {CF}$_2$ and {CF} radicals in a {CF}$_4$ {RF} plasma},}\ }\href@noop {} {\bibfield  {journal} {\bibinfo  {journal} {Journal of applied physics}\ }\textbf {\bibinfo {volume} {66}},\ \bibinfo {pages} {5251--5257} (\bibinfo {year} {1989})}\BibitemShut {NoStop}%
\bibitem [{\citenamefont {King}, \citenamefont {Schenck},\ and\ \citenamefont {Stephenson}(1979)}]{king1979spectroscopy}%
  \BibitemOpen
  \bibfield  {author} {\bibinfo {author} {\bibfnamefont {D.~S.}\ \bibnamefont {King}}, \bibinfo {author} {\bibfnamefont {P.~K.}\ \bibnamefont {Schenck}}, \ and\ \bibinfo {author} {\bibfnamefont {J.~C.}\ \bibnamefont {Stephenson}},\ }\bibfield  {title} {\enquote {\bibinfo {title} {{Spectroscopy and photophysics of the CF$_2$ $\tilde{A}^{1}$B$_1$-$\tilde{X}^{1}$A$_1$ system}},}\ }\href@noop {} {\bibfield  {journal} {\bibinfo  {journal} {Journal of Molecular Spectroscopy}\ }\textbf {\bibinfo {volume} {78}},\ \bibinfo {pages} {1--15} (\bibinfo {year} {1979})}\BibitemShut {NoStop}%
\bibitem [{\citenamefont {McMillin}\ and\ \citenamefont {Zachariah}(1997)}]{mcmillin1997two}%
  \BibitemOpen
  \bibfield  {author} {\bibinfo {author} {\bibfnamefont {B.~K.}\ \bibnamefont {McMillin}}\ and\ \bibinfo {author} {\bibfnamefont {M.~R.}\ \bibnamefont {Zachariah}},\ }\bibfield  {title} {\enquote {\bibinfo {title} {{Two-dimensional imaging of CF$_2$ density by laser-induced fluorescence in CF$_4$ etching plasmas in the gaseous electronics conference reference cell}},}\ }\href@noop {} {\bibfield  {journal} {\bibinfo  {journal} {Journal of Vacuum Science \& Technology A: Vacuum, Surfaces, and Films}\ }\textbf {\bibinfo {volume} {15}},\ \bibinfo {pages} {230--237} (\bibinfo {year} {1997})}\BibitemShut {NoStop}%
\bibitem [{\citenamefont {Rubio}, \citenamefont {Santos},\ and\ \citenamefont {Torresano}(2001)}]{rubio2001laser}%
  \BibitemOpen
  \bibfield  {author} {\bibinfo {author} {\bibfnamefont {L.}~\bibnamefont {Rubio}}, \bibinfo {author} {\bibfnamefont {M.}~\bibnamefont {Santos}}, \ and\ \bibinfo {author} {\bibfnamefont {J.}~\bibnamefont {Torresano}},\ }\bibfield  {title} {\enquote {\bibinfo {title} {{Laser induced fluorescence detection of CF, CF$_2$ and CF$_3$ in the infrared multiphoton dissociation of C$_3$F$_6$}},}\ }\href@noop {} {\bibfield  {journal} {\bibinfo  {journal} {Journal of Photochemistry and Photobiology A: Chemistry}\ }\textbf {\bibinfo {volume} {146}},\ \bibinfo {pages} {1--8} (\bibinfo {year} {2001})}\BibitemShut {NoStop}%
\bibitem [{\citenamefont {Yang}, \citenamefont {Williamson},\ and\ \citenamefont {Miller}(1997)}]{yang1997rotational}%
  \BibitemOpen
  \bibfield  {author} {\bibinfo {author} {\bibfnamefont {M.-C.}\ \bibnamefont {Yang}}, \bibinfo {author} {\bibfnamefont {J.}~\bibnamefont {Williamson}}, \ and\ \bibinfo {author} {\bibfnamefont {T.~A.}\ \bibnamefont {Miller}},\ }\bibfield  {title} {\enquote {\bibinfo {title} {Rotational analyses of the laser induced fluorescence excitation spectra of jet-cooled {CF}$_3${O} and {CF}$_3${S}},}\ }\href@noop {} {\bibfield  {journal} {\bibinfo  {journal} {Journal of molecular spectroscopy}\ }\textbf {\bibinfo {volume} {186}},\ \bibinfo {pages} {1--14} (\bibinfo {year} {1997})}\BibitemShut {NoStop}%
\bibitem [{\citenamefont {Li}\ and\ \citenamefont {Francisco}(1991)}]{li1991laser}%
  \BibitemOpen
  \bibfield  {author} {\bibinfo {author} {\bibfnamefont {Z.}~\bibnamefont {Li}}\ and\ \bibinfo {author} {\bibfnamefont {J.}~\bibnamefont {Francisco}},\ }\bibfield  {title} {\enquote {\bibinfo {title} {{Laser-induced fluorescence spectroscopic study of the $\tilde{A}^{2}$A$_{1}$ -- $\tilde{X}^{2}$E transition of trifluoromethoxy radical}},}\ }\href@noop {} {\bibfield  {journal} {\bibinfo  {journal} {Chemical physics letters}\ }\textbf {\bibinfo {volume} {186}},\ \bibinfo {pages} {336--342} (\bibinfo {year} {1991})}\BibitemShut {NoStop}%
\bibitem [{\citenamefont {Purdy}\ and\ \citenamefont {Thrush}(1980)}]{purdy1980laser}%
  \BibitemOpen
  \bibfield  {author} {\bibinfo {author} {\bibfnamefont {J.}~\bibnamefont {Purdy}}\ and\ \bibinfo {author} {\bibfnamefont {B.}~\bibnamefont {Thrush}},\ }\bibfield  {title} {\enquote {\bibinfo {title} {{Laser-induced fluorescence of CFBr in the gas phase}},}\ }\href@noop {} {\bibfield  {journal} {\bibinfo  {journal} {Chemical Physics Letters}\ }\textbf {\bibinfo {volume} {73}},\ \bibinfo {pages} {228--230} (\bibinfo {year} {1980})}\BibitemShut {NoStop}%
\bibitem [{\citenamefont {Truscott}, \citenamefont {Elliott},\ and\ \citenamefont {Western}(2009)}]{truscott2009reanalysis}%
  \BibitemOpen
  \bibfield  {author} {\bibinfo {author} {\bibfnamefont {B.~S.}\ \bibnamefont {Truscott}}, \bibinfo {author} {\bibfnamefont {N.~L.}\ \bibnamefont {Elliott}}, \ and\ \bibinfo {author} {\bibfnamefont {C.~M.}\ \bibnamefont {Western}},\ }\bibfield  {title} {\enquote {\bibinfo {title} {{A reanalysis of the $\tilde{A}^{1}$A$^{''}$ -- $\tilde{X}^{''}$A$^{''}$ transition of CFBr}},}\ }\href@noop {} {\bibfield  {journal} {\bibinfo  {journal} {The Journal of chemical physics}\ }\textbf {\bibinfo {volume} {130}} (\bibinfo {year} {2009})}\BibitemShut {NoStop}%
\bibitem [{\citenamefont {Guss}, \citenamefont {Votava},\ and\ \citenamefont {Kable}(2001)}]{guss2001electronic}%
  \BibitemOpen
  \bibfield  {author} {\bibinfo {author} {\bibfnamefont {J.~S.}\ \bibnamefont {Guss}}, \bibinfo {author} {\bibfnamefont {O.}~\bibnamefont {Votava}}, \ and\ \bibinfo {author} {\bibfnamefont {S.~H.}\ \bibnamefont {Kable}},\ }\bibfield  {title} {\enquote {\bibinfo {title} {{Electronic spectroscopy of jet-cooled CFCl: Laser-induced fluorescence, dispersed fluorescence, lifetimes, and C--Cl dissociation barrier}},}\ }\href@noop {} {\bibfield  {journal} {\bibinfo  {journal} {The Journal of Chemical Physics}\ }\textbf {\bibinfo {volume} {115}},\ \bibinfo {pages} {11118--11130} (\bibinfo {year} {2001})}\BibitemShut {NoStop}%
\bibitem [{\citenamefont {Raiche}\ and\ \citenamefont {Jeffries}(1993)}]{raiche1993laser}%
  \BibitemOpen
  \bibfield  {author} {\bibinfo {author} {\bibfnamefont {G.~A.}\ \bibnamefont {Raiche}}\ and\ \bibinfo {author} {\bibfnamefont {J.~B.}\ \bibnamefont {Jeffries}},\ }\bibfield  {title} {\enquote {\bibinfo {title} {Laser-induced fluorescence temperature measurements in a dc arcjet used for diamond deposition},}\ }\href@noop {} {\bibfield  {journal} {\bibinfo  {journal} {Applied optics}\ }\textbf {\bibinfo {volume} {32}},\ \bibinfo {pages} {4629--4635} (\bibinfo {year} {1993})}\BibitemShut {NoStop}%
\bibitem [{\citenamefont {Luque}, \citenamefont {Juchmann},\ and\ \citenamefont {Jeffries}(1997{\natexlab{b}})}]{luque1997absolute}%
  \BibitemOpen
  \bibfield  {author} {\bibinfo {author} {\bibfnamefont {J.}~\bibnamefont {Luque}}, \bibinfo {author} {\bibfnamefont {W.}~\bibnamefont {Juchmann}}, \ and\ \bibinfo {author} {\bibfnamefont {J.}~\bibnamefont {Jeffries}},\ }\bibfield  {title} {\enquote {\bibinfo {title} {{Absolute concentration measurements of CH radicals in a diamond-depositing dc-arcjet reactor}},}\ }\href@noop {} {\bibfield  {journal} {\bibinfo  {journal} {Applied optics}\ }\textbf {\bibinfo {volume} {36}},\ \bibinfo {pages} {3261--3270} (\bibinfo {year} {1997}{\natexlab{b}})}\BibitemShut {NoStop}%
\bibitem [{\citenamefont {Engelhard}\ \emph {et~al.}(1995)\citenamefont {Engelhard}, \citenamefont {Jacob}, \citenamefont {M{\"o}ller},\ and\ \citenamefont {Koch}}]{engelhard1995new}%
  \BibitemOpen
  \bibfield  {author} {\bibinfo {author} {\bibfnamefont {M.}~\bibnamefont {Engelhard}}, \bibinfo {author} {\bibfnamefont {W.}~\bibnamefont {Jacob}}, \bibinfo {author} {\bibfnamefont {W.}~\bibnamefont {M{\"o}ller}}, \ and\ \bibinfo {author} {\bibfnamefont {A.}~\bibnamefont {Koch}},\ }\bibfield  {title} {\enquote {\bibinfo {title} {{New calibration method for the determination of the absolute density of CH radicals through laser-induced fluorescence}},}\ }\href@noop {} {\bibfield  {journal} {\bibinfo  {journal} {Applied optics}\ }\textbf {\bibinfo {volume} {34}},\ \bibinfo {pages} {4542--4551} (\bibinfo {year} {1995})}\BibitemShut {NoStop}%
\bibitem [{\citenamefont {Luque}\ \emph {et~al.}(2002)\citenamefont {Luque}, \citenamefont {Klein-Douwel}, \citenamefont {Jeffries}, \citenamefont {Smith},\ and\ \citenamefont {Crosley}}]{luque2002quantitative}%
  \BibitemOpen
  \bibfield  {author} {\bibinfo {author} {\bibfnamefont {J.}~\bibnamefont {Luque}}, \bibinfo {author} {\bibfnamefont {R.}~\bibnamefont {Klein-Douwel}}, \bibinfo {author} {\bibfnamefont {J.}~\bibnamefont {Jeffries}}, \bibinfo {author} {\bibfnamefont {G.}~\bibnamefont {Smith}}, \ and\ \bibinfo {author} {\bibfnamefont {D.}~\bibnamefont {Crosley}},\ }\bibfield  {title} {\enquote {\bibinfo {title} {{Quantitative laser-induced fluorescence of CH in atmospheric pressure flames}},}\ }\href@noop {} {\bibfield  {journal} {\bibinfo  {journal} {Applied Physics B}\ }\textbf {\bibinfo {volume} {75}},\ \bibinfo {pages} {779--790} (\bibinfo {year} {2002})}\BibitemShut {NoStop}%
\bibitem [{\citenamefont {Hakuta}(1984)}]{hakuta1984vibration}%
  \BibitemOpen
  \bibfield  {author} {\bibinfo {author} {\bibfnamefont {K.}~\bibnamefont {Hakuta}},\ }\bibfield  {title} {\enquote {\bibinfo {title} {Vibration-rotation spectrum of {HCF} ({X$^1$A$^{\prime}$}) by laser-induced fluorescence},}\ }\href@noop {} {\bibfield  {journal} {\bibinfo  {journal} {Journal of Molecular Spectroscopy}\ }\textbf {\bibinfo {volume} {106}},\ \bibinfo {pages} {56--63} (\bibinfo {year} {1984})}\BibitemShut {NoStop}%
\bibitem [{\citenamefont {L'Esp{\'e}rance}, \citenamefont {Williams},\ and\ \citenamefont {Fleming}(1997)}]{l1997detection}%
  \BibitemOpen
  \bibfield  {author} {\bibinfo {author} {\bibfnamefont {D.}~\bibnamefont {L'Esp{\'e}rance}}, \bibinfo {author} {\bibfnamefont {B.~A.}\ \bibnamefont {Williams}}, \ and\ \bibinfo {author} {\bibfnamefont {J.~W.}\ \bibnamefont {Fleming}},\ }\bibfield  {title} {\enquote {\bibinfo {title} {Detection of fluorocarbon intermediates in low-pressure premixed flames by laser-induced fluorescence},}\ }\href@noop {} {\bibfield  {journal} {\bibinfo  {journal} {Chemical physics letters}\ }\textbf {\bibinfo {volume} {280}},\ \bibinfo {pages} {113--118} (\bibinfo {year} {1997})}\BibitemShut {NoStop}%
\bibitem [{\citenamefont {Kakimoto}, \citenamefont {Saito},\ and\ \citenamefont {Hirota}(1981)}]{kakimoto1981doppler}%
  \BibitemOpen
  \bibfield  {author} {\bibinfo {author} {\bibfnamefont {M.}~\bibnamefont {Kakimoto}}, \bibinfo {author} {\bibfnamefont {S.}~\bibnamefont {Saito}}, \ and\ \bibinfo {author} {\bibfnamefont {E.}~\bibnamefont {Hirota}},\ }\bibfield  {title} {\enquote {\bibinfo {title} {{Doppler-limited dye laser excitation spectroscopy of HCF}},}\ }\href@noop {} {\bibfield  {journal} {\bibinfo  {journal} {Journal of Molecular Spectroscopy}\ }\textbf {\bibinfo {volume} {88}},\ \bibinfo {pages} {300--310} (\bibinfo {year} {1981})}\BibitemShut {NoStop}%
\bibitem [{\citenamefont {Qiu}, \citenamefont {Zhou},\ and\ \citenamefont {Shi}(1987)}]{qiu1987laser}%
  \BibitemOpen
  \bibfield  {author} {\bibinfo {author} {\bibfnamefont {Y.}~\bibnamefont {Qiu}}, \bibinfo {author} {\bibfnamefont {S.}~\bibnamefont {Zhou}}, \ and\ \bibinfo {author} {\bibfnamefont {J.}~\bibnamefont {Shi}},\ }\bibfield  {title} {\enquote {\bibinfo {title} {{Laser-induced fluorescence of HCF and HCCl}},}\ }\href@noop {} {\bibfield  {journal} {\bibinfo  {journal} {Chemical physics letters}\ }\textbf {\bibinfo {volume} {136}},\ \bibinfo {pages} {93--96} (\bibinfo {year} {1987})}\BibitemShut {NoStop}%
\bibitem [{\citenamefont {Ashfold}\ \emph {et~al.}(1980)\citenamefont {Ashfold}, \citenamefont {Castano}, \citenamefont {Hancock},\ and\ \citenamefont {Ketley}}]{ashfold1980laser}%
  \BibitemOpen
  \bibfield  {author} {\bibinfo {author} {\bibfnamefont {M.~N.}\ \bibnamefont {Ashfold}}, \bibinfo {author} {\bibfnamefont {F.}~\bibnamefont {Castano}}, \bibinfo {author} {\bibfnamefont {G.}~\bibnamefont {Hancock}}, \ and\ \bibinfo {author} {\bibfnamefont {G.}~\bibnamefont {Ketley}},\ }\bibfield  {title} {\enquote {\bibinfo {title} {{Laser-induced fluorescence of the CHF radical}},}\ }\href@noop {} {\bibfield  {journal} {\bibinfo  {journal} {Chemical Physics Letters}\ }\textbf {\bibinfo {volume} {73}},\ \bibinfo {pages} {421--424} (\bibinfo {year} {1980})}\BibitemShut {NoStop}%
\bibitem [{\citenamefont {Hancock}\ and\ \citenamefont {Ketley}(1982)}]{hancock1982chf}%
  \BibitemOpen
  \bibfield  {author} {\bibinfo {author} {\bibfnamefont {G.}~\bibnamefont {Hancock}}\ and\ \bibinfo {author} {\bibfnamefont {G.~W.}\ \bibnamefont {Ketley}},\ }\bibfield  {title} {\enquote {\bibinfo {title} {{CHF (X$^1$A$^{'}$) radical kinetics. Part 1.--Reaction with NO and O$_2$}},}\ }\href@noop {} {\bibfield  {journal} {\bibinfo  {journal} {Journal of the Chemical Society, Faraday Transactions 2: Molecular and Chemical Physics}\ }\textbf {\bibinfo {volume} {78}},\ \bibinfo {pages} {1283--1291} (\bibinfo {year} {1982})}\BibitemShut {NoStop}%
\bibitem [{\citenamefont {Schmidt}, \citenamefont {Bacskay},\ and\ \citenamefont {Kable}(1999)}]{schmidt1999characterization}%
  \BibitemOpen
  \bibfield  {author} {\bibinfo {author} {\bibfnamefont {T.~W.}\ \bibnamefont {Schmidt}}, \bibinfo {author} {\bibfnamefont {G.~B.}\ \bibnamefont {Bacskay}}, \ and\ \bibinfo {author} {\bibfnamefont {S.~H.}\ \bibnamefont {Kable}},\ }\bibfield  {title} {\enquote {\bibinfo {title} {{Characterization of the $\tilde{A}(^{1}A^{''})$ state of HCF by laser induced fluorescence spectroscopy}},}\ }\href@noop {} {\bibfield  {journal} {\bibinfo  {journal} {The Journal of chemical physics}\ }\textbf {\bibinfo {volume} {110}},\ \bibinfo {pages} {11277--11285} (\bibinfo {year} {1999})}\BibitemShut {NoStop}%
\bibitem [{\citenamefont {Danon}\ \emph {et~al.}(1978)\citenamefont {Danon}, \citenamefont {Filseth}, \citenamefont {Feldmann}, \citenamefont {Zacharias}, \citenamefont {Dugan},\ and\ \citenamefont {Welge}}]{danon1978laser}%
  \BibitemOpen
  \bibfield  {author} {\bibinfo {author} {\bibfnamefont {J.}~\bibnamefont {Danon}}, \bibinfo {author} {\bibfnamefont {S.}~\bibnamefont {Filseth}}, \bibinfo {author} {\bibfnamefont {D.}~\bibnamefont {Feldmann}}, \bibinfo {author} {\bibfnamefont {H.}~\bibnamefont {Zacharias}}, \bibinfo {author} {\bibfnamefont {C.}~\bibnamefont {Dugan}}, \ and\ \bibinfo {author} {\bibfnamefont {K.}~\bibnamefont {Welge}},\ }\bibfield  {title} {\enquote {\bibinfo {title} {{Laser induced fluorescence of CH$_2$ ($\tilde{a}^1$A$_1$) produced in the photodissociation of ketene at 337 nm. The CH$_2$ ($\tilde{a}^1$A$_1$ - $\tilde{X}^3$B$_1$) energy separation}},}\ }\href@noop {} {\bibfield  {journal} {\bibinfo  {journal} {Chemical Physics}\ }\textbf {\bibinfo {volume} {29}},\ \bibinfo {pages} {345--351} (\bibinfo {year} {1978})}\BibitemShut {NoStop}%
\bibitem [{\citenamefont {Brackmann}\ \emph {et~al.}(2005)\citenamefont {Brackmann}, \citenamefont {Li}, \citenamefont {Rupinski}, \citenamefont {Docquier}, \citenamefont {Pengloan},\ and\ \citenamefont {Ald{\'e}n}}]{brackmann2005strategies}%
  \BibitemOpen
  \bibfield  {author} {\bibinfo {author} {\bibfnamefont {C.}~\bibnamefont {Brackmann}}, \bibinfo {author} {\bibfnamefont {Z.}~\bibnamefont {Li}}, \bibinfo {author} {\bibfnamefont {M.}~\bibnamefont {Rupinski}}, \bibinfo {author} {\bibfnamefont {N.}~\bibnamefont {Docquier}}, \bibinfo {author} {\bibfnamefont {G.}~\bibnamefont {Pengloan}}, \ and\ \bibinfo {author} {\bibfnamefont {M.}~\bibnamefont {Ald{\'e}n}},\ }\bibfield  {title} {\enquote {\bibinfo {title} {{Strategies for formaldehyde detection in flames and engines using a single-mode Nd: YAG/OPO laser system}},}\ }\href@noop {} {\bibfield  {journal} {\bibinfo  {journal} {Applied spectroscopy}\ }\textbf {\bibinfo {volume} {59}},\ \bibinfo {pages} {763--768} (\bibinfo {year} {2005})}\BibitemShut {NoStop}%
\bibitem [{\citenamefont {Harrington}\ and\ \citenamefont {Smyth}(1993)}]{harrington1993laser}%
  \BibitemOpen
  \bibfield  {author} {\bibinfo {author} {\bibfnamefont {J.~E.}\ \bibnamefont {Harrington}}\ and\ \bibinfo {author} {\bibfnamefont {K.~C.}\ \bibnamefont {Smyth}},\ }\bibfield  {title} {\enquote {\bibinfo {title} {Laser-induced fluorescence measurements of formaldehyde in a methane/air diffusion flame},}\ }\href@noop {} {\bibfield  {journal} {\bibinfo  {journal} {Chemical Physics Letters}\ }\textbf {\bibinfo {volume} {202}},\ \bibinfo {pages} {196--202} (\bibinfo {year} {1993})}\BibitemShut {NoStop}%
\bibitem [{\citenamefont {Gutman}, \citenamefont {Sanders},\ and\ \citenamefont {Butler}(1982)}]{gutman1982kinetics}%
  \BibitemOpen
  \bibfield  {author} {\bibinfo {author} {\bibfnamefont {D.}~\bibnamefont {Gutman}}, \bibinfo {author} {\bibfnamefont {N.}~\bibnamefont {Sanders}}, \ and\ \bibinfo {author} {\bibfnamefont {J.}~\bibnamefont {Butler}},\ }\bibfield  {title} {\enquote {\bibinfo {title} {Kinetics of the reactions of methoxy and ethoxy radicals with oxygen},}\ }\href@noop {} {\bibfield  {journal} {\bibinfo  {journal} {The Journal of Physical Chemistry}\ }\textbf {\bibinfo {volume} {86}},\ \bibinfo {pages} {66--70} (\bibinfo {year} {1982})}\BibitemShut {NoStop}%
\bibitem [{\citenamefont {Ebata}\ \emph {et~al.}(1982)\citenamefont {Ebata}, \citenamefont {Yanagishita}, \citenamefont {Obi},\ and\ \citenamefont {Tanaka}}]{ebata1982a}%
  \BibitemOpen
  \bibfield  {author} {\bibinfo {author} {\bibfnamefont {T.}~\bibnamefont {Ebata}}, \bibinfo {author} {\bibfnamefont {H.}~\bibnamefont {Yanagishita}}, \bibinfo {author} {\bibfnamefont {K.}~\bibnamefont {Obi}}, \ and\ \bibinfo {author} {\bibfnamefont {I.}~\bibnamefont {Tanaka}},\ }\bibfield  {title} {\enquote {\bibinfo {title} {{$\tilde{A}$ $\rightarrow$ $\tilde{X}$ fluorescence spectra of CH$_3$O and C$_2$H$_5$O generated by the ArF laser photolysis of alkyl nitrites}},}\ }\href@noop {} {\bibfield  {journal} {\bibinfo  {journal} {Chemical Physics}\ }\textbf {\bibinfo {volume} {69}},\ \bibinfo {pages} {27--33} (\bibinfo {year} {1982})}\BibitemShut {NoStop}%
\bibitem [{\citenamefont {Inoue}, \citenamefont {Akimoto},\ and\ \citenamefont {Okuda}(1979)}]{inoue1979laser}%
  \BibitemOpen
  \bibfield  {author} {\bibinfo {author} {\bibfnamefont {G.}~\bibnamefont {Inoue}}, \bibinfo {author} {\bibfnamefont {H.}~\bibnamefont {Akimoto}}, \ and\ \bibinfo {author} {\bibfnamefont {M.}~\bibnamefont {Okuda}},\ }\bibfield  {title} {\enquote {\bibinfo {title} {{Laser-induced fluorescence spectra of CH$_3$O}},}\ }\href@noop {} {\bibfield  {journal} {\bibinfo  {journal} {Chemical Physics Letters}\ }\textbf {\bibinfo {volume} {63}},\ \bibinfo {pages} {213--216} (\bibinfo {year} {1979})}\BibitemShut {NoStop}%
\bibitem [{\citenamefont {Inoue}, \citenamefont {Akimoto},\ and\ \citenamefont {Okuda}(1980)}]{inoue1980spectroscopy}%
  \BibitemOpen
  \bibfield  {author} {\bibinfo {author} {\bibfnamefont {G.}~\bibnamefont {Inoue}}, \bibinfo {author} {\bibfnamefont {H.}~\bibnamefont {Akimoto}}, \ and\ \bibinfo {author} {\bibfnamefont {M.}~\bibnamefont {Okuda}},\ }\bibfield  {title} {\enquote {\bibinfo {title} {{Spectroscopy of the CH$_3$O A$^2$A$_1$--X$^2$E system by laser-excited fluorescence method}},}\ }\href@noop {} {\bibfield  {journal} {\bibinfo  {journal} {The Journal of Chemical Physics}\ }\textbf {\bibinfo {volume} {72}},\ \bibinfo {pages} {1769--1775} (\bibinfo {year} {1980})}\BibitemShut {NoStop}%
\bibitem [{\citenamefont {Kappert}\ and\ \citenamefont {Temps}(1989)}]{kappert1989rotationally}%
  \BibitemOpen
  \bibfield  {author} {\bibinfo {author} {\bibfnamefont {J.}~\bibnamefont {Kappert}}\ and\ \bibinfo {author} {\bibfnamefont {F.}~\bibnamefont {Temps}},\ }\bibfield  {title} {\enquote {\bibinfo {title} {{Rotationally resolved laser-induced fluorescence excitation studies of CH$_3$O}},}\ }\href@noop {} {\bibfield  {journal} {\bibinfo  {journal} {Chemical Physics}\ }\textbf {\bibinfo {volume} {132}},\ \bibinfo {pages} {197--208} (\bibinfo {year} {1989})}\BibitemShut {NoStop}%
\bibitem [{\citenamefont {Zu}\ \emph {et~al.}(2004)\citenamefont {Zu}, \citenamefont {Liu}, \citenamefont {Tarczay}, \citenamefont {Dupr{\'e}},\ and\ \citenamefont {Miller}}]{zu2004jet}%
  \BibitemOpen
  \bibfield  {author} {\bibinfo {author} {\bibfnamefont {L.}~\bibnamefont {Zu}}, \bibinfo {author} {\bibfnamefont {J.}~\bibnamefont {Liu}}, \bibinfo {author} {\bibfnamefont {G.}~\bibnamefont {Tarczay}}, \bibinfo {author} {\bibfnamefont {P.}~\bibnamefont {Dupr{\'e}}}, \ and\ \bibinfo {author} {\bibfnamefont {T.~A.}\ \bibnamefont {Miller}},\ }\bibfield  {title} {\enquote {\bibinfo {title} {Jet-cooled laser spectroscopy of the cyclohexoxy radical},}\ }\href@noop {} {\bibfield  {journal} {\bibinfo  {journal} {The Journal of chemical physics}\ }\textbf {\bibinfo {volume} {120}},\ \bibinfo {pages} {10579--10593} (\bibinfo {year} {2004})}\BibitemShut {NoStop}%
\bibitem [{\citenamefont {Suzuki}, \citenamefont {Inoue},\ and\ \citenamefont {Akimoto}(1984)}]{suzuki1984laser}%
  \BibitemOpen
  \bibfield  {author} {\bibinfo {author} {\bibfnamefont {M.}~\bibnamefont {Suzuki}}, \bibinfo {author} {\bibfnamefont {G.}~\bibnamefont {Inoue}}, \ and\ \bibinfo {author} {\bibfnamefont {H.}~\bibnamefont {Akimoto}},\ }\bibfield  {title} {\enquote {\bibinfo {title} {{Laser induced fluorescence of CH$_3$S and CD$_3$S radicals}},}\ }\href@noop {} {\bibfield  {journal} {\bibinfo  {journal} {The Journal of chemical physics}\ }\textbf {\bibinfo {volume} {81}},\ \bibinfo {pages} {5405--5412} (\bibinfo {year} {1984})}\BibitemShut {NoStop}%
\bibitem [{\citenamefont {Black}\ and\ \citenamefont {Jusinski}(1986)}]{black1986laser}%
  \BibitemOpen
  \bibfield  {author} {\bibinfo {author} {\bibfnamefont {G.}~\bibnamefont {Black}}\ and\ \bibinfo {author} {\bibfnamefont {L.~E.}\ \bibnamefont {Jusinski}},\ }\bibfield  {title} {\enquote {\bibinfo {title} {Laser-induced fluorescence studies of the {CH}$_3${S} radical},}\ }\href@noop {} {\bibfield  {journal} {\bibinfo  {journal} {Journal of the Chemical Society, Faraday Transactions 2: Molecular and Chemical Physics}\ }\textbf {\bibinfo {volume} {82}},\ \bibinfo {pages} {2143--2151} (\bibinfo {year} {1986})}\BibitemShut {NoStop}%
\bibitem [{\citenamefont {Conley}\ \emph {et~al.}(1980)\citenamefont {Conley}, \citenamefont {Halpern}, \citenamefont {Wood}, \citenamefont {Vaughn},\ and\ \citenamefont {Jackson}}]{conley1980laser}%
  \BibitemOpen
  \bibfield  {author} {\bibinfo {author} {\bibfnamefont {C.}~\bibnamefont {Conley}}, \bibinfo {author} {\bibfnamefont {J.~B.}\ \bibnamefont {Halpern}}, \bibinfo {author} {\bibfnamefont {J.}~\bibnamefont {Wood}}, \bibinfo {author} {\bibfnamefont {C.}~\bibnamefont {Vaughn}}, \ and\ \bibinfo {author} {\bibfnamefont {W.~M.}\ \bibnamefont {Jackson}},\ }\bibfield  {title} {\enquote {\bibinfo {title} {{Laser excitation of the CN B$^{2}\Sigma^{+}$ $\leftarrow$ A$^{2}\Pi$ 0--0 and 1--0 bands}},}\ }\href@noop {} {\bibfield  {journal} {\bibinfo  {journal} {Chemical Physics Letters}\ }\textbf {\bibinfo {volume} {73}},\ \bibinfo {pages} {224--227} (\bibinfo {year} {1980})}\BibitemShut {NoStop}%
\bibitem [{\citenamefont {Cody}, \citenamefont {Sabety-Dzvonik},\ and\ \citenamefont {Jackson}(1977)}]{cody1977laser}%
  \BibitemOpen
  \bibfield  {author} {\bibinfo {author} {\bibfnamefont {R.}~\bibnamefont {Cody}}, \bibinfo {author} {\bibfnamefont {M.~J.}\ \bibnamefont {Sabety-Dzvonik}}, \ and\ \bibinfo {author} {\bibfnamefont {W.}~\bibnamefont {Jackson}},\ }\bibfield  {title} {\enquote {\bibinfo {title} {{Laser-induced fluorescence of CN(X$^{2}\Sigma^{+}$) produced by photolysis of C$_2$N$_2$ at 160 nm}},}\ }\href@noop {} {\bibfield  {journal} {\bibinfo  {journal} {The Journal of Chemical Physics}\ }\textbf {\bibinfo {volume} {66}},\ \bibinfo {pages} {2145--2152} (\bibinfo {year} {1977})}\BibitemShut {NoStop}%
\bibitem [{\citenamefont {Bonczyk}\ and\ \citenamefont {Shirley}(1979)}]{bonczyk1979measurement}%
  \BibitemOpen
  \bibfield  {author} {\bibinfo {author} {\bibfnamefont {P.}~\bibnamefont {Bonczyk}}\ and\ \bibinfo {author} {\bibfnamefont {J.}~\bibnamefont {Shirley}},\ }\bibfield  {title} {\enquote {\bibinfo {title} {{Measurement of CH and CN concentration in flames by laser-induced saturated fluorescence}},}\ }\href@noop {} {\bibfield  {journal} {\bibinfo  {journal} {Combustion and flame}\ }\textbf {\bibinfo {volume} {34}},\ \bibinfo {pages} {253--264} (\bibinfo {year} {1979})}\BibitemShut {NoStop}%
\bibitem [{\citenamefont {Sutton}, \citenamefont {Williams},\ and\ \citenamefont {Fleming}(2008)}]{sutton2008laser}%
  \BibitemOpen
  \bibfield  {author} {\bibinfo {author} {\bibfnamefont {J.~A.}\ \bibnamefont {Sutton}}, \bibinfo {author} {\bibfnamefont {B.~A.}\ \bibnamefont {Williams}}, \ and\ \bibinfo {author} {\bibfnamefont {J.~W.}\ \bibnamefont {Fleming}},\ }\bibfield  {title} {\enquote {\bibinfo {title} {{Laser-induced fluorescence measurements of NCN in low-pressure CH$_4$/O$_2$/N$_2$ flames and its role in prompt NO formation}},}\ }\href@noop {} {\bibfield  {journal} {\bibinfo  {journal} {Combustion and Flame}\ }\textbf {\bibinfo {volume} {153}},\ \bibinfo {pages} {465--478} (\bibinfo {year} {2008})}\BibitemShut {NoStop}%
\bibitem [{\citenamefont {Smith}, \citenamefont {Copeland},\ and\ \citenamefont {Crosley}(1989)}]{smith1989electronic}%
  \BibitemOpen
  \bibfield  {author} {\bibinfo {author} {\bibfnamefont {G.~P.}\ \bibnamefont {Smith}}, \bibinfo {author} {\bibfnamefont {R.~A.}\ \bibnamefont {Copeland}}, \ and\ \bibinfo {author} {\bibfnamefont {D.~R.}\ \bibnamefont {Crosley}},\ }\bibfield  {title} {\enquote {\bibinfo {title} {{Electronic quenching, fluorescence lifetime, and spectroscopy of the A$^{3}\Pi_{u}$ state of NCN}},}\ }\href@noop {} {\bibfield  {journal} {\bibinfo  {journal} {The Journal of chemical physics}\ }\textbf {\bibinfo {volume} {91}},\ \bibinfo {pages} {1987--1993} (\bibinfo {year} {1989})}\BibitemShut {NoStop}%
\bibitem [{\citenamefont {Curtis}, \citenamefont {Levick},\ and\ \citenamefont {Sarre}(1988)}]{curtis1988laser}%
  \BibitemOpen
  \bibfield  {author} {\bibinfo {author} {\bibfnamefont {M.}~\bibnamefont {Curtis}}, \bibinfo {author} {\bibfnamefont {A.}~\bibnamefont {Levick}}, \ and\ \bibinfo {author} {\bibfnamefont {P.}~\bibnamefont {Sarre}},\ }\bibfield  {title} {\enquote {\bibinfo {title} {{Laser-induced-fluorescence spectrum of the CNN molecule}},}\ }\href@noop {} {\bibfield  {journal} {\bibinfo  {journal} {Laser chemistry}\ }\textbf {\bibinfo {volume} {9}},\ \bibinfo {pages} {359--368} (\bibinfo {year} {1988})}\BibitemShut {NoStop}%
\bibitem [{\citenamefont {Dally}\ \emph {et~al.}(2003)\citenamefont {Dally}, \citenamefont {Masri}, \citenamefont {Barlow},\ and\ \citenamefont {Fiechtner}}]{dally2003two}%
  \BibitemOpen
  \bibfield  {author} {\bibinfo {author} {\bibfnamefont {B.}~\bibnamefont {Dally}}, \bibinfo {author} {\bibfnamefont {A.}~\bibnamefont {Masri}}, \bibinfo {author} {\bibfnamefont {R.}~\bibnamefont {Barlow}}, \ and\ \bibinfo {author} {\bibfnamefont {G.}~\bibnamefont {Fiechtner}},\ }\bibfield  {title} {\enquote {\bibinfo {title} {Two-photon laser-induced fluorescence measurement of {CO} in turbulent non-premixed bluff body flames},}\ }\href@noop {} {\bibfield  {journal} {\bibinfo  {journal} {Combustion and flame}\ }\textbf {\bibinfo {volume} {132}},\ \bibinfo {pages} {272--274} (\bibinfo {year} {2003})}\BibitemShut {NoStop}%
\bibitem [{\citenamefont {Zetterberg}\ \emph {et~al.}(2012)\citenamefont {Zetterberg}, \citenamefont {Blomberg}, \citenamefont {Gustafson}, \citenamefont {Sun}, \citenamefont {Li}, \citenamefont {Lundgren},\ and\ \citenamefont {Ald{\'e}n}}]{zetterberg2012situ}%
  \BibitemOpen
  \bibfield  {author} {\bibinfo {author} {\bibfnamefont {J.}~\bibnamefont {Zetterberg}}, \bibinfo {author} {\bibfnamefont {S.}~\bibnamefont {Blomberg}}, \bibinfo {author} {\bibfnamefont {J.}~\bibnamefont {Gustafson}}, \bibinfo {author} {\bibfnamefont {Z.}~\bibnamefont {Sun}}, \bibinfo {author} {\bibfnamefont {Z.}~\bibnamefont {Li}}, \bibinfo {author} {\bibfnamefont {E.}~\bibnamefont {Lundgren}}, \ and\ \bibinfo {author} {\bibfnamefont {M.}~\bibnamefont {Ald{\'e}n}},\ }\bibfield  {title} {\enquote {\bibinfo {title} {{An in situ set up for the detection of CO$_{2}$ from catalytic CO oxidation by using planar laser-induced fluorescence}},}\ }\href@noop {} {\bibfield  {journal} {\bibinfo  {journal} {Review of scientific instruments}\ }\textbf {\bibinfo {volume} {83}} (\bibinfo {year} {2012})}\BibitemShut {NoStop}%
\bibitem [{\citenamefont {Alwahabi}\ \emph {et~al.}(2007)\citenamefont {Alwahabi}, \citenamefont {Zetterberg}, \citenamefont {Li},\ and\ \citenamefont {Ald{\'e}n}}]{alwahabi2007high}%
  \BibitemOpen
  \bibfield  {author} {\bibinfo {author} {\bibfnamefont {Z.}~\bibnamefont {Alwahabi}}, \bibinfo {author} {\bibfnamefont {J.}~\bibnamefont {Zetterberg}}, \bibinfo {author} {\bibfnamefont {Z.}~\bibnamefont {Li}}, \ and\ \bibinfo {author} {\bibfnamefont {M.}~\bibnamefont {Ald{\'e}n}},\ }\bibfield  {title} {\enquote {\bibinfo {title} {{High resolution polarization spectroscopy and laser induced fluorescence of CO$_{2}$ around 2$\mu$m}},}\ }\href@noop {} {\bibfield  {journal} {\bibinfo  {journal} {The European Physical Journal D}\ }\textbf {\bibinfo {volume} {42}},\ \bibinfo {pages} {41--47} (\bibinfo {year} {2007})}\BibitemShut {NoStop}%
\bibitem [{\citenamefont {Clough}\ and\ \citenamefont {Johnston}(1980)}]{clough1980laser}%
  \BibitemOpen
  \bibfield  {author} {\bibinfo {author} {\bibfnamefont {P.}~\bibnamefont {Clough}}\ and\ \bibinfo {author} {\bibfnamefont {J.}~\bibnamefont {Johnston}},\ }\bibfield  {title} {\enquote {\bibinfo {title} {{Laser-induced fluorescence measurement of the vibrational distribution of CS formed in the reaction O+ CS$_2$ $\rightarrow$ CS + SO}},}\ }\href@noop {} {\bibfield  {journal} {\bibinfo  {journal} {Chemical Physics Letters}\ }\textbf {\bibinfo {volume} {71}},\ \bibinfo {pages} {253--257} (\bibinfo {year} {1980})}\BibitemShut {NoStop}%
\bibitem [{\citenamefont {Hynes}\ and\ \citenamefont {Brophy}(1979)}]{hynes1979laser}%
  \BibitemOpen
  \bibfield  {author} {\bibinfo {author} {\bibfnamefont {A.~J.}\ \bibnamefont {Hynes}}\ and\ \bibinfo {author} {\bibfnamefont {J.~H.}\ \bibnamefont {Brophy}},\ }\bibfield  {title} {\enquote {\bibinfo {title} {{Laser-induced fluorescence study of radiative lifetimes and quenching rates in CS A$^{1}\Pi$ ($\upsilon$= 0)}},}\ }\href@noop {} {\bibfield  {journal} {\bibinfo  {journal} {Chemical physics letters}\ }\textbf {\bibinfo {volume} {63}},\ \bibinfo {pages} {93--96} (\bibinfo {year} {1979})}\BibitemShut {NoStop}%
\bibitem [{\citenamefont {Martin}\ and\ \citenamefont {Donovan}(1982)}]{martin1982two}%
  \BibitemOpen
  \bibfield  {author} {\bibinfo {author} {\bibfnamefont {M.}~\bibnamefont {Martin}}\ and\ \bibinfo {author} {\bibfnamefont {R.~J.}\ \bibnamefont {Donovan}},\ }\bibfield  {title} {\enquote {\bibinfo {title} {{Two-photon dissociation of CS$_2$ with a KrF laser ($\lambda$= 248 nm)}},}\ }\href@noop {} {\bibfield  {journal} {\bibinfo  {journal} {Journal of Photochemistry}\ }\textbf {\bibinfo {volume} {18}},\ \bibinfo {pages} {245--250} (\bibinfo {year} {1982})}\BibitemShut {NoStop}%
\bibitem [{\citenamefont {Ochi}\ \emph {et~al.}(1987)\citenamefont {Ochi}, \citenamefont {Watanabe}, \citenamefont {Tsuchiya},\ and\ \citenamefont {Koda}}]{ochi1987rotationally}%
  \BibitemOpen
  \bibfield  {author} {\bibinfo {author} {\bibfnamefont {N.}~\bibnamefont {Ochi}}, \bibinfo {author} {\bibfnamefont {H.}~\bibnamefont {Watanabe}}, \bibinfo {author} {\bibfnamefont {S.}~\bibnamefont {Tsuchiya}}, \ and\ \bibinfo {author} {\bibfnamefont {S.}~\bibnamefont {Koda}},\ }\bibfield  {title} {\enquote {\bibinfo {title} {{Rotationally resolved laser-induced fluorescence and zeeman quantum beat spectroscopy of the V $^{1}$B$_{2}$ state of jet-cooled CS$_{2}$}},}\ }\href@noop {} {\bibfield  {journal} {\bibinfo  {journal} {Chemical physics}\ }\textbf {\bibinfo {volume} {113}},\ \bibinfo {pages} {271--285} (\bibinfo {year} {1987})}\BibitemShut {NoStop}%
\bibitem [{\citenamefont {Weyh}\ and\ \citenamefont {Demtr{\"o}der}(1996)}]{weyh1996lifetime}%
  \BibitemOpen
  \bibfield  {author} {\bibinfo {author} {\bibfnamefont {T.}~\bibnamefont {Weyh}}\ and\ \bibinfo {author} {\bibfnamefont {W.}~\bibnamefont {Demtr{\"o}der}},\ }\bibfield  {title} {\enquote {\bibinfo {title} {{Lifetime measurements of selectively excited rovibrational levels of the V$^{1}$B$_{2}$ state of CS$_{2}$}},}\ }\href@noop {} {\bibfield  {journal} {\bibinfo  {journal} {The Journal of chemical physics}\ }\textbf {\bibinfo {volume} {104}},\ \bibinfo {pages} {6938--6948} (\bibinfo {year} {1996})}\BibitemShut {NoStop}%
\bibitem [{\citenamefont {Mart{\'\i}nez}\ \emph {et~al.}(1997)\citenamefont {Mart{\'\i}nez}, \citenamefont {L{\'o}pez}, \citenamefont {Albaladejo},\ and\ \citenamefont {Poblete}}]{martinez1997laser}%
  \BibitemOpen
  \bibfield  {author} {\bibinfo {author} {\bibfnamefont {E.}~\bibnamefont {Mart{\'\i}nez}}, \bibinfo {author} {\bibfnamefont {M.}~\bibnamefont {L{\'o}pez}}, \bibinfo {author} {\bibfnamefont {J.}~\bibnamefont {Albaladejo}}, \ and\ \bibinfo {author} {\bibfnamefont {F.}~\bibnamefont {Poblete}},\ }\bibfield  {title} {\enquote {\bibinfo {title} {{Laser-induced fluorescence from selected excited states of the CS$_2$ molecule}},}\ }\href@noop {} {\bibfield  {journal} {\bibinfo  {journal} {Journal of molecular structure}\ }\textbf {\bibinfo {volume} {408}},\ \bibinfo {pages} {553--556} (\bibinfo {year} {1997})}\BibitemShut {NoStop}%
\bibitem [{\citenamefont {Sisk}\ \emph {et~al.}(1999)\citenamefont {Sisk}, \citenamefont {Sarkar}, \citenamefont {Ikeda},\ and\ \citenamefont {Hayashi}}]{sisk1999influence}%
  \BibitemOpen
  \bibfield  {author} {\bibinfo {author} {\bibfnamefont {W.~N.}\ \bibnamefont {Sisk}}, \bibinfo {author} {\bibfnamefont {N.}~\bibnamefont {Sarkar}}, \bibinfo {author} {\bibfnamefont {S.}~\bibnamefont {Ikeda}}, \ and\ \bibinfo {author} {\bibfnamefont {H.}~\bibnamefont {Hayashi}},\ }\bibfield  {title} {\enquote {\bibinfo {title} {{Influence of large magnetic fields on fluorescence of gaseous CS$_2$ excited through several V bands}},}\ }\href@noop {} {\bibfield  {journal} {\bibinfo  {journal} {The Journal of Physical Chemistry A}\ }\textbf {\bibinfo {volume} {103}},\ \bibinfo {pages} {7179--7185} (\bibinfo {year} {1999})}\BibitemShut {NoStop}%
\bibitem [{\citenamefont {Sunahori}, \citenamefont {Zhang},\ and\ \citenamefont {Clouthier}(2006)}]{sunahori2006electronic}%
  \BibitemOpen
  \bibfield  {author} {\bibinfo {author} {\bibfnamefont {F.~X.}\ \bibnamefont {Sunahori}}, \bibinfo {author} {\bibfnamefont {X.}~\bibnamefont {Zhang}}, \ and\ \bibinfo {author} {\bibfnamefont {D.~J.}\ \bibnamefont {Clouthier}},\ }\bibfield  {title} {\enquote {\bibinfo {title} {The electronic spectrum of jet-cooled copper hydrosulfide (cush)},}\ }\href@noop {} {\bibfield  {journal} {\bibinfo  {journal} {The Journal of chemical physics}\ }\textbf {\bibinfo {volume} {125}} (\bibinfo {year} {2006})}\BibitemShut {NoStop}%
\bibitem [{\citenamefont {Dreyfus}(1991)}]{dreyfus1991cu0}%
  \BibitemOpen
  \bibfield  {author} {\bibinfo {author} {\bibfnamefont {R.}~\bibnamefont {Dreyfus}},\ }\bibfield  {title} {\enquote {\bibinfo {title} {{Cu$^0$, Cu$^{+}$, and Cu$_{2}$ from excimer-ablated copper}},}\ }\href@noop {} {\bibfield  {journal} {\bibinfo  {journal} {Journal of applied physics}\ }\textbf {\bibinfo {volume} {69}},\ \bibinfo {pages} {1721--1729} (\bibinfo {year} {1991})}\BibitemShut {NoStop}%
\bibitem [{\citenamefont {Vekselman}\ \emph {et~al.}(2018)\citenamefont {Vekselman}, \citenamefont {Khrabry}, \citenamefont {Kaganovich}, \citenamefont {Stratton}, \citenamefont {Selinsky},\ and\ \citenamefont {Raitses}}]{vekselman2018quantitative}%
  \BibitemOpen
  \bibfield  {author} {\bibinfo {author} {\bibfnamefont {V.}~\bibnamefont {Vekselman}}, \bibinfo {author} {\bibfnamefont {A.}~\bibnamefont {Khrabry}}, \bibinfo {author} {\bibfnamefont {I.}~\bibnamefont {Kaganovich}}, \bibinfo {author} {\bibfnamefont {B.}~\bibnamefont {Stratton}}, \bibinfo {author} {\bibfnamefont {R.}~\bibnamefont {Selinsky}}, \ and\ \bibinfo {author} {\bibfnamefont {Y.}~\bibnamefont {Raitses}},\ }\bibfield  {title} {\enquote {\bibinfo {title} {Quantitative imaging of carbon dimer precursor for nanomaterial synthesis in the carbon arc},}\ }\href@noop {} {\bibfield  {journal} {\bibinfo  {journal} {Plasma Sources Science and Technology}\ }\textbf {\bibinfo {volume} {27}},\ \bibinfo {pages} {025008} (\bibinfo {year} {2018})}\BibitemShut {NoStop}%
\bibitem [{\citenamefont {Reisler}, \citenamefont {Mangir},\ and\ \citenamefont {Wittig}(1980)}]{reisler1980kinetics}%
  \BibitemOpen
  \bibfield  {author} {\bibinfo {author} {\bibfnamefont {H.}~\bibnamefont {Reisler}}, \bibinfo {author} {\bibfnamefont {M.}~\bibnamefont {Mangir}}, \ and\ \bibinfo {author} {\bibfnamefont {C.}~\bibnamefont {Wittig}},\ }\bibfield  {title} {\enquote {\bibinfo {title} {{The kinetics of free radicals generated by IR laser photolysis. II. Reactions of C$_2$(X$^{1}\Sigma^{+}_{g}$), C$_2$(a$^{3}\Pi_{u}$), C$_{3}(\tilde{X}^{1}\Sigma^{+}_{g})$ and CN(X$^{2}\Sigma^{+}$) with O$_2$}},}\ }\href@noop {} {\bibfield  {journal} {\bibinfo  {journal} {Chemical Physics}\ }\textbf {\bibinfo {volume} {47}},\ \bibinfo {pages} {49--58} (\bibinfo {year} {1980})}\BibitemShut {NoStop}%
\bibitem [{\citenamefont {Pitts}, \citenamefont {Pasternack},\ and\ \citenamefont {McDonald}(1982)}]{pitts1982temperature}%
  \BibitemOpen
  \bibfield  {author} {\bibinfo {author} {\bibfnamefont {W.~M.}\ \bibnamefont {Pitts}}, \bibinfo {author} {\bibfnamefont {L.}~\bibnamefont {Pasternack}}, \ and\ \bibinfo {author} {\bibfnamefont {J.}~\bibnamefont {McDonald}},\ }\bibfield  {title} {\enquote {\bibinfo {title} {{Temperature dependence of the C$_2$ (X$^{1}\Sigma^{+}_{g}$) reaction with H$_2$ and CH$_4$ and C$_2$ (X$^{1}\Sigma_{g}^{+}$ and a$^{3}\Pi_{u}$ equilibrated states) with O$_2$}},}\ }\href@noop {} {\bibfield  {journal} {\bibinfo  {journal} {Chemical Physics}\ }\textbf {\bibinfo {volume} {68}},\ \bibinfo {pages} {417--422} (\bibinfo {year} {1982})}\BibitemShut {NoStop}%
\bibitem [{\citenamefont {Jones}\ and\ \citenamefont {Mackie}(1976)}]{jones1976evaluation}%
  \BibitemOpen
  \bibfield  {author} {\bibinfo {author} {\bibfnamefont {D.}~\bibnamefont {Jones}}\ and\ \bibinfo {author} {\bibfnamefont {J.}~\bibnamefont {Mackie}},\ }\bibfield  {title} {\enquote {\bibinfo {title} {{Evaluation of C$_2$ resonance fluorescence as a technique for transient flame studies}},}\ }\href@noop {} {\bibfield  {journal} {\bibinfo  {journal} {Combustion and Flame}\ }\textbf {\bibinfo {volume} {27}},\ \bibinfo {pages} {143--146} (\bibinfo {year} {1976})}\BibitemShut {NoStop}%
\bibitem [{\citenamefont {Tatarczyk}, \citenamefont {Fink},\ and\ \citenamefont {Becker}(1976)}]{tatarczyk1976lifetime}%
  \BibitemOpen
  \bibfield  {author} {\bibinfo {author} {\bibfnamefont {T.}~\bibnamefont {Tatarczyk}}, \bibinfo {author} {\bibfnamefont {E.}~\bibnamefont {Fink}}, \ and\ \bibinfo {author} {\bibfnamefont {K.}~\bibnamefont {Becker}},\ }\bibfield  {title} {\enquote {\bibinfo {title} {{Lifetime measurements on single vibrational levels of C$_2$ (d$^{3}\Pi_{g}$) by laser fluorescence excitation}},}\ }\href@noop {} {\bibfield  {journal} {\bibinfo  {journal} {Chemical Physics Letters}\ }\textbf {\bibinfo {volume} {40}},\ \bibinfo {pages} {126--130} (\bibinfo {year} {1976})}\BibitemShut {NoStop}%
\bibitem [{\citenamefont {Williams}\ and\ \citenamefont {Fleming}(2002)}]{williams2002laser}%
  \BibitemOpen
  \bibfield  {author} {\bibinfo {author} {\bibfnamefont {B.}~\bibnamefont {Williams}}\ and\ \bibinfo {author} {\bibfnamefont {J.}~\bibnamefont {Fleming}},\ }\bibfield  {title} {\enquote {\bibinfo {title} {Laser-induced fluorescence detection of acetylene in low-pressure propane and methane flames},}\ }\href@noop {} {\bibfield  {journal} {\bibinfo  {journal} {Applied Physics B}\ }\textbf {\bibinfo {volume} {75}},\ \bibinfo {pages} {883--890} (\bibinfo {year} {2002})}\BibitemShut {NoStop}%
\bibitem [{\citenamefont {Inoue}\ and\ \citenamefont {Akimoto}(1981)}]{inoue1981laserC2H3O}%
  \BibitemOpen
  \bibfield  {author} {\bibinfo {author} {\bibfnamefont {G.}~\bibnamefont {Inoue}}\ and\ \bibinfo {author} {\bibfnamefont {H.}~\bibnamefont {Akimoto}},\ }\bibfield  {title} {\enquote {\bibinfo {title} {{Laser-induced fluorescence of the C$_2$H$_3$O radical}},}\ }\href@noop {} {\bibfield  {journal} {\bibinfo  {journal} {The Journal of Chemical Physics}\ }\textbf {\bibinfo {volume} {74}},\ \bibinfo {pages} {425--433} (\bibinfo {year} {1981})}\BibitemShut {NoStop}%
\bibitem [{\citenamefont {Kleinermanns}\ and\ \citenamefont {Luntz}(1981)}]{kleinermanns1981laser}%
  \BibitemOpen
  \bibfield  {author} {\bibinfo {author} {\bibfnamefont {K.}~\bibnamefont {Kleinermanns}}\ and\ \bibinfo {author} {\bibfnamefont {A.}~\bibnamefont {Luntz}},\ }\bibfield  {title} {\enquote {\bibinfo {title} {{Laser-induced fluorescence of CH$_2$CHO produced in the crossed molecular beam reactions of O($^3$P) with Olefins}},}\ }\href@noop {} {\bibfield  {journal} {\bibinfo  {journal} {The Journal of Physical Chemistry}\ }\textbf {\bibinfo {volume} {85}},\ \bibinfo {pages} {1966--1968} (\bibinfo {year} {1981})}\BibitemShut {NoStop}%
\bibitem [{\citenamefont {DiMauro}, \citenamefont {Heaven},\ and\ \citenamefont {Miller}(1984)}]{dimauro1984laser}%
  \BibitemOpen
  \bibfield  {author} {\bibinfo {author} {\bibfnamefont {L.}~\bibnamefont {DiMauro}}, \bibinfo {author} {\bibfnamefont {M.}~\bibnamefont {Heaven}}, \ and\ \bibinfo {author} {\bibfnamefont {T.~A.}\ \bibnamefont {Miller}},\ }\bibfield  {title} {\enquote {\bibinfo {title} {{Laser induced fluorescence study of the $\tilde{B}^{2}$A$^{''}$ $\rightarrow$ $\tilde{X}^{2}$A$^{''}$ transition of the vinoxy radical in a supersonic free jet expansion}},}\ }\href@noop {} {\bibfield  {journal} {\bibinfo  {journal} {The Journal of chemical physics}\ }\textbf {\bibinfo {volume} {81}},\ \bibinfo {pages} {2339--2346} (\bibinfo {year} {1984})}\BibitemShut {NoStop}%
\bibitem [{\citenamefont {Nakajima}\ \emph {et~al.}(2007{\natexlab{b}})\citenamefont {Nakajima}, \citenamefont {Miyoshi}, \citenamefont {Sumiyoshi},\ and\ \citenamefont {Endo}}]{nakajima2007laser}%
  \BibitemOpen
  \bibfield  {author} {\bibinfo {author} {\bibfnamefont {M.}~\bibnamefont {Nakajima}}, \bibinfo {author} {\bibfnamefont {A.}~\bibnamefont {Miyoshi}}, \bibinfo {author} {\bibfnamefont {Y.}~\bibnamefont {Sumiyoshi}}, \ and\ \bibinfo {author} {\bibfnamefont {Y.}~\bibnamefont {Endo}},\ }\bibfield  {title} {\enquote {\bibinfo {title} {{Laser-induced fluorescence and pure rotational spectroscopy of the CH$_2$CHS (vinylthio) radical}},}\ }\href@noop {} {\bibfield  {journal} {\bibinfo  {journal} {The Journal of chemical physics}\ }\textbf {\bibinfo {volume} {126}} (\bibinfo {year} {2007}{\natexlab{b}})}\BibitemShut {NoStop}%
\bibitem [{\citenamefont {Inoue}, \citenamefont {Okuda},\ and\ \citenamefont {Akimoto}(1981)}]{inoue1981laser}%
  \BibitemOpen
  \bibfield  {author} {\bibinfo {author} {\bibfnamefont {G.}~\bibnamefont {Inoue}}, \bibinfo {author} {\bibfnamefont {M.}~\bibnamefont {Okuda}}, \ and\ \bibinfo {author} {\bibfnamefont {H.}~\bibnamefont {Akimoto}},\ }\bibfield  {title} {\enquote {\bibinfo {title} {{Laser-induced fluorescence of the C$_2$H$_5$O radical}},}\ }\href@noop {} {\bibfield  {journal} {\bibinfo  {journal} {The Journal of Chemical Physics}\ }\textbf {\bibinfo {volume} {75}},\ \bibinfo {pages} {2060--2065} (\bibinfo {year} {1981})}\BibitemShut {NoStop}%
\bibitem [{\citenamefont {Black}\ and\ \citenamefont {Jusinski}(1987{\natexlab{a}})}]{black1987laser}%
  \BibitemOpen
  \bibfield  {author} {\bibinfo {author} {\bibfnamefont {G.}~\bibnamefont {Black}}\ and\ \bibinfo {author} {\bibfnamefont {L.~E.}\ \bibnamefont {Jusinski}},\ }\bibfield  {title} {\enquote {\bibinfo {title} {{Laser-induced fluorescence of C$_{2}$H$_{5}$S radicals}},}\ }\href@noop {} {\bibfield  {journal} {\bibinfo  {journal} {Chemical physics letters}\ }\textbf {\bibinfo {volume} {136}},\ \bibinfo {pages} {241--246} (\bibinfo {year} {1987}{\natexlab{a}})}\BibitemShut {NoStop}%
\bibitem [{\citenamefont {Hung}\ \emph {et~al.}(1996)\citenamefont {Hung}, \citenamefont {Shen}, \citenamefont {Yu},\ and\ \citenamefont {Lee}}]{hung1996vibronic}%
  \BibitemOpen
  \bibfield  {author} {\bibinfo {author} {\bibfnamefont {W.-C.}\ \bibnamefont {Hung}}, \bibinfo {author} {\bibfnamefont {M.-y.}\ \bibnamefont {Shen}}, \bibinfo {author} {\bibfnamefont {C.-h.}\ \bibnamefont {Yu}}, \ and\ \bibinfo {author} {\bibfnamefont {Y.-P.}\ \bibnamefont {Lee}},\ }\bibfield  {title} {\enquote {\bibinfo {title} {{Vibronic analysis of the $\tilde{B}^{2}$A$^{''}$ -- $\tilde{X}^{2}$A$^{''}$ laser-induced fluorescence of jet-cooled C$_{2}$H$_{5}$S}},}\ }\href@noop {} {\bibfield  {journal} {\bibinfo  {journal} {The Journal of chemical physics}\ }\textbf {\bibinfo {volume} {105}},\ \bibinfo {pages} {5722--5730} (\bibinfo {year} {1996})}\BibitemShut {NoStop}%
\bibitem [{\citenamefont {Kakimoto}\ and\ \citenamefont {Kasuya}(1982)}]{kakimoto1982doppler}%
  \BibitemOpen
  \bibfield  {author} {\bibinfo {author} {\bibfnamefont {M.}~\bibnamefont {Kakimoto}}\ and\ \bibinfo {author} {\bibfnamefont {T.}~\bibnamefont {Kasuya}},\ }\bibfield  {title} {\enquote {\bibinfo {title} {{Doppler-limited dye laser excitation spectroscopy of the CCN radical}},}\ }\href@noop {} {\bibfield  {journal} {\bibinfo  {journal} {Journal of Molecular Spectroscopy}\ }\textbf {\bibinfo {volume} {94}},\ \bibinfo {pages} {380--392} (\bibinfo {year} {1982})}\BibitemShut {NoStop}%
\bibitem [{\citenamefont {Hakuta}\ and\ \citenamefont {Uehara}(1983)}]{hakuta1983laser}%
  \BibitemOpen
  \bibfield  {author} {\bibinfo {author} {\bibfnamefont {K.}~\bibnamefont {Hakuta}}\ and\ \bibinfo {author} {\bibfnamefont {H.}~\bibnamefont {Uehara}},\ }\bibfield  {title} {\enquote {\bibinfo {title} {{Laser-induced fluorescence spectrum of the CCN radical with an Ar$^{+}$ laser}},}\ }\href@noop {} {\bibfield  {journal} {\bibinfo  {journal} {The Journal of Chemical Physics}\ }\textbf {\bibinfo {volume} {78}},\ \bibinfo {pages} {6484--6489} (\bibinfo {year} {1983})}\BibitemShut {NoStop}%
\bibitem [{\citenamefont {Pitts}\ \emph {et~al.}(1981)\citenamefont {Pitts}, \citenamefont {Donnelley}, \citenamefont {Baronavski},\ and\ \citenamefont {McDonald}}]{pitts1981c2o}%
  \BibitemOpen
  \bibfield  {author} {\bibinfo {author} {\bibfnamefont {W.}~\bibnamefont {Pitts}}, \bibinfo {author} {\bibfnamefont {V.}~\bibnamefont {Donnelley}}, \bibinfo {author} {\bibfnamefont {A.}~\bibnamefont {Baronavski}}, \ and\ \bibinfo {author} {\bibfnamefont {J.}~\bibnamefont {McDonald}},\ }\bibfield  {title} {\enquote {\bibinfo {title} {{C$_2$O ($\tilde{A}^{3}\Pi_{i}$ $\leftrightarrow$ $\tilde{X}^{3}\Sigma^{-}$): laser induced excitation and fluorescence spectra}},}\ }\href@noop {} {\bibfield  {journal} {\bibinfo  {journal} {Chemical Physics}\ }\textbf {\bibinfo {volume} {61}},\ \bibinfo {pages} {451--464} (\bibinfo {year} {1981})}\BibitemShut {NoStop}%
\bibitem [{\citenamefont {Sunahori}, \citenamefont {Wei},\ and\ \citenamefont {Clouthier}(2008)}]{sunahori2008electronic}%
  \BibitemOpen
  \bibfield  {author} {\bibinfo {author} {\bibfnamefont {F.~X.}\ \bibnamefont {Sunahori}}, \bibinfo {author} {\bibfnamefont {J.}~\bibnamefont {Wei}}, \ and\ \bibinfo {author} {\bibfnamefont {D.~J.}\ \bibnamefont {Clouthier}},\ }\bibfield  {title} {\enquote {\bibinfo {title} {{The electronic spectrum of the C$_2$P free radical and a Renner--Teller analysis of the $^{2}\Delta$ and $\tilde{X}^{2}\Pi$ electronic states}},}\ }\href@noop {} {\bibfield  {journal} {\bibinfo  {journal} {The Journal of chemical physics}\ }\textbf {\bibinfo {volume} {128}} (\bibinfo {year} {2008})}\BibitemShut {NoStop}%
\bibitem [{\citenamefont {Schoeffler}\ \emph {et~al.}(2001)\citenamefont {Schoeffler}, \citenamefont {Kohguchi}, \citenamefont {Hoshina}, \citenamefont {Ohshima},\ and\ \citenamefont {Endo}}]{schoeffler2001laser}%
  \BibitemOpen
  \bibfield  {author} {\bibinfo {author} {\bibfnamefont {A.}~\bibnamefont {Schoeffler}}, \bibinfo {author} {\bibfnamefont {H.}~\bibnamefont {Kohguchi}}, \bibinfo {author} {\bibfnamefont {K.}~\bibnamefont {Hoshina}}, \bibinfo {author} {\bibfnamefont {Y.}~\bibnamefont {Ohshima}}, \ and\ \bibinfo {author} {\bibfnamefont {Y.}~\bibnamefont {Endo}},\ }\bibfield  {title} {\enquote {\bibinfo {title} {Laser induced fluorescence spectroscopy of the $\tilde{A}_{3}{\Pi}_{i}$ ← $\tilde{X}^{3}{\Pi}^{-}$ transition of the {CCS} radical},}\ }\href@noop {} {\bibfield  {journal} {\bibinfo  {journal} {The Journal of Chemical Physics}\ }\textbf {\bibinfo {volume} {114}},\ \bibinfo {pages} {6142--6150} (\bibinfo {year} {2001})}\BibitemShut {NoStop}%
\bibitem [{\citenamefont {Ikegami}\ \emph {et~al.}(2001)\citenamefont {Ikegami}, \citenamefont {Ishibashi}, \citenamefont {Yamagata}, \citenamefont {Ebihara}, \citenamefont {Thareja},\ and\ \citenamefont {Narayan}}]{ikegami2001spatial}%
  \BibitemOpen
  \bibfield  {author} {\bibinfo {author} {\bibfnamefont {T.}~\bibnamefont {Ikegami}}, \bibinfo {author} {\bibfnamefont {S.}~\bibnamefont {Ishibashi}}, \bibinfo {author} {\bibfnamefont {Y.}~\bibnamefont {Yamagata}}, \bibinfo {author} {\bibfnamefont {K.}~\bibnamefont {Ebihara}}, \bibinfo {author} {\bibfnamefont {R.}~\bibnamefont {Thareja}}, \ and\ \bibinfo {author} {\bibfnamefont {J.}~\bibnamefont {Narayan}},\ }\bibfield  {title} {\enquote {\bibinfo {title} {Spatial distribution of carbon species in laser ablation of graphite target},}\ }\href@noop {} {\bibfield  {journal} {\bibinfo  {journal} {Journal of Vacuum Science \& Technology A: Vacuum, Surfaces, and Films}\ }\textbf {\bibinfo {volume} {19}},\ \bibinfo {pages} {1304--1307} (\bibinfo {year} {2001})}\BibitemShut {NoStop}%
\bibitem [{\citenamefont {Black}\ and\ \citenamefont {Jusinski}(1987{\natexlab{b}})}]{black1987laserC3H7S}%
  \BibitemOpen
  \bibfield  {author} {\bibinfo {author} {\bibfnamefont {G.}~\bibnamefont {Black}}\ and\ \bibinfo {author} {\bibfnamefont {L.~E.}\ \bibnamefont {Jusinski}},\ }\bibfield  {title} {\enquote {\bibinfo {title} {{Laser-induced fluorescence of i--C$_3$H$_7$S radicals}},}\ }\href@noop {} {\bibfield  {journal} {\bibinfo  {journal} {Chemical physics letters}\ }\textbf {\bibinfo {volume} {139}},\ \bibinfo {pages} {431--436} (\bibinfo {year} {1987}{\natexlab{b}})}\BibitemShut {NoStop}%
\bibitem [{\citenamefont {Hoshina}\ and\ \citenamefont {Endo}(2007)}]{hoshina2007laser}%
  \BibitemOpen
  \bibfield  {author} {\bibinfo {author} {\bibfnamefont {K.}~\bibnamefont {Hoshina}}\ and\ \bibinfo {author} {\bibfnamefont {Y.}~\bibnamefont {Endo}},\ }\bibfield  {title} {\enquote {\bibinfo {title} {{Laser induced fluorescence spectroscopy of the C$_3$N radical}},}\ }\href@noop {} {\bibfield  {journal} {\bibinfo  {journal} {The Journal of chemical physics}\ }\textbf {\bibinfo {volume} {127}} (\bibinfo {year} {2007})}\BibitemShut {NoStop}%
\bibitem [{\citenamefont {Hoshina}\ \emph {et~al.}(1998)\citenamefont {Hoshina}, \citenamefont {Kohguchi}, \citenamefont {Ohshima},\ and\ \citenamefont {Endo}}]{hoshina1998laser}%
  \BibitemOpen
  \bibfield  {author} {\bibinfo {author} {\bibfnamefont {K.}~\bibnamefont {Hoshina}}, \bibinfo {author} {\bibfnamefont {H.}~\bibnamefont {Kohguchi}}, \bibinfo {author} {\bibfnamefont {Y.}~\bibnamefont {Ohshima}}, \ and\ \bibinfo {author} {\bibfnamefont {Y.}~\bibnamefont {Endo}},\ }\bibfield  {title} {\enquote {\bibinfo {title} {{Laser-induced fluorescence spectroscopy of the C$_4$H and C$_4$D radicals in a supersonic jet}},}\ }\href@noop {} {\bibfield  {journal} {\bibinfo  {journal} {The Journal of chemical physics}\ }\textbf {\bibinfo {volume} {108}},\ \bibinfo {pages} {3465--3478} (\bibinfo {year} {1998})}\BibitemShut {NoStop}%
\bibitem [{\citenamefont {Williams}\ and\ \citenamefont {Fleming}(1997)}]{williams1997laser}%
  \BibitemOpen
  \bibfield  {author} {\bibinfo {author} {\bibfnamefont {B.~A.}\ \bibnamefont {Williams}}\ and\ \bibinfo {author} {\bibfnamefont {J.~W.}\ \bibnamefont {Fleming}},\ }\bibfield  {title} {\enquote {\bibinfo {title} {{Laser-induced fluorescence spectrum of the FCO radical}},}\ }\href@noop {} {\bibfield  {journal} {\bibinfo  {journal} {The Journal of chemical physics}\ }\textbf {\bibinfo {volume} {106}},\ \bibinfo {pages} {4376--4382} (\bibinfo {year} {1997})}\BibitemShut {NoStop}%
\bibitem [{\citenamefont {Sunahori}\ and\ \citenamefont {Clouthier}(2009)}]{sunahori2009electronic}%
  \BibitemOpen
  \bibfield  {author} {\bibinfo {author} {\bibfnamefont {F.~X.}\ \bibnamefont {Sunahori}}\ and\ \bibinfo {author} {\bibfnamefont {D.~J.}\ \bibnamefont {Clouthier}},\ }\bibfield  {title} {\enquote {\bibinfo {title} {The electronic spectrum of the fluoroborane free radical. {II}. analysis of laser-induced fluorescence and single vibronic level emission spectra},}\ }\href@noop {} {\bibfield  {journal} {\bibinfo  {journal} {The Journal of chemical physics}\ }\textbf {\bibinfo {volume} {130}} (\bibinfo {year} {2009})}\BibitemShut {NoStop}%
\bibitem [{\citenamefont {Callaghan}\ \emph {et~al.}(1989)\citenamefont {Callaghan}, \citenamefont {Huang}, \citenamefont {Arepalli},\ and\ \citenamefont {Gordon}}]{callaghan1989single}%
  \BibitemOpen
  \bibfield  {author} {\bibinfo {author} {\bibfnamefont {R.}~\bibnamefont {Callaghan}}, \bibinfo {author} {\bibfnamefont {Y.-L.}\ \bibnamefont {Huang}}, \bibinfo {author} {\bibfnamefont {S.}~\bibnamefont {Arepalli}}, \ and\ \bibinfo {author} {\bibfnamefont {R.~J.}\ \bibnamefont {Gordon}},\ }\bibfield  {title} {\enquote {\bibinfo {title} {{Single-photon VUV laser-induced fluorescence spectra of HCl and HBr}},}\ }\href@noop {} {\bibfield  {journal} {\bibinfo  {journal} {Chemical physics letters}\ }\textbf {\bibinfo {volume} {158}},\ \bibinfo {pages} {531--534} (\bibinfo {year} {1989})}\BibitemShut {NoStop}%
\bibitem [{\citenamefont {K{\"o}nig}\ and\ \citenamefont {Lademann}(1983)}]{konig1983laser}%
  \BibitemOpen
  \bibfield  {author} {\bibinfo {author} {\bibfnamefont {R.}~\bibnamefont {K{\"o}nig}}\ and\ \bibinfo {author} {\bibfnamefont {J.}~\bibnamefont {Lademann}},\ }\bibfield  {title} {\enquote {\bibinfo {title} {{Laser-induced fluorescence detection of HCO produced by laser photolysis of formaldehyde}},}\ }\href@noop {} {\bibfield  {journal} {\bibinfo  {journal} {Chemical Physics Letters}\ }\textbf {\bibinfo {volume} {94}},\ \bibinfo {pages} {152--155} (\bibinfo {year} {1983})}\BibitemShut {NoStop}%
\bibitem [{\citenamefont {Inoue}\ and\ \citenamefont {Suzuki}(1986)}]{inoue1986laser}%
  \BibitemOpen
  \bibfield  {author} {\bibinfo {author} {\bibfnamefont {G.}~\bibnamefont {Inoue}}\ and\ \bibinfo {author} {\bibfnamefont {M.}~\bibnamefont {Suzuki}},\ }\bibfield  {title} {\enquote {\bibinfo {title} {{Laser induced fluorescence of HCCO (DCCO) radical formed in O+ C$_2$H$_2$ (C$_2$D$_2$) reaction}},}\ }\href@noop {} {\bibfield  {journal} {\bibinfo  {journal} {The Journal of chemical physics}\ }\textbf {\bibinfo {volume} {84}},\ \bibinfo {pages} {3709--3716} (\bibinfo {year} {1986})}\BibitemShut {NoStop}%
\bibitem [{\citenamefont {Nakajima}\ \emph {et~al.}(2005)\citenamefont {Nakajima}, \citenamefont {Yoneda}, \citenamefont {Toyoshima}, \citenamefont {Sumiyoshi},\ and\ \citenamefont {Endo}}]{nakajima2005gas}%
  \BibitemOpen
  \bibfield  {author} {\bibinfo {author} {\bibfnamefont {M.}~\bibnamefont {Nakajima}}, \bibinfo {author} {\bibfnamefont {Y.}~\bibnamefont {Yoneda}}, \bibinfo {author} {\bibfnamefont {H.}~\bibnamefont {Toyoshima}}, \bibinfo {author} {\bibfnamefont {Y.}~\bibnamefont {Sumiyoshi}}, \ and\ \bibinfo {author} {\bibfnamefont {Y.}~\bibnamefont {Endo}},\ }\bibfield  {title} {\enquote {\bibinfo {title} {{Gas phase electronic spectrum of the HSCCS radical by laser-induced fluorescence spectroscopy}},}\ }\href@noop {} {\bibfield  {journal} {\bibinfo  {journal} {Journal of Molecular Spectroscopy}\ }\textbf {\bibinfo {volume} {232}},\ \bibinfo {pages} {255--263} (\bibinfo {year} {2005})}\BibitemShut {NoStop}%
\bibitem [{\citenamefont {Nakajima}, \citenamefont {Sumiyoshi},\ and\ \citenamefont {Endo}(2002{\natexlab{a}})}]{nakajima2002laser}%
  \BibitemOpen
  \bibfield  {author} {\bibinfo {author} {\bibfnamefont {M.}~\bibnamefont {Nakajima}}, \bibinfo {author} {\bibfnamefont {Y.}~\bibnamefont {Sumiyoshi}}, \ and\ \bibinfo {author} {\bibfnamefont {Y.}~\bibnamefont {Endo}},\ }\bibfield  {title} {\enquote {\bibinfo {title} {{Laser induced fluorescence spectroscopy of the HC$_4$S and DC$_4$S radicals}},}\ }\href@noop {} {\bibfield  {journal} {\bibinfo  {journal} {Chemical physics letters}\ }\textbf {\bibinfo {volume} {351}},\ \bibinfo {pages} {359--364} (\bibinfo {year} {2002}{\natexlab{a}})}\BibitemShut {NoStop}%
\bibitem [{\citenamefont {Reilly}\ \emph {et~al.}(2006)\citenamefont {Reilly}, \citenamefont {Cupitt}, \citenamefont {Kable},\ and\ \citenamefont {Schmidt}}]{reilly2006experimental}%
  \BibitemOpen
  \bibfield  {author} {\bibinfo {author} {\bibfnamefont {N.}~\bibnamefont {Reilly}}, \bibinfo {author} {\bibfnamefont {G.}~\bibnamefont {Cupitt}}, \bibinfo {author} {\bibfnamefont {S.}~\bibnamefont {Kable}}, \ and\ \bibinfo {author} {\bibfnamefont {T.}~\bibnamefont {Schmidt}},\ }\bibfield  {title} {\enquote {\bibinfo {title} {{Experimental and theoretical investigation of the dispersed fluorescence spectroscopy of HC$_4$S}},}\ }\href@noop {} {\bibfield  {journal} {\bibinfo  {journal} {The Journal of chemical physics}\ }\textbf {\bibinfo {volume} {124}} (\bibinfo {year} {2006})}\BibitemShut {NoStop}%
\bibitem [{\citenamefont {Nakajima}, \citenamefont {Sumiyoshi},\ and\ \citenamefont {Endo}(2002{\natexlab{b}})}]{nakajima2002laserHC6S}%
  \BibitemOpen
  \bibfield  {author} {\bibinfo {author} {\bibfnamefont {M.}~\bibnamefont {Nakajima}}, \bibinfo {author} {\bibfnamefont {Y.}~\bibnamefont {Sumiyoshi}}, \ and\ \bibinfo {author} {\bibfnamefont {Y.}~\bibnamefont {Endo}},\ }\bibfield  {title} {\enquote {\bibinfo {title} {{Laser induced fluorescence spectroscopy of the HC$_6$S radical}},}\ }\href@noop {} {\bibfield  {journal} {\bibinfo  {journal} {Chemical physics letters}\ }\textbf {\bibinfo {volume} {355}},\ \bibinfo {pages} {116--122} (\bibinfo {year} {2002}{\natexlab{b}})}\BibitemShut {NoStop}%
\bibitem [{\citenamefont {Crosley}(1986)}]{crosley1986laser}%
  \BibitemOpen
  \bibfield  {author} {\bibinfo {author} {\bibfnamefont {D.~R.}\ \bibnamefont {Crosley}},\ }\bibfield  {title} {\enquote {\bibinfo {title} {Laser-induced fluorescence measurement of combustion chemistry intermediates},}\ }\href@noop {} {\bibfield  {journal} {\bibinfo  {journal} {High Temperature Materials and Processes}\ }\textbf {\bibinfo {volume} {7}},\ \bibinfo {pages} {41--54} (\bibinfo {year} {1986})}\BibitemShut {NoStop}%
\bibitem [{\citenamefont {Mayama}, \citenamefont {Egashira},\ and\ \citenamefont {Obi}(1989)}]{mayama1989laser}%
  \BibitemOpen
  \bibfield  {author} {\bibinfo {author} {\bibfnamefont {S.}~\bibnamefont {Mayama}}, \bibinfo {author} {\bibfnamefont {K.}~\bibnamefont {Egashira}}, \ and\ \bibinfo {author} {\bibfnamefont {K.}~\bibnamefont {Obi}},\ }\bibfield  {title} {\enquote {\bibinfo {title} {{Laser induced fluorescence of HNO and DNO $\tilde{A}^{1}$A$^{''}$ -- $\tilde{X}^{1}$A$^{''}$ in a supersonic free jet}},}\ }\href@noop {} {\bibfield  {journal} {\bibinfo  {journal} {Research on chemical intermediates}\ }\textbf {\bibinfo {volume} {12}},\ \bibinfo {pages} {285--302} (\bibinfo {year} {1989})}\BibitemShut {NoStop}%
\bibitem [{\citenamefont {Taylor}\ and\ \citenamefont {Lemmer}(2024)}]{taylor2024laser}%
  \BibitemOpen
  \bibfield  {author} {\bibinfo {author} {\bibfnamefont {N.~R.}\ \bibnamefont {Taylor}}\ and\ \bibinfo {author} {\bibfnamefont {K.~M.}\ \bibnamefont {Lemmer}},\ }\bibfield  {title} {\enquote {\bibinfo {title} {{Laser-Induced Fluorescence Detection of Nitroxyl (HNO) Formed from the Thermal Decomposition of Hydroxylammonium Nitrate Vapor}},}\ }\href@noop {} {\bibfield  {journal} {\bibinfo  {journal} {Journal of Ionic Liquids}\ ,\ \bibinfo {pages} {100084}} (\bibinfo {year} {2024})}\BibitemShut {NoStop}%
\bibitem [{\citenamefont {Harjanto}, \citenamefont {Harper},\ and\ \citenamefont {Clouthier}(1996)}]{harjanto1996resolution}%
  \BibitemOpen
  \bibfield  {author} {\bibinfo {author} {\bibfnamefont {H.}~\bibnamefont {Harjanto}}, \bibinfo {author} {\bibfnamefont {W.~W.}\ \bibnamefont {Harper}}, \ and\ \bibinfo {author} {\bibfnamefont {D.~J.}\ \bibnamefont {Clouthier}},\ }\bibfield  {title} {\enquote {\bibinfo {title} {{Resolution of anomalies in the geometry and vibrational frequencies of monobromosilylene (HSiBr) by pulsed discharge jet spectroscopy}},}\ }\href@noop {} {\bibfield  {journal} {\bibinfo  {journal} {The Journal of chemical physics}\ }\textbf {\bibinfo {volume} {105}},\ \bibinfo {pages} {10189--10200} (\bibinfo {year} {1996})}\BibitemShut {NoStop}%
\bibitem [{\citenamefont {Ho}, \citenamefont {Breiland},\ and\ \citenamefont {Carr}(1986)}]{ho1986kinetics}%
  \BibitemOpen
  \bibfield  {author} {\bibinfo {author} {\bibfnamefont {P.}~\bibnamefont {Ho}}, \bibinfo {author} {\bibfnamefont {W.~G.}\ \bibnamefont {Breiland}}, \ and\ \bibinfo {author} {\bibfnamefont {R.~W.}\ \bibnamefont {Carr}},\ }\bibfield  {title} {\enquote {\bibinfo {title} {{Kinetics of the reactions of HSiCl with SiH$_4$ and SiH$_2$Cl$_2$}},}\ }\href@noop {} {\bibfield  {journal} {\bibinfo  {journal} {Chemical physics letters}\ }\textbf {\bibinfo {volume} {132}},\ \bibinfo {pages} {422--426} (\bibinfo {year} {1986})}\BibitemShut {NoStop}%
\bibitem [{\citenamefont {de~Nalda}\ \emph {et~al.}(2000)\citenamefont {de~Nalda}, \citenamefont {Mavromanolakis}, \citenamefont {Couris},\ and\ \citenamefont {Castillejo}}]{de2000induced}%
  \BibitemOpen
  \bibfield  {author} {\bibinfo {author} {\bibfnamefont {R.}~\bibnamefont {de~Nalda}}, \bibinfo {author} {\bibfnamefont {A.}~\bibnamefont {Mavromanolakis}}, \bibinfo {author} {\bibfnamefont {S.}~\bibnamefont {Couris}}, \ and\ \bibinfo {author} {\bibfnamefont {M.}~\bibnamefont {Castillejo}},\ }\bibfield  {title} {\enquote {\bibinfo {title} {{Induced HSiCl emission in the UV photodissociation of 2-chloroethenylsilane}},}\ }\href@noop {} {\bibfield  {journal} {\bibinfo  {journal} {Chemical Physics Letters}\ }\textbf {\bibinfo {volume} {316}},\ \bibinfo {pages} {449--454} (\bibinfo {year} {2000})}\BibitemShut {NoStop}%
\bibitem [{\citenamefont {Harper}\ and\ \citenamefont {Clouthier}(1997)}]{harper1997reinvestigation}%
  \BibitemOpen
  \bibfield  {author} {\bibinfo {author} {\bibfnamefont {W.~W.}\ \bibnamefont {Harper}}\ and\ \bibinfo {author} {\bibfnamefont {D.~J.}\ \bibnamefont {Clouthier}},\ }\bibfield  {title} {\enquote {\bibinfo {title} {{Reinvestigation of the HSiCl electronic spectrum: Experimental reevaluation of the geometry, rotational constants, and vibrational frequencies}},}\ }\href@noop {} {\bibfield  {journal} {\bibinfo  {journal} {The Journal of chemical physics}\ }\textbf {\bibinfo {volume} {106}},\ \bibinfo {pages} {9461--9473} (\bibinfo {year} {1997})}\BibitemShut {NoStop}%
\bibitem [{\citenamefont {Lee}\ and\ \citenamefont {Deneufville}(1983)}]{lee1983laser}%
  \BibitemOpen
  \bibfield  {author} {\bibinfo {author} {\bibfnamefont {H.~U.}\ \bibnamefont {Lee}}\ and\ \bibinfo {author} {\bibfnamefont {J.~P.}\ \bibnamefont {Deneufville}},\ }\bibfield  {title} {\enquote {\bibinfo {title} {Laser-induced fluorescence of the {HSiF} radical},}\ }\href@noop {} {\bibfield  {journal} {\bibinfo  {journal} {Chemical physics letters}\ }\textbf {\bibinfo {volume} {99}},\ \bibinfo {pages} {394--398} (\bibinfo {year} {1983})}\BibitemShut {NoStop}%
\bibitem [{\citenamefont {Dixon}\ and\ \citenamefont {Wright}(1985)}]{dixon1985rotational}%
  \BibitemOpen
  \bibfield  {author} {\bibinfo {author} {\bibfnamefont {R.}~\bibnamefont {Dixon}}\ and\ \bibinfo {author} {\bibfnamefont {N.}~\bibnamefont {Wright}},\ }\bibfield  {title} {\enquote {\bibinfo {title} {{The rotational analysis of a Doppler-limited $\tilde{A}$ $^{1}$A$^{\prime\prime}$ -- $\tilde{X}$ $^{1}$A$^{\prime}$ fluorescence excitation spectrum of HSiF at 430 nm}},}\ }\href@noop {} {\bibfield  {journal} {\bibinfo  {journal} {Chemical physics letters}\ }\textbf {\bibinfo {volume} {117}},\ \bibinfo {pages} {280--285} (\bibinfo {year} {1985})}\BibitemShut {NoStop}%
\bibitem [{\citenamefont {Evans}\ and\ \citenamefont {Dover}(2009)}]{evans2009spectroscopic}%
  \BibitemOpen
  \bibfield  {author} {\bibinfo {author} {\bibfnamefont {C.~J.}\ \bibnamefont {Evans}}\ and\ \bibinfo {author} {\bibfnamefont {M.~R.}\ \bibnamefont {Dover}},\ }\bibfield  {title} {\enquote {\bibinfo {title} {{Spectroscopic Investigation of the Electronic $\tilde{A}^{1}$A$^{''}$ -- $\tilde{X}^{1}$A$^{'}$ Transition of HSiNC}},}\ }\href@noop {} {\bibfield  {journal} {\bibinfo  {journal} {The Journal of Physical Chemistry A}\ }\textbf {\bibinfo {volume} {113}},\ \bibinfo {pages} {8533--8539} (\bibinfo {year} {2009})}\BibitemShut {NoStop}%
\bibitem [{\citenamefont {Dover}, \citenamefont {Evans},\ and\ \citenamefont {Western}(2009)}]{dover2009spectroscopic}%
  \BibitemOpen
  \bibfield  {author} {\bibinfo {author} {\bibfnamefont {M.~R.}\ \bibnamefont {Dover}}, \bibinfo {author} {\bibfnamefont {C.~J.}\ \bibnamefont {Evans}}, \ and\ \bibinfo {author} {\bibfnamefont {C.~M.}\ \bibnamefont {Western}},\ }\bibfield  {title} {\enquote {\bibinfo {title} {{Spectroscopic investigation of the $\tilde{A}^{1}$A$^{''}$ -- $\tilde{X}^{1}$A$^{'}$ electronic transition of HSiNCO}},}\ }\href@noop {} {\bibfield  {journal} {\bibinfo  {journal} {The Journal of chemical physics}\ }\textbf {\bibinfo {volume} {131}} (\bibinfo {year} {2009})}\BibitemShut {NoStop}%
\bibitem [{\citenamefont {Lee}, \citenamefont {Lee},\ and\ \citenamefont {Wang}(1994)}]{lee1994kinetics}%
  \BibitemOpen
  \bibfield  {author} {\bibinfo {author} {\bibfnamefont {Y.-Y.}\ \bibnamefont {Lee}}, \bibinfo {author} {\bibfnamefont {Y.-P.}\ \bibnamefont {Lee}}, \ and\ \bibinfo {author} {\bibfnamefont {N.~S.}\ \bibnamefont {Wang}},\ }\bibfield  {title} {\enquote {\bibinfo {title} {{Kinetics of the reaction of HSO with O$_3$ at temperatures 273--423 K}},}\ }\href@noop {} {\bibfield  {journal} {\bibinfo  {journal} {The Journal of chemical physics}\ }\textbf {\bibinfo {volume} {100}},\ \bibinfo {pages} {387--392} (\bibinfo {year} {1994})}\BibitemShut {NoStop}%
\bibitem [{\citenamefont {Kawasaki}(1987)}]{kawasaki1987laser}%
  \BibitemOpen
  \bibfield  {author} {\bibinfo {author} {\bibfnamefont {M.}~\bibnamefont {Kawasaki}},\ }\bibfield  {title} {\enquote {\bibinfo {title} {{Laser-induced Fluorescence of Unstable Intermediates in Combustion: HSO and H$_2$CS}},}\ }\href@noop {} {\bibfield  {journal} {\bibinfo  {journal} {Laser Diagnostics and Modeling of Combustion}\ ,\ \bibinfo {pages} {203--210}} (\bibinfo {year} {1987})}\BibitemShut {NoStop}%
\bibitem [{\citenamefont {Yoshikawa}\ \emph {et~al.}(2009)\citenamefont {Yoshikawa}, \citenamefont {Watanabe}, \citenamefont {Sumiyoshi},\ and\ \citenamefont {Endo}}]{yoshikawa2009laser}%
  \BibitemOpen
  \bibfield  {author} {\bibinfo {author} {\bibfnamefont {T.}~\bibnamefont {Yoshikawa}}, \bibinfo {author} {\bibfnamefont {A.}~\bibnamefont {Watanabe}}, \bibinfo {author} {\bibfnamefont {Y.}~\bibnamefont {Sumiyoshi}}, \ and\ \bibinfo {author} {\bibfnamefont {Y.}~\bibnamefont {Endo}},\ }\bibfield  {title} {\enquote {\bibinfo {title} {{Laser spectroscopy of the $\tilde{A}$A$^{''}$ -- $\tilde{X}$A$^{''}$ system for the HSO radical}},}\ }\href@noop {} {\bibfield  {journal} {\bibinfo  {journal} {Journal of Molecular Spectroscopy}\ }\textbf {\bibinfo {volume} {254}},\ \bibinfo {pages} {119--125} (\bibinfo {year} {2009})}\BibitemShut {NoStop}%
\bibitem [{\citenamefont {Rothschopf}, \citenamefont {Smith},\ and\ \citenamefont {Clouthier}(2022)}]{rothschopf2022barely}%
  \BibitemOpen
  \bibfield  {author} {\bibinfo {author} {\bibfnamefont {G.}~\bibnamefont {Rothschopf}}, \bibinfo {author} {\bibfnamefont {T.~C.}\ \bibnamefont {Smith}}, \ and\ \bibinfo {author} {\bibfnamefont {D.~J.}\ \bibnamefont {Clouthier}},\ }\bibfield  {title} {\enquote {\bibinfo {title} {{Barely fluorescent molecules. I. Twin-discharge jet laser-induced fluorescence spectroscopy of HSnCl and DSnCl}},}\ }\href@noop {} {\bibfield  {journal} {\bibinfo  {journal} {The Journal of Chemical Physics}\ }\textbf {\bibinfo {volume} {156}} (\bibinfo {year} {2022})}\BibitemShut {NoStop}%
\bibitem [{\citenamefont {Lempert}\ \emph {et~al.}(1991)\citenamefont {Lempert}, \citenamefont {Diskin}, \citenamefont {Kumar}, \citenamefont {Glesk},\ and\ \citenamefont {Miles}}]{lempert1991two}%
  \BibitemOpen
  \bibfield  {author} {\bibinfo {author} {\bibfnamefont {W.}~\bibnamefont {Lempert}}, \bibinfo {author} {\bibfnamefont {G.}~\bibnamefont {Diskin}}, \bibinfo {author} {\bibfnamefont {V.}~\bibnamefont {Kumar}}, \bibinfo {author} {\bibfnamefont {I.}~\bibnamefont {Glesk}}, \ and\ \bibinfo {author} {\bibfnamefont {R.}~\bibnamefont {Miles}},\ }\bibfield  {title} {\enquote {\bibinfo {title} {{Two-dimensional imaging of molecular hydrogen in H$_{2}$--air diffusion flames using two-photon laser-induced fluorescence}},}\ }\href@noop {} {\bibfield  {journal} {\bibinfo  {journal} {Optics letters}\ }\textbf {\bibinfo {volume} {16}},\ \bibinfo {pages} {660--662} (\bibinfo {year} {1991})}\BibitemShut {NoStop}%
\bibitem [{\citenamefont {Northrup}\ \emph {et~al.}(1984)\citenamefont {Northrup}, \citenamefont {Polanyi}, \citenamefont {Wallace},\ and\ \citenamefont {Williamson}}]{northrup1984vuv}%
  \BibitemOpen
  \bibfield  {author} {\bibinfo {author} {\bibfnamefont {F.}~\bibnamefont {Northrup}}, \bibinfo {author} {\bibfnamefont {J.}~\bibnamefont {Polanyi}}, \bibinfo {author} {\bibfnamefont {S.}~\bibnamefont {Wallace}}, \ and\ \bibinfo {author} {\bibfnamefont {J.}~\bibnamefont {Williamson}},\ }\bibfield  {title} {\enquote {\bibinfo {title} {{VUV laser-induced fluorescence of molecular hydrogen}},}\ }\href@noop {} {\bibfield  {journal} {\bibinfo  {journal} {Chemical physics letters}\ }\textbf {\bibinfo {volume} {105}},\ \bibinfo {pages} {34--37} (\bibinfo {year} {1984})}\BibitemShut {NoStop}%
\bibitem [{\citenamefont {Mosbach}, \citenamefont {Katsch},\ and\ \citenamefont {D{\"o}bele}(2000)}]{mosbach2000situ}%
  \BibitemOpen
  \bibfield  {author} {\bibinfo {author} {\bibfnamefont {T.}~\bibnamefont {Mosbach}}, \bibinfo {author} {\bibfnamefont {H.-M.}\ \bibnamefont {Katsch}}, \ and\ \bibinfo {author} {\bibfnamefont {H.}~\bibnamefont {D{\"o}bele}},\ }\bibfield  {title} {\enquote {\bibinfo {title} {In situ diagnostics in plasmas of electronic-ground-state hydrogen molecules in high vibrational and rotational states by laser-induced fluorescence with vacuum-ultraviolet radiation},}\ }\href@noop {} {\bibfield  {journal} {\bibinfo  {journal} {Physical Review Letters}\ }\textbf {\bibinfo {volume} {85}},\ \bibinfo {pages} {3420} (\bibinfo {year} {2000})}\BibitemShut {NoStop}%
\bibitem [{\citenamefont {Gabriel}, \citenamefont {Schram},\ and\ \citenamefont {Engeln}(2008)}]{gabriel2008formation}%
  \BibitemOpen
  \bibfield  {author} {\bibinfo {author} {\bibfnamefont {O.}~\bibnamefont {Gabriel}}, \bibinfo {author} {\bibfnamefont {D.}~\bibnamefont {Schram}}, \ and\ \bibinfo {author} {\bibfnamefont {R.}~\bibnamefont {Engeln}},\ }\bibfield  {title} {\enquote {\bibinfo {title} {{Formation and relaxation of rovibrationally excited H$_2$ molecules due to plasma-surface interaction}},}\ }\href@noop {} {\bibfield  {journal} {\bibinfo  {journal} {Physical Review E}\ }\textbf {\bibinfo {volume} {78}},\ \bibinfo {pages} {016407} (\bibinfo {year} {2008})}\BibitemShut {NoStop}%
\bibitem [{\citenamefont {Turnipseed}\ \emph {et~al.}(1995)\citenamefont {Turnipseed}, \citenamefont {Gilles}, \citenamefont {Burkholder},\ and\ \citenamefont {Ravishankara}}]{turnipseed1995lif}%
  \BibitemOpen
  \bibfield  {author} {\bibinfo {author} {\bibfnamefont {A.~A.}\ \bibnamefont {Turnipseed}}, \bibinfo {author} {\bibfnamefont {M.~K.}\ \bibnamefont {Gilles}}, \bibinfo {author} {\bibfnamefont {J.~B.}\ \bibnamefont {Burkholder}}, \ and\ \bibinfo {author} {\bibfnamefont {A.}~\bibnamefont {Ravishankara}},\ }\bibfield  {title} {\enquote {\bibinfo {title} {{LIF detection of IO and the rate coefficients for I+ O$_3$ and IO+ NO reactions}},}\ }\href@noop {} {\bibfield  {journal} {\bibinfo  {journal} {Chemical physics letters}\ }\textbf {\bibinfo {volume} {242}},\ \bibinfo {pages} {427--434} (\bibinfo {year} {1995})}\BibitemShut {NoStop}%
\bibitem [{\citenamefont {Bloss}\ \emph {et~al.}(2003)\citenamefont {Bloss}, \citenamefont {Gravestock}, \citenamefont {Heard}, \citenamefont {Ingham}, \citenamefont {Johnson},\ and\ \citenamefont {Lee}}]{bloss2003application}%
  \BibitemOpen
  \bibfield  {author} {\bibinfo {author} {\bibfnamefont {W.~J.}\ \bibnamefont {Bloss}}, \bibinfo {author} {\bibfnamefont {T.~J.}\ \bibnamefont {Gravestock}}, \bibinfo {author} {\bibfnamefont {D.~E.}\ \bibnamefont {Heard}}, \bibinfo {author} {\bibfnamefont {T.}~\bibnamefont {Ingham}}, \bibinfo {author} {\bibfnamefont {G.~P.}\ \bibnamefont {Johnson}}, \ and\ \bibinfo {author} {\bibfnamefont {J.~D.}\ \bibnamefont {Lee}},\ }\bibfield  {title} {\enquote {\bibinfo {title} {{Application of a compact all solid-state laser system to the in situ detection of atmospheric OH, HO$_2$, NO and IO by laser-induced fluorescence}},}\ }\href@noop {} {\bibfield  {journal} {\bibinfo  {journal} {Journal of Environmental Monitoring}\ }\textbf {\bibinfo {volume} {5}},\ \bibinfo {pages} {21--28} (\bibinfo {year} {2003})}\BibitemShut {NoStop}%
\bibitem [{\citenamefont {Hiller}\ and\ \citenamefont {Hanson}(1990)}]{hiller1990properties}%
  \BibitemOpen
  \bibfield  {author} {\bibinfo {author} {\bibfnamefont {B.}~\bibnamefont {Hiller}}\ and\ \bibinfo {author} {\bibfnamefont {R.}~\bibnamefont {Hanson}},\ }\bibfield  {title} {\enquote {\bibinfo {title} {Properties of the iodine molecule relevant to laser-induced fluorescence experiments in gas flows},}\ }\href@noop {} {\bibfield  {journal} {\bibinfo  {journal} {Experiments in fluids}\ }\textbf {\bibinfo {volume} {10}},\ \bibinfo {pages} {1--11} (\bibinfo {year} {1990})}\BibitemShut {NoStop}%
\bibitem [{\citenamefont {Ezekiel}\ and\ \citenamefont {Weiss}(1968)}]{ezekiel1968laser}%
  \BibitemOpen
  \bibfield  {author} {\bibinfo {author} {\bibfnamefont {S.}~\bibnamefont {Ezekiel}}\ and\ \bibinfo {author} {\bibfnamefont {R.}~\bibnamefont {Weiss}},\ }\bibfield  {title} {\enquote {\bibinfo {title} {Laser-induced fluorescence in a molecular beam of iodine},}\ }\href@noop {} {\bibfield  {journal} {\bibinfo  {journal} {Physical Review Letters}\ }\textbf {\bibinfo {volume} {20}},\ \bibinfo {pages} {91} (\bibinfo {year} {1968})}\BibitemShut {NoStop}%
\bibitem [{\citenamefont {Zucco}, \citenamefont {Robertsson},\ and\ \citenamefont {Wallerand}(2013)}]{zucco2013laser}%
  \BibitemOpen
  \bibfield  {author} {\bibinfo {author} {\bibfnamefont {M.}~\bibnamefont {Zucco}}, \bibinfo {author} {\bibfnamefont {L.}~\bibnamefont {Robertsson}}, \ and\ \bibinfo {author} {\bibfnamefont {J.}~\bibnamefont {Wallerand}},\ }\bibfield  {title} {\enquote {\bibinfo {title} {Laser-induced fluorescence as a tool to verify the reproducibility of iodine-based laser standards: a study of 96 iodine cells},}\ }\href@noop {} {\bibfield  {journal} {\bibinfo  {journal} {Metrologia}\ }\textbf {\bibinfo {volume} {50}},\ \bibinfo {pages} {402} (\bibinfo {year} {2013})}\BibitemShut {NoStop}%
\bibitem [{\citenamefont {Lin}\ and\ \citenamefont {Chang}(1989)}]{lin1989state}%
  \BibitemOpen
  \bibfield  {author} {\bibinfo {author} {\bibfnamefont {K.-C.}\ \bibnamefont {Lin}}\ and\ \bibinfo {author} {\bibfnamefont {H.-C.}\ \bibnamefont {Chang}},\ }\bibfield  {title} {\enquote {\bibinfo {title} {{State-selective reaction of excited potassium atom with hydrogen molecule. K$^{*}$ + H$_2$ $\rightarrow$ KH + H}},}\ }\href@noop {} {\bibfield  {journal} {\bibinfo  {journal} {The Journal of Chemical Physics}\ }\textbf {\bibinfo {volume} {90}},\ \bibinfo {pages} {6151--6156} (\bibinfo {year} {1989})}\BibitemShut {NoStop}%
\bibitem [{\citenamefont {Tango}, \citenamefont {Link},\ and\ \citenamefont {Zare}(1968)}]{tango1968spectroscopy}%
  \BibitemOpen
  \bibfield  {author} {\bibinfo {author} {\bibfnamefont {W.~J.}\ \bibnamefont {Tango}}, \bibinfo {author} {\bibfnamefont {J.~K.}\ \bibnamefont {Link}}, \ and\ \bibinfo {author} {\bibfnamefont {R.~N.}\ \bibnamefont {Zare}},\ }\bibfield  {title} {\enquote {\bibinfo {title} {Spectroscopy of {K}$_2$ using laser-induced fluorescence},}\ }\href@noop {} {\bibfield  {journal} {\bibinfo  {journal} {The Journal of Chemical Physics}\ }\textbf {\bibinfo {volume} {49}},\ \bibinfo {pages} {4264--4268} (\bibinfo {year} {1968})}\BibitemShut {NoStop}%
\bibitem [{\citenamefont {Milosevic}, \citenamefont {Kowalczyk},\ and\ \citenamefont {Pichler}(1987)}]{milosevic1987study}%
  \BibitemOpen
  \bibfield  {author} {\bibinfo {author} {\bibfnamefont {S.}~\bibnamefont {Milosevic}}, \bibinfo {author} {\bibfnamefont {P.}~\bibnamefont {Kowalczyk}}, \ and\ \bibinfo {author} {\bibfnamefont {G.}~\bibnamefont {Pichler}},\ }\bibfield  {title} {\enquote {\bibinfo {title} {{A study of structured continua in K$_2$ excited by the 457.9 nm Ar-ion laser line}},}\ }\href@noop {} {\bibfield  {journal} {\bibinfo  {journal} {Journal of Physics B: Atomic and Molecular Physics}\ }\textbf {\bibinfo {volume} {20}},\ \bibinfo {pages} {2231} (\bibinfo {year} {1987})}\BibitemShut {NoStop}%
\bibitem [{\citenamefont {Meiwes}\ and\ \citenamefont {Engelke}(1982)}]{meiwes1982predissociation}%
  \BibitemOpen
  \bibfield  {author} {\bibinfo {author} {\bibfnamefont {K.}~\bibnamefont {Meiwes}}\ and\ \bibinfo {author} {\bibfnamefont {F.}~\bibnamefont {Engelke}},\ }\bibfield  {title} {\enquote {\bibinfo {title} {{Predissociation of K2: molecular beam-laser-induced fluorescence spectroscopy of the C$^{1}\Pi_{u}$ -- X$^{1}\Sigma_{g}^{+}$ band system}},}\ }\href@noop {} {\bibfield  {journal} {\bibinfo  {journal} {Chemical Physics Letters}\ }\textbf {\bibinfo {volume} {85}},\ \bibinfo {pages} {409--414} (\bibinfo {year} {1982})}\BibitemShut {NoStop}%
\bibitem [{\citenamefont {Ross}\ \emph {et~al.}(1986)\citenamefont {Ross}, \citenamefont {Effantin}, \citenamefont {d'Incan},\ and\ \citenamefont {Barrow}}]{ross1986laser}%
  \BibitemOpen
  \bibfield  {author} {\bibinfo {author} {\bibfnamefont {A.}~\bibnamefont {Ross}}, \bibinfo {author} {\bibfnamefont {C.}~\bibnamefont {Effantin}}, \bibinfo {author} {\bibfnamefont {J.}~\bibnamefont {d'Incan}}, \ and\ \bibinfo {author} {\bibfnamefont {R.}~\bibnamefont {Barrow}},\ }\bibfield  {title} {\enquote {\bibinfo {title} {{Laser-induced fluorescence of NaK: the b(1) $^{3}\Pi$ state}},}\ }\href@noop {} {\bibfield  {journal} {\bibinfo  {journal} {Journal of Physics B: Atomic and Molecular Physics}\ }\textbf {\bibinfo {volume} {19}},\ \bibinfo {pages} {1449} (\bibinfo {year} {1986})}\BibitemShut {NoStop}%
\bibitem [{\citenamefont {Demtr{\"o}der}, \citenamefont {McClintock},\ and\ \citenamefont {Zare}(1969)}]{demtroder1969spectroscopy}%
  \BibitemOpen
  \bibfield  {author} {\bibinfo {author} {\bibfnamefont {W.}~\bibnamefont {Demtr{\"o}der}}, \bibinfo {author} {\bibfnamefont {M.}~\bibnamefont {McClintock}}, \ and\ \bibinfo {author} {\bibfnamefont {R.}~\bibnamefont {Zare}},\ }\bibfield  {title} {\enquote {\bibinfo {title} {Spectroscopy of {Na}$_2$ using laser-induced fluorescence},}\ }\href@noop {} {\bibfield  {journal} {\bibinfo  {journal} {The Journal of Chemical Physics}\ }\textbf {\bibinfo {volume} {51}},\ \bibinfo {pages} {5495--5508} (\bibinfo {year} {1969})}\BibitemShut {NoStop}%
\bibitem [{\citenamefont {Copeland}, \citenamefont {Crosley},\ and\ \citenamefont {Smith}(1985)}]{copeland1985laser}%
  \BibitemOpen
  \bibfield  {author} {\bibinfo {author} {\bibfnamefont {R.~A.}\ \bibnamefont {Copeland}}, \bibinfo {author} {\bibfnamefont {D.~R.}\ \bibnamefont {Crosley}}, \ and\ \bibinfo {author} {\bibfnamefont {G.~P.}\ \bibnamefont {Smith}},\ }\bibfield  {title} {\enquote {\bibinfo {title} {Laser-induced fluorescence spectroscopy of {NCO} and {NH}$_2$ in atmospheric pressure flames},}\ }\href@noop {} {\bibfield  {journal} {\bibinfo  {journal} {Symposium (International) on Combustion}\ }\textbf {\bibinfo {volume} {20}},\ \bibinfo {pages} {1195--1203} (\bibinfo {year} {1985})}\BibitemShut {NoStop}%
\bibitem [{\citenamefont {Yoshikawa}\ \emph {et~al.}(2008)\citenamefont {Yoshikawa}, \citenamefont {Sumiyoshi}, \citenamefont {Takada}, \citenamefont {Hoshina},\ and\ \citenamefont {Endo}}]{yoshikawa2008laser}%
  \BibitemOpen
  \bibfield  {author} {\bibinfo {author} {\bibfnamefont {T.}~\bibnamefont {Yoshikawa}}, \bibinfo {author} {\bibfnamefont {Y.}~\bibnamefont {Sumiyoshi}}, \bibinfo {author} {\bibfnamefont {H.}~\bibnamefont {Takada}}, \bibinfo {author} {\bibfnamefont {K.}~\bibnamefont {Hoshina}}, \ and\ \bibinfo {author} {\bibfnamefont {Y.}~\bibnamefont {Endo}},\ }\bibfield  {title} {\enquote {\bibinfo {title} {{Laser induced fluorescence spectroscopy of NC$_{3}$O}},}\ }\href@noop {} {\bibfield  {journal} {\bibinfo  {journal} {The Journal of chemical physics}\ }\textbf {\bibinfo {volume} {128}} (\bibinfo {year} {2008})}\BibitemShut {NoStop}%
\bibitem [{\citenamefont {Nakajima}\ \emph {et~al.}(2004)\citenamefont {Nakajima}, \citenamefont {Yoneda}, \citenamefont {Sumiyoshi},\ and\ \citenamefont {Endo}}]{nakajima2004laser}%
  \BibitemOpen
  \bibfield  {author} {\bibinfo {author} {\bibfnamefont {M.}~\bibnamefont {Nakajima}}, \bibinfo {author} {\bibfnamefont {Y.}~\bibnamefont {Yoneda}}, \bibinfo {author} {\bibfnamefont {Y.}~\bibnamefont {Sumiyoshi}}, \ and\ \bibinfo {author} {\bibfnamefont {Y.}~\bibnamefont {Endo}},\ }\bibfield  {title} {\enquote {\bibinfo {title} {{Laser-induced fluorescence spectroscopy of NC$_3$S}},}\ }\href@noop {} {\bibfield  {journal} {\bibinfo  {journal} {The Journal of chemical physics}\ }\textbf {\bibinfo {volume} {120}},\ \bibinfo {pages} {2662--2666} (\bibinfo {year} {2004})}\BibitemShut {NoStop}%
\bibitem [{\citenamefont {Heidner~III}\ \emph {et~al.}(1989)\citenamefont {Heidner~III}, \citenamefont {Helvajian}, \citenamefont {Holloway},\ and\ \citenamefont {Koffend}}]{heidner1989direct}%
  \BibitemOpen
  \bibfield  {author} {\bibinfo {author} {\bibfnamefont {R.}~\bibnamefont {Heidner~III}}, \bibinfo {author} {\bibfnamefont {H.}~\bibnamefont {Helvajian}}, \bibinfo {author} {\bibfnamefont {J.}~\bibnamefont {Holloway}}, \ and\ \bibinfo {author} {\bibfnamefont {J.~B.}\ \bibnamefont {Koffend}},\ }\bibfield  {title} {\enquote {\bibinfo {title} {{Direct observation of NF(X) using laser-induced fluorescence: kinetics of the NF $^3{\Sigma}^-$ ground state}},}\ }\href@noop {} {\bibfield  {journal} {\bibinfo  {journal} {The Journal of Physical Chemistry}\ }\textbf {\bibinfo {volume} {93}},\ \bibinfo {pages} {7813--7818} (\bibinfo {year} {1989})}\BibitemShut {NoStop}%
\bibitem [{\citenamefont {Wang}, \citenamefont {Wang},\ and\ \citenamefont {Guiberti}(2023)}]{wang2023simultaneous}%
  \BibitemOpen
  \bibfield  {author} {\bibinfo {author} {\bibfnamefont {G.}~\bibnamefont {Wang}}, \bibinfo {author} {\bibfnamefont {S.}~\bibnamefont {Wang}}, \ and\ \bibinfo {author} {\bibfnamefont {T.~F.}\ \bibnamefont {Guiberti}},\ }\bibfield  {title} {\enquote {\bibinfo {title} {{Simultaneous planar laser-induced fluorescence measurement of reactant NH$_3$, radical NH, and pollutant NO in ammonia-hydrogen flames using a single dye laser}},}\ }\href@noop {} {\bibfield  {journal} {\bibinfo  {journal} {Combustion and Flame}\ }\textbf {\bibinfo {volume} {256}},\ \bibinfo {pages} {112981} (\bibinfo {year} {2023})}\BibitemShut {NoStop}%
\bibitem [{\citenamefont {Copeland}\ \emph {et~al.}(1989)\citenamefont {Copeland}, \citenamefont {Wise}, \citenamefont {Rensberger},\ and\ \citenamefont {Crosley}}]{copeland1989time}%
  \BibitemOpen
  \bibfield  {author} {\bibinfo {author} {\bibfnamefont {R.~A.}\ \bibnamefont {Copeland}}, \bibinfo {author} {\bibfnamefont {M.~L.}\ \bibnamefont {Wise}}, \bibinfo {author} {\bibfnamefont {K.~J.}\ \bibnamefont {Rensberger}}, \ and\ \bibinfo {author} {\bibfnamefont {D.~R.}\ \bibnamefont {Crosley}},\ }\bibfield  {title} {\enquote {\bibinfo {title} {Time resolved laser induced fluorescence of the {NH} radical in low pressure {N}$_{2}${O} flames},}\ }\href@noop {} {\bibfield  {journal} {\bibinfo  {journal} {Applied optics}\ }\textbf {\bibinfo {volume} {28}},\ \bibinfo {pages} {3199--3205} (\bibinfo {year} {1989})}\BibitemShut {NoStop}%
\bibitem [{\citenamefont {Umemoto}\ and\ \citenamefont {Matsumoto}(1996)}]{umemoto1996production}%
  \BibitemOpen
  \bibfield  {author} {\bibinfo {author} {\bibfnamefont {H.}~\bibnamefont {Umemoto}}\ and\ \bibinfo {author} {\bibfnamefont {K.-i.}\ \bibnamefont {Matsumoto}},\ }\bibfield  {title} {\enquote {\bibinfo {title} {{Production of NH (ND) Radicals in the Reactions of N (2$^2$D) with H$_2$ (D$_2$): Nascent Vibrational Distributions of NH (X $^3\Sigma^{-}$) and ND (X$^{3}\Sigma^{-}$)}},}\ }\href@noop {} {\bibfield  {journal} {\bibinfo  {journal} {The Journal of chemical physics}\ }\textbf {\bibinfo {volume} {104}},\ \bibinfo {pages} {9640--9643} (\bibinfo {year} {1996})}\BibitemShut {NoStop}%
\bibitem [{\citenamefont {Halpern}\ \emph {et~al.}(1975)\citenamefont {Halpern}, \citenamefont {Hancock}, \citenamefont {Lenzi},\ and\ \citenamefont {Welge}}]{halpern1975laser}%
  \BibitemOpen
  \bibfield  {author} {\bibinfo {author} {\bibfnamefont {J.}~\bibnamefont {Halpern}}, \bibinfo {author} {\bibfnamefont {G.}~\bibnamefont {Hancock}}, \bibinfo {author} {\bibfnamefont {M.}~\bibnamefont {Lenzi}}, \ and\ \bibinfo {author} {\bibfnamefont {K.}~\bibnamefont {Welge}},\ }\bibfield  {title} {\enquote {\bibinfo {title} {{Laser induced fluorescence from NH$_2$($^{2}$A$_1$). State selected radiative lifetimes and collisional de-excitation rates}},}\ }\href@noop {} {\bibfield  {journal} {\bibinfo  {journal} {The Journal of Chemical Physics}\ }\textbf {\bibinfo {volume} {63}},\ \bibinfo {pages} {4808--4816} (\bibinfo {year} {1975})}\BibitemShut {NoStop}%
\bibitem [{\citenamefont {Hancock}\ \emph {et~al.}(1975)\citenamefont {Hancock}, \citenamefont {Lange}, \citenamefont {Lenzi},\ and\ \citenamefont {Welge}}]{hancock1975laser}%
  \BibitemOpen
  \bibfield  {author} {\bibinfo {author} {\bibfnamefont {G.}~\bibnamefont {Hancock}}, \bibinfo {author} {\bibfnamefont {W.}~\bibnamefont {Lange}}, \bibinfo {author} {\bibfnamefont {M.}~\bibnamefont {Lenzi}}, \ and\ \bibinfo {author} {\bibfnamefont {K.}~\bibnamefont {Welge}},\ }\bibfield  {title} {\enquote {\bibinfo {title} {{Laser fluorescence of NH2 and rate constant measurement of NH2+ NO}},}\ }\href@noop {} {\bibfield  {journal} {\bibinfo  {journal} {Chemical Physics Letters}\ }\textbf {\bibinfo {volume} {33}},\ \bibinfo {pages} {168--172} (\bibinfo {year} {1975})}\BibitemShut {NoStop}%
\bibitem [{\citenamefont {Westblom}\ and\ \citenamefont {Alden}(1990)}]{westblom1990laser}%
  \BibitemOpen
  \bibfield  {author} {\bibinfo {author} {\bibfnamefont {U.}~\bibnamefont {Westblom}}\ and\ \bibinfo {author} {\bibfnamefont {M.}~\bibnamefont {Alden}},\ }\bibfield  {title} {\enquote {\bibinfo {title} {Laser-induced fluorescence detection of {NH}$_{3}$ in flames with the use of two-photon excitation},}\ }\href@noop {} {\bibfield  {journal} {\bibinfo  {journal} {Applied spectroscopy}\ }\textbf {\bibinfo {volume} {44}},\ \bibinfo {pages} {881--886} (\bibinfo {year} {1990})}\BibitemShut {NoStop}%
\bibitem [{\citenamefont {Beaman}\ \emph {et~al.}(1987)\citenamefont {Beaman}, \citenamefont {Nelson}, \citenamefont {Richards},\ and\ \citenamefont {Setser}}]{beaman1987observation}%
  \BibitemOpen
  \bibfield  {author} {\bibinfo {author} {\bibfnamefont {R.}~\bibnamefont {Beaman}}, \bibinfo {author} {\bibfnamefont {T.}~\bibnamefont {Nelson}}, \bibinfo {author} {\bibfnamefont {D.}~\bibnamefont {Richards}}, \ and\ \bibinfo {author} {\bibfnamefont {D.}~\bibnamefont {Setser}},\ }\bibfield  {title} {\enquote {\bibinfo {title} {Observation of azido radical by laser-induced fluorescence},}\ }\href@noop {} {\bibfield  {journal} {\bibinfo  {journal} {Journal of Physical Chemistry}\ }\textbf {\bibinfo {volume} {91}},\ \bibinfo {pages} {6090--6092} (\bibinfo {year} {1987})}\BibitemShut {NoStop}%
\bibitem [{\citenamefont {Brackmann}\ \emph {et~al.}(2017)\citenamefont {Brackmann}, \citenamefont {Bood}, \citenamefont {Naucler}, \citenamefont {Konnov},\ and\ \citenamefont {Alden}}]{brackmann2017quantitative}%
  \BibitemOpen
  \bibfield  {author} {\bibinfo {author} {\bibfnamefont {C.}~\bibnamefont {Brackmann}}, \bibinfo {author} {\bibfnamefont {J.}~\bibnamefont {Bood}}, \bibinfo {author} {\bibfnamefont {J.~D.}\ \bibnamefont {Naucler}}, \bibinfo {author} {\bibfnamefont {A.~A.}\ \bibnamefont {Konnov}}, \ and\ \bibinfo {author} {\bibfnamefont {M.}~\bibnamefont {Alden}},\ }\bibfield  {title} {\enquote {\bibinfo {title} {Quantitative picosecond laser-induced fluorescence measurements of nitric oxide in flames},}\ }\href@noop {} {\bibfield  {journal} {\bibinfo  {journal} {Proceedings of the Combustion Institute}\ }\textbf {\bibinfo {volume} {36}},\ \bibinfo {pages} {4533--4540} (\bibinfo {year} {2017})}\BibitemShut {NoStop}%
\bibitem [{\citenamefont {Van~Gessel}\ \emph {et~al.}(2013)\citenamefont {Van~Gessel}, \citenamefont {Hrycak}, \citenamefont {Jasi{\'n}ski}, \citenamefont {Mizeraczyk}, \citenamefont {Van Der~Mullen},\ and\ \citenamefont {Bruggeman}}]{van2013temperature}%
  \BibitemOpen
  \bibfield  {author} {\bibinfo {author} {\bibfnamefont {A.}~\bibnamefont {Van~Gessel}}, \bibinfo {author} {\bibfnamefont {B.}~\bibnamefont {Hrycak}}, \bibinfo {author} {\bibfnamefont {M.}~\bibnamefont {Jasi{\'n}ski}}, \bibinfo {author} {\bibfnamefont {J.}~\bibnamefont {Mizeraczyk}}, \bibinfo {author} {\bibfnamefont {J.}~\bibnamefont {Van Der~Mullen}}, \ and\ \bibinfo {author} {\bibfnamefont {P.}~\bibnamefont {Bruggeman}},\ }\bibfield  {title} {\enquote {\bibinfo {title} {{Temperature and NO density measurements by LIF and OES on an atmospheric pressure plasma jet}},}\ }\href@noop {} {\bibfield  {journal} {\bibinfo  {journal} {Journal of Physics D: Applied Physics}\ }\textbf {\bibinfo {volume} {46}},\ \bibinfo {pages} {095201} (\bibinfo {year} {2013})}\BibitemShut {NoStop}%
\bibitem [{\citenamefont {Lee}, \citenamefont {McMillin},\ and\ \citenamefont {Hanson}(1993)}]{lee1993temperature}%
  \BibitemOpen
  \bibfield  {author} {\bibinfo {author} {\bibfnamefont {M.~P.}\ \bibnamefont {Lee}}, \bibinfo {author} {\bibfnamefont {B.~K.}\ \bibnamefont {McMillin}}, \ and\ \bibinfo {author} {\bibfnamefont {R.~K.}\ \bibnamefont {Hanson}},\ }\bibfield  {title} {\enquote {\bibinfo {title} {{Temperature measurements in gases by use of planar laser-induced fluorescence imaging of NO}},}\ }\href@noop {} {\bibfield  {journal} {\bibinfo  {journal} {Applied Optics}\ }\textbf {\bibinfo {volume} {32}},\ \bibinfo {pages} {5379--5396} (\bibinfo {year} {1993})}\BibitemShut {NoStop}%
\bibitem [{\citenamefont {Wodtke}\ \emph {et~al.}(1988)\citenamefont {Wodtke}, \citenamefont {Huwel}, \citenamefont {Schluter}, \citenamefont {Meijer}, \citenamefont {Andersen},\ and\ \citenamefont {Voges}}]{wodtke1988high}%
  \BibitemOpen
  \bibfield  {author} {\bibinfo {author} {\bibfnamefont {A.}~\bibnamefont {Wodtke}}, \bibinfo {author} {\bibfnamefont {L.}~\bibnamefont {Huwel}}, \bibinfo {author} {\bibfnamefont {H.}~\bibnamefont {Schluter}}, \bibinfo {author} {\bibfnamefont {G.}~\bibnamefont {Meijer}}, \bibinfo {author} {\bibfnamefont {P.}~\bibnamefont {Andersen}}, \ and\ \bibinfo {author} {\bibfnamefont {H.}~\bibnamefont {Voges}},\ }\bibfield  {title} {\enquote {\bibinfo {title} {{High-sensitivity detection of NO in a flame using a tunable ArF laser}},}\ }\href@noop {} {\bibfield  {journal} {\bibinfo  {journal} {Optics letters}\ }\textbf {\bibinfo {volume} {13}},\ \bibinfo {pages} {910--912} (\bibinfo {year} {1988})}\BibitemShut {NoStop}%
\bibitem [{\citenamefont {Delon}\ and\ \citenamefont {Jost}(1991)}]{delon1991laser}%
  \BibitemOpen
  \bibfield  {author} {\bibinfo {author} {\bibfnamefont {A.}~\bibnamefont {Delon}}\ and\ \bibinfo {author} {\bibfnamefont {R.}~\bibnamefont {Jost}},\ }\bibfield  {title} {\enquote {\bibinfo {title} {{Laser induced dispersed fluorescence spectra of jet cooled NO$_2$: The complete set of vibrational levels up to 10 000 cm$^{-1}$ and the onset of the $\tilde{X}^{2}$A$_{1}$ -- $\tilde{A}^{2}$B$_{2}$ vibronic interaction}},}\ }\href@noop {} {\bibfield  {journal} {\bibinfo  {journal} {The Journal of chemical physics}\ }\textbf {\bibinfo {volume} {95}},\ \bibinfo {pages} {5686--5700} (\bibinfo {year} {1991})}\BibitemShut {NoStop}%
\bibitem [{\citenamefont {Kim}, \citenamefont {Hunter},\ and\ \citenamefont {Johnston}(1992)}]{kim1992no3}%
  \BibitemOpen
  \bibfield  {author} {\bibinfo {author} {\bibfnamefont {B.}~\bibnamefont {Kim}}, \bibinfo {author} {\bibfnamefont {P.~L.}\ \bibnamefont {Hunter}}, \ and\ \bibinfo {author} {\bibfnamefont {H.}~\bibnamefont {Johnston}},\ }\bibfield  {title} {\enquote {\bibinfo {title} {{NO$_3$ radical studied by laser-induced fluorescence}},}\ }\href@noop {} {\bibfield  {journal} {\bibinfo  {journal} {The Journal of chemical physics}\ }\textbf {\bibinfo {volume} {96}},\ \bibinfo {pages} {4057--4067} (\bibinfo {year} {1992})}\BibitemShut {NoStop}%
\bibitem [{\citenamefont {Fukushima}(2022)}]{fukushima2022laser}%
  \BibitemOpen
  \bibfield  {author} {\bibinfo {author} {\bibfnamefont {M.}~\bibnamefont {Fukushima}},\ }\bibfield  {title} {\enquote {\bibinfo {title} {{Laser induced fluorescence spectra of the $\tilde{B}^{2}$E$^{'}$ -- $\tilde{X}^{2}$A$^{'}_{2}$ transition of jet cooled $^{14}$NO$_3$ and $^{15}$NO$_3$ I: $\nu_{4}$ progressions in the ground $\tilde{X}^{2}$A$^{'}_{2}$ state}},}\ }\href@noop {} {\bibfield  {journal} {\bibinfo  {journal} {Journal of Molecular Spectroscopy}\ }\textbf {\bibinfo {volume} {387}},\ \bibinfo {pages} {111646} (\bibinfo {year} {2022})}\BibitemShut {NoStop}%
\bibitem [{\citenamefont {Jeffries}\ and\ \citenamefont {Crosley}(1986)}]{jeffries1986laser}%
  \BibitemOpen
  \bibfield  {author} {\bibinfo {author} {\bibfnamefont {J.~B.}\ \bibnamefont {Jeffries}}\ and\ \bibinfo {author} {\bibfnamefont {D.~R.}\ \bibnamefont {Crosley}},\ }\bibfield  {title} {\enquote {\bibinfo {title} {Laser-induced fluorescence detection of the {NS} radical in sulfur and nitrogen doped methane flames},}\ }\href@noop {} {\bibfield  {journal} {\bibinfo  {journal} {Combustion and flame}\ }\textbf {\bibinfo {volume} {64}},\ \bibinfo {pages} {55--64} (\bibinfo {year} {1986})}\BibitemShut {NoStop}%
\bibitem [{\citenamefont {Ongstad}, \citenamefont {Lawconnell},\ and\ \citenamefont {Henshaw}(1992)}]{ongstad1992photodissociation}%
  \BibitemOpen
  \bibfield  {author} {\bibinfo {author} {\bibfnamefont {A.~P.}\ \bibnamefont {Ongstad}}, \bibinfo {author} {\bibfnamefont {R.~I.}\ \bibnamefont {Lawconnell}}, \ and\ \bibinfo {author} {\bibfnamefont {T.~L.}\ \bibnamefont {Henshaw}},\ }\bibfield  {title} {\enquote {\bibinfo {title} {Photodissociation dynamics of {S}$_4${N}$_4$ at 222 and 248 nm},}\ }\href@noop {} {\bibfield  {journal} {\bibinfo  {journal} {The Journal of chemical physics}\ }\textbf {\bibinfo {volume} {97}},\ \bibinfo {pages} {1053--1064} (\bibinfo {year} {1992})}\BibitemShut {NoStop}%
\bibitem [{\citenamefont {Goldsmith}\ and\ \citenamefont {Anderson}(1986)}]{goldsmith1986laser}%
  \BibitemOpen
  \bibfield  {author} {\bibinfo {author} {\bibfnamefont {J.}~\bibnamefont {Goldsmith}}\ and\ \bibinfo {author} {\bibfnamefont {R.}~\bibnamefont {Anderson}},\ }\bibfield  {title} {\enquote {\bibinfo {title} {Laser-induced fluorescence spectroscopy and imaging of molecular oxygen in flames},}\ }\href@noop {} {\bibfield  {journal} {\bibinfo  {journal} {Optics letters}\ }\textbf {\bibinfo {volume} {11}},\ \bibinfo {pages} {67--69} (\bibinfo {year} {1986})}\BibitemShut {NoStop}%
\bibitem [{\citenamefont {Irvine}\ \emph {et~al.}(1990)\citenamefont {Irvine}, \citenamefont {Smith}, \citenamefont {Tuckett},\ and\ \citenamefont {Yang}}]{irvine1990laser}%
  \BibitemOpen
  \bibfield  {author} {\bibinfo {author} {\bibfnamefont {A.~M.}\ \bibnamefont {Irvine}}, \bibinfo {author} {\bibfnamefont {I.~W.}\ \bibnamefont {Smith}}, \bibinfo {author} {\bibfnamefont {R.~P.}\ \bibnamefont {Tuckett}}, \ and\ \bibinfo {author} {\bibfnamefont {X.-F.}\ \bibnamefont {Yang}},\ }\bibfield  {title} {\enquote {\bibinfo {title} {{A laser-induced fluorescence determination of the complete internal state distribution of OH produced in the reaction: H+ NO$_2$ $\rightarrow$ OH+ NO}},}\ }\href@noop {} {\bibfield  {journal} {\bibinfo  {journal} {The Journal of chemical physics}\ }\textbf {\bibinfo {volume} {93}},\ \bibinfo {pages} {3177--3186} (\bibinfo {year} {1990})}\BibitemShut {NoStop}%
\bibitem [{\citenamefont {Zhu}\ \emph {et~al.}(2022)\citenamefont {Zhu}, \citenamefont {Wang}, \citenamefont {Chen}, \citenamefont {Chen}, \citenamefont {Yang}, \citenamefont {Yin},\ and\ \citenamefont {Liu}}]{zhu2022fine}%
  \BibitemOpen
  \bibfield  {author} {\bibinfo {author} {\bibfnamefont {C.}~\bibnamefont {Zhu}}, \bibinfo {author} {\bibfnamefont {H.}~\bibnamefont {Wang}}, \bibinfo {author} {\bibfnamefont {B.}~\bibnamefont {Chen}}, \bibinfo {author} {\bibfnamefont {Y.}~\bibnamefont {Chen}}, \bibinfo {author} {\bibfnamefont {T.}~\bibnamefont {Yang}}, \bibinfo {author} {\bibfnamefont {J.}~\bibnamefont {Yin}}, \ and\ \bibinfo {author} {\bibfnamefont {J.}~\bibnamefont {Liu}},\ }\bibfield  {title} {\enquote {\bibinfo {title} {{Fine and hyperfine interactions of PbF studied by laser-induced fluorescence spectroscopy}},}\ }\href@noop {} {\bibfield  {journal} {\bibinfo  {journal} {The Journal of chemical physics}\ }\textbf {\bibinfo {volume} {157}} (\bibinfo {year} {2022})}\BibitemShut {NoStop}%
\bibitem [{\citenamefont {Chen}\ \emph {et~al.}(2022)\citenamefont {Chen}, \citenamefont {Chen}, \citenamefont {Pan}, \citenamefont {Yin},\ and\ \citenamefont {Wang}}]{chen2022spectroscopic}%
  \BibitemOpen
  \bibfield  {author} {\bibinfo {author} {\bibfnamefont {B.}~\bibnamefont {Chen}}, \bibinfo {author} {\bibfnamefont {Y.-N.}\ \bibnamefont {Chen}}, \bibinfo {author} {\bibfnamefont {J.-N.}\ \bibnamefont {Pan}}, \bibinfo {author} {\bibfnamefont {J.-P.}\ \bibnamefont {Yin}}, \ and\ \bibinfo {author} {\bibfnamefont {H.-L.}\ \bibnamefont {Wang}},\ }\bibfield  {title} {\enquote {\bibinfo {title} {{Spectroscopic study of B$^{2}\Sigma^{+}$ -- X$^{2}\Pi_{1/2}$ transition of electron electric dipole moment candidate PbF}},}\ }\href@noop {} {\bibfield  {journal} {\bibinfo  {journal} {Chinese Physics B}\ }\textbf {\bibinfo {volume} {31}},\ \bibinfo {pages} {093301} (\bibinfo {year} {2022})}\BibitemShut {NoStop}%
\bibitem [{\citenamefont {Shestakov}\ \emph {et~al.}(1992)\citenamefont {Shestakov}, \citenamefont {Pravilov}, \citenamefont {Demes},\ and\ \citenamefont {Fink}}]{shestakov1992radiative}%
  \BibitemOpen
  \bibfield  {author} {\bibinfo {author} {\bibfnamefont {O.}~\bibnamefont {Shestakov}}, \bibinfo {author} {\bibfnamefont {A.}~\bibnamefont {Pravilov}}, \bibinfo {author} {\bibfnamefont {H.}~\bibnamefont {Demes}}, \ and\ \bibinfo {author} {\bibfnamefont {E.}~\bibnamefont {Fink}},\ }\bibfield  {title} {\enquote {\bibinfo {title} {{Radiative lifetime and quenching of the A$^{2}\Sigma^{+}$ and X$_{2}^{2}\Pi_{3/2}$ states of PbF}},}\ }\href@noop {} {\bibfield  {journal} {\bibinfo  {journal} {Chemical physics}\ }\textbf {\bibinfo {volume} {165}},\ \bibinfo {pages} {415--427} (\bibinfo {year} {1992})}\BibitemShut {NoStop}%
\bibitem [{\citenamefont {Chen}\ \emph {et~al.}(1994)\citenamefont {Chen}, \citenamefont {Zhang}, \citenamefont {Zhang}, \citenamefont {Chen}, \citenamefont {Yu},\ and\ \citenamefont {Ma}}]{chen1994laser}%
  \BibitemOpen
  \bibfield  {author} {\bibinfo {author} {\bibfnamefont {Y.}~\bibnamefont {Chen}}, \bibinfo {author} {\bibfnamefont {Q.}~\bibnamefont {Zhang}}, \bibinfo {author} {\bibfnamefont {D.}~\bibnamefont {Zhang}}, \bibinfo {author} {\bibfnamefont {C.}~\bibnamefont {Chen}}, \bibinfo {author} {\bibfnamefont {S.}~\bibnamefont {Yu}}, \ and\ \bibinfo {author} {\bibfnamefont {X.}~\bibnamefont {Ma}},\ }\bibfield  {title} {\enquote {\bibinfo {title} {Laser-induced fluorescence spectrum of {PH}$_2$ cooled in a supersonic jet},}\ }\href@noop {} {\bibfield  {journal} {\bibinfo  {journal} {Chemical physics letters}\ }\textbf {\bibinfo {volume} {223}},\ \bibinfo {pages} {104--109} (\bibinfo {year} {1994})}\BibitemShut {NoStop}%
\bibitem [{\citenamefont {Huie}, \citenamefont {Long},\ and\ \citenamefont {Thrush}(1978)}]{huie1978laser}%
  \BibitemOpen
  \bibfield  {author} {\bibinfo {author} {\bibfnamefont {R.~E.}\ \bibnamefont {Huie}}, \bibinfo {author} {\bibfnamefont {N.~J.}\ \bibnamefont {Long}}, \ and\ \bibinfo {author} {\bibfnamefont {B.~A.}\ \bibnamefont {Thrush}},\ }\bibfield  {title} {\enquote {\bibinfo {title} {{Laser induced fluorescence of the PH$_2$ radical}},}\ }\href@noop {} {\bibfield  {journal} {\bibinfo  {journal} {Journal of the Chemical Society, Faraday Transactions 2: Molecular and Chemical Physics}\ }\textbf {\bibinfo {volume} {74}},\ \bibinfo {pages} {1253--1262} (\bibinfo {year} {1978})}\BibitemShut {NoStop}%
\bibitem [{\citenamefont {Xuan}\ and\ \citenamefont {Margani}(1990)}]{xuan1990dynamics}%
  \BibitemOpen
  \bibfield  {author} {\bibinfo {author} {\bibfnamefont {C.~N.}\ \bibnamefont {Xuan}}\ and\ \bibinfo {author} {\bibfnamefont {A.}~\bibnamefont {Margani}},\ }\bibfield  {title} {\enquote {\bibinfo {title} {{Dynamics and spectroscopy of PH$_{2}$ ($\tilde{A}^{2}$A$_{1}$)}},}\ }\href@noop {} {\bibfield  {journal} {\bibinfo  {journal} {The Journal of chemical physics}\ }\textbf {\bibinfo {volume} {93}},\ \bibinfo {pages} {136--146} (\bibinfo {year} {1990})}\BibitemShut {NoStop}%
\bibitem [{\citenamefont {Clyne}\ and\ \citenamefont {Heaven}(1981)}]{clyne1981laser}%
  \BibitemOpen
  \bibfield  {author} {\bibinfo {author} {\bibfnamefont {M.~A.}\ \bibnamefont {Clyne}}\ and\ \bibinfo {author} {\bibfnamefont {M.~C.}\ \bibnamefont {Heaven}},\ }\bibfield  {title} {\enquote {\bibinfo {title} {{Laser-induced fluorescence of the PO radical}},}\ }\href@noop {} {\bibfield  {journal} {\bibinfo  {journal} {Chemical Physics}\ }\textbf {\bibinfo {volume} {58}},\ \bibinfo {pages} {145--150} (\bibinfo {year} {1981})}\BibitemShut {NoStop}%
\bibitem [{\citenamefont {Wong}, \citenamefont {Anderson},\ and\ \citenamefont {Kotlar}(1986)}]{wong1986radiative}%
  \BibitemOpen
  \bibfield  {author} {\bibinfo {author} {\bibfnamefont {K.~N.}\ \bibnamefont {Wong}}, \bibinfo {author} {\bibfnamefont {W.~R.}\ \bibnamefont {Anderson}}, \ and\ \bibinfo {author} {\bibfnamefont {A.~J.}\ \bibnamefont {Kotlar}},\ }\bibfield  {title} {\enquote {\bibinfo {title} {{Radiative processes following laser excitation of the A$^2\Sigma^{+}$ state of PO}},}\ }\href@noop {} {\bibfield  {journal} {\bibinfo  {journal} {The Journal of chemical physics}\ }\textbf {\bibinfo {volume} {85}},\ \bibinfo {pages} {2406--2413} (\bibinfo {year} {1986})}\BibitemShut {NoStop}%
\bibitem [{\citenamefont {Wang}, \citenamefont {Ng},\ and\ \citenamefont {Cheung}(2012)}]{wang2012laser}%
  \BibitemOpen
  \bibfield  {author} {\bibinfo {author} {\bibfnamefont {N.}~\bibnamefont {Wang}}, \bibinfo {author} {\bibfnamefont {Y.}~\bibnamefont {Ng}}, \ and\ \bibinfo {author} {\bibfnamefont {A.-C.}\ \bibnamefont {Cheung}},\ }\bibfield  {title} {\enquote {\bibinfo {title} {Laser induced fluorescence spectroscopy of ruthenium monoboride},}\ }\href@noop {} {\bibfield  {journal} {\bibinfo  {journal} {Chemical Physics Letters}\ }\textbf {\bibinfo {volume} {547}},\ \bibinfo {pages} {21--23} (\bibinfo {year} {2012})}\BibitemShut {NoStop}%
\bibitem [{\citenamefont {DaBell}, \citenamefont {Meyer},\ and\ \citenamefont {Morse}(2001)}]{dabell2001electronic}%
  \BibitemOpen
  \bibfield  {author} {\bibinfo {author} {\bibfnamefont {R.~S.}\ \bibnamefont {DaBell}}, \bibinfo {author} {\bibfnamefont {R.~G.}\ \bibnamefont {Meyer}}, \ and\ \bibinfo {author} {\bibfnamefont {M.~D.}\ \bibnamefont {Morse}},\ }\bibfield  {title} {\enquote {\bibinfo {title} {{Electronic structure of the 4d transition metal carbides: Dispersed fluorescence spectroscopy of MoC, RuC, and PdC}},}\ }\href@noop {} {\bibfield  {journal} {\bibinfo  {journal} {The Journal of Chemical Physics}\ }\textbf {\bibinfo {volume} {114}},\ \bibinfo {pages} {2938--2954} (\bibinfo {year} {2001})}\BibitemShut {NoStop}%
\bibitem [{\citenamefont {Zarringhalam}(2022)}]{zarringhalam2022high}%
  \BibitemOpen
  \bibfield  {author} {\bibinfo {author} {\bibfnamefont {H.}~\bibnamefont {Zarringhalam}},\ }\emph {\bibinfo {title} {High-resolution laser and far-infrared Fourier transform synchrotron-based spectroscopy of selected molecules}},\ \href@noop {} {Ph.D. thesis},\ \bibinfo  {school} {University of New Brunswick} (\bibinfo {year} {2022})\BibitemShut {NoStop}%
\bibitem [{\citenamefont {Steimle}\ and\ \citenamefont {Virgo}(2003)}]{steimle2003permanent}%
  \BibitemOpen
  \bibfield  {author} {\bibinfo {author} {\bibfnamefont {T.~C.}\ \bibnamefont {Steimle}}\ and\ \bibinfo {author} {\bibfnamefont {W.}~\bibnamefont {Virgo}},\ }\bibfield  {title} {\enquote {\bibinfo {title} {{The permanent electric dipole moments and magnetic hyperfine interactions of ruthenium mononitride, RuN}},}\ }\href@noop {} {\bibfield  {journal} {\bibinfo  {journal} {The Journal of chemical physics}\ }\textbf {\bibinfo {volume} {119}},\ \bibinfo {pages} {12965--12972} (\bibinfo {year} {2003})}\BibitemShut {NoStop}%
\bibitem [{\citenamefont {Tiee}\ \emph {et~al.}(1981)\citenamefont {Tiee}, \citenamefont {Wampler}, \citenamefont {Oldenborg},\ and\ \citenamefont {Rice}}]{tiee1981spectroscopy}%
  \BibitemOpen
  \bibfield  {author} {\bibinfo {author} {\bibfnamefont {J.}~\bibnamefont {Tiee}}, \bibinfo {author} {\bibfnamefont {F.}~\bibnamefont {Wampler}}, \bibinfo {author} {\bibfnamefont {R.}~\bibnamefont {Oldenborg}}, \ and\ \bibinfo {author} {\bibfnamefont {W.}~\bibnamefont {Rice}},\ }\bibfield  {title} {\enquote {\bibinfo {title} {{Spectroscopy and reaction kinetics of HS radicals}},}\ }\href@noop {} {\bibfield  {journal} {\bibinfo  {journal} {Chemical physics letters}\ }\textbf {\bibinfo {volume} {82}},\ \bibinfo {pages} {80--84} (\bibinfo {year} {1981})}\BibitemShut {NoStop}%
\bibitem [{\citenamefont {Hawkins}\ and\ \citenamefont {Houston}(1980)}]{hawkins1980193}%
  \BibitemOpen
  \bibfield  {author} {\bibinfo {author} {\bibfnamefont {W.}~\bibnamefont {Hawkins}}\ and\ \bibinfo {author} {\bibfnamefont {P.}~\bibnamefont {Houston}},\ }\bibfield  {title} {\enquote {\bibinfo {title} {{193 nm photodissociation of H$_2$S: The SH internal energy distribution}},}\ }\href@noop {} {\bibfield  {journal} {\bibinfo  {journal} {The Journal of Chemical Physics}\ }\textbf {\bibinfo {volume} {73}},\ \bibinfo {pages} {297--302} (\bibinfo {year} {1980})}\BibitemShut {NoStop}%
\bibitem [{\citenamefont {Becker}, \citenamefont {HAAKS},\ and\ \citenamefont {TATARCZYK}(1974)}]{becker1974lifetime}%
  \BibitemOpen
  \bibfield  {author} {\bibinfo {author} {\bibfnamefont {K.}~\bibnamefont {Becker}}, \bibinfo {author} {\bibfnamefont {D.}~\bibnamefont {HAAKS}}, \ and\ \bibinfo {author} {\bibfnamefont {T.}~\bibnamefont {TATARCZYK}},\ }\bibfield  {title} {\enquote {\bibinfo {title} {Lifetime measurements at selectively excited-states of diatomic hydrides},}\ }\href@noop {} {\bibfield  {journal} {\bibinfo  {journal} {Berichte Der Bunsen-Gesellschaft-Physical Chemistry Chemical Physics}\ }\textbf {\bibinfo {volume} {78}},\ \bibinfo {pages} {1157--1160} (\bibinfo {year} {1974})}\BibitemShut {NoStop}%
\bibitem [{\citenamefont {Herman}\ \emph {et~al.}(1996)\citenamefont {Herman}, \citenamefont {Donnelly}, \citenamefont {Cheng},\ and\ \citenamefont {Guinn}}]{herman1996surface}%
  \BibitemOpen
  \bibfield  {author} {\bibinfo {author} {\bibfnamefont {I.~P.}\ \bibnamefont {Herman}}, \bibinfo {author} {\bibfnamefont {V.~M.}\ \bibnamefont {Donnelly}}, \bibinfo {author} {\bibfnamefont {C.-C.}\ \bibnamefont {Cheng}}, \ and\ \bibinfo {author} {\bibfnamefont {K.~V.}\ \bibnamefont {Guinn}},\ }\bibfield  {title} {\enquote {\bibinfo {title} {Surface analysis during plasma etching by laser-induced thermal desorption},}\ }\href@noop {} {\bibfield  {journal} {\bibinfo  {journal} {Japanese journal of applied physics}\ }\textbf {\bibinfo {volume} {35}},\ \bibinfo {pages} {2410} (\bibinfo {year} {1996})}\BibitemShut {NoStop}%
\bibitem [{\citenamefont {Cheng}\ \emph {et~al.}(1995)\citenamefont {Cheng}, \citenamefont {Guinn}, \citenamefont {Herman},\ and\ \citenamefont {Donnelly}}]{cheng1995competitive}%
  \BibitemOpen
  \bibfield  {author} {\bibinfo {author} {\bibfnamefont {C.}~\bibnamefont {Cheng}}, \bibinfo {author} {\bibfnamefont {K.}~\bibnamefont {Guinn}}, \bibinfo {author} {\bibfnamefont {I.}~\bibnamefont {Herman}}, \ and\ \bibinfo {author} {\bibfnamefont {V.}~\bibnamefont {Donnelly}},\ }\bibfield  {title} {\enquote {\bibinfo {title} {{Competitive halogenation of silicon surfaces in HBr/Cl$_{2}$ plasmas studied with x-ray photoelectron spectroscopy and in situ, real-time, pulsed laser-induced thermal desorption}},}\ }\href@noop {} {\bibfield  {journal} {\bibinfo  {journal} {Journal of Vacuum Science \& Technology A: Vacuum, Surfaces, and Films}\ }\textbf {\bibinfo {volume} {13}},\ \bibinfo {pages} {1970--1976} (\bibinfo {year} {1995})}\BibitemShut {NoStop}%
\bibitem [{\citenamefont {Rothschopf}, \citenamefont {Smith},\ and\ \citenamefont {Clouthier}(2019)}]{rothschopf2019high}%
  \BibitemOpen
  \bibfield  {author} {\bibinfo {author} {\bibfnamefont {G.}~\bibnamefont {Rothschopf}}, \bibinfo {author} {\bibfnamefont {T.~C.}\ \bibnamefont {Smith}}, \ and\ \bibinfo {author} {\bibfnamefont {D.~J.}\ \bibnamefont {Clouthier}},\ }\bibfield  {title} {\enquote {\bibinfo {title} {{The high-resolution LIF spectrum of the SiCCl free radical: Probing the silicon-carbon triple bond}},}\ }\href@noop {} {\bibfield  {journal} {\bibinfo  {journal} {Journal of Molecular Spectroscopy}\ }\textbf {\bibinfo {volume} {359}},\ \bibinfo {pages} {22--30} (\bibinfo {year} {2019})}\BibitemShut {NoStop}%
\bibitem [{\citenamefont {Rothschopf}, \citenamefont {Smith},\ and\ \citenamefont {Clouthier}(2018)}]{rothschopf2018laser}%
  \BibitemOpen
  \bibfield  {author} {\bibinfo {author} {\bibfnamefont {G.}~\bibnamefont {Rothschopf}}, \bibinfo {author} {\bibfnamefont {T.~C.}\ \bibnamefont {Smith}}, \ and\ \bibinfo {author} {\bibfnamefont {D.~J.}\ \bibnamefont {Clouthier}},\ }\bibfield  {title} {\enquote {\bibinfo {title} {{Laser-induced fluorescence detection of the elusive SiCF free radical}},}\ }\href@noop {} {\bibfield  {journal} {\bibinfo  {journal} {The Journal of Chemical Physics}\ }\textbf {\bibinfo {volume} {149}} (\bibinfo {year} {2018})}\BibitemShut {NoStop}%
\bibitem [{\citenamefont {Smith}\ \emph {et~al.}(2000)\citenamefont {Smith}, \citenamefont {Li}, \citenamefont {Clouthier}, \citenamefont {Kingston},\ and\ \citenamefont {Merer}}]{smith2000electronic}%
  \BibitemOpen
  \bibfield  {author} {\bibinfo {author} {\bibfnamefont {T.~C.}\ \bibnamefont {Smith}}, \bibinfo {author} {\bibfnamefont {H.}~\bibnamefont {Li}}, \bibinfo {author} {\bibfnamefont {D.~J.}\ \bibnamefont {Clouthier}}, \bibinfo {author} {\bibfnamefont {C.~T.}\ \bibnamefont {Kingston}}, \ and\ \bibinfo {author} {\bibfnamefont {A.~J.}\ \bibnamefont {Merer}},\ }\bibfield  {title} {\enquote {\bibinfo {title} {{The electronic spectrum of silicon methylidyne (SiCH), a molecule with a silicon--carbon triple bond in the excited state}},}\ }\href@noop {} {\bibfield  {journal} {\bibinfo  {journal} {The Journal of Chemical Physics}\ }\textbf {\bibinfo {volume} {112}},\ \bibinfo {pages} {3662--3670} (\bibinfo {year} {2000})}\BibitemShut {NoStop}%
\bibitem [{\citenamefont {Singleton}\ \emph {et~al.}(1992)\citenamefont {Singleton}, \citenamefont {McKendrick}, \citenamefont {Copeland},\ and\ \citenamefont {Jeffries}}]{singleton1992vibrational}%
  \BibitemOpen
  \bibfield  {author} {\bibinfo {author} {\bibfnamefont {S.}~\bibnamefont {Singleton}}, \bibinfo {author} {\bibfnamefont {K.~G.}\ \bibnamefont {McKendrick}}, \bibinfo {author} {\bibfnamefont {R.~A.}\ \bibnamefont {Copeland}}, \ and\ \bibinfo {author} {\bibfnamefont {J.~B.}\ \bibnamefont {Jeffries}},\ }\bibfield  {title} {\enquote {\bibinfo {title} {{Vibrational transition probabilities in the B--X and B$^{'}$--X systems of the chlorosilylidyne radical}},}\ }\href@noop {} {\bibfield  {journal} {\bibinfo  {journal} {The Journal of Physical Chemistry}\ }\textbf {\bibinfo {volume} {96}},\ \bibinfo {pages} {9703--9709} (\bibinfo {year} {1992})}\BibitemShut {NoStop}%
\bibitem [{\citenamefont {Suzuki}, \citenamefont {Washida},\ and\ \citenamefont {Inoue}(1986)}]{suzuki1986laser}%
  \BibitemOpen
  \bibfield  {author} {\bibinfo {author} {\bibfnamefont {M.}~\bibnamefont {Suzuki}}, \bibinfo {author} {\bibfnamefont {N.}~\bibnamefont {Washida}}, \ and\ \bibinfo {author} {\bibfnamefont {G.}~\bibnamefont {Inoue}},\ }\bibfield  {title} {\enquote {\bibinfo {title} {{Laser-induced fluorescence of the SiCl$_{2}$ radical}},}\ }\href@noop {} {\bibfield  {journal} {\bibinfo  {journal} {Chemical Physics Letters}\ }\textbf {\bibinfo {volume} {131}},\ \bibinfo {pages} {24--30} (\bibinfo {year} {1986})}\BibitemShut {NoStop}%
\bibitem [{\citenamefont {Smith}\ and\ \citenamefont {Clouthier}(2020)}]{smith2020identification}%
  \BibitemOpen
  \bibfield  {author} {\bibinfo {author} {\bibfnamefont {T.~C.}\ \bibnamefont {Smith}}\ and\ \bibinfo {author} {\bibfnamefont {D.~J.}\ \bibnamefont {Clouthier}},\ }\bibfield  {title} {\enquote {\bibinfo {title} {{Identification of the Jahn--Teller active trichlorosiloxy (SiCl$_{3}$O) free radical in the gas phase}},}\ }\href@noop {} {\bibfield  {journal} {\bibinfo  {journal} {The Journal of Chemical Physics}\ }\textbf {\bibinfo {volume} {152}} (\bibinfo {year} {2020})}\BibitemShut {NoStop}%
\bibitem [{\citenamefont {Umeki}, \citenamefont {Nakajima},\ and\ \citenamefont {Endo}(2015)}]{umeki2015laser}%
  \BibitemOpen
  \bibfield  {author} {\bibinfo {author} {\bibfnamefont {H.}~\bibnamefont {Umeki}}, \bibinfo {author} {\bibfnamefont {M.}~\bibnamefont {Nakajima}}, \ and\ \bibinfo {author} {\bibfnamefont {Y.}~\bibnamefont {Endo}},\ }\bibfield  {title} {\enquote {\bibinfo {title} {{Laser spectroscopy of the $\tilde{A}^{2}\Sigma^{+}$ -- $\tilde{X}^{2}\Pi_{i}$ band system of l-SiC$_{3}$H}},}\ }\href@noop {} {\bibfield  {journal} {\bibinfo  {journal} {The Journal of Chemical Physics}\ }\textbf {\bibinfo {volume} {143}} (\bibinfo {year} {2015})}\BibitemShut {NoStop}%
\bibitem [{\citenamefont {Hebner}(2001)}]{hebner2001spatially}%
  \BibitemOpen
  \bibfield  {author} {\bibinfo {author} {\bibfnamefont {G.}~\bibnamefont {Hebner}},\ }\bibfield  {title} {\enquote {\bibinfo {title} {Spatially resolved {SiF} and {SiF}$_2$ densities in inductively driven discharges containing {C$_2$F$_6$} and {C$_4$F$_8$}},}\ }\href@noop {} {\bibfield  {journal} {\bibinfo  {journal} {Journal of Applied Physics}\ }\textbf {\bibinfo {volume} {90}},\ \bibinfo {pages} {4938--4945} (\bibinfo {year} {2001})}\BibitemShut {NoStop}%
\bibitem [{\citenamefont {Nozaki}\ \emph {et~al.}(2000)\citenamefont {Nozaki}, \citenamefont {Kongo}, \citenamefont {Miyazaki}, \citenamefont {Kitazoe}, \citenamefont {Horii}, \citenamefont {Umemoto}, \citenamefont {Masuda},\ and\ \citenamefont {Matsumura}}]{nozaki2000identification}%
  \BibitemOpen
  \bibfield  {author} {\bibinfo {author} {\bibfnamefont {Y.}~\bibnamefont {Nozaki}}, \bibinfo {author} {\bibfnamefont {K.}~\bibnamefont {Kongo}}, \bibinfo {author} {\bibfnamefont {T.}~\bibnamefont {Miyazaki}}, \bibinfo {author} {\bibfnamefont {M.}~\bibnamefont {Kitazoe}}, \bibinfo {author} {\bibfnamefont {K.}~\bibnamefont {Horii}}, \bibinfo {author} {\bibfnamefont {H.}~\bibnamefont {Umemoto}}, \bibinfo {author} {\bibfnamefont {A.}~\bibnamefont {Masuda}}, \ and\ \bibinfo {author} {\bibfnamefont {H.}~\bibnamefont {Matsumura}},\ }\bibfield  {title} {\enquote {\bibinfo {title} {{Identification of Si and SiH in catalytic chemical vapor deposition of SiH$_4$ by laser induced fluorescence spectroscopy}},}\ }\href@noop {} {\bibfield  {journal} {\bibinfo  {journal} {Journal of Applied Physics}\ }\textbf {\bibinfo {volume} {88}},\ \bibinfo {pages} {5437--5443} (\bibinfo {year} {2000})}\BibitemShut {NoStop}%
\bibitem [{\citenamefont {Hertl}\ \emph {et~al.}(1998)\citenamefont {Hertl}, \citenamefont {Dorval}, \citenamefont {Leroy}, \citenamefont {Jolly},\ and\ \citenamefont {P{\'e}alat}}]{hertl1998laser}%
  \BibitemOpen
  \bibfield  {author} {\bibinfo {author} {\bibfnamefont {M.}~\bibnamefont {Hertl}}, \bibinfo {author} {\bibfnamefont {N.}~\bibnamefont {Dorval}}, \bibinfo {author} {\bibfnamefont {O.}~\bibnamefont {Leroy}}, \bibinfo {author} {\bibfnamefont {J.}~\bibnamefont {Jolly}}, \ and\ \bibinfo {author} {\bibfnamefont {M.}~\bibnamefont {P{\'e}alat}},\ }\bibfield  {title} {\enquote {\bibinfo {title} {{Laser-induced fluorescence measurements of absolute SiH densities in-RF discharges and comparison with a numerical model}},}\ }\href@noop {} {\bibfield  {journal} {\bibinfo  {journal} {Plasma Sources Science and Technology}\ }\textbf {\bibinfo {volume} {7}},\ \bibinfo {pages} {130} (\bibinfo {year} {1998})}\BibitemShut {NoStop}%
\bibitem [{\citenamefont {Matsumi}\ \emph {et~al.}(1986)\citenamefont {Matsumi}, \citenamefont {Hayashi}, \citenamefont {Yoshikawa},\ and\ \citenamefont {Komiya}}]{matsumi1986laser}%
  \BibitemOpen
  \bibfield  {author} {\bibinfo {author} {\bibfnamefont {Y.}~\bibnamefont {Matsumi}}, \bibinfo {author} {\bibfnamefont {T.}~\bibnamefont {Hayashi}}, \bibinfo {author} {\bibfnamefont {H.}~\bibnamefont {Yoshikawa}}, \ and\ \bibinfo {author} {\bibfnamefont {S.}~\bibnamefont {Komiya}},\ }\bibfield  {title} {\enquote {\bibinfo {title} {{Laser diagnostics of a silane plasma—SiH radicals in an a-Si: H chemical vapor deposition system}},}\ }\href@noop {} {\bibfield  {journal} {\bibinfo  {journal} {Journal of Vacuum Science \& Technology A: Vacuum, Surfaces, and Films}\ }\textbf {\bibinfo {volume} {4}},\ \bibinfo {pages} {1786--1790} (\bibinfo {year} {1986})}\BibitemShut {NoStop}%
\bibitem [{\citenamefont {Hertl}\ and\ \citenamefont {Jolly}(2000)}]{hertl2000laser}%
  \BibitemOpen
  \bibfield  {author} {\bibinfo {author} {\bibfnamefont {M.}~\bibnamefont {Hertl}}\ and\ \bibinfo {author} {\bibfnamefont {J.}~\bibnamefont {Jolly}},\ }\bibfield  {title} {\enquote {\bibinfo {title} {Laser-induced fluorescence detection and kinetics of {SiH}$_2$ radicals in {Ar}/{H}$_2$/{SiH}$_4$ {RF} discharges},}\ }\href@noop {} {\bibfield  {journal} {\bibinfo  {journal} {Journal of Physics D: Applied Physics}\ }\textbf {\bibinfo {volume} {33}},\ \bibinfo {pages} {381} (\bibinfo {year} {2000})}\BibitemShut {NoStop}%
\bibitem [{\citenamefont {Kono}\ \emph {et~al.}(1993)\citenamefont {Kono}, \citenamefont {Koike}, \citenamefont {Okuda},\ and\ \citenamefont {Goto}}]{kono1993laser}%
  \BibitemOpen
  \bibfield  {author} {\bibinfo {author} {\bibfnamefont {A.}~\bibnamefont {Kono}}, \bibinfo {author} {\bibfnamefont {N.}~\bibnamefont {Koike}}, \bibinfo {author} {\bibfnamefont {K.}~\bibnamefont {Okuda}}, \ and\ \bibinfo {author} {\bibfnamefont {T.~G.~T.}\ \bibnamefont {Goto}},\ }\bibfield  {title} {\enquote {\bibinfo {title} {{Laser-induced-fluorescence detection of SiH$_2$ radicals in a radio-frequency silane plasma}},}\ }\href@noop {} {\bibfield  {journal} {\bibinfo  {journal} {Japanese journal of applied physics}\ }\textbf {\bibinfo {volume} {32}},\ \bibinfo {pages} {L543} (\bibinfo {year} {1993})}\BibitemShut {NoStop}%
\bibitem [{\citenamefont {Kono}\ \emph {et~al.}(1995)\citenamefont {Kono}, \citenamefont {Koike}, \citenamefont {Nomura},\ and\ \citenamefont {Goto}}]{kono1995laser}%
  \BibitemOpen
  \bibfield  {author} {\bibinfo {author} {\bibfnamefont {A.}~\bibnamefont {Kono}}, \bibinfo {author} {\bibfnamefont {N.}~\bibnamefont {Koike}}, \bibinfo {author} {\bibfnamefont {H.}~\bibnamefont {Nomura}}, \ and\ \bibinfo {author} {\bibfnamefont {T.~G.~T.}\ \bibnamefont {Goto}},\ }\bibfield  {title} {\enquote {\bibinfo {title} {{Laser-induced-fluorescence study of the SiH2 density in RF SiH$_4$ plasmas with Xe, Ar, He, and H$_2$ dilution gases}},}\ }\href@noop {} {\bibfield  {journal} {\bibinfo  {journal} {Japanese journal of applied physics}\ }\textbf {\bibinfo {volume} {34}},\ \bibinfo {pages} {307} (\bibinfo {year} {1995})}\BibitemShut {NoStop}%
\bibitem [{\citenamefont {Walkup}\ \emph {et~al.}(1984)\citenamefont {Walkup}, \citenamefont {Avouris}, \citenamefont {Dreyfus}, \citenamefont {Jasinski},\ and\ \citenamefont {Selwyn}}]{walkup1984laser}%
  \BibitemOpen
  \bibfield  {author} {\bibinfo {author} {\bibfnamefont {R.}~\bibnamefont {Walkup}}, \bibinfo {author} {\bibfnamefont {P.}~\bibnamefont {Avouris}}, \bibinfo {author} {\bibfnamefont {R.}~\bibnamefont {Dreyfus}}, \bibinfo {author} {\bibfnamefont {J.}~\bibnamefont {Jasinski}}, \ and\ \bibinfo {author} {\bibfnamefont {G.}~\bibnamefont {Selwyn}},\ }\bibfield  {title} {\enquote {\bibinfo {title} {Laser detection of diatomic products of plasma sputtering and etching},}\ }\href@noop {} {\bibfield  {journal} {\bibinfo  {journal} {Applied physics letters}\ }\textbf {\bibinfo {volume} {45}},\ \bibinfo {pages} {372--374} (\bibinfo {year} {1984})}\BibitemShut {NoStop}%
\bibitem [{\citenamefont {Chrystie}\ \emph {et~al.}(2017)\citenamefont {Chrystie}, \citenamefont {Feroughi}, \citenamefont {Dreier},\ and\ \citenamefont {Schulz}}]{chrystie2017sio}%
  \BibitemOpen
  \bibfield  {author} {\bibinfo {author} {\bibfnamefont {R.~S.}\ \bibnamefont {Chrystie}}, \bibinfo {author} {\bibfnamefont {O.~M.}\ \bibnamefont {Feroughi}}, \bibinfo {author} {\bibfnamefont {T.}~\bibnamefont {Dreier}}, \ and\ \bibinfo {author} {\bibfnamefont {C.}~\bibnamefont {Schulz}},\ }\bibfield  {title} {\enquote {\bibinfo {title} {{SiO multi-line laser-induced fluorescence for quantitative temperature imaging in flame-synthesis of nanoparticles}},}\ }\href@noop {} {\bibfield  {journal} {\bibinfo  {journal} {Applied Physics B}\ }\textbf {\bibinfo {volume} {123}},\ \bibinfo {pages} {1--12} (\bibinfo {year} {2017})}\BibitemShut {NoStop}%
\bibitem [{\citenamefont {Chrystie}\ \emph {et~al.}(2019)\citenamefont {Chrystie}, \citenamefont {Ebertz}, \citenamefont {Dreier},\ and\ \citenamefont {Schulz}}]{chrystie2019absolute}%
  \BibitemOpen
  \bibfield  {author} {\bibinfo {author} {\bibfnamefont {R.~S.}\ \bibnamefont {Chrystie}}, \bibinfo {author} {\bibfnamefont {F.~L.}\ \bibnamefont {Ebertz}}, \bibinfo {author} {\bibfnamefont {T.}~\bibnamefont {Dreier}}, \ and\ \bibinfo {author} {\bibfnamefont {C.}~\bibnamefont {Schulz}},\ }\bibfield  {title} {\enquote {\bibinfo {title} {{Absolute SiO concentration imaging in low-pressure nanoparticle-synthesis flames via laser-induced fluorescence}},}\ }\href@noop {} {\bibfield  {journal} {\bibinfo  {journal} {Applied Physics B}\ }\textbf {\bibinfo {volume} {125}},\ \bibinfo {pages} {1--15} (\bibinfo {year} {2019})}\BibitemShut {NoStop}%
\bibitem [{\citenamefont {Greenberg}\ and\ \citenamefont {Hargis~Jr}(1990)}]{greenberg1990laser}%
  \BibitemOpen
  \bibfield  {author} {\bibinfo {author} {\bibfnamefont {K.}~\bibnamefont {Greenberg}}\ and\ \bibinfo {author} {\bibfnamefont {P.}~\bibnamefont {Hargis~Jr}},\ }\bibfield  {title} {\enquote {\bibinfo {title} {Laser-induced-fluorescence detection of {SO} and {SO}$_2$ in {SF}$_6$/{O}$_2$ plasma-etching discharges},}\ }\href@noop {} {\bibfield  {journal} {\bibinfo  {journal} {Journal of applied physics}\ }\textbf {\bibinfo {volume} {68}},\ \bibinfo {pages} {505--511} (\bibinfo {year} {1990})}\BibitemShut {NoStop}%
\bibitem [{\citenamefont {Weng}, \citenamefont {Aldén},\ and\ \citenamefont {Li}(2019)}]{weng2019quantitative}%
  \BibitemOpen
  \bibfield  {author} {\bibinfo {author} {\bibfnamefont {W.}~\bibnamefont {Weng}}, \bibinfo {author} {\bibfnamefont {M.}~\bibnamefont {Aldén}}, \ and\ \bibinfo {author} {\bibfnamefont {Z.}~\bibnamefont {Li}},\ }\bibfield  {title} {\enquote {\bibinfo {title} {Quantitative {SO}$_2$ detection in combustion environments using broad band ultraviolet absorption and laser-induced fluorescence},}\ }\href@noop {} {\bibfield  {journal} {\bibinfo  {journal} {Analytical chemistry}\ }\textbf {\bibinfo {volume} {91}},\ \bibinfo {pages} {10849--10855} (\bibinfo {year} {2019})}\BibitemShut {NoStop}%
\bibitem [{\citenamefont {Smith}\ and\ \citenamefont {Hopkins}(1981)}]{smith1981fluorescence}%
  \BibitemOpen
  \bibfield  {author} {\bibinfo {author} {\bibfnamefont {A.~L.}\ \bibnamefont {Smith}}\ and\ \bibinfo {author} {\bibfnamefont {J.~B.}\ \bibnamefont {Hopkins}},\ }\bibfield  {title} {\enquote {\bibinfo {title} {{Fluorescence of S$_2$(B--X) excited by fixed frequency ultraviolet lasers}},}\ }\href@noop {} {\bibfield  {journal} {\bibinfo  {journal} {The Journal of Chemical Physics}\ }\textbf {\bibinfo {volume} {75}},\ \bibinfo {pages} {2080--2084} (\bibinfo {year} {1981})}\BibitemShut {NoStop}%
\bibitem [{\citenamefont {Zhang}\ \emph {et~al.}(1995)\citenamefont {Zhang}, \citenamefont {Dupr{\'e}}, \citenamefont {Grzybowski},\ and\ \citenamefont {Vaccaro}}]{zhang1995laser}%
  \BibitemOpen
  \bibfield  {author} {\bibinfo {author} {\bibfnamefont {Q.}~\bibnamefont {Zhang}}, \bibinfo {author} {\bibfnamefont {P.}~\bibnamefont {Dupr{\'e}}}, \bibinfo {author} {\bibfnamefont {B.}~\bibnamefont {Grzybowski}}, \ and\ \bibinfo {author} {\bibfnamefont {P.~H.}\ \bibnamefont {Vaccaro}},\ }\bibfield  {title} {\enquote {\bibinfo {title} {{Laser-induced fluorescence studies of jet-cooled S$_2$O: Axis-switching and predissociation effects}},}\ }\href@noop {} {\bibfield  {journal} {\bibinfo  {journal} {The Journal of chemical physics}\ }\textbf {\bibinfo {volume} {103}},\ \bibinfo {pages} {67--79} (\bibinfo {year} {1995})}\BibitemShut {NoStop}%
\bibitem [{\citenamefont {Nakhate}, \citenamefont {Mukund},\ and\ \citenamefont {Bhattacharyya}(2021)}]{nakhate2021jet}%
  \BibitemOpen
  \bibfield  {author} {\bibinfo {author} {\bibfnamefont {S.}~\bibnamefont {Nakhate}}, \bibinfo {author} {\bibfnamefont {S.}~\bibnamefont {Mukund}}, \ and\ \bibinfo {author} {\bibfnamefont {S.}~\bibnamefont {Bhattacharyya}},\ }\bibfield  {title} {\enquote {\bibinfo {title} {Jet cooled laser-induced fluorescence spectroscopy of tantalum monocarbide: Observation of the ground state vibrations and the low-energy states},}\ }\href@noop {} {\bibfield  {journal} {\bibinfo  {journal} {Journal of Molecular Structure}\ }\textbf {\bibinfo {volume} {1243}},\ \bibinfo {pages} {130888} (\bibinfo {year} {2021})}\BibitemShut {NoStop}%
\bibitem [{\citenamefont {Nakhate}, \citenamefont {Mukund},\ and\ \citenamefont {Bhattacharyya}(2017)}]{nakhate2017laser}%
  \BibitemOpen
  \bibfield  {author} {\bibinfo {author} {\bibfnamefont {S.}~\bibnamefont {Nakhate}}, \bibinfo {author} {\bibfnamefont {S.}~\bibnamefont {Mukund}}, \ and\ \bibinfo {author} {\bibfnamefont {S.}~\bibnamefont {Bhattacharyya}},\ }\bibfield  {title} {\enquote {\bibinfo {title} {{Laser-induced fluorescence spectroscopy of jet-cooled TiC: Observation of low-lying $^{1}\Sigma^{+}$ states}},}\ }\href@noop {} {\bibfield  {journal} {\bibinfo  {journal} {Chemical Physics Letters}\ }\textbf {\bibinfo {volume} {680}},\ \bibinfo {pages} {51--55} (\bibinfo {year} {2017})}\BibitemShut {NoStop}%
\bibitem [{\citenamefont {Zhang}\ \emph {et~al.}(2009)\citenamefont {Zhang}, \citenamefont {Guo}, \citenamefont {Yu}, \citenamefont {Zhen},\ and\ \citenamefont {Chen}}]{zhang2009laser}%
  \BibitemOpen
  \bibfield  {author} {\bibinfo {author} {\bibfnamefont {Z.}~\bibnamefont {Zhang}}, \bibinfo {author} {\bibfnamefont {J.}~\bibnamefont {Guo}}, \bibinfo {author} {\bibfnamefont {X.}~\bibnamefont {Yu}}, \bibinfo {author} {\bibfnamefont {J.}~\bibnamefont {Zhen}}, \ and\ \bibinfo {author} {\bibfnamefont {Y.}~\bibnamefont {Chen}},\ }\bibfield  {title} {\enquote {\bibinfo {title} {{The laser-induced fluorescence spectroscopy of TiF in the ultraviolet region}},}\ }\href@noop {} {\bibfield  {journal} {\bibinfo  {journal} {Journal of Molecular Spectroscopy}\ }\textbf {\bibinfo {volume} {253}},\ \bibinfo {pages} {112--115} (\bibinfo {year} {2009})}\BibitemShut {NoStop}%
\bibitem [{\citenamefont {Zhao}\ \emph {et~al.}(2019)\citenamefont {Zhao}, \citenamefont {Lei}, \citenamefont {Li}, \citenamefont {Li}, \citenamefont {Ma}, \citenamefont {Zhang}, \citenamefont {Guo},\ and\ \citenamefont {Lu}}]{zhao2019experimental}%
  \BibitemOpen
  \bibfield  {author} {\bibinfo {author} {\bibfnamefont {N.}~\bibnamefont {Zhao}}, \bibinfo {author} {\bibfnamefont {D.}~\bibnamefont {Lei}}, \bibinfo {author} {\bibfnamefont {X.}~\bibnamefont {Li}}, \bibinfo {author} {\bibfnamefont {J.}~\bibnamefont {Li}}, \bibinfo {author} {\bibfnamefont {Q.}~\bibnamefont {Ma}}, \bibinfo {author} {\bibfnamefont {Q.}~\bibnamefont {Zhang}}, \bibinfo {author} {\bibfnamefont {L.}~\bibnamefont {Guo}}, \ and\ \bibinfo {author} {\bibfnamefont {Y.}~\bibnamefont {Lu}},\ }\bibfield  {title} {\enquote {\bibinfo {title} {Experimental investigation of laser-induced breakdown spectroscopy assisted with laser-induced fluorescence for trace aluminum detection in steatite ceramics},}\ }\href@noop {} {\bibfield  {journal} {\bibinfo  {journal} {Applied optics}\ }\textbf {\bibinfo {volume} {58}},\ \bibinfo {pages} {1895--1899} (\bibinfo {year} {2019})}\BibitemShut {NoStop}%
\bibitem [{\citenamefont {Dreyfus}, \citenamefont {Kelly},\ and\ \citenamefont {Walkup}(1986)}]{dreyfus1986laser}%
  \BibitemOpen
  \bibfield  {author} {\bibinfo {author} {\bibfnamefont {R.}~\bibnamefont {Dreyfus}}, \bibinfo {author} {\bibfnamefont {R.}~\bibnamefont {Kelly}}, \ and\ \bibinfo {author} {\bibfnamefont {R.}~\bibnamefont {Walkup}},\ }\bibfield  {title} {\enquote {\bibinfo {title} {Laser-induced fluorescence studies of excimer laser ablation of {Al}$_2${O}$_3$},}\ }\href@noop {} {\bibfield  {journal} {\bibinfo  {journal} {Applied physics letters}\ }\textbf {\bibinfo {volume} {49}},\ \bibinfo {pages} {1478--1480} (\bibinfo {year} {1986})}\BibitemShut {NoStop}%
\bibitem [{\citenamefont {Selwyn}(1987)}]{selwyn1987atomic}%
  \BibitemOpen
  \bibfield  {author} {\bibinfo {author} {\bibfnamefont {G.~S.}\ \bibnamefont {Selwyn}},\ }\bibfield  {title} {\enquote {\bibinfo {title} {Atomic arsenic detection by {ArF} laser-induced fluorescence},}\ }\href@noop {} {\bibfield  {journal} {\bibinfo  {journal} {Applied physics letters}\ }\textbf {\bibinfo {volume} {51}},\ \bibinfo {pages} {167--168} (\bibinfo {year} {1987})}\BibitemShut {NoStop}%
\bibitem [{\citenamefont {Li}\ \emph {et~al.}(2016)\citenamefont {Li}, \citenamefont {Hao}, \citenamefont {Zou}, \citenamefont {Zhou}, \citenamefont {Li}, \citenamefont {Guo}, \citenamefont {Li}, \citenamefont {Lu},\ and\ \citenamefont {Zeng}}]{li2016determinations}%
  \BibitemOpen
  \bibfield  {author} {\bibinfo {author} {\bibfnamefont {C.}~\bibnamefont {Li}}, \bibinfo {author} {\bibfnamefont {Z.}~\bibnamefont {Hao}}, \bibinfo {author} {\bibfnamefont {Z.}~\bibnamefont {Zou}}, \bibinfo {author} {\bibfnamefont {R.}~\bibnamefont {Zhou}}, \bibinfo {author} {\bibfnamefont {J.}~\bibnamefont {Li}}, \bibinfo {author} {\bibfnamefont {L.}~\bibnamefont {Guo}}, \bibinfo {author} {\bibfnamefont {X.}~\bibnamefont {Li}}, \bibinfo {author} {\bibfnamefont {Y.}~\bibnamefont {Lu}}, \ and\ \bibinfo {author} {\bibfnamefont {X.}~\bibnamefont {Zeng}},\ }\bibfield  {title} {\enquote {\bibinfo {title} {Determinations of trace boron in superalloys and steels using laser-induced breakdown spectroscopy assisted with laser-induced fluorescence},}\ }\href@noop {} {\bibfield  {journal} {\bibinfo  {journal} {Optics Express}\ }\textbf {\bibinfo {volume} {24}},\ \bibinfo {pages} {7850--7857} (\bibinfo {year} {2016})}\BibitemShut {NoStop}%
\bibitem [{\citenamefont {Aubert}\ \emph {et~al.}(2017)\citenamefont {Aubert}, \citenamefont {Duluard}, \citenamefont {Sadeghi},\ and\ \citenamefont {Gicquel}}]{aubert2017comparison}%
  \BibitemOpen
  \bibfield  {author} {\bibinfo {author} {\bibfnamefont {X.}~\bibnamefont {Aubert}}, \bibinfo {author} {\bibfnamefont {C.}~\bibnamefont {Duluard}}, \bibinfo {author} {\bibfnamefont {N.}~\bibnamefont {Sadeghi}}, \ and\ \bibinfo {author} {\bibfnamefont {A.}~\bibnamefont {Gicquel}},\ }\bibfield  {title} {\enquote {\bibinfo {title} {Comparison of three optical diagnostic techniques for the measurement of boron atom density in a {H}$_2$/{B}$_2${H}$_6$ microwave plasma},}\ }\href@noop {} {\bibfield  {journal} {\bibinfo  {journal} {Plasma Sources Science and Technology}\ }\textbf {\bibinfo {volume} {26}},\ \bibinfo {pages} {115011} (\bibinfo {year} {2017})}\BibitemShut {NoStop}%
\bibitem [{\citenamefont {Sirse}\ \emph {et~al.}(2014)\citenamefont {Sirse}, \citenamefont {Foucher}, \citenamefont {Chabert},\ and\ \citenamefont {Booth}}]{sirse2014ground}%
  \BibitemOpen
  \bibfield  {author} {\bibinfo {author} {\bibfnamefont {N.}~\bibnamefont {Sirse}}, \bibinfo {author} {\bibfnamefont {M.}~\bibnamefont {Foucher}}, \bibinfo {author} {\bibfnamefont {P.}~\bibnamefont {Chabert}}, \ and\ \bibinfo {author} {\bibfnamefont {J.-P.}\ \bibnamefont {Booth}},\ }\bibfield  {title} {\enquote {\bibinfo {title} {Ground state bromine atom density measurements by two-photon absorption laser-induced fluorescence},}\ }\href@noop {} {\bibfield  {journal} {\bibinfo  {journal} {Plasma Sources Science and Technology}\ }\textbf {\bibinfo {volume} {23}},\ \bibinfo {pages} {062003} (\bibinfo {year} {2014})}\BibitemShut {NoStop}%
\bibitem [{\citenamefont {Ald{\'e}n}, \citenamefont {Bengtsson},\ and\ \citenamefont {Westblom}(1989)}]{alden1989detection}%
  \BibitemOpen
  \bibfield  {author} {\bibinfo {author} {\bibfnamefont {M.}~\bibnamefont {Ald{\'e}n}}, \bibinfo {author} {\bibfnamefont {P.-E.}\ \bibnamefont {Bengtsson}}, \ and\ \bibinfo {author} {\bibfnamefont {U.}~\bibnamefont {Westblom}},\ }\bibfield  {title} {\enquote {\bibinfo {title} {Detection of carbon atoms in flames using stimulated emission induced by two-photon laser excitation},}\ }\href@noop {} {\bibfield  {journal} {\bibinfo  {journal} {Optics Communications}\ }\textbf {\bibinfo {volume} {71}},\ \bibinfo {pages} {263--268} (\bibinfo {year} {1989})}\BibitemShut {NoStop}%
\bibitem [{\citenamefont {Booth}\ \emph {et~al.}(2012)\citenamefont {Booth}, \citenamefont {Azamoum}, \citenamefont {Sirse},\ and\ \citenamefont {Chabert}}]{booth2012absolute}%
  \BibitemOpen
  \bibfield  {author} {\bibinfo {author} {\bibfnamefont {J.}~\bibnamefont {Booth}}, \bibinfo {author} {\bibfnamefont {Y.}~\bibnamefont {Azamoum}}, \bibinfo {author} {\bibfnamefont {N.}~\bibnamefont {Sirse}}, \ and\ \bibinfo {author} {\bibfnamefont {P.}~\bibnamefont {Chabert}},\ }\bibfield  {title} {\enquote {\bibinfo {title} {{Absolute atomic chlorine densities in a Cl$_2$ inductively coupled plasma determined by two-photon laser-induced fluorescence with a new calibration method}},}\ }\href@noop {} {\bibfield  {journal} {\bibinfo  {journal} {Journal of Physics D: Applied Physics}\ }\textbf {\bibinfo {volume} {45}},\ \bibinfo {pages} {195201} (\bibinfo {year} {2012})}\BibitemShut {NoStop}%
\bibitem [{\citenamefont {Heaven}\ \emph {et~al.}(1982)\citenamefont {Heaven}, \citenamefont {Miller}, \citenamefont {Freeman}, \citenamefont {White},\ and\ \citenamefont {Bokor}}]{heaven1982two}%
  \BibitemOpen
  \bibfield  {author} {\bibinfo {author} {\bibfnamefont {M.}~\bibnamefont {Heaven}}, \bibinfo {author} {\bibfnamefont {T.~A.}\ \bibnamefont {Miller}}, \bibinfo {author} {\bibfnamefont {R.~R.}\ \bibnamefont {Freeman}}, \bibinfo {author} {\bibfnamefont {J.}~\bibnamefont {White}}, \ and\ \bibinfo {author} {\bibfnamefont {J.}~\bibnamefont {Bokor}},\ }\bibfield  {title} {\enquote {\bibinfo {title} {{Two-photon absorption, laser-induced fluorescence detection of Cl atoms}},}\ }\href@noop {} {\bibfield  {journal} {\bibinfo  {journal} {Chemical Physics Letters}\ }\textbf {\bibinfo {volume} {86}},\ \bibinfo {pages} {458--462} (\bibinfo {year} {1982})}\BibitemShut {NoStop}%
\bibitem [{\citenamefont {Sdorra}, \citenamefont {Quentmeier},\ and\ \citenamefont {Niemax}(1989)}]{sdorra1989basic}%
  \BibitemOpen
  \bibfield  {author} {\bibinfo {author} {\bibfnamefont {W.}~\bibnamefont {Sdorra}}, \bibinfo {author} {\bibfnamefont {A.}~\bibnamefont {Quentmeier}}, \ and\ \bibinfo {author} {\bibfnamefont {K.}~\bibnamefont {Niemax}},\ }\bibfield  {title} {\enquote {\bibinfo {title} {{Basic investigations for laser microanalysis: II. Laser-induced fluorescence in laser-produced sample plumes}},}\ }\href@noop {} {\bibfield  {journal} {\bibinfo  {journal} {Microchimica Acta}\ }\textbf {\bibinfo {volume} {98}},\ \bibinfo {pages} {201--218} (\bibinfo {year} {1989})}\BibitemShut {NoStop}%
\bibitem [{\citenamefont {Gellert}(1988)}]{gellert1988measurement}%
  \BibitemOpen
  \bibfield  {author} {\bibinfo {author} {\bibfnamefont {B.}~\bibnamefont {Gellert}},\ }\bibfield  {title} {\enquote {\bibinfo {title} {Measurement of high copper vapour densities by laser-induced fluorescence},}\ }\href@noop {} {\bibfield  {journal} {\bibinfo  {journal} {Journal of Physics D: Applied Physics}\ }\textbf {\bibinfo {volume} {21}},\ \bibinfo {pages} {710} (\bibinfo {year} {1988})}\BibitemShut {NoStop}%
\bibitem [{\citenamefont {Bolshov}, \citenamefont {Zybin},\ and\ \citenamefont {Smirenkina}(1981)}]{bolshov1981atomic}%
  \BibitemOpen
  \bibfield  {author} {\bibinfo {author} {\bibfnamefont {M.}~\bibnamefont {Bolshov}}, \bibinfo {author} {\bibfnamefont {A.}~\bibnamefont {Zybin}}, \ and\ \bibinfo {author} {\bibfnamefont {I.}~\bibnamefont {Smirenkina}},\ }\bibfield  {title} {\enquote {\bibinfo {title} {Atomic fluorescence spectrometry with laser excitation},}\ }\href@noop {} {\bibfield  {journal} {\bibinfo  {journal} {Spectrochimica Acta Part B: Atomic Spectroscopy}\ }\textbf {\bibinfo {volume} {36}},\ \bibinfo {pages} {1143--1152} (\bibinfo {year} {1981})}\BibitemShut {NoStop}%
\bibitem [{\citenamefont {Falk}\ \emph {et~al.}(1988)\citenamefont {Falk}, \citenamefont {Paetzold}, \citenamefont {Schmidt},\ and\ \citenamefont {Tilch}}]{falk1988analytical}%
  \BibitemOpen
  \bibfield  {author} {\bibinfo {author} {\bibfnamefont {H.}~\bibnamefont {Falk}}, \bibinfo {author} {\bibfnamefont {H.-J.}\ \bibnamefont {Paetzold}}, \bibinfo {author} {\bibfnamefont {K.}~\bibnamefont {Schmidt}}, \ and\ \bibinfo {author} {\bibfnamefont {J.}~\bibnamefont {Tilch}},\ }\bibfield  {title} {\enquote {\bibinfo {title} {Analytical application of laser excited atomic fluorescence using a graphite cup atomizer},}\ }\href@noop {} {\bibfield  {journal} {\bibinfo  {journal} {Spectrochimica Acta Part B: Atomic Spectroscopy}\ }\textbf {\bibinfo {volume} {43}},\ \bibinfo {pages} {1101--1109} (\bibinfo {year} {1988})}\BibitemShut {NoStop}%
\bibitem [{\citenamefont {MA}\ \emph {et~al.}(2002)\citenamefont {MA}, \citenamefont {TAO}, \citenamefont {ZHANG},\ and\ \citenamefont {CHEN}}]{ma2002determination}%
  \BibitemOpen
  \bibfield  {author} {\bibinfo {author} {\bibfnamefont {W.}~\bibnamefont {MA}}, \bibinfo {author} {\bibfnamefont {S.}~\bibnamefont {TAO}}, \bibinfo {author} {\bibfnamefont {D.}~\bibnamefont {ZHANG}}, \ and\ \bibinfo {author} {\bibfnamefont {D.}~\bibnamefont {CHEN}},\ }\bibfield  {title} {\enquote {\bibinfo {title} {Determination of trace gallium in rock and sediment samples by laser-induced fluorescence spectrometry},}\ }\href@noop {} {\bibfield  {journal} {\bibinfo  {journal} {Analytical Sciences/Supplements}\ }\textbf {\bibinfo {volume} {17}},\ \bibinfo {pages} {a215--a218} (\bibinfo {year} {2002})}\BibitemShut {NoStop}%
\bibitem [{\citenamefont {Brewer}\ \emph {et~al.}(1983)\citenamefont {Brewer}, \citenamefont {Das}, \citenamefont {Ondrey},\ and\ \citenamefont {Bersohn}}]{brewer1983measurement}%
  \BibitemOpen
  \bibfield  {author} {\bibinfo {author} {\bibfnamefont {P.}~\bibnamefont {Brewer}}, \bibinfo {author} {\bibfnamefont {P.}~\bibnamefont {Das}}, \bibinfo {author} {\bibfnamefont {G.}~\bibnamefont {Ondrey}}, \ and\ \bibinfo {author} {\bibfnamefont {R.}~\bibnamefont {Bersohn}},\ }\bibfield  {title} {\enquote {\bibinfo {title} {Measurement of the relative populations of {I} ($^2${P}$^0_{1/2}$) and {I}($^2${P}$^0_{3/2}$) by laser induced vacuum ultraviolet fluorescence},}\ }\href@noop {} {\bibfield  {journal} {\bibinfo  {journal} {The Journal of Chemical Physics}\ }\textbf {\bibinfo {volume} {79}},\ \bibinfo {pages} {720--723} (\bibinfo {year} {1983})}\BibitemShut {NoStop}%
\bibitem [{\citenamefont {Zhu}\ \emph {et~al.}(2020)\citenamefont {Zhu}, \citenamefont {Barkley}, \citenamefont {Dedic}, \citenamefont {Sippel},\ and\ \citenamefont {Michael}}]{zhu2020two}%
  \BibitemOpen
  \bibfield  {author} {\bibinfo {author} {\bibfnamefont {K.}~\bibnamefont {Zhu}}, \bibinfo {author} {\bibfnamefont {S.~J.}\ \bibnamefont {Barkley}}, \bibinfo {author} {\bibfnamefont {C.~E.}\ \bibnamefont {Dedic}}, \bibinfo {author} {\bibfnamefont {T.~R.}\ \bibnamefont {Sippel}}, \ and\ \bibinfo {author} {\bibfnamefont {J.~B.}\ \bibnamefont {Michael}},\ }\bibfield  {title} {\enquote {\bibinfo {title} {Two-photon laser-induced fluorescence of sodium in multiphase combustion},}\ }\href@noop {} {\bibfield  {journal} {\bibinfo  {journal} {Applied Optics}\ }\textbf {\bibinfo {volume} {59}},\ \bibinfo {pages} {5632--5641} (\bibinfo {year} {2020})}\BibitemShut {NoStop}%
\bibitem [{\citenamefont {Daily}\ and\ \citenamefont {Chan}(1978)}]{daily1978laser}%
  \BibitemOpen
  \bibfield  {author} {\bibinfo {author} {\bibfnamefont {J.~W.}\ \bibnamefont {Daily}}\ and\ \bibinfo {author} {\bibfnamefont {C.}~\bibnamefont {Chan}},\ }\bibfield  {title} {\enquote {\bibinfo {title} {Laser-induced fluorescence measurement of sodium in flames},}\ }\href@noop {} {\bibfield  {journal} {\bibinfo  {journal} {Combustion and Flame}\ }\textbf {\bibinfo {volume} {33}},\ \bibinfo {pages} {47--53} (\bibinfo {year} {1978})}\BibitemShut {NoStop}%
\bibitem [{\citenamefont {Erdmann}, \citenamefont {Figger},\ and\ \citenamefont {Walther}(1972)}]{erdmann1972lifetime}%
  \BibitemOpen
  \bibfield  {author} {\bibinfo {author} {\bibfnamefont {T.}~\bibnamefont {Erdmann}}, \bibinfo {author} {\bibfnamefont {H.}~\bibnamefont {Figger}}, \ and\ \bibinfo {author} {\bibfnamefont {H.}~\bibnamefont {Walther}},\ }\bibfield  {title} {\enquote {\bibinfo {title} {Lifetime measurements with a tunable flashlamp pumped dye laser},}\ }\href@noop {} {\bibfield  {journal} {\bibinfo  {journal} {Optics Communications}\ }\textbf {\bibinfo {volume} {6}},\ \bibinfo {pages} {166--168} (\bibinfo {year} {1972})}\BibitemShut {NoStop}%
\bibitem [{\citenamefont {Laville}\ \emph {et~al.}(2009)\citenamefont {Laville}, \citenamefont {Goueguel}, \citenamefont {Loudyi}, \citenamefont {Vidal}, \citenamefont {Chaker},\ and\ \citenamefont {Sabsabi}}]{laville2009laser}%
  \BibitemOpen
  \bibfield  {author} {\bibinfo {author} {\bibfnamefont {S.}~\bibnamefont {Laville}}, \bibinfo {author} {\bibfnamefont {C.}~\bibnamefont {Goueguel}}, \bibinfo {author} {\bibfnamefont {H.}~\bibnamefont {Loudyi}}, \bibinfo {author} {\bibfnamefont {F.}~\bibnamefont {Vidal}}, \bibinfo {author} {\bibfnamefont {M.}~\bibnamefont {Chaker}}, \ and\ \bibinfo {author} {\bibfnamefont {M.}~\bibnamefont {Sabsabi}},\ }\bibfield  {title} {\enquote {\bibinfo {title} {Laser-induced fluorescence detection of lead atoms in a laser-induced plasma: An experimental analytical optimization study},}\ }\href@noop {} {\bibfield  {journal} {\bibinfo  {journal} {Spectrochimica Acta Part B: Atomic Spectroscopy}\ }\textbf {\bibinfo {volume} {64}},\ \bibinfo {pages} {347--353} (\bibinfo {year} {2009})}\BibitemShut {NoStop}%
\bibitem [{\citenamefont {Brewer}, \citenamefont {Van~Veen},\ and\ \citenamefont {Bersohn}(1982)}]{brewer1982two}%
  \BibitemOpen
  \bibfield  {author} {\bibinfo {author} {\bibfnamefont {P.}~\bibnamefont {Brewer}}, \bibinfo {author} {\bibfnamefont {N.}~\bibnamefont {Van~Veen}}, \ and\ \bibinfo {author} {\bibfnamefont {R.}~\bibnamefont {Bersohn}},\ }\bibfield  {title} {\enquote {\bibinfo {title} {Two-photon induced fluorescence and resonance-enhanced ionization of sulfur atoms},}\ }\href@noop {} {\bibfield  {journal} {\bibinfo  {journal} {Chemical Physics Letters}\ }\textbf {\bibinfo {volume} {91}},\ \bibinfo {pages} {126--129} (\bibinfo {year} {1982})}\BibitemShut {NoStop}%
\bibitem [{\citenamefont {Roth}, \citenamefont {Spears},\ and\ \citenamefont {Wong}(1984)}]{roth1984spatial}%
  \BibitemOpen
  \bibfield  {author} {\bibinfo {author} {\bibfnamefont {R.}~\bibnamefont {Roth}}, \bibinfo {author} {\bibfnamefont {K.}~\bibnamefont {Spears}}, \ and\ \bibinfo {author} {\bibfnamefont {G.}~\bibnamefont {Wong}},\ }\bibfield  {title} {\enquote {\bibinfo {title} {Spatial concentrations of silicon atoms by laser-induced fluorescence in a silane glow discharge},}\ }\href@noop {} {\bibfield  {journal} {\bibinfo  {journal} {Applied physics letters}\ }\textbf {\bibinfo {volume} {45}},\ \bibinfo {pages} {28--30} (\bibinfo {year} {1984})}\BibitemShut {NoStop}%
\bibitem [{\citenamefont {Britun}, \citenamefont {Gaillard},\ and\ \citenamefont {Han}(2008)}]{britun2008laser}%
  \BibitemOpen
  \bibfield  {author} {\bibinfo {author} {\bibfnamefont {N.}~\bibnamefont {Britun}}, \bibinfo {author} {\bibfnamefont {M.}~\bibnamefont {Gaillard}}, \ and\ \bibinfo {author} {\bibfnamefont {J.}~\bibnamefont {Han}},\ }\bibfield  {title} {\enquote {\bibinfo {title} {{Laser induced fluorescence for Ti and Ti$^+$ density characterization in a magnetron discharge}},}\ }\href@noop {} {\bibfield  {journal} {\bibinfo  {journal} {Journal of Physics D: Applied Physics}\ }\textbf {\bibinfo {volume} {41}},\ \bibinfo {pages} {185201} (\bibinfo {year} {2008})}\BibitemShut {NoStop}%
\bibitem [{\citenamefont {Ljung}\ \emph {et~al.}(1997)\citenamefont {Ljung}, \citenamefont {Nystr{\"o}m}, \citenamefont {Enger}, \citenamefont {Ljungberg},\ and\ \citenamefont {Axner}}]{ljung1997detection}%
  \BibitemOpen
  \bibfield  {author} {\bibinfo {author} {\bibfnamefont {P.}~\bibnamefont {Ljung}}, \bibinfo {author} {\bibfnamefont {E.}~\bibnamefont {Nystr{\"o}m}}, \bibinfo {author} {\bibfnamefont {J.}~\bibnamefont {Enger}}, \bibinfo {author} {\bibfnamefont {P.}~\bibnamefont {Ljungberg}}, \ and\ \bibinfo {author} {\bibfnamefont {O.}~\bibnamefont {Axner}},\ }\bibfield  {title} {\enquote {\bibinfo {title} {{Detection of titanium in electrothermal atomizers by laser-induced fluorescence. Part 1. Determination of optimum excitation and detection wavelengths}},}\ }\href@noop {} {\bibfield  {journal} {\bibinfo  {journal} {Spectrochimica Acta Part B: Atomic Spectroscopy}\ }\textbf {\bibinfo {volume} {52}},\ \bibinfo {pages} {675--701} (\bibinfo {year} {1997})}\BibitemShut {NoStop}%
\bibitem [{\citenamefont {Hadrath}\ \emph {et~al.}(2005)\citenamefont {Hadrath}, \citenamefont {Ehlbeck}, \citenamefont {Lieder},\ and\ \citenamefont {Sigeneger}}]{hadrath2005determination}%
  \BibitemOpen
  \bibfield  {author} {\bibinfo {author} {\bibfnamefont {S.}~\bibnamefont {Hadrath}}, \bibinfo {author} {\bibfnamefont {J.}~\bibnamefont {Ehlbeck}}, \bibinfo {author} {\bibfnamefont {G.}~\bibnamefont {Lieder}}, \ and\ \bibinfo {author} {\bibfnamefont {F.}~\bibnamefont {Sigeneger}},\ }\bibfield  {title} {\enquote {\bibinfo {title} {Determination of absolute population densities of eroded tungsten in hollow cathode lamps and fluorescent lamps by laser-induced fluorescence},}\ }\href@noop {} {\bibfield  {journal} {\bibinfo  {journal} {Journal of Physics D: Applied Physics}\ }\textbf {\bibinfo {volume} {38}},\ \bibinfo {pages} {3285} (\bibinfo {year} {2005})}\BibitemShut {NoStop}%
\bibitem [{\citenamefont {Georgiev}, \citenamefont {Blagoev},\ and\ \citenamefont {Pashov}(2019)}]{georgiev2019detection}%
  \BibitemOpen
  \bibfield  {author} {\bibinfo {author} {\bibfnamefont {A.}~\bibnamefont {Georgiev}}, \bibinfo {author} {\bibfnamefont {A.}~\bibnamefont {Blagoev}}, \ and\ \bibinfo {author} {\bibfnamefont {A.}~\bibnamefont {Pashov}},\ }\bibfield  {title} {\enquote {\bibinfo {title} {Detection of neutral tungsten by laser-induced fluorescence},}\ }\href@noop {} {\bibfield  {journal} {\bibinfo  {journal} {AIP Conference Proceedings}\ }\textbf {\bibinfo {volume} {2075}} (\bibinfo {year} {2019})}\BibitemShut {NoStop}%
\bibitem [{\citenamefont {Lunt}, \citenamefont {Fussmann},\ and\ \citenamefont {Waldmann}(2008)}]{lunt2008experimental}%
  \BibitemOpen
  \bibfield  {author} {\bibinfo {author} {\bibfnamefont {T.}~\bibnamefont {Lunt}}, \bibinfo {author} {\bibfnamefont {G.}~\bibnamefont {Fussmann}}, \ and\ \bibinfo {author} {\bibfnamefont {O.}~\bibnamefont {Waldmann}},\ }\bibfield  {title} {\enquote {\bibinfo {title} {Experimental investigation of the plasma-wall transition},}\ }\href@noop {} {\bibfield  {journal} {\bibinfo  {journal} {Physical review letters}\ }\textbf {\bibinfo {volume} {100}},\ \bibinfo {pages} {175004} (\bibinfo {year} {2008})}\BibitemShut {NoStop}%
\bibitem [{\citenamefont {Tanida}, \citenamefont {Kuwahara},\ and\ \citenamefont {Shinohara}(2016)}]{tanida2016spatial}%
  \BibitemOpen
  \bibfield  {author} {\bibinfo {author} {\bibfnamefont {Y.}~\bibnamefont {Tanida}}, \bibinfo {author} {\bibfnamefont {D.}~\bibnamefont {Kuwahara}}, \ and\ \bibinfo {author} {\bibfnamefont {S.}~\bibnamefont {Shinohara}},\ }\bibfield  {title} {\enquote {\bibinfo {title} {Spatial profile of ion velocity distribution function in helicon high-density plasma by laser induced fluorescence method},}\ }\href@noop {} {\bibfield  {journal} {\bibinfo  {journal} {Transactions of the Japan Society for Aeronatutical and Space Sciences, Aerospace Technology Japan}\ }\textbf {\bibinfo {volume} {14}},\ \bibinfo {pages} {Pb\_7--Pb\_12} (\bibinfo {year} {2016})}\BibitemShut {NoStop}%
\bibitem [{\citenamefont {Thakur}\ \emph {et~al.}(2016)\citenamefont {Thakur}, \citenamefont {Gosselin}, \citenamefont {McKee}, \citenamefont {Scime}, \citenamefont {Sears},\ and\ \citenamefont {Tynan}}]{thakur2016development}%
  \BibitemOpen
  \bibfield  {author} {\bibinfo {author} {\bibfnamefont {S.~C.}\ \bibnamefont {Thakur}}, \bibinfo {author} {\bibfnamefont {J.}~\bibnamefont {Gosselin}}, \bibinfo {author} {\bibfnamefont {J.}~\bibnamefont {McKee}}, \bibinfo {author} {\bibfnamefont {E.}~\bibnamefont {Scime}}, \bibinfo {author} {\bibfnamefont {S.}~\bibnamefont {Sears}}, \ and\ \bibinfo {author} {\bibfnamefont {G.}~\bibnamefont {Tynan}},\ }\bibfield  {title} {\enquote {\bibinfo {title} {Development of core ion temperature gradients and edge sheared flows in a helicon plasma device investigated by laser induced fluorescence measurements},}\ }\href@noop {} {\bibfield  {journal} {\bibinfo  {journal} {Physics of Plasmas}\ }\textbf {\bibinfo {volume} {23}} (\bibinfo {year} {2016})}\BibitemShut {NoStop}%
\bibitem [{\citenamefont {Biloiu}\ \emph {et~al.}(2005)\citenamefont {Biloiu}, \citenamefont {Sun}, \citenamefont {Choueiri}, \citenamefont {Doss}, \citenamefont {Scime}, \citenamefont {Heard}, \citenamefont {Spektor},\ and\ \citenamefont {Ventura}}]{biloiu2005evolution}%
  \BibitemOpen
  \bibfield  {author} {\bibinfo {author} {\bibfnamefont {C.}~\bibnamefont {Biloiu}}, \bibinfo {author} {\bibfnamefont {X.}~\bibnamefont {Sun}}, \bibinfo {author} {\bibfnamefont {E.}~\bibnamefont {Choueiri}}, \bibinfo {author} {\bibfnamefont {F.}~\bibnamefont {Doss}}, \bibinfo {author} {\bibfnamefont {E.}~\bibnamefont {Scime}}, \bibinfo {author} {\bibfnamefont {J.}~\bibnamefont {Heard}}, \bibinfo {author} {\bibfnamefont {R.}~\bibnamefont {Spektor}}, \ and\ \bibinfo {author} {\bibfnamefont {D.}~\bibnamefont {Ventura}},\ }\bibfield  {title} {\enquote {\bibinfo {title} {Evolution of the parallel and perpendicular ion velocity distribution functions in pulsed helicon plasma sources obtained by time resolved laser induced fluorescence},}\ }\href@noop {} {\bibfield  {journal} {\bibinfo  {journal} {Plasma Sources Science and Technology}\ }\textbf {\bibinfo {volume} {14}},\ \bibinfo {pages} {766} (\bibinfo {year} {2005})}\BibitemShut {NoStop}%
\bibitem [{\citenamefont {Severn}\ \emph {et~al.}(2006)\citenamefont {Severn}, \citenamefont {Wang}, \citenamefont {Ko}, \citenamefont {Hershkowitz}, \citenamefont {Turner},\ and\ \citenamefont {McWilliams}}]{severn2006ion}%
  \BibitemOpen
  \bibfield  {author} {\bibinfo {author} {\bibfnamefont {G.}~\bibnamefont {Severn}}, \bibinfo {author} {\bibfnamefont {X.}~\bibnamefont {Wang}}, \bibinfo {author} {\bibfnamefont {E.}~\bibnamefont {Ko}}, \bibinfo {author} {\bibfnamefont {N.}~\bibnamefont {Hershkowitz}}, \bibinfo {author} {\bibfnamefont {M.}~\bibnamefont {Turner}}, \ and\ \bibinfo {author} {\bibfnamefont {R.}~\bibnamefont {McWilliams}},\ }\bibfield  {title} {\enquote {\bibinfo {title} {Ion flow and sheath physics studies in multiple ion species plasmas using diode laser based laser-induced fluorescence},}\ }\href@noop {} {\bibfield  {journal} {\bibinfo  {journal} {Thin Solid Films}\ }\textbf {\bibinfo {volume} {506}},\ \bibinfo {pages} {674--678} (\bibinfo {year} {2006})}\BibitemShut {NoStop}%
\bibitem [{\citenamefont {Bieber}\ \emph {et~al.}(2011)\citenamefont {Bieber}, \citenamefont {Bardin}, \citenamefont {De~Poucques}, \citenamefont {Brochard}, \citenamefont {Hugon}, \citenamefont {Vasseur},\ and\ \citenamefont {Bougdira}}]{bieber2011measurements}%
  \BibitemOpen
  \bibfield  {author} {\bibinfo {author} {\bibfnamefont {T.}~\bibnamefont {Bieber}}, \bibinfo {author} {\bibfnamefont {S.}~\bibnamefont {Bardin}}, \bibinfo {author} {\bibfnamefont {L.}~\bibnamefont {De~Poucques}}, \bibinfo {author} {\bibfnamefont {F.}~\bibnamefont {Brochard}}, \bibinfo {author} {\bibfnamefont {R.}~\bibnamefont {Hugon}}, \bibinfo {author} {\bibfnamefont {J.}~\bibnamefont {Vasseur}}, \ and\ \bibinfo {author} {\bibfnamefont {J.}~\bibnamefont {Bougdira}},\ }\bibfield  {title} {\enquote {\bibinfo {title} {Measurements on argon ion by tunable diode-laser induced fluorescence in a low magnetic field helicon configuration reactor},}\ }\href@noop {} {\bibfield  {journal} {\bibinfo  {journal} {Plasma Sources Science and Technology}\ }\textbf {\bibinfo {volume} {20}},\ \bibinfo {pages} {015023} (\bibinfo {year} {2011})}\BibitemShut {NoStop}%
\bibitem [{\citenamefont {Keesee}, \citenamefont {Scime},\ and\ \citenamefont {Boivin}(2004)}]{keesee2004laser}%
  \BibitemOpen
  \bibfield  {author} {\bibinfo {author} {\bibfnamefont {A.~M.}\ \bibnamefont {Keesee}}, \bibinfo {author} {\bibfnamefont {E.~E.}\ \bibnamefont {Scime}}, \ and\ \bibinfo {author} {\bibfnamefont {R.~F.}\ \bibnamefont {Boivin}},\ }\bibfield  {title} {\enquote {\bibinfo {title} {Laser-induced fluorescence measurements of three plasma species with a tunable diode laser},}\ }\href@noop {} {\bibfield  {journal} {\bibinfo  {journal} {Review of Scientific Instruments}\ }\textbf {\bibinfo {volume} {75}},\ \bibinfo {pages} {4091--4093} (\bibinfo {year} {2004})}\BibitemShut {NoStop}%
\bibitem [{\citenamefont {Kuwahara}\ \emph {et~al.}(2015)\citenamefont {Kuwahara}, \citenamefont {Tanida}, \citenamefont {Watanabe}, \citenamefont {Teshigahara}, \citenamefont {Yamagata},\ and\ \citenamefont {Shinohara}}]{kuwahara2015development}%
  \BibitemOpen
  \bibfield  {author} {\bibinfo {author} {\bibfnamefont {D.}~\bibnamefont {Kuwahara}}, \bibinfo {author} {\bibfnamefont {Y.}~\bibnamefont {Tanida}}, \bibinfo {author} {\bibfnamefont {M.}~\bibnamefont {Watanabe}}, \bibinfo {author} {\bibfnamefont {N.}~\bibnamefont {Teshigahara}}, \bibinfo {author} {\bibfnamefont {Y.}~\bibnamefont {Yamagata}}, \ and\ \bibinfo {author} {\bibfnamefont {S.}~\bibnamefont {Shinohara}},\ }\bibfield  {title} {\enquote {\bibinfo {title} {Development of {Ar} {I} and {Ar} {II} measuring system using laser-induced fluorescence methods in high-density helicon plasma},}\ }\href@noop {} {\bibfield  {journal} {\bibinfo  {journal} {Plasma and Fusion Research}\ }\textbf {\bibinfo {volume} {10}},\ \bibinfo {pages} {3401057--3401057} (\bibinfo {year} {2015})}\BibitemShut {NoStop}%
\bibitem [{\citenamefont {Severn}, \citenamefont {Edrich},\ and\ \citenamefont {McWilliams}(1998)}]{severn1998argon}%
  \BibitemOpen
  \bibfield  {author} {\bibinfo {author} {\bibfnamefont {G.}~\bibnamefont {Severn}}, \bibinfo {author} {\bibfnamefont {D.}~\bibnamefont {Edrich}}, \ and\ \bibinfo {author} {\bibfnamefont {R.}~\bibnamefont {McWilliams}},\ }\bibfield  {title} {\enquote {\bibinfo {title} {Argon ion laser-induced fluorescence with diode lasers},}\ }\href@noop {} {\bibfield  {journal} {\bibinfo  {journal} {Review of Scientific Instruments}\ }\textbf {\bibinfo {volume} {69}},\ \bibinfo {pages} {10--15} (\bibinfo {year} {1998})}\BibitemShut {NoStop}%
\bibitem [{\citenamefont {Harris}, \citenamefont {Eland},\ and\ \citenamefont {Tuckett}(1983)}]{harris1983ax}%
  \BibitemOpen
  \bibfield  {author} {\bibinfo {author} {\bibfnamefont {T.}~\bibnamefont {Harris}}, \bibinfo {author} {\bibfnamefont {J.}~\bibnamefont {Eland}}, \ and\ \bibinfo {author} {\bibfnamefont {R.}~\bibnamefont {Tuckett}},\ }\bibfield  {title} {\enquote {\bibinfo {title} {{The AX system of Br$_{2}^{+}$ radical cations}},}\ }\href@noop {} {\bibfield  {journal} {\bibinfo  {journal} {Journal of Molecular Spectroscopy}\ }\textbf {\bibinfo {volume} {98}},\ \bibinfo {pages} {269--281} (\bibinfo {year} {1983})}\BibitemShut {NoStop}%
\bibitem [{\citenamefont {Kumagai}\ \emph {et~al.}(1999)\citenamefont {Kumagai}, \citenamefont {Sasaki}, \citenamefont {Koyanagi},\ and\ \citenamefont {Hane}}]{kumagai1999detection}%
  \BibitemOpen
  \bibfield  {author} {\bibinfo {author} {\bibfnamefont {S.~K.~S.}\ \bibnamefont {Kumagai}}, \bibinfo {author} {\bibfnamefont {M.~S.~M.}\ \bibnamefont {Sasaki}}, \bibinfo {author} {\bibfnamefont {M.~K.~M.}\ \bibnamefont {Koyanagi}}, \ and\ \bibinfo {author} {\bibfnamefont {K.~H.~K.}\ \bibnamefont {Hane}},\ }\bibfield  {title} {\enquote {\bibinfo {title} {Detection of metastable chlorine ions in time-modulated plasma by time resolved laser-induced fluorescence},}\ }\href@noop {} {\bibfield  {journal} {\bibinfo  {journal} {Japanese Journal of Applied Physics}\ }\textbf {\bibinfo {volume} {38}},\ \bibinfo {pages} {7126} (\bibinfo {year} {1999})}\BibitemShut {NoStop}%
\bibitem [{\citenamefont {Malyshev}\ \emph {et~al.}(1999)\citenamefont {Malyshev}, \citenamefont {Fuller}, \citenamefont {Bogart}, \citenamefont {Donnelly},\ and\ \citenamefont {Herman}}]{malyshev1999laser}%
  \BibitemOpen
  \bibfield  {author} {\bibinfo {author} {\bibfnamefont {M.}~\bibnamefont {Malyshev}}, \bibinfo {author} {\bibfnamefont {N.}~\bibnamefont {Fuller}}, \bibinfo {author} {\bibfnamefont {K.}~\bibnamefont {Bogart}}, \bibinfo {author} {\bibfnamefont {V.}~\bibnamefont {Donnelly}}, \ and\ \bibinfo {author} {\bibfnamefont {I.~P.}\ \bibnamefont {Herman}},\ }\bibfield  {title} {\enquote {\bibinfo {title} {Laser-induced fluorescence and langmuir probe determination of {Cl$_2^+$} and {Cl$^+$} absolute densities in transformer-coupled chlorine plasmas},}\ }\href@noop {} {\bibfield  {journal} {\bibinfo  {journal} {Applied physics letters}\ }\textbf {\bibinfo {volume} {74}},\ \bibinfo {pages} {1666--1668} (\bibinfo {year} {1999})}\BibitemShut {NoStop}%
\bibitem [{\citenamefont {Bondybey}, \citenamefont {English},\ and\ \citenamefont {Miller}(1979)}]{bondybey1979laser}%
  \BibitemOpen
  \bibfield  {author} {\bibinfo {author} {\bibfnamefont {V.}~\bibnamefont {Bondybey}}, \bibinfo {author} {\bibfnamefont {J.}~\bibnamefont {English}}, \ and\ \bibinfo {author} {\bibfnamefont {T.~A.}\ \bibnamefont {Miller}},\ }\bibfield  {title} {\enquote {\bibinfo {title} {{Laser induced fluorescence spectrum of matrix isolated CS$^{+}_{2}$}},}\ }\href@noop {} {\bibfield  {journal} {\bibinfo  {journal} {The Journal of Chemical Physics}\ }\textbf {\bibinfo {volume} {70}},\ \bibinfo {pages} {1621--1625} (\bibinfo {year} {1979})}\BibitemShut {NoStop}%
\bibitem [{\citenamefont {Zen}\ and\ \citenamefont {Lee}(1995)}]{zen1995laser}%
  \BibitemOpen
  \bibfield  {author} {\bibinfo {author} {\bibfnamefont {C.-C.}\ \bibnamefont {Zen}}\ and\ \bibinfo {author} {\bibfnamefont {Y.-P.}\ \bibnamefont {Lee}},\ }\bibfield  {title} {\enquote {\bibinfo {title} {{Laser-induced fluorescence of the A$^{2}\Pi_{u}$--X$^{2}\Pi_{g}$ transition of CS$_{2}^{+}$ in solid Ne. Reanalysis of vibronic spectra}},}\ }\href@noop {} {\bibfield  {journal} {\bibinfo  {journal} {Chemical physics letters}\ }\textbf {\bibinfo {volume} {244}},\ \bibinfo {pages} {177--182} (\bibinfo {year} {1995})}\BibitemShut {NoStop}%
\bibitem [{\citenamefont {Nakajima}\ \emph {et~al.}(2003)\citenamefont {Nakajima}, \citenamefont {Yoneda}, \citenamefont {Sumiyoshi}, \citenamefont {Nagata},\ and\ \citenamefont {Endo}}]{nakajima2003laser}%
  \BibitemOpen
  \bibfield  {author} {\bibinfo {author} {\bibfnamefont {M.}~\bibnamefont {Nakajima}}, \bibinfo {author} {\bibfnamefont {Y.}~\bibnamefont {Yoneda}}, \bibinfo {author} {\bibfnamefont {Y.}~\bibnamefont {Sumiyoshi}}, \bibinfo {author} {\bibfnamefont {T.}~\bibnamefont {Nagata}}, \ and\ \bibinfo {author} {\bibfnamefont {Y.}~\bibnamefont {Endo}},\ }\bibfield  {title} {\enquote {\bibinfo {title} {{Laser-induced fluorescence and fluorescence depletion spectroscopy of SCCS$^{-}$}},}\ }\href@noop {} {\bibfield  {journal} {\bibinfo  {journal} {The Journal of chemical physics}\ }\textbf {\bibinfo {volume} {119}},\ \bibinfo {pages} {7805--7813} (\bibinfo {year} {2003})}\BibitemShut {NoStop}%
\bibitem [{\citenamefont {Steinberger}\ and\ \citenamefont {Scime}(2018)}]{steinberger2018laser}%
  \BibitemOpen
  \bibfield  {author} {\bibinfo {author} {\bibfnamefont {T.~E.}\ \bibnamefont {Steinberger}}\ and\ \bibinfo {author} {\bibfnamefont {E.~E.}\ \bibnamefont {Scime}},\ }\bibfield  {title} {\enquote {\bibinfo {title} {Laser-induced fluorescence of singly ionized iodine},}\ }\href@noop {} {\bibfield  {journal} {\bibinfo  {journal} {Journal of Propulsion and Power}\ }\textbf {\bibinfo {volume} {34}},\ \bibinfo {pages} {1235--1239} (\bibinfo {year} {2018})}\BibitemShut {NoStop}%
\bibitem [{\citenamefont {Lejeune}, \citenamefont {Bourgeois},\ and\ \citenamefont {Mazouffre}(2012)}]{lejeune2012kr}%
  \BibitemOpen
  \bibfield  {author} {\bibinfo {author} {\bibfnamefont {A.}~\bibnamefont {Lejeune}}, \bibinfo {author} {\bibfnamefont {G.}~\bibnamefont {Bourgeois}}, \ and\ \bibinfo {author} {\bibfnamefont {S.}~\bibnamefont {Mazouffre}},\ }\bibfield  {title} {\enquote {\bibinfo {title} {{Kr} {II} and {Xe} {II} axial velocity distribution functions in a cross-field ion source},}\ }\href@noop {} {\bibfield  {journal} {\bibinfo  {journal} {Physics of Plasmas}\ }\textbf {\bibinfo {volume} {19}} (\bibinfo {year} {2012})}\BibitemShut {NoStop}%
\bibitem [{\citenamefont {Hargus}, \citenamefont {Azarnia},\ and\ \citenamefont {Nakles}(2011)}]{hargus2011demonstration}%
  \BibitemOpen
  \bibfield  {author} {\bibinfo {author} {\bibfnamefont {W.}~\bibnamefont {Hargus}}, \bibinfo {author} {\bibfnamefont {G.~M.}\ \bibnamefont {Azarnia}}, \ and\ \bibinfo {author} {\bibfnamefont {M.~R.}\ \bibnamefont {Nakles}},\ }\bibfield  {title} {\enquote {\bibinfo {title} {Demonstration of laser-induced fluorescence on a krypton hall effect thruster},}\ }in\ \href@noop {} {\emph {\bibinfo {booktitle} {32nd International Electric Propulsion Conference}}}\ (\bibinfo {year} {2011})\ pp.\ \bibinfo {pages} {11--15}\BibitemShut {NoStop}%
\bibitem [{\citenamefont {Tremblay}, \citenamefont {Smith},\ and\ \citenamefont {Winefordner}(1987)}]{tremblay1987laser}%
  \BibitemOpen
  \bibfield  {author} {\bibinfo {author} {\bibfnamefont {M.}~\bibnamefont {Tremblay}}, \bibinfo {author} {\bibfnamefont {B.~W.}\ \bibnamefont {Smith}}, \ and\ \bibinfo {author} {\bibfnamefont {J.~D.}\ \bibnamefont {Winefordner}},\ }\bibfield  {title} {\enquote {\bibinfo {title} {Laser-excited ionic fluorescence spectrometry of rare-earth elements in the inductively-coupled plasma},}\ }\href@noop {} {\bibfield  {journal} {\bibinfo  {journal} {Analytica chimica acta}\ }\textbf {\bibinfo {volume} {199}},\ \bibinfo {pages} {111--118} (\bibinfo {year} {1987})}\BibitemShut {NoStop}%
\bibitem [{\citenamefont {Woodcock}\ \emph {et~al.}(1997)\citenamefont {Woodcock}, \citenamefont {Busby}, \citenamefont {Freegarde},\ and\ \citenamefont {Hancock}}]{woodcock1997doppler}%
  \BibitemOpen
  \bibfield  {author} {\bibinfo {author} {\bibfnamefont {B.}~\bibnamefont {Woodcock}}, \bibinfo {author} {\bibfnamefont {J.}~\bibnamefont {Busby}}, \bibinfo {author} {\bibfnamefont {T.}~\bibnamefont {Freegarde}}, \ and\ \bibinfo {author} {\bibfnamefont {G.}~\bibnamefont {Hancock}},\ }\bibfield  {title} {\enquote {\bibinfo {title} {Doppler spectroscopic measurements of sheath ion velocities in radio-frequency plasmas},}\ }\href@noop {} {\bibfield  {journal} {\bibinfo  {journal} {Journal of applied physics}\ }\textbf {\bibinfo {volume} {81}},\ \bibinfo {pages} {5945--5949} (\bibinfo {year} {1997})}\BibitemShut {NoStop}%
\bibitem [{\citenamefont {Gharaibeh}\ and\ \citenamefont {Clouthier}(2012)}]{gharaibeh2012laser}%
  \BibitemOpen
  \bibfield  {author} {\bibinfo {author} {\bibfnamefont {M.~A.}\ \bibnamefont {Gharaibeh}}\ and\ \bibinfo {author} {\bibfnamefont {D.~J.}\ \bibnamefont {Clouthier}},\ }\bibfield  {title} {\enquote {\bibinfo {title} {{A laser-induced fluorescence study of the jet-cooled nitrous oxide cation (N$_{2}$O$^{+}$)}},}\ }\href@noop {} {\bibfield  {journal} {\bibinfo  {journal} {The Journal of Chemical Physics}\ }\textbf {\bibinfo {volume} {136}} (\bibinfo {year} {2012})}\BibitemShut {NoStop}%
\bibitem [{\citenamefont {Li}, \citenamefont {Bierbaum},\ and\ \citenamefont {Leone}(2000)}]{li2000laser}%
  \BibitemOpen
  \bibfield  {author} {\bibinfo {author} {\bibfnamefont {J.}~\bibnamefont {Li}}, \bibinfo {author} {\bibfnamefont {V.~M.}\ \bibnamefont {Bierbaum}}, \ and\ \bibinfo {author} {\bibfnamefont {S.~R.}\ \bibnamefont {Leone}},\ }\bibfield  {title} {\enquote {\bibinfo {title} {{Laser-induced fluorescence study of two weak vibrational bands of the O$_{2}^{+}$ A$^2{\Pi}_u$-X$^2{\Pi}_g$ system}},}\ }\href@noop {} {\bibfield  {journal} {\bibinfo  {journal} {Chemical Physics Letters}\ }\textbf {\bibinfo {volume} {330}},\ \bibinfo {pages} {331--338} (\bibinfo {year} {2000})}\BibitemShut {NoStop}%
\bibitem [{\citenamefont {Matsuo}\ \emph {et~al.}(1997)\citenamefont {Matsuo}, \citenamefont {Nakajima}, \citenamefont {Kobayashi},\ and\ \citenamefont {Takami}}]{matsuo1997formation}%
  \BibitemOpen
  \bibfield  {author} {\bibinfo {author} {\bibfnamefont {Y.}~\bibnamefont {Matsuo}}, \bibinfo {author} {\bibfnamefont {T.}~\bibnamefont {Nakajima}}, \bibinfo {author} {\bibfnamefont {T.}~\bibnamefont {Kobayashi}}, \ and\ \bibinfo {author} {\bibfnamefont {M.}~\bibnamefont {Takami}},\ }\bibfield  {title} {\enquote {\bibinfo {title} {Formation and laser-induced-fluorescence study of {SiO}$^+$ ions produced by laser ablation of {Si} in oxygen gas},}\ }\href@noop {} {\bibfield  {journal} {\bibinfo  {journal} {Applied physics letters}\ }\textbf {\bibinfo {volume} {71}},\ \bibinfo {pages} {996--998} (\bibinfo {year} {1997})}\BibitemShut {NoStop}%
\bibitem [{\citenamefont {Smith}\ \emph {et~al.}(2005)\citenamefont {Smith}, \citenamefont {Ngom}, \citenamefont {Linnell},\ and\ \citenamefont {Gallimore}}]{smith2005diode}%
  \BibitemOpen
  \bibfield  {author} {\bibinfo {author} {\bibfnamefont {T.}~\bibnamefont {Smith}}, \bibinfo {author} {\bibfnamefont {B.}~\bibnamefont {Ngom}}, \bibinfo {author} {\bibfnamefont {J.}~\bibnamefont {Linnell}}, \ and\ \bibinfo {author} {\bibfnamefont {A.}~\bibnamefont {Gallimore}},\ }\bibfield  {title} {\enquote {\bibinfo {title} {Diode laser-induced fluorescence of xenon ion velocity distributions},}\ }in\ \href@noop {} {\emph {\bibinfo {booktitle} {41st AIAA/ASME/SAE/ASEE Joint Propulsion Conference \& Exhibit}}}\ (\bibinfo {year} {2005})\ p.\ \bibinfo {pages} {4406}\BibitemShut {NoStop}%
\bibitem [{\citenamefont {Smith}\ \emph {et~al.}(2004)\citenamefont {Smith}, \citenamefont {Herman}, \citenamefont {Gallimore},\ and\ \citenamefont {Williams}}]{smith2004laser}%
  \BibitemOpen
  \bibfield  {author} {\bibinfo {author} {\bibfnamefont {T.}~\bibnamefont {Smith}}, \bibinfo {author} {\bibfnamefont {D.}~\bibnamefont {Herman}}, \bibinfo {author} {\bibfnamefont {A.}~\bibnamefont {Gallimore}}, \ and\ \bibinfo {author} {\bibfnamefont {G.}~\bibnamefont {Williams}},\ }\bibfield  {title} {\enquote {\bibinfo {title} {{Laser-induced fluorescence velocimetry of Xe II in the 30-cm NSTAR-type ion engine plume}},}\ }in\ \href@noop {} {\emph {\bibinfo {booktitle} {40th AIAA/ASME/SAE/ASEE Joint Propulsion Conference and Exhibit}}}\ (\bibinfo {year} {2004})\ p.\ \bibinfo {pages} {3963}\BibitemShut {NoStop}%
\bibitem [{\citenamefont {Sadeghi}\ \emph {et~al.}(1997)\citenamefont {Sadeghi}, \citenamefont {Van De~Grift}, \citenamefont {Vender}, \citenamefont {Kroesen},\ and\ \citenamefont {De~Hoog}}]{sadeghi1997transport}%
  \BibitemOpen
  \bibfield  {author} {\bibinfo {author} {\bibfnamefont {N.}~\bibnamefont {Sadeghi}}, \bibinfo {author} {\bibfnamefont {M.}~\bibnamefont {Van De~Grift}}, \bibinfo {author} {\bibfnamefont {D.}~\bibnamefont {Vender}}, \bibinfo {author} {\bibfnamefont {G.}~\bibnamefont {Kroesen}}, \ and\ \bibinfo {author} {\bibfnamefont {F.}~\bibnamefont {De~Hoog}},\ }\bibfield  {title} {\enquote {\bibinfo {title} {Transport of argon ions in an inductively coupled high-density plasma reactor},}\ }\href@noop {} {\bibfield  {journal} {\bibinfo  {journal} {Applied physics letters}\ }\textbf {\bibinfo {volume} {70}},\ \bibinfo {pages} {835--837} (\bibinfo {year} {1997})}\BibitemShut {NoStop}%
\bibitem [{\citenamefont {Engeln}\ \emph {et~al.}(2001)\citenamefont {Engeln}, \citenamefont {Mazouffre}, \citenamefont {Vankan}, \citenamefont {Schram},\ and\ \citenamefont {Sadeghi}}]{engeln2001flow}%
  \BibitemOpen
  \bibfield  {author} {\bibinfo {author} {\bibfnamefont {R.}~\bibnamefont {Engeln}}, \bibinfo {author} {\bibfnamefont {S.}~\bibnamefont {Mazouffre}}, \bibinfo {author} {\bibfnamefont {P.}~\bibnamefont {Vankan}}, \bibinfo {author} {\bibfnamefont {D.}~\bibnamefont {Schram}}, \ and\ \bibinfo {author} {\bibfnamefont {N.}~\bibnamefont {Sadeghi}},\ }\bibfield  {title} {\enquote {\bibinfo {title} {Flow dynamics and invasion by background gas of a supersonically expanding thermal plasma},}\ }\href@noop {} {\bibfield  {journal} {\bibinfo  {journal} {Plasma Sources Science and Technology}\ }\textbf {\bibinfo {volume} {10}},\ \bibinfo {pages} {595} (\bibinfo {year} {2001})}\BibitemShut {NoStop}%
\bibitem [{\citenamefont {Short}\ \emph {et~al.}(2016)\citenamefont {Short}, \citenamefont {Siddiqui}, \citenamefont {Henriquez}, \citenamefont {McKee},\ and\ \citenamefont {Scime}}]{short2016novel}%
  \BibitemOpen
  \bibfield  {author} {\bibinfo {author} {\bibfnamefont {Z.~D.}\ \bibnamefont {Short}}, \bibinfo {author} {\bibfnamefont {M.~U.}\ \bibnamefont {Siddiqui}}, \bibinfo {author} {\bibfnamefont {M.~F.}\ \bibnamefont {Henriquez}}, \bibinfo {author} {\bibfnamefont {J.~S.}\ \bibnamefont {McKee}}, \ and\ \bibinfo {author} {\bibfnamefont {E.~E.}\ \bibnamefont {Scime}},\ }\bibfield  {title} {\enquote {\bibinfo {title} {A novel laser-induced fluorescence scheme for {Ar}-{I} in a plasma},}\ }\href@noop {} {\bibfield  {journal} {\bibinfo  {journal} {Review of Scientific Instruments}\ }\textbf {\bibinfo {volume} {87}} (\bibinfo {year} {2016})}\BibitemShut {NoStop}%
\bibitem [{\citenamefont {Bergert}\ \emph {et~al.}(2021)\citenamefont {Bergert}, \citenamefont {Isberner}, \citenamefont {Mitic},\ and\ \citenamefont {Thoma}}]{bergert2021quantitative}%
  \BibitemOpen
  \bibfield  {author} {\bibinfo {author} {\bibfnamefont {R.}~\bibnamefont {Bergert}}, \bibinfo {author} {\bibfnamefont {L.~W.}\ \bibnamefont {Isberner}}, \bibinfo {author} {\bibfnamefont {S.}~\bibnamefont {Mitic}}, \ and\ \bibinfo {author} {\bibfnamefont {M.~H.}\ \bibnamefont {Thoma}},\ }\bibfield  {title} {\enquote {\bibinfo {title} {Quantitative evaluation of laser-induced fluorescence in magnetized low-pressure argon plasma},}\ }\href@noop {} {\bibfield  {journal} {\bibinfo  {journal} {Physics of Plasmas}\ }\textbf {\bibinfo {volume} {28}} (\bibinfo {year} {2021})}\BibitemShut {NoStop}%
\bibitem [{\citenamefont {Hansen}\ \emph {et~al.}(1990)\citenamefont {Hansen}, \citenamefont {Luckman}, \citenamefont {Nieman},\ and\ \citenamefont {Colson}}]{hansen1990formation}%
  \BibitemOpen
  \bibfield  {author} {\bibinfo {author} {\bibfnamefont {S.}~\bibnamefont {Hansen}}, \bibinfo {author} {\bibfnamefont {G.}~\bibnamefont {Luckman}}, \bibinfo {author} {\bibfnamefont {G.~C.}\ \bibnamefont {Nieman}}, \ and\ \bibinfo {author} {\bibfnamefont {S.~D.}\ \bibnamefont {Colson}},\ }\bibfield  {title} {\enquote {\bibinfo {title} {Formation and decay of metastable fluorine atoms in pulsed fluorocarbon/oxygen discharges monitored by laser-induced fluorescence},}\ }\href@noop {} {\bibfield  {journal} {\bibinfo  {journal} {Applied physics letters}\ }\textbf {\bibinfo {volume} {56}},\ \bibinfo {pages} {719--721} (\bibinfo {year} {1990})}\BibitemShut {NoStop}%
\bibitem [{\citenamefont {Frost}, \citenamefont {Himmelmann},\ and\ \citenamefont {Palmer}(2001)}]{frost2001laser}%
  \BibitemOpen
  \bibfield  {author} {\bibinfo {author} {\bibfnamefont {M.~J.}\ \bibnamefont {Frost}}, \bibinfo {author} {\bibfnamefont {S.}~\bibnamefont {Himmelmann}}, \ and\ \bibinfo {author} {\bibfnamefont {D.}~\bibnamefont {Palmer}},\ }\bibfield  {title} {\enquote {\bibinfo {title} {{Laser-induced fluorescence studies of elementary processes in a helium plasma following He 3$^{3}$P--2$^{3}$S and 3$^{1}$P--2$^{1}$S excitation}},}\ }\href@noop {} {\bibfield  {journal} {\bibinfo  {journal} {Journal of Physics B: Atomic, Molecular and Optical Physics}\ }\textbf {\bibinfo {volume} {34}},\ \bibinfo {pages} {1569} (\bibinfo {year} {2001})}\BibitemShut {NoStop}%
\bibitem [{\citenamefont {Rellergert}\ \emph {et~al.}(2008)\citenamefont {Rellergert}, \citenamefont {Cahn}, \citenamefont {Garvan}, \citenamefont {Hanson}, \citenamefont {Lippincott}, \citenamefont {Nikkel},\ and\ \citenamefont {McKinsey}}]{rellergert2008detection}%
  \BibitemOpen
  \bibfield  {author} {\bibinfo {author} {\bibfnamefont {W.}~\bibnamefont {Rellergert}}, \bibinfo {author} {\bibfnamefont {S.}~\bibnamefont {Cahn}}, \bibinfo {author} {\bibfnamefont {A.}~\bibnamefont {Garvan}}, \bibinfo {author} {\bibfnamefont {J.}~\bibnamefont {Hanson}}, \bibinfo {author} {\bibfnamefont {W.}~\bibnamefont {Lippincott}}, \bibinfo {author} {\bibfnamefont {J.}~\bibnamefont {Nikkel}}, \ and\ \bibinfo {author} {\bibfnamefont {D.}~\bibnamefont {McKinsey}},\ }\bibfield  {title} {\enquote {\bibinfo {title} {{Detection and imaging of He$_{2}$ molecules in superfluid helium}},}\ }\href@noop {} {\bibfield  {journal} {\bibinfo  {journal} {Physical review letters}\ }\textbf {\bibinfo {volume} {100}},\ \bibinfo {pages} {025301} (\bibinfo {year} {2008})}\BibitemShut {NoStop}%
\bibitem [{\citenamefont {Hargus}(2010)}]{hargus2010preliminary}%
  \BibitemOpen
  \bibfield  {author} {\bibinfo {author} {\bibfnamefont {W.}~\bibnamefont {Hargus}},\ }\bibfield  {title} {\enquote {\bibinfo {title} {A preliminary study of krypton laser-induced fluorescence},}\ }in\ \href@noop {} {\emph {\bibinfo {booktitle} {46th AIAA/ASME/SAE/ASEE Joint Propulsion Conference \& Exhibit}}}\ (\bibinfo {year} {2010})\ p.\ \bibinfo {pages} {6524}\BibitemShut {NoStop}%
\bibitem [{\citenamefont {Mustafa}\ \emph {et~al.}(2017)\citenamefont {Mustafa}, \citenamefont {Hunt}, \citenamefont {Parziale}, \citenamefont {Smith},\ and\ \citenamefont {Marineau}}]{mustafa2017krypton}%
  \BibitemOpen
  \bibfield  {author} {\bibinfo {author} {\bibfnamefont {M.}~\bibnamefont {Mustafa}}, \bibinfo {author} {\bibfnamefont {M.~B.}\ \bibnamefont {Hunt}}, \bibinfo {author} {\bibfnamefont {N.~J.}\ \bibnamefont {Parziale}}, \bibinfo {author} {\bibfnamefont {M.~S.}\ \bibnamefont {Smith}}, \ and\ \bibinfo {author} {\bibfnamefont {E.~C.}\ \bibnamefont {Marineau}},\ }\bibfield  {title} {\enquote {\bibinfo {title} {{Krypton tagging velocimetry (KTV) investigation of shock-wave/turbulent boundary-layer interaction}},}\ }in\ \href@noop {} {\emph {\bibinfo {booktitle} {55th AIAA Aerospace Sciences Meeting}}}\ (\bibinfo {year} {2017})\ p.\ \bibinfo {pages} {0025}\BibitemShut {NoStop}%
\bibitem [{\citenamefont {Nemschokmichal}\ \emph {et~al.}(2011)\citenamefont {Nemschokmichal}, \citenamefont {Bernhardt}, \citenamefont {Krames},\ and\ \citenamefont {Meichsner}}]{nemschokmichal2011laser}%
  \BibitemOpen
  \bibfield  {author} {\bibinfo {author} {\bibfnamefont {S.}~\bibnamefont {Nemschokmichal}}, \bibinfo {author} {\bibfnamefont {F.}~\bibnamefont {Bernhardt}}, \bibinfo {author} {\bibfnamefont {B.}~\bibnamefont {Krames}}, \ and\ \bibinfo {author} {\bibfnamefont {J.}~\bibnamefont {Meichsner}},\ }\bibfield  {title} {\enquote {\bibinfo {title} {{Laser-induced fluorescence spectroscopy of {N}$_{2}$({A}$^3\Sigma_{u}^{+}$) and absolute density calibration by Rayleigh scattering in capacitively coupled RF discharges}},}\ }\href@noop {} {\bibfield  {journal} {\bibinfo  {journal} {Journal of Physics D: Applied Physics}\ }\textbf {\bibinfo {volume} {44}},\ \bibinfo {pages} {205201} (\bibinfo {year} {2011})}\BibitemShut {NoStop}%
\bibitem [{\citenamefont {Ono}\ \emph {et~al.}(2009)\citenamefont {Ono}, \citenamefont {Tobaru}, \citenamefont {Teramoto},\ and\ \citenamefont {Oda}}]{ono2009laser}%
  \BibitemOpen
  \bibfield  {author} {\bibinfo {author} {\bibfnamefont {R.}~\bibnamefont {Ono}}, \bibinfo {author} {\bibfnamefont {C.}~\bibnamefont {Tobaru}}, \bibinfo {author} {\bibfnamefont {Y.}~\bibnamefont {Teramoto}}, \ and\ \bibinfo {author} {\bibfnamefont {T.}~\bibnamefont {Oda}},\ }\bibfield  {title} {\enquote {\bibinfo {title} {{Laser-induced fluorescence of {N}$_{2}$({A}$^3\Sigma_{u}^{+}$) metastable in N$_2$ pulsed positive corona discharge}},}\ }\href@noop {} {\bibfield  {journal} {\bibinfo  {journal} {Plasma Sources Science and Technology}\ }\textbf {\bibinfo {volume} {18}},\ \bibinfo {pages} {025006} (\bibinfo {year} {2009})}\BibitemShut {NoStop}%
\bibitem [{\citenamefont {Shen}\ \emph {et~al.}(2009)\citenamefont {Shen}, \citenamefont {Wang}, \citenamefont {Xie}, \citenamefont {Gao}, \citenamefont {Ling},\ and\ \citenamefont {Lu}}]{shen2009detection}%
  \BibitemOpen
  \bibfield  {author} {\bibinfo {author} {\bibfnamefont {X.}~\bibnamefont {Shen}}, \bibinfo {author} {\bibfnamefont {H.}~\bibnamefont {Wang}}, \bibinfo {author} {\bibfnamefont {Z.}~\bibnamefont {Xie}}, \bibinfo {author} {\bibfnamefont {Y.}~\bibnamefont {Gao}}, \bibinfo {author} {\bibfnamefont {H.}~\bibnamefont {Ling}}, \ and\ \bibinfo {author} {\bibfnamefont {Y.}~\bibnamefont {Lu}},\ }\bibfield  {title} {\enquote {\bibinfo {title} {Detection of trace phosphorus in steel using laser-induced breakdown spectroscopy combined with laser-induced fluorescence},}\ }\href@noop {} {\bibfield  {journal} {\bibinfo  {journal} {Applied optics}\ }\textbf {\bibinfo {volume} {48}},\ \bibinfo {pages} {2551--2558} (\bibinfo {year} {2009})}\BibitemShut {NoStop}%
\bibitem [{\citenamefont {Kondo}, \citenamefont {Hamada},\ and\ \citenamefont {Wagatsuma}(2009)}]{kondo2009determination}%
  \BibitemOpen
  \bibfield  {author} {\bibinfo {author} {\bibfnamefont {H.}~\bibnamefont {Kondo}}, \bibinfo {author} {\bibfnamefont {N.}~\bibnamefont {Hamada}}, \ and\ \bibinfo {author} {\bibfnamefont {K.}~\bibnamefont {Wagatsuma}},\ }\bibfield  {title} {\enquote {\bibinfo {title} {Determination of phosphorus in steel by the combined technique of laser induced breakdown spectrometry with laser induced fluorescence spectrometry},}\ }\href@noop {} {\bibfield  {journal} {\bibinfo  {journal} {Spectrochimica Acta Part B: Atomic Spectroscopy}\ }\textbf {\bibinfo {volume} {64}},\ \bibinfo {pages} {884--890} (\bibinfo {year} {2009})}\BibitemShut {NoStop}%
\bibitem [{\citenamefont {Mazouffre}\ \emph {et~al.}(2011)\citenamefont {Mazouffre}, \citenamefont {Bourgeois}, \citenamefont {Garrigues},\ and\ \citenamefont {Pawelec}}]{mazouffre2011comprehensive}%
  \BibitemOpen
  \bibfield  {author} {\bibinfo {author} {\bibfnamefont {S.}~\bibnamefont {Mazouffre}}, \bibinfo {author} {\bibfnamefont {G.}~\bibnamefont {Bourgeois}}, \bibinfo {author} {\bibfnamefont {L.}~\bibnamefont {Garrigues}}, \ and\ \bibinfo {author} {\bibfnamefont {E.}~\bibnamefont {Pawelec}},\ }\bibfield  {title} {\enquote {\bibinfo {title} {A comprehensive study on the atom flow in the cross-field discharge of a hall thruster},}\ }\href@noop {} {\bibfield  {journal} {\bibinfo  {journal} {Journal of Physics D: Applied Physics}\ }\textbf {\bibinfo {volume} {44}},\ \bibinfo {pages} {105203} (\bibinfo {year} {2011})}\BibitemShut {NoStop}%
\bibitem [{\citenamefont {Cedolin}\ \emph {et~al.}(1997)\citenamefont {Cedolin}, \citenamefont {Hargus~Jr}, \citenamefont {Storm}, \citenamefont {Hanson},\ and\ \citenamefont {Cappelli}}]{cedolin1997laser}%
  \BibitemOpen
  \bibfield  {author} {\bibinfo {author} {\bibfnamefont {R.}~\bibnamefont {Cedolin}}, \bibinfo {author} {\bibfnamefont {W.}~\bibnamefont {Hargus~Jr}}, \bibinfo {author} {\bibfnamefont {P.}~\bibnamefont {Storm}}, \bibinfo {author} {\bibfnamefont {R.}~\bibnamefont {Hanson}}, \ and\ \bibinfo {author} {\bibfnamefont {M.}~\bibnamefont {Cappelli}},\ }\bibfield  {title} {\enquote {\bibinfo {title} {Laser-induced fluorescence study of a xenon hall thruster},}\ }\href@noop {} {\bibfield  {journal} {\bibinfo  {journal} {Applied Physics B}\ }\textbf {\bibinfo {volume} {65}},\ \bibinfo {pages} {459--469} (\bibinfo {year} {1997})}\BibitemShut {NoStop}%
\end{thebibliography}%


\end{document}